\documentclass[11pt,fleqn,a4paper]{book} % Default font size and left-justified equations

\title{PhD: On Filaments within Molecular Clouds and Their Connection to Star Formation}

\usepackage[english]{babel}

\usepackage{amsfonts,amsmath,amssymb,amstext}

\usepackage[bottom=3cm]{geometry}

\usepackage{xcolor} % Required for specifying colors by name

\definecolor{ocre}{RGB}{199,102,25} % Define the orange color used for highlighting throughout the book

\usepackage{avant} % Use the Avantgarde font for headings
\usepackage{mathptmx} % Use the Adobe Times Roman as the default text font together with math symbols from the Symbol, Chancery and Computer Modern fonts

\usepackage{microtype} % Slightly tweak font spacing for aesthetics
\usepackage[utf8]{inputenc} % Required for including letters with accents
\usepackage[T1]{fontenc} % Use 8-bit encoding that has 256 glyphs

\usepackage[margin=10pt,font=small,labelfont={footnotesize,bf},textfont=footnotesize,format=plain,figureposition=bottom]{caption}
\usepackage{natbib}
	\bibpunct{(}{)}{,}{a}{,}{,} % to follow the A&A style
\def\apjl{{Astrophys. J. Lett.} }

\def\aap{{Astron. Astrophys.} }

\def\apjs {{Astrophys. J. Suppl.}}

\usepackage{rotating,pdflscape,subcaption,wrapfig,chngcntr,setspace,fancyhdr}
\usepackage[pdftex,
            pdfauthor={Roxana-Adela Chira},
            pdftitle={On Filaments within Molecular Clouds and their Connection to Star Formation},
            pdfsubject={PhD Thesis},
            pdfkeywords={ISM: clouds – ISM: structure – ISM: kinematics and dynamics – Stars: formation},
            pdfproducer={Latex with hyperref},
            pdfcreator={pdflatex}]{hyperref}
            
\newenvironment{mysidewaysfigure}
	{\begin{sidewaysfigure*}\vspace*{.0\textwidth}\centering}{\end{sidewaysfigure*}}
    
\newcommand*\xbar[1]{{\hbox{\vbox{\hrule height 0.5pt \kern0.5ex \hbox{\kern-0.1em \ensuremath{#1} \kern-0.1em}}}}}

\pdfoutput=1

\usepackage{titlesec} % Allows customization of titles

\usepackage{graphicx} % Required for including pictures

\usepackage{lipsum} % Inserts dummy text

\usepackage{tikz} % Required for drawing custom shapes

\usepackage[english]{babel} % English language/hyphenation

\usepackage{enumitem} % Customize lists
\setlist{nolistsep} % Reduce spacing between bullet points and numbered lists

\usepackage{booktabs} % Required for nicer horizontal rules in tables

\usepackage{eso-pic} % Required for specifying an image background in the title page

\usepackage{titletoc} % Required for manipulating the table of contents

\contentsmargin{0cm} % Removes the default margin
\titlecontents{chapter}[1.25cm] % Indentation
{\addvspace{15pt}\large\sffamily\bfseries} % Spacing and font options for chapters
{\color{ocre!60}\contentslabel[\Large\thecontentslabel]{1.25cm}\color{ocre}} % Chapter number
{}  
{\color{ocre!60}\normalsize\sffamily\bfseries\;\titlerule*[.5pc]{.}\;\thecontentspage} % Page number
\titlecontents{section}[1.25cm] % Indentation
{\addvspace{5pt}\sffamily\bfseries} % Spacing and font options for sections
{\contentslabel[\thecontentslabel]{1.25cm}} % Section number
{}
{\sffamily\hfill\color{black}\thecontentspage} % Page number
[]
\titlecontents{subsection}[1.25cm] % Indentation
{\addvspace{1pt}\sffamily\small} % Spacing and font options for subsections
{\contentslabel[\thecontentslabel]{1.25cm}} % Subsection number
{}
{\sffamily\;\titlerule*[.5pc]{.}\;\thecontentspage} % Page number
[] 

\titlecontents{lsection}[0em] % Indendating
{\footnotesize\sffamily} % Font settings
{}
{}
{}

\titlecontents{lsubsection}[.5em] % Indentation
{\normalfont\footnotesize\sffamily} % Font settings
{}
{}
{}
 
\usepackage{fancyhdr} % Required for header and footer configuration

\fancypagestyle{plain}{\fancyhead{}} % Style for when a plain pagestyle is specified

\makeatletter
\renewcommand{\cleardoublepage}{
\clearpage\ifodd\c@page\else
\hbox{}
\vspace*{\fill}
\thispagestyle{empty}
\newpage
\fi}

\usepackage{amsmath,amsfonts,amssymb,amsthm} % For math equations, theorems, symbols, etc

\newtheoremstyle{ocrenumbox}% % Theorem style name
{0pt}% Space above
{0pt}% Space below
{\normalfont}% % Body font
{}% Indent amount
{\small\bf\sffamily\color{ocre}}% % Theorem head font
{\;}% Punctuation after theorem head
{0.25em}% Space after theorem head
{\small\sffamily\color{ocre}\thmname{#1}\nobreakspace\thmnumber{\@ifnotempty{#1}{}\@upn{#2}}% Theorem text (e.g. Theorem 2.1)
\thmnote{\nobreakspace\the\thm@notefont\sffamily\bfseries\color{black}---\nobreakspace#3.}} % Optional theorem note
\newtheoremstyle{blacknumex}% Theorem style name
{5pt}% Space above
{5pt}% Space below
{\normalfont}% Body font
{} % Indent amount
{\small\bf\sffamily}% Theorem head font
{\;}% Punctuation after theorem head
{0.25em}% Space after theorem head
{\small\sffamily{\tiny\ensuremath{\blacksquare}}\nobreakspace\thmname{#1}\nobreakspace\thmnumber{\@ifnotempty{#1}{}\@upn{#2}}% Theorem text (e.g. Theorem 2.1)
\thmnote{\nobreakspace\the\thm@notefont\sffamily\bfseries---\nobreakspace#3.}}% Optional theorem note

\newtheoremstyle{blacknumbox} % Theorem style name
{0pt}% Space above
{0pt}% Space below
{\normalfont}% Body font
{}% Indent amount
{\small\bf\sffamily}% Theorem head font
{\;}% Punctuation after theorem head
{0.25em}% Space after theorem head
{\small\sffamily\thmname{#1}\nobreakspace\thmnumber{\@ifnotempty{#1}{}\@upn{#2}}% Theorem text (e.g. Theorem 2.1)
\thmnote{\nobreakspace\the\thm@notefont\sffamily\bfseries---\nobreakspace#3.}}% Optional theorem note

\newtheoremstyle{ocrenum}% % Theorem style name
{5pt}% Space above
{5pt}% Space below
{\normalfont}% % Body font
{}% Indent amount
{\small\bf\sffamily\color{ocre}}% % Theorem head font
{\;}% Punctuation after theorem head
{0.25em}% Space after theorem head
{\small\sffamily\color{ocre}\thmname{#1}\nobreakspace\thmnumber{\@ifnotempty{#1}{}\@upn{#2}}% Theorem text (e.g. Theorem 2.1)
\thmnote{\nobreakspace\the\thm@notefont\sffamily\bfseries\color{black}---\nobreakspace#3.}} % Optional theorem note
\makeatother

\newcounter{dummy} 
\numberwithin{dummy}{section}
\theoremstyle{ocrenumbox}
\newtheorem{theoremeT}[dummy]{Theorem}

\newtheorem{exerciseT}{Exercise}[chapter]
\theoremstyle{blacknumex}
\newtheorem{exampleT}{Example}[chapter]
\theoremstyle{blacknumbox}

\newtheorem{definitionT}{Definition}[section]
\newtheorem{corollaryT}[dummy]{Corollary}
\theoremstyle{ocrenum}

\RequirePackage[framemethod=default]{mdframed} % Required for creating the theorem, definition, exercise and corollary boxes

\newmdenv[skipabove=7pt,
skipbelow=7pt,
backgroundcolor=black!5,
linecolor=ocre,
innerleftmargin=5pt,
innerrightmargin=5pt,
innertopmargin=5pt,
leftmargin=0cm,
rightmargin=0cm,
innerbottommargin=5pt]{tBox}

\newmdenv[skipabove=7pt,
skipbelow=7pt,
rightline=false,
leftline=true,
topline=false,
bottomline=false,
backgroundcolor=ocre!10,
linecolor=ocre,
innerleftmargin=5pt,
innerrightmargin=5pt,
innertopmargin=5pt,
innerbottommargin=5pt,
leftmargin=0cm,
rightmargin=0cm,
linewidth=4pt]{eBox}	

\newmdenv[skipabove=7pt,
skipbelow=7pt,
rightline=false,
leftline=true,
topline=false,
bottomline=false,
linecolor=ocre,
innerleftmargin=5pt,
innerrightmargin=5pt,
innertopmargin=0pt,
leftmargin=0cm,
rightmargin=0cm,
linewidth=4pt,
innerbottommargin=0pt]{dBox}	

\newmdenv[skipabove=7pt,
skipbelow=7pt,
rightline=false,
leftline=true,
topline=false,
bottomline=false,
linecolor=gray,
backgroundcolor=black!5,
innerleftmargin=5pt,
innerrightmargin=5pt,
innertopmargin=5pt,
leftmargin=0cm,
rightmargin=0cm,
linewidth=4pt,
innerbottommargin=5pt]{cBox}

\makeatletter
\renewcommand{\@seccntformat}[1]{\llap{\textcolor{ocre}{\csname the#1\endcsname}\hspace{1em}}}                    
\renewcommand{\section}{\@startsection{section}{1}{\z@}
{-4ex \@plus -1ex \@minus -.4ex}
{1ex \@plus.2ex }
{\normalfont\large\sffamily\bfseries}}
\renewcommand{\subsection}{\@startsection {subsection}{2}{\z@}
{-3ex \@plus -0.1ex \@minus -.4ex}
{0.5ex \@plus.2ex }
{\normalfont\sffamily\bfseries}}
\renewcommand{\subsubsection}{\@startsection {subsubsection}{3}{\z@}
{-2ex \@plus -0.1ex \@minus -.2ex}
{.2ex \@plus.2ex }
{\normalfont\small\sffamily\bfseries}}                        
\renewcommand\paragraph{\@startsection{paragraph}{4}{\z@}
{-2ex \@plus-.2ex \@minus .2ex}
{.1ex}
{\normalfont\small\sffamily\bfseries}}

\usepackage{hyperref}
\hypersetup{hidelinks,backref=true,pagebackref=true,hyperindex=true,colorlinks=false,breaklinks=true,urlcolor= ocre,bookmarks=true,bookmarksopen=false,pdftitle={Title},pdfauthor={Author}}

\newcommand{\thechapterimage}{}
\newcommand{\chapterimage}[1]{\renewcommand{\thechapterimage}{#1}}

\def\thechapter{\arabic{chapter}}
\def\@makechapterhead#1{
\thispagestyle{empty}
{\centering \normalfont\sffamily
\ifnum \c@secnumdepth >\m@ne
\if@mainmatter
\startcontents
\begin{tikzpicture}[remember picture,overlay]
\node at (current page.north west)
{\begin{tikzpicture}[remember picture,overlay]
\node[anchor=north west,inner sep=0pt] at (0,0) {\includegraphics[width=\paperwidth]{\thechapterimage}};
\draw[anchor=west] (5cm,-9cm) node [rounded corners=20pt,fill=ocre!10!white,text opacity=1,draw=ocre,draw opacity=1,line width=1.5pt,fill opacity=.6,inner sep=12pt]{\huge\sffamily\bfseries\textcolor{black}{\thechapter. #1\strut\makebox[22cm]{}}};
\end{tikzpicture}};
\end{tikzpicture}}
\par\vspace*{230\p@}
\fi
\fi}

\def\@makeschapterhead#1{
\thispagestyle{empty}
{\centering \normalfont\sffamily
\ifnum \c@secnumdepth >\m@ne
\if@mainmatter
\begin{tikzpicture}[remember picture,overlay]
\node at (current page.north west)
{\begin{tikzpicture}[remember picture,overlay]
\node[anchor=north west,inner sep=0pt] at (0,0) {\includegraphics[width=\paperwidth]{\thechapterimage}};
\draw[anchor=west] (5cm,-9cm) node [rounded corners=20pt,fill=ocre!10!white,fill opacity=.6,inner sep=12pt,text opacity=1,draw=ocre,draw opacity=1,line width=1.5pt]{\huge\sffamily\bfseries\textcolor{black}{#1\strut\makebox[22cm]{}}};
\end{tikzpicture}};
\end{tikzpicture}}
\par\vspace*{230\p@}
\fi
\fi
}
\makeatother

\fancypagestyle{bibliography}{%
  \fancyhead[RO,LE]{\slshape Bibliography}%
  \fancyhead[LO,RE]{\slshape \ifnum\value{page}<2\relax\else\thepage\fi}
}
\begin{document}

%---------------------------------------------------------------

%% this will generate title pages similar to the template provided
%% by the Department of Physics and Astronomy Heidelberg
%%
%% More information:
%% http://www.physik.uni-heidelberg.de/aktuelles/studium/
%% (PDF link: ...studium/download/145/Vorlage_Diplomarbeit_Formular.pdf)

%% Titleintro
\thispagestyle{empty}
\begin{center}
	\renewcommand{\baselinestretch}{2.00}
    {\large\sffamily
		Dissertation \\
		submitted to the \\
		Combined Faculties of the Natural Sciences and Mathematics \\ 
		of the Ruperto-Carola-University of Heidelberg, Germany \\
        for the degree of \\
		Doctor of Natural Sciences
    }
        
	\par\vfill\normalsize
    Put forward by \\
	Roxana-Adela Chira\\
	born in: Temeschburg\\
	Oral examination: January 22, 2018
\end{center}
\newpage

%% Titlepage
\thispagestyle{empty}
\cleardoublepage

\thispagestyle{empty}
\begin{center}
	\renewcommand{\baselinestretch}{2.00}
	\Large\bfseries\sffamily
		On Filaments within Molecular Clouds \\
        and their Connection to Star Formation
	\par
	\vfill
		%% additionally insert second supervisor here if carrying out an
		%% external diploma thesis. Reduce vspace in L. 44 accordingly.
\end{center}\par

\large\normalfont
\begin{flushleft}
Referees:\\
Prof.~Dr.~Thomas Henning \\
Prof.~Dr.~Cornelis P.~Dullemond
\end{flushleft}
% \vspace{5\baselineskip}

% reset baselinestretch
\renewcommand{\baselinestretch}{1.00}\normalsize

\newpage
%\thispagestyle{empty}
%\cleardoublepage

~\vfill
\thispagestyle{empty}

%TeX Template Copyright \copyright\ 2014 Andrea Hidalgo\\ % Copyright notice

\noindent \textit{Second edition, Januar 2018} % Printing/edition date

\noindent Diese Dissertation wurde durchgeführt von Roxana-Adela Chira, geboren in Temeschburg, unter der Betreuung von Prof.~Dr.~Thomas Henning am Max-Planck-Institut für Astronomie, Heidelberg, und Dr.~Ralf Siebenmorgen an der European Southern Observatory, Garching bei München. 

\noindent This PhD thesis has been carried out by Roxana-Adela Chira, born in Temeschburg, under the supervision of Prof.~Dr.~Thomas Henning at the Max-Planck-Institute for Astronomie, Heidelberg, and Dr.~Ralf Siebenmorgen at the European Southern Observatory, Garching bei München.

\newpage

\thispagestyle{empty}
%\onehalfspacing
\vspace*{0.5cm}

\begin{center}
    \textbf{Abstract}
\end{center}

\noindent In recent years, there have been many studies on the omnipresence and structures of filaments in star-forming regions, as well as the role of their fragmentation in the process of star formation. 
However, only a few comprehensive studies have analysed the evolution of filaments and their distribution with the Galactic disk where the filaments form self-consistently as part of large-scale molecular cloud evolution.
In this thesis, I study the effect of inclination on dust observations of filaments to evaluate whether the variations would enable the identification of further filaments in existing dust surveys.
I address the early evolution of pc-scale filaments that form within individual clouds and focus on the questions how and when the filaments fragment, and how the fragmentation relates to typically used observables of the filaments.
I perform dust radiative transfer calculations on models of cylinders and reconstructions of observed star-forming regions.
For evaluating the equilibrium state of filaments and the nature of their fragmentation I examine three simulated molecular clouds formed in kpc-scale numerical simulations modelling a self-gravitating, magnetised, stratified, supernova-driven interstellar medium.
I find that the observables of filaments in dust emission are on average on small scales influenced by inclination; yet the variations strongly depend on the structure of the object.
The first fragments appear when the line masses of the simulated filaments lie well below the critical line mass of Ostriker’s isolated hydrostatic equilibrium solution.
This indicate that, although the turbulence of the entire clouds is mostly driven by gravitational contraction, fragmentation does not occur do to gravitational instability, but is supported by colliding flow motions.
I conclude that there is no single quantity in my analysis that can uniquely trace the inclination and 3D structure of a filament based on dust observations alone. 
A simple model of an isolated, isothermal cylinder may not provide a good approach for fragmentation analysis, independently of the dominant driving source of the parental cloud.

\begin{center}
    \textbf{Zusammenfassung}
\end{center}

\noindent In den letzten Jahren wiesen viele Studien sowohl auf die Omnipräsenz und Strukturen von Filamenten in Sternentstehungsregionen, als auch auf die Rolle ihrer Fragmentierung in der Sternentstehung hin. 
Jedoch gibt es nur wenige umfassende Studien, die einheitlich Filamente in großskaligen Simulationen von Molekülwolken entwickeln lassen und deren Entwicklung und Verteilung innerhalb der galaktischen Scheibe analysieren.
In dieser Arbeit untersuche ich den Effekt von Inklination auf Staubbeobachtungen von Filamenten and evaluiere, ob die Variationen die Identifizierung weiterer Filamente in bereits existierenden Staubstudien ermöglichen.
I führe Staubstrahlungstransportberechnungen für zylindrische Modelle und Rekonstruktionen von beobachteten Sternentstehungsregionen durch.
Um das Gleichgewichtsstadium und die Fragmentierung der Filamente zu beurteilen, examiniere ich drei simulierte Molekülwolken, die in kpc-großen numerischen Simulationen, die ein eigengravitierendes, magnetisiertes, geschichtetes, durch Supernoven getriebenes interstellares Medium modellieren, entstanden sind.
Dabei finde ich, dass die Observablen in Staubemission nur gering von der Inklination der Filamente beeinflusst werden; wobei die Variationen stark von der Struktur der Objekte selbst abhängen.
Die ersten Fragmente erscheinen, wenn die Linienmasse der simulierten Filamente unterhalb des kritischen Wertes, der durch Ostrikers isolierte, hydrostatische Gleichgewichtslösung gegeben ist.
Das zeigt, dass, obwohl die Turbulenz der ganzen Wolke hauptsächlich durch gravitative Kontraktionen getrieben wird, die Fragmentierung nicht aufgrund gravitativer Instabilität geschieht, sondern durch kollidierende Gasflüsse angeregt wird.
I schließe daraus, dass es keine Messgröße in meiner Analyse, die alleinstehend und eindeutig die Inklination und 3D Strukturen eines Filaments widerspiegeln kann. 
Außerdem kann ein simples Model eines isolierten, isothermen Zylinders keine gute Näherung für die Fragmentierung bieten, unabhängig von der dominanten Turbulenzquelle der übergeordneten Wolke.

%---------------------------------------------------------------
%	TABLE OF CONTENTS
%---------------------------------------------------------------

\chapterimage{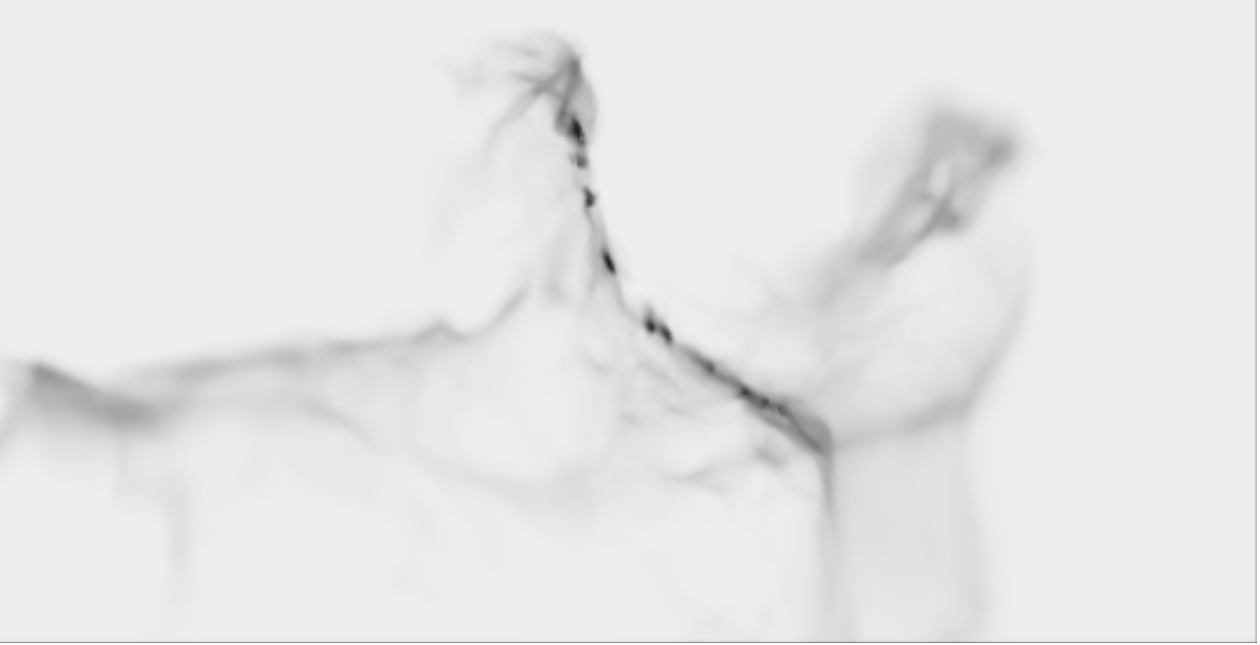} % Table of contents heading image

\pagestyle{empty} % No headers

\tableofcontents % Print the table of contents itself

\cleardoublepage % Forces the first chapter to start on an odd page so it's on the right

\pagestyle{fancy} % Print headers again

%---------------------------------------------------------------
%	CHAPTERS
%---------------------------------------------------------------

\chapter[A short Story about Filaments]{A short Story about Filaments\footnotemark}\label{intro}

\footnotetext{Parts of this chapter are already published in \citet{Chira2016,Chira2017}}

Stars are essential components of the Universe.
They emit light over a wide range of wavelengths and enrich the gas content of the interstellar and intergalactic medium by producing heavier elements through nuclear fusion.
High-mass stars play a special role, not only because they are luminous enough to be observable across the Universe, but also because they return gas to the Interstellar Medium (ISM) and drive the turbulence within the ISM by exploding as supernovae (SNe).
Therefore, the question of how stars form and evolve has been constantly investigated for many decades, both observationally and theoretically. 

It has become clear that stars form in the densest and coldest phases of the ISM, the so-called molecular clouds \citep[among others,][]{Lada2003,Tielens2010,Draine2011,Andre2014}.
These clouds consist predominantly of molecular hydrogen with average number densities on the order of 10$^2$--10$^6$~cm$^{-3}$ and gas temperatures of 10--15~K \citep{Draine2011}.
These conditions are supported by the coupling of the gas with dust.
Although dust provides only 1-2\% of mass within the ISM, it is a crucial ingredient for many physical and chemical processes.
Furthermore, dust is an important element for all heating and cooling processes within molecular clouds as it is a very efficient coolant and protects the molecular gas to be dissociated by high-energy photons and cosmic rays.

Unlike often assumed, the gas is not homogeneously distributed within the molecular clouds.
Instead, one sees a highly hierarchical network of substructures.
The most prominent features are filaments, elongated streams of gas that connect clumps and cores, the very sites of star formation, with each other \citep{Bergin2007,Williams2000,Menshchikov2010,Arzoumanian2011,Peretto2012,Andre2014}.
Additionally, these filaments build the bridges that transfer gas from the large scales of the entire cloud onto those cores \citep{Schmalzl2010,Myers2011,Schneider2012,Koenyves2015}.

Due to their omnipresence in the ISM, filamentary structures have been investigated for many years.
Much effort has been invested in studying their morphology \citep{Barnard1927,Schneider1979}, properties \citep{Schmalzl2010,Hacar2013,Andre2014,Smith2014b,Kainulainen2017}, distribution within the Milky Way \citep{Molinari2010,Ragan2014,Abreu-Vicente2016,Li2016}, and formation \citep{Rivera2015,Federrath2016,Smith2016}.
The large number of studies is not only motivated by the potential role of filaments in star formation, but also by the accessibility to instruments and methods that are able to examine filaments in detail.

Yet, little is really understood about their origin and their exact role in the star formation process.
One difficulty is that there is no unique, universal and physically motivated definition of what a filament actually is.
In general, filaments are detected as structures of condensed gas with an aspect ratio of at least 1:5$\sim$10 \citep[meaning that one axis is at least five to ten times longer than the other axis,][]{Andre2014}.
However, other studies focus on more specific subsamples of filaments that require different properties (e.g., an aspect ratio of 1:50, \citeauthor{Zucker2015}, \citeyear{Zucker2015}, or velocity coherence, \citeauthor{Hacar2013}, \citeyear{Hacar2013}). 
It is due to this lack of definition that there are many ways to detect and identify filamentary structures with a large variety of properties and on all scales all across the Universe.

In the context of this thesis, the most prominent groups of filaments are seen embedded within molecular clouds \citep{Schneider1979,Bally1987,Beuther2011a,Arzoumanian2011,Hacar2013,Kainulainen2013c,Andre2014,Molinari2014,Stutz2015} or as isolated objects across the Milky Way \citep{Jackson2010,Ragan2014,Wang2015,Zucker2015,Wang2016}.
The properties of those filaments range over several orders of magnitude, with length of a few to hundreds of parsecs and masses on the orders 10$^3$--10$^5$~M$_\odot$.
Yet, they all have in common that the filaments contain gas and dust at high densities (with 10$^4$~cm$^{-3}$ and higher) and low temperatures (below 30~K), as well as supersonic gas flows. 
These are essential ingredients for the onset of star formation, based on our current understanding.

Indeed, both kinds of filaments show signs of star formation which indicate that they are indeed involved in the process.
However, it widens the range of forces that possibly form filaments and set their properties. 
The most prominent contracting force, that acts on all scales from molecular clouds to individual stars, is gravity. 
As \citet{Federrath2016} has demonstrated, gravity is able to form filamentary structures, though their properties do not match those of observed filaments. 
In this example, the filaments that have been produced in the 'gravity only' runs have widths half as large as those filaments that are normally observed in comparable molecular clouds \citep[e.g.,][]{Arzoumanian2011}.
Thus, counteracting forces are essential for a complete picture of filaments, and of the initial conditions of star formation.

Traditionally, one assumes that the internal thermal pressure of the gas opposes those collapsing motions.
This is the basis of commonly used theoretical predictions, such as the Jeans analysis \citep{Jeans1902} or the cylindrical model described by \citet{Ostriker1964b}.
However, there have been many studies that show evidence that thermal pressure alone cannot prevent molecular clouds from collapsing onto a single object \citep{Klein1999,Toci2015}.

Consequently, other forces need to be considered, as well.
On scales of molecular clouds, the most promising candidates are turbulence, feedback and magnetic fields.
Isolated filaments may also feel the impact of the Galactic shear, spiral density waves, the interstellar radiation field, and colliding flows due to the expansion of (super-)shells, that have been formed by SNe and HII regions \citep{Battisti2013}. 

Since these forces are not directly observable, one way to test their influence is to use numerical simulations that form filaments self-consistently and compare the properties of the resulting structures with those observed. 
Yet, only a few comprehensive studies have analysed the evolution of filaments as part of large-scale molecular cloud evolution \citep{Smith2014a,Federrath2016}.

However, often this comparison cannot occur on a one-to-one basis as observations preliminarily show the two-dimensional (2D) projection of the three-dimensional (3D) objects. 
With precise and highly-resolved line observations one is able to produce so-called position-position-velocity (PPV) diagrams that reconstruct the third dimension along the line-of-sight (LoS) using the local standard of rest velocity and models of the Galactic rotation curve \citep{Reid2009}.
However, it is not clear how reliably this approach really resembles the third dimension, especially in the context of large filaments on galactic scales \citep{Ragan2014,Zucker2015,Abreu-Vicente2016}.
Furthermore, there are many LoS effects, such as optical depth, freezing-out or overlap effects, which increase the uncertainties of the observations \citep{Ballesteros2002,ZamoraAviles2017}.

Other studies use dust observations to reconstruct the 3D geometry of complex structures \citep{Kainulainen2014,Schmiedeke2016}.
For example, \citet{Schmiedeke2016} deduced the 3D structure of Sagittarius B2 by fitting synthetic observations prepared with \texttt{RADMC-3D}\footnote{The code is  available  with  the  permission  of the main  author,
Cornelis Dullemond, at the webpage \url{http://www.ita.uni-heidelberg.de/dullemond/software/RADMC-3D/}. There is also a manual on the website including more detailed explanations of the different functions and parameters.} on dust observations conducted with \textit{Herschel}, SMA, VLA, and ATLASGAL.
The framework \texttt{pandora} that the authors implemented for their investigations can utilise any kind of input distribution and, thus, will be able to reconstruct more 3D structures of observed objects, such as filaments.

As dust does not transfer any velocity information, line observations are still required to confirm the 3D coherence of a filament.
Yet, dust traces more orders of magnitude of density than gas.
This makes it a favoured tool in theoretical studies and a starting point of many observational studies and catalogues of filaments \citep{Lenfestey2013,Li2016}.
From the theoretical point of view, dust radiative transfer (RT) is the easiest method to project the data simulated in 3D onto 2D synthetic observations. 
Besides instrumental limitations of the telescopes (such as resolution, instrument noise, etc.), this requires RT calculations that include all the physical processes and chemical conditions along the entire LoS.
A complete realisation of line RT is computationally very expensive, whereas dust RT can well be approximated.

Studies like those by \citet{Juvela2012a} and \citet{Smith2014b} have applied RT codes to filaments formed within their (magneto-)hydrodynamical (MHD) models.
They have found that the column density profiles of resolved filaments obtained from their synthetic images were similar to observed ones, suggesting that RT effects are negligible.
This is no longer the case when the filament is located further away and becomes poorly-resolved.
However, they have also shown that filamentary structures in dust column density maps may be the result of superpositions along the LoS.
This emphasises the importance of detecting counterparting filaments in line observations.
Yet, such theoretical studies used to follow the custom to project the simulated filaments are projected in a way that the synthetic LoS is perpendicular to the filaments' axis.
This highlights that a more generalised view on filaments is necessary to understand their entire nature and importance in star formation.

Thus, the questions I address in this thesis are:
Chapter~\ref{viewing} asks whether one would recognise inclined filaments in dust observations.
What are the signatures of inclined filaments in the far-infrared (FIR) and sub-mm observations?
Are variations along different LoS ssignificant enough on average to be detectable?
And can one use these variations to learn more about the distribution of filaments within the Milky Way based on already existing data?
In Chapters~\ref{frag} and~\ref{turb} I investigate how filaments evolve and fragment in numerical simulations.
Is the fragmentation picture seen in simulations in agreement with the quasi-static analytic framework of gravitational fragmentation?
Or are additional forces, such as colliding flows, required to create fragments?
Are the fragmentation modes related to the global turbulence modes of the entire parental cloud?

%\chapterimage{head02.pdf}

\chapter[Project I: Filaments at Different Viewing Angles]{Filaments at Different Viewing Angles\footnotemark[3]}\label{viewing}

\footnotetext[3]{The content and results presented in this chapter are published in \citet{Chira2016}.}

As described in Chapter~\ref{intro}, two of the major foci of studies on the early stages of star formation have been the filamentary structures of the ISM and molecular clouds and the role of these filaments in star-forming activities \citep[e.g.,][]{Menshchikov2010,Arzoumanian2011,Peretto2012,Andre2014}.
As filaments are rather easily identified as elongated structures in dust observations most observational studies are entirely based on dust data or at least use samples of filaments that have been selected based on dust surveys \citep[for example,][]{Ragan2014,Abreu-Vicente2016,Wang2015}.
However, since dust observations do not provide any kinetic information, one lacks any details about the inclination of the filament axis relative to the line-of-sight (LoS).
Unless there are complementary observations of molecules or atoms only a detailed reconstruction, including synthetic observations, of the observed filament can reduce these uncertainties.

However, this effort is, if at all, only done for filaments that have already been observed.
Those samples are biased in a way that the long axes of the filaments are typically aligned parallel to the plane of the sky.
Otherwise, the filaments would not have been seen as elongated structures in the first place and, thus, not included in the sample of observed objects.
As a consequence, the reconstructed inclinations of these filaments are automatically small \citep{Arzoumanian2011,Fischera2012a}.
Furthermore, assuming that one can find all filaments by looking for elongated, high-density structures only implies that the filaments have preferential locations and directions within the Galactic disk.
This is not necessarily the case and may exclude a notable fraction of filaments \citep{Smith2014b}.

The questions I attempt to answer in this chapter are:
Would one recognise inclined filaments in dust observations?
What are the signatures of inclined filaments in the far-infrared (FIR) and sub-mm observations?
Are variations along different LoSs significant enough on average to be detectable?

%%%%%%%%%%%%%%%%%%%%%%%%%%%%%%%%%%%%%%%%%%%%%%%%%
\pagebreak

\section{Dust Radiative Transfer}\label{viewing:radtrans}

In this section, I briefly summarise the basics of dust radiative transfer (RT) calculations and the so-called Monte Carlo method (Sect.~\ref{viewing:radtrans_basics}). 
The descriptions are adapted from the lecture notes by Prof.~Dr.~Cornelis~Dullemond\footnote[4]{Lecture notes on "Radiative Transfer in Astrophysics \url{http://www.ita.uni-heidelberg.de/~dullemond/lectures/radtrans_2012/index.shtml}"}.
I refer to the lecture notes for more detailed review of the topic.
In Sect.~\ref{viewing:mc} I describe the code I apply to create the synthetic observations I analyse in Sect.~\ref{viewing:fila}, as well as my contribution to it. 

\subsection{The Basics and Monte Carlo Methods}\label{viewing:radtrans_basics}

The interest in interstellar dust has increased continuously during the last years, for a few reasons.
First, despite its small contribution to the total mass, dust has chemical properties that are critical for shaping the structures and properties of the ISM, and especially of molecular clouds, since the grains' surfaces enable and catalyse chemical reactions.
Second, dust grains also protect the molecules in the interior of molecular clouds due to the efficient absorption of high-energy photons and cosmic rays over broad ranges of wavelengths.
Third, dust efficiently controls heating and cooling processes within molecular clouds, especially in the densest regions where dust and gas are thermally coupled.

The wide range over which dust is observable is accounted for by the very characteristic way in which dust grains process radiation.
Contrary to molecules, the atoms within dust grains are too tightly connected with each other. 
Thus, a grain does not need a photon with a specific energy to excite a certain transition within the grain. 
Instead the grain can absorb any high-energy photon and distribute its energy the over several atoms or couplings. 
This broadens the range of absorbable photons from discrete energies to a continuum from optical to X-ray wavelengths.

At longer wavelengths, that are much larger than the grains' diameter, the interaction between dust particles and photons is less effected by the geometry of the grains, but their volume mass.
Thereby, the dielectric elements within the dust grains react to the electromagnetic field of the radiation field. 
The longer the wavelength, the deeper the photon can propagate into the interior of the grains. 
This may cause that the dielectrics within the grains create their own electromagnetic field that amplifies the interaction between radiation field and dust. 

Which range of photons that interact with a specific dust grain is primarily defined by the properties of the grain, such as the opacity, $\kappa_\nu$, of the grain.
The opacity is one of the key parameters in the context of RT and thermal processes in dusty environments because it quantifies which fraction of incoming photons interacts with the dust. 
Assuming that the dust species are homogeneously distributed along the light path, the ratio of out-going intensity, $I(s)$, to in-coming intensity, $I_0$, is given by the source-free RT equation,
\begin{equation}
	\ln \left[ \frac{I(s)}{I_0} \right] = - \int_0^s \kappa_\nu(x) \rho(x) dx = - \int_0^s \tau_\nu(x) dx ,
    \label{equ:viewing_radtrans_intratio}
\end{equation}
with $\kappa_\nu(x)$ and $\tau_\nu(x) = \kappa_\nu(x) \rho(x)$ being the total dust opacity and total optical depth, respectively, at frequency $\nu$ and the location $x$, $\rho$ the density, $s$ the path length of the light.
With $\sigma_\nu$ being the cross-section, meaning the surface within which the grain interacts with a photon, and $m$ the mass of the dust grain, the total dust opacity is defined by 
\begin{equation}
	\kappa_\nu(x) = \frac{\sigma_\nu(x)}{m(x)} .
    \label{equ:viewing_radtrans_kappa_def}
\end{equation}
\pagebreak

\noindent In general, most of the quantities in Equations~(\ref{equ:viewing_radtrans_intratio})~and~(\ref{equ:viewing_radtrans_kappa_def}) are not considered to be strongly dependent on the actual location. 
This approach ignores that dust particles, contrary to molecules, do not have unique sizes, shapes, or composition.
Thereby, those properties are essential for determining the mass of the individual grain, its cross section and, in a more global context, also the density of the dusty medium. 

Obviously, considering all these factors is computationally too expensive.
Thus, dust models \citep[e.g.,][]{MRN1977,Ossenkopf1994} introduce assumptions that reduce the number of free parameters by grouping the dust particles into categories roughly representing our nature.
For example, dust particles are mainly classified as silicates (mainly Si-O bearing grains), carbonates (amorphous carbon, graphite, PAHs, organics) and dust grains that are covered by layers of ice (water ice, CO ice, etc.). 
In my studies, I focus on silicates and amorphous carbons.
The respective quantities are correspondingly labelled with Si and aC hereafter.

Another pair of significant assumptions are that the assumptions that dust particles are spherically symmetric and only optically interacting with photons. 
These reduce Eq.~(\ref{equ:viewing_radtrans_kappa_def}) to
\begin{equation}
	\kappa_\nu(x,r) = \frac{\sigma_\mathrm{geo}}{m(x)} = \frac{\pi r^2}{m(x)} ,
    \label{equ:viewing_radtrans_kappa_geo}
\end{equation}
with $\sigma_\mathrm{geo} = \pi r^2$ being the geometrical cross section of the grain and $r$ the grain's radius.
Although this assumption is commonly used in many studies, there are also investigations that demonstrate that a more proper treatment of the geometry of the dust grains may lead to a significant difference in the grains' optical behaviour \citep[e.g.,][]{Ossenkopf1992,Jaeger2003,Jones2013}.

Another common assumption is that the dust is supposed to be homogeneously distributed along the light path.
Consequently, Equations~(\ref{equ:viewing_radtrans_intratio})~and~(\ref{equ:viewing_radtrans_kappa_geo}) are normally given as
\begin{align}
	\ln \left[ \frac{I(s)}{I_0} \right] & = - \kappa_\nu \rho s = - \tau_\nu s \mbox{\hspace*{1em}} \Rightarrow I(s) = I_0~e^{- \kappa_\nu \rho s} = I_0~e^{- \tau_\nu s} \\
    \kappa_\nu(r) & = \frac{\pi r^2}{m}.
    \label{equ:viewing_radtrans_simply}
\end{align}
This assumption is difficult on all scales. 
On large scales it ignores that the density within the objects of interest (e.g., molecular clouds) is not constantly distributed, as in reality the density of an object is very structured and non-uniform.
This way, the optical depths along the path of photons varies over as many orders of magnitude as the density.
Although I do not explicitly apply this assumption in my work, the nature of gridded data implies it.
Since I cannot resolve the density within a grid cell, it can be treated as constant in a zero-order approximation; and consequently the opacities, as well. 
Yet, this caveat needs to be kept in mind.

The total opacity described above summarises all interactions between the dust grains and photons.
Those interactions can be classified as either scattering or absorption. 
When a photon is scattered its propagation direction is changed.
If a photon is absorbed, the dust particle consumes its entire energy, ending the propagation of the photon completely. 
Mathematically speaking, this means:
\begin{align}
	\kappa_\nu = \kappa_\nu^\mathrm{sca} + \kappa_\nu^\mathrm{abs} , \qquad
    \kappa_\nu^\mathrm{sca} = \eta_\nu \kappa_\nu , \qquad
    \kappa_\nu^\mathrm{abs} = (1 - \eta_\nu) \kappa_\nu ,
    \label{equ:viewing_radtrans_def_sca_abs}
\end{align}
where $\kappa_\nu^\mathrm{sca}$ and $\kappa_\nu^\mathrm{abs}$ are the opacities referring to scattering and absorption processes, respectively. 
$\eta_\nu$ is the albedo that reflects the ratio of scattering events relative to all dust-photon-interactions in total.
\pagebreak

\noindent In the context of RT calculation, absorption events are important since those events are able to change the thermal state of the dust and its radiation behaviour at long wavelengths (FIR, sub-mm, radio). 
In RT codes this is taken into account by always connecting an absorption event with the re-emission of a thermal photon (package).
This, of course, adds an additional source term to the RT equations.
Fortunately, due to the properties of the particles the emission behaviour of dust can be described by as a black body, following the Planck radiation given as
\begin{equation}
	B_\nu(T_\mathrm{d}) = \frac{2 h \nu^3}{c^2} \left( e^{\frac{h \nu}{k_\mathrm{B} T_\mathrm{d}}} - 1 \right)^{-1} ,
    \label{equ:viewing_radtrans_planck}
\end{equation}
where $h$ is the Planck constant, $c$ the light speed, $k_\mathrm{B}$ the Boltzmann constant and $T_\mathrm{d}$ the dust temperature.
Note that this relation only then reflects the entire thermal dust emission if the respective dust is isothermal. 
If that should not be case, as in molecular clouds where dust used to be colder in the interior compared to the outer skirts, the actual total dust emission is a density-weighted superposition of the individual contributors.
An example of this effect is the spectral energy distribution (SED) of a protostar that is surrounded by a disk. 
Such a SED normally consists of two main components, namely a hotter blackbody emitted by the protostar and a colder one reflecting the disk. 
%\medskip

Of course, there are more properties and physical events that should actually taken into account when solving the RT equation.
However, my examples already demonstrate that an entire and proper solution would be very expensive and often not possible analytically.
Thus, numerical methods are essential in astronomical studies, especially wherever RT routines are only embedded in larger simulations for self-consistently deriving temperatures.

One commonly used numerical implementation is the Monte Carlo methods.
The methods are based on simulating random events and, therefore, very efficient in imitating scattering events.
The code I have worked on and use for the investigations presented in Sect.~\ref{viewing:fila} uses Monte Carlo methods to perform the dust RT and create SEDs and intensity maps. 
In Sect.~\ref{viewing:mc}, I outline what I contributed to the code. 
For the rest of this section, I briefly summarise the basics of how photon packages propagate and interact with the dust within the code.

Monte Carlo provides more of a statistical approximation than actual solution to the RT equations.
Yet, it is intuitive to understand since the processes are reminiscent of the actual propagation of photons. 
To make Monte Carlo more computationally efficient the methods, however, do not propagate individual photons, but so-called photon packages representing a high number of individual photons with the same frequency.
Therefore, the spectrum of the heating source(s) is divided into $n_f$ bins, each containing the same amount of energy.
Although preparing such a binning might be difficult depending on the source spectrum, the advantage of this approach is that energy per photon package, $\epsilon_\mathrm{pak}$, and with it most of the RT calculations, are independent of the initial frequency of the photon package.

It is important to emphasise that Monte Carlo methods are not completely frequency independent. 
An essential part of Monte Carlo is that the decision whether an event occurs and which kind it is (scattering or absorption/re-emission) strongly depends on the dust opacity in the respective grid cell. 
\pagebreak

\noindent The RT processes of the code can, thereby, be summarised in the following way \citep{Lucy1999,Bjorkman2001}:
\begin{enumerate}
\item The photon packages propagate through the data cube after being ejected from the radiation source. After each incremental step along its path, a random number, $\xi_1$ is generated and compared to the local optical depth, $\tau_\nu$. Therefore, $\tau_\nu$ represents the connection between the dust and the radiation field as it depends on both the local density and the frequency of the photon package.
If $\tau_\nu < -\ln(\xi_1)$ there is no interaction between the dust and the photon package continues to propagate forward. If not, the photon package interacts with the dust.
\item The nature of the interaction is given by a second random number, $\xi_2$. If $\xi_2 < \eta_\nu$ the photon packages will be scattered, and otherwise absorbed and re-emitted.
\item The code distinguishes the interactions of individual dust species (Si and aC, in my case). This way the user can easily disentangle the contributions of the species in the final SED or image. In the Monte Carlo routine this is realised by introducing a new random number, $\xi_3$, that is compared to the fraction of extinction due to silicates relative to the total extinction. If $\xi_3$ is bigger than this ratio the photon package interacts with the amorphous carbons, and otherwise with the silicates.
\item If the dust has absorbed the photon package, its temperature is updated accounting to the newly gained energy. The code, thereby, separates the temperatures of the individual dust species. In order to fulfil energy conservation, a new photon package needs to be thermally re-emitted, at a frequency that depends on the updated dust temperature.
\item For both the scattered and re-emitted photon packages a new propagation direction is computed.
\end{enumerate}

\noindent This routine is repeated for each photon package as long as it is moving within the data cube. 
When the photon packages leave the cube, they are counted, taking into account to the final frequency and the dust species they have interacted with last.
Based on these photon counts, the user can easily create a SED. 
I apply this method in Sect.~\ref{viewing:mc_benchmarks} for benchmarking my extensions on the code.
The simulated dust images in Sect.~\ref{viewing:fila} are produced by additional ray-tracing calculations \citep[see][]{Heymann2012}.

\subsection{The Code}\label{viewing:mc}

Since the focus of this chapter is to investigate how filaments are observable in thermal dust emission at infrared and (sub-)mm wavelengths, I use and extend the work on a 3D vectorised Monte Carlo dust RT code.
The code, which I call \texttt{MC} for simplicity hereafter, has been introduced by \citet{Heymann2012} and is based on the one-dimensional RT code by \citet{Kruegel2008}.
It solves 3D dust RT equation and includes many techniques that optimise the computations, e.g.~ by running it on GPUs.
By specialising the code to dust RT, \texttt{MC} offers many dust-specific features that makes it advantageous for studies of interstellar dust.
For example, besides utilising the standard opacity tables, \texttt{MC} offers an option that enables the user to use their own opacity tables, or only parts of it if the range of used grain sizes shall be reduced.

Initially, the code was only able to generate symmetrical density distributions that were heated from single internal point sources; for example, a sphere containing a protostar in the middle, or a disk surrounding an active galactic nucleus (AGN).
For my investigations, however, I need to consider externally heated filaments that have arbitrary density structures and are sterile, meaning that they do not contain any internal heating sources. 
This would not have been possible with the initial version of the code.
\pagebreak

\noindent In what follows, I describe the extensions I implemented to be able to conduct my studies.
First, I describe how the arbitrary density distributions are read.
Next, I illustrate how the user can heat his object of interest externally by using an isotropic radiation field, such as the interstellar radiation field (ISRF).
I present the benchmark tests and introduce how the flux errors are estimated when spectral energy distributions (SEDs) are generated by counting photons.

\subsubsection{Arbitrary Density Distributions}\label{viewing:mc_densdistri}

So far, \texttt{MC} has only generated symmetric density distributions, e.g.~spheres, discs and tori. 
Filaments, however, have a roughly cylindrically symmetry, if any at all. 
Thus, I implement a new function that reads in arbitrary density distributions from binary files and deactivates any symmetry-motivated assumptions or simplifications within the code.
This not only means that several steps in the normal routine of the code need to be skipped.
It also implies a re-ordering of the entire code and re-defining of substantial subroutines as they have used to be too specified for my purposes.
During my work, I generalised the actual RT subroutines so that the RT calculations act independently of how the studied density distribution has been created.

Next, I have implemented new subroutines that read in the desired density distribution from a binary file and automatically transform it into the arrays and parameters the code needs for conducting the RT calculations.
The density distribution, therefore, needs to be given as Cartesian grid, with the options for subgrids within individual cells similar to the structure of adaptive mesh refinement (AMR) data.
This means that the code is limited to specific formats of input. 
However, most observational and simulated data can be mapped onto such a grid, which automatically broadens the range of possible applications without changing \texttt{MC} and its included RT routines anymore.

\subsubsection{External Heating}\label{viewing:mc_externalheating}

Besides the symmetry of studied density distribution, the user had initially been limited to a single point source when applying \texttt{MC}.
Therefore, one has been able to choose the kind of source (protostar, black body radiation, AGN), its effective temperature (in the case of black bodies) or total luminosities (in the case of AGNs), and its position within the data cube (centre or origin of the cube).
These settings imply many symmetry-related assumptions.
However, this handling has not been applicable for my studies on sterile filaments that are solely heated by the interstellar radiation field (ISRF).
Therefore, a proper treatment of external heating is crucial for calculating correct emission and temperature distributions of such filaments.
Thus, I connect the generalisation of \texttt{MC}, when implementing the arbitrary density option, with a widening of possible heating sources.
In particular, I focus on the implementation of an external, isotropic heating field that is able to resemble the ISRF.

In a first step, I add a new parameter to the routine treating arbitrary density distributions.
With this parameter, $f_{\mathrm{isrf}}$ hereafter, the user can set the nature of heating source.
If $f_{\mathrm{isrf}}$~=~0, no heating source is used.
If $f_{\mathrm{isrf}}$ is negative, the code considers a single point source, as in the initial version of \texttt{MC}.
If the user chooses a positive value for $f_{\mathrm{isrf}}$ an external heating field is applied.

The external radiation field is implemented as follows:
The frequency bins are not generated by \texttt{MC} itself anymore, as in the case of the point sources, but are read in from a new input file. 
This means, of course, that the user needs to prepare the sequence of frequency bins themselves, similar to the grid structures of the arbitrary density distribution.
Therefore, the user needs to ensure that all frequency bins conform with the requirements of the Monte Carlo method, meaning that each bin contains the same amount of energy that is propagated by the photon packages.
The advantage of this ansatz, however, is that the user is free to use their favoured radiation field model. 
\pagebreak

\noindent The total source luminosity, $L_\mathrm{source}$, of the read-in external radiation field is then scaled with $f_{\mathrm{isrf}}$, as follows:
\begin{equation}
	L_\mathrm{source}~=~f_\mathrm{isrf} \, \cdot \, \underbrace{(2 \, a)^2 \cdot \left( n_\mathrm{x} \, n_\mathrm{y} + n_\mathrm{x} \, n_\mathrm{z} + n_\mathrm{y} \, n_\mathrm{z} \right)}_{\mbox{\small total outer surface } c_{\mathrm{c3}} \cdot S_\mathrm{out}} \, \cdot \, l_\mathrm{isrf}  ,
    \label{equ:viewing_mc_def_lisrf}
\end{equation}
with $l_\mathrm{isrf}$~=~0.036~erg~s$^{-1}$~cm$^{-2}$ representing the surface luminosity density of the ISRF \citep{Kruegel2003}, $a$ the edge length of a single cell of the grid (without refinement levels), and $n_\mathrm{x}$, $n_\mathrm{y}$, and $n_\mathrm{z}$ the number of grid cells in x-, y-, and z-direction. 
Following the standard routines of \texttt{MC}, the energy per photon package is then given by,
\begin{equation}
	\epsilon_\mathrm{pak}~=~\frac{L_\mathrm{source}}{n_\mathrm{f} \cdot n_\mathrm{zyk} \cdot c_{\mathrm{c3}}} ,
	\label{equ:viewing_mc_def_epak}
\end{equation}
with $n_\mathrm{f}$ being the number of frequency bins of the heating source and $n_\mathrm{zyk}$ being the number of photon packages that are ejected within each frequency bin.
$c_{\mathrm{c3}}$ is a scalar factor taking the symmetric structure of the model into account.
\texttt{MC} offers the possibility to reduce the computational effort by assuming that the studied object is axisymmetric. 
This means that the input grid represents only 1/8 of the actual object that shall be studied. 
I call this practice \textit{axial mode}, hereafter.
$c_{\mathrm{c3}}$, then, corrects for this simplification so that the outputs refer to the entire object (\textit{3D mode}, henceforth).

The photon packages are ejected from the outer surface, $S_\mathrm{out}$, into the cube. 
The surface is, therefore, defined as
\begin{equation}
	S_\mathrm{out}~=~\begin{cases}
\{ \, (i,j,k) \, | \, i~=~n_\mathrm{x} \vee j~=~n_\mathrm{y} \vee k~=~n_\mathrm{z} \} \footnotemark[5] & \mbox{ axial mode} \\
\{ \, (i,j,k) \, | \, i~=~1 \vee i~=~n_\mathrm{x} \vee j~=~1 \vee j~=~n_\mathrm{y} \vee k~=~1 \vee k~=~n_\mathrm{z} \} & \mbox{ 3D mode} 
          \end{cases} .
	\label{equ:viewing_mc_def_surfout}
\end{equation}
\footnotetext[5]{Note that the code uses \texttt{FORTRAN} where indices of an array run from 1 to $n_\mathrm{x/y/z}$, instead of 0 to $n_\mathrm{x/y/z}$-1.}

\vspace{-\baselineskip}
\noindent The starting cell, $\vec{a}$, from which the the respective photon package is ejected, is an element of $S_\mathrm{out}$.
It is calculated for each photon package individually by using three random numbers, $\xi_1$, $\xi_2$, and $\xi_3$.
The first random number, $\xi_1$, determines the precise plane of $S_\mathrm{out}$.
For example, in the axial mode $\xi_1$ decided whether $\vec{a} = \left( n_x , f(\xi_2) , f(\xi_3) \right)^T$, $\vec{a} = \left( f(\xi_3) , n_y , f(\xi_2) \right)^T$, or $\vec{a} = \left( f(\xi_2) , f(\xi_3) , n_z \right)^T$. 
As indicated in the example, the other two random numbers define the two missing coordinates of $\vec{a}$. 
If $\vec{a}$ is too close to the corners of the data cube, the calculations are repeated with three new random numbers.

The direction vector, $\vec{d}$, along which the photon package starts propagating into the cube, is similarly computed with two new random numbers, $\xi_4$ and $\xi_5$, and normalised to a unit vector.
To ensure that the resulting radiation field is isotropic, $\vec{d}$ is given by
\begin{align}
	a_x = 1 \mbox{ or } a_x = n_x: ~ & \vec{d} = \left( \pm \mu , \sin(\theta) \cdot \cos(\phi) , \sin(\theta) \cdot \sin(\phi) \right)^T \\
	a_y = 1 \mbox{ or } a_y = n_y: ~ & \vec{d} = \left(  \sin(\theta) \cdot \sin(\phi) , \pm \mu , \sin(\theta) \cdot \cos(\phi) \right)^T \\
	a_z = 1 \mbox{ or } a_z = n_z: ~ & \vec{d} = \left( \sin(\theta) \cdot \cos(\phi) , \sin(\theta) \cdot \sin(\phi) , \pm \mu \right)^T
    \label{equ:viewing_mc_def_dir}
\end{align}
with $\mu$~=~$\sqrt[]{\xi_4}$, $\sin(\theta)$~=~$\sqrt[]{1 - \mu^2}$ and $\phi$~=~$2 \pi \xi_5$.
In the case $\vec{a}$ is located on one of the outer surfaces (meaning where either $a_x$~=~$n_\mathrm{x}$, $a_y$~=~$n_\mathrm{y}$, or $a_z$~=~$n_\mathrm{z}$), the signs of $\vec{d}$ are set in a way that the photon package is going \textit{into} the cube.

\noindent Of course, this only results in an isotropic radiation field if the number of packages is large enough. 
According to my tests, this requires at least 10$^3$ photon packages per frequency bins.
Fewer photon packages would be distributed randomly, as well, but there is no guarantee that they are equally ejected within all viewing angles, which is essential for obtaining an isotropic radiation field. 
Instead the radiation field would correspond to a surrounding cluster of point sources. 
However, for sampling the SED correctly with the photon counting method (see below) the user needs a high number of photon packages to reduce the Monte Carlo noise, so that this problem does not arise.

\subsubsection{Benchmark Tests on External Heating}\label{viewing:mc_benchmarks}

In order to verify that the new implementations provide correct results, I benchmark the new routines against \texttt{HII} \citep{Kruegel2003}.
\texttt{HII} is a one-dimensional RT code that solves the RT equations analytically for dusty spheres and rings.
This makes it a reliable object of comparison.

For the benchmark runs, I choose homogeneous spheres with an inner radius, $R_{\mathrm{in}}$, and outer radius, $R_\mathrm{out}$.
The volume within $R_{\mathrm{in}}$ is empty, as well as the volume outside $R_\mathrm{out}$.
The volume between $R_{\mathrm{in}}$ and $R_\mathrm{out}$ is homogeneously filled with dust of constant density, $\rho$.

The 3D model is divided into $n_\mathrm{x} \times n_\mathrm{y} \times n_\mathrm{z}$ cubes in x-, y- and z-direction, respectively, that have a physical edge length, $a$.
The total extinction, $A_{\mathrm{V}}$, along the LoS between the centre of the sphere and the observer is given by
\begin{equation}
	A_{\mathrm{V}}~=~\int_{R_{\mathrm{in}}}^{R_\mathrm{out}} dr \, \rho(r) \, \kappa_{\mathrm{V}}~=~\rho \cdot \kappa_{\mathrm{V}} \cdot \left( R_\mathrm{out} - R_{\mathrm{in}} \right) 
    \label{equ:viewing_mc_def_av}
\end{equation}
where $\kappa_{\mathrm{V}}$ is the total dust opacity in the V band ($\lambda_{\mathrm{V}}$~=~0.55 $\mu$m). 
Table~\ref{tab:viewing_mc_bench_setup} summarises the values I use for the benchmark spheres.

\begin{table}
	\centering
	\begin{tabular}{l|c|c}
		Quantity & Symbol & Value \\ \hline
		inner radius & $R_{\mathrm{in}}$ & 1.0 $\times$ 10$^{14}$ cm \\
		outer radius & $R_{\mathrm{out}}$ & 3.4 $\times$ 10$^{15}$ cm \\
		edge length of major grids & $\Delta GW$ & 1.0 $\times$ 10$^{14}$ cm \\
		major grids in x/y/z direction & $n_{\mathrm{x,y,z}}$ & 35 \\
		total luminosity & $L_{\mathrm{tot}}$ & 1.0 L$\odot$ \\
		black body temperature & $T_{\mathrm{BB}}$ & 2,500 K \\
		dilution factor & $w$ & 5.9 $\times$ 10$^{-9}$
	\end{tabular}
    
	\caption[Summary of parameters used for benchmark models]{Summary of parameters used for benchmark models. I use the given numbers for defining the spherical volumes as well as the radiation field for both codes. The radiation field is given by a diluted black body.
	}
	\label{tab:viewing_mc_bench_setup}
\end{table}

\texttt{HII}, on the contrary, does not automatically create a equidistant grid as \texttt{MC}, but generates it such that the difference in optical depth in V band, $\tau_{\mathrm{V}}$, between two neighbouring grid points is smaller than 0.1.
Thus, the radial grid depends on the density distribution.
Since the density is constant in my models the grid is equidistant and directly comparable to the results by \texttt{MC}.

In both codes, I fill the spheres with dust which is a mixture of amorphous carbon (aC) and silicates (Si).
Respectively, the grain sizes range from 16 to 128 nm and 32 to 256 nm.
The grains of both species are spherical and follow the size distribution by \citet{MRN1977} with number density $n \, \propto \, a^{-3.5} da$.
\citet{Siebenmorgen2014} provide the cross-sections and opacities. 
The abundances are chosen to be $\chi_{\mathrm{aC}}$~=~2.0~$\times$~10$^{-4}$ and $\chi_{\mathrm{Si}}$~=~3.1~$\times$~10$^{-5}$ relative to H$_2$, respectively.

Both codes heat the dust with an external, isotropic radiation field.
For simplicity, I use a black body that is diluted by the factor $w$.
\pagebreak

\noindent My sample consists of six test cases (see Table~\ref{tab:viewing_mc_bench_av} for more details):
\begin{itemize}
	\item an empty sphere ($A_{\mathrm{V}}$~=~0 mag) \\
		this model verifies that the propagation of photon packages works correctly by reproducing the input radiation field;
	\item three (very) optically thin models ($A_{\mathrm{V}}$~=~0.01, 0.1, 1 mag); and
	\item two models with optically thick dust ($A_{\mathrm{V}}$~=~10, 100 mag)
\end{itemize}

\begin{table}
	\centering
	\begin{tabular}{c|c|c}
		Total extinction & Dust density & Dust mass \\ \hline
		$A_{\mathrm{V}}$ [mag] & $\rho$ [g cm$^{-3}$] & $M_{\mathrm{d}}$ [M$_\odot$] \\ \hline
		0  & 0.0 & 0.0 \\
		$10^{-2}$ & $10^{-22}$ & 8.53 $\times$ 10$^{-9}$ \\
		$10^{-1}$ & $10^{-21}$ & 8.53 $\times$ 10$^{-8}$ \\
		$10^{0}$  & $10^{-20}$ & 8.53 $\times$ 10$^{-7}$ \\
		$10^{1}$  & $10^{-19}$ & 8.53 $\times$ 10$^{-6}$ \\
		$10^{2}$  & $10^{-18}$ & 8.53 $\times$ 10$^{-5}$ \\
	\end{tabular}
	\caption[Summary of parameters used for setting-up the benchmark]{Summary of parameters used for setting-up the benchmark.
		The table lists the total extinction, $A_{\mathrm{V}}$, of the dust along the line-of-sight through the spheres, the dust density, $\rho$, and the dust mass, $M_{\mathrm{d}}$.
	}
	\label{tab:viewing_mc_bench_av}
\end{table}

\noindent Fig.~\ref{pic:viewing_mc_bench_av} illustrates the results of the benchmark runs.
The top panel shows the normalised SEDs as computed by \texttt{HII} (black solid line) and \texttt{MC} (red dotted line), whereby the latter uses the photon counting methods to do so.
That means that the code counts the number of photon packages per frequency bin when those packages are leaving the data cube.
Since the photon packages in \texttt{MC} have the same energy by definition the total number of photon packages within each frequency bin is equivalent to the energy within the same frequency bin (see below).
In order to reduce the Monte Carlo noise and ensure isotropy I use 10$^5$ photon packages for each of the 8192 initial frequency bins.

The residuals shown in the bottom panel of Fig.~\ref{pic:viewing_mc_bench_av} are within $\leq$ 5\% in relative errors, meaning that the SEDs of both codes are almost identical.
Only the model with $A_{\mathrm{V}}$~=~100 mag includes larger differences.
There are two effects responsible for this:
\begin{enumerate}[label=\alph*]
	\item \texttt{HII} and \texttt{MC} use different frequency grids. Thus, when comparing the results I can only use grid points that are at least very close to each other. That automatically introduces a certain level of error due to the steep gradient of the SED.
	\item The spatial grids of both codes differ from each other, as well. If the dust within the cells is optically thick, the grid may not sufficiently resolve regions with high temperature gradients. If the resolution is too low, the variations are averaged which leads to profiles that are shallower than they should be. \texttt{HII} automatically takes that into account when setting-up its spatial grid, but not \texttt{MC}. This problem can be fixed by, for example, using finer grids with more and smaller cells. This would increase the spatial resolution significantly, but also the computational effort. The user needs to find a compromise between the both factors depending on the structures that shall be studied.
\end{enumerate}

\begin{mysidewaysfigure}
	\includegraphics[width=\textwidth]{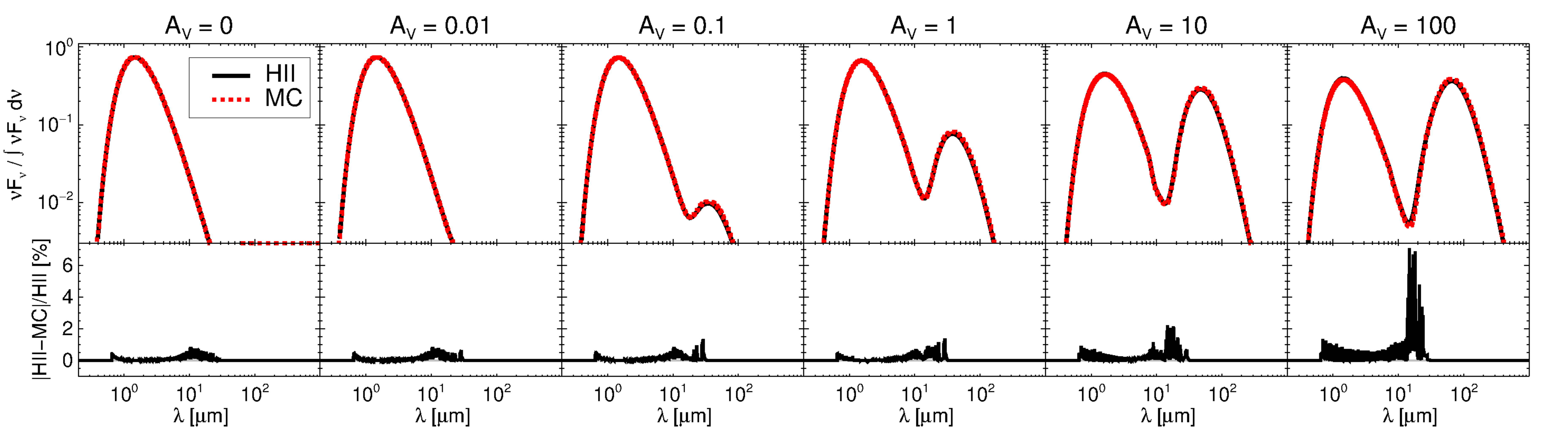}
	\caption[Results of the benchmark tests]{Results of the benchmark tests. 
	\textit{Top:} SEDs of dust spheres that are heated externally by a diluted black body for various total extinctions (as labelled).
	The SEDs are normalised to the total energy.
	Results are shown for \texttt{HII} (black solid) and \texttt{MC} (red dashed).
	\textit{Bottom:} Residuals are given as absolute relative error of \texttt{MC} compared to HII.
	one sees that the results derived by both methods agree well with each other.}
	\label{pic:viewing_mc_bench_av}
\end{mysidewaysfigure}

\pagebreak

\subsubsection{Flux Errors}\label{viewing:mc_fluxerror}

There are two ways how \texttt{MC} computes SEDs.
The first way is by ray-tracing, the second by photon counting.
In the latter case, the fluxes within each solid angle $\Omega$ are computed by counting the photon packages that leave the data cube within that angle. 
$\Omega$ is here defined as the area between two azimuthal angles $\theta_{\mathrm{j}}$ and $\theta_{\mathrm{j+1}}$ for all inclinations. 
The flux is then given by
\begin{equation}
	F^\mathrm{i} \left( k, \theta_{\mathrm{j}} \right) \,~=~\, \frac{c_{\mathrm{c3}} \times \mbox{10$^{23}$}}{4 \pi d^2} \, \frac{n_\mathrm{ph}^\mathrm{i}(k,\theta_{\mathrm{j}}) \, \epsilon_\mathrm{pak}}{\Delta \nu(k)} \, \mathrm{Jy} ,
    \label{equ:viewing_mc_def_flux}
\end{equation}
where $i$ distinguishes the emission from the dust ($i$~=~d) and the original heating field ($i$~=~s).
Thus, $F^\mathrm{i}$ is the flux of the $i\mathrm{th}$ emission contributor and $n_\mathrm{ph}^\mathrm{i}$ the number of photon packages within the current azimuthal angle $\theta_{\mathrm{j}}$ and current frequency bin $\Delta \nu(k)$.
$d$ represents the distance between the object and the observer.
The factor 10$^{23}$ converts the flux values from erg~cm$^{-2}$~s$^{-1}$~Hz$^{-1}$ into Janskys. 

Since the distance and energy per photon package are set by the user, the only quantity in Eq.~(\ref{equ:viewing_mc_def_flux}) that contains errors is $n_\mathrm{ph}^\mathrm{i}$. 
Following Poisson statistics, the errors of counted photons, $\Delta n_\mathrm{ph}^\mathrm{i}$, is given by
\begin{equation}
	\Delta n_\mathrm{ph}^\mathrm{i} \,~=~\, \sqrt{n_\mathrm{ph}^\mathrm{i}} .
    \label{equ:viewing_mc_photerr}
\end{equation}
The upper and lower errors of the final fluxes of dust and stellar contributor, that is a first order interpolation of Eq.~(\ref{equ:viewing_mc_def_flux}), and the total flux are computed by adding or subtracting the the corresponding energy according Eq.~(\ref{equ:viewing_mc_def_flux}) to the emission terms of each component. 
The code applies the same computations as before to the limit terms and returns the differences to the measured flux value in the output file.

%%%%%%%%%%%%%%%%%%%%%%%%%%%%%%%%%%%%%%%%%%%%%%%%%%%%%%%%%%%%%%%%%%

\section{Cylindrical Models}\label{viewing:cyl}

In this section, I investigate the question whether filaments show specific signatures in FIR and sub-mm observations with which one can derive their inclination.
For this, I analyse three analytic models of isothermal cylinders and explore how the mean flux density varies as a function of the viewing angle.
Since the dust is isothermal and optically thin at FIR and sub-mm	 wavelengths (see Table~\ref{tab:viewing_cyl_details}), the mean column density is expected to vary similarly with the mean flux density.
These simplified models provide insight into the behaviour of the mean column density based on the inclination alone, before I consider more complex models in Sect.~\ref{viewing:fila}.

Doing so, one needs to keep in mind that the flux density can only be observed relative to the fore- and background emission.
Furthermore, it is influenced by other parameters, like dust properties, observational noise or, essentially, by the way observational data are treated (see Sect.~\ref{viewing:g11}).
This is the reason why I focus on variations in mean flux densities relative to the initial LoS only in this section instead of analysing absolute values.
\pagebreak

\subsection{Model description}\label{viewing:cyl_models}

I consider three cylindrical models \citep[analogues to those models examined by][and others]{Ostriker1964b}:
\begin{enumerate}[label=(\alph*)]
	\item a homogeneous, constant density cylinder (\textit{homcyl}) that represents a coreless filament; 
	\item a cylinder with a power-law radial density profile (\textit{cyl}) that reflects a more realistic filament without a core; and 
	\item a cylinder that includes a sphere with both having a power-law radial density profile (\textit{cylsph}) that mimics a fragmented filament.
\end{enumerate}
%\pagebreak

\noindent The radial density profiles are based on Plummer functions \citep{Plummer1911,Nutter2008} and are given by
\begin{equation}
	\rho(r)~=~\underbrace{\frac{\rho_{\mathrm{c}}}{\left[ 1 + \left( \frac{r}{R_\mathrm{flat}} \right)^2 \right]^{\frac{p_{\mathrm{c}}}{2}}}}_\mathrm{cylinder~component} +  \underbrace{\frac{\rho_{\mathrm{s}}}{\left[ 1 + \left( \frac{\vec{r}-\vec{r}_0}{R_\mathrm{flat}} \right)^2 \right]^{\frac{p_{\mathrm{s}}}{2}}}}_\mathrm{sphere~component} \, ,
	\label{equ:viewing_cyl_def_plummer}
\end{equation}
where $\rho_{\mathrm{c}}$ and $\rho_{\mathrm{s}}$ are the central gas densities and $p_{\mathrm{c}}$ and $p_{\mathrm{s}}$ the power-law exponents of the density profile of the cylinder and sphere component, respectively, and of $R_\mathrm{flat}$ defines the region where the density in the inner part of the cylinder is relatively constant. 
For the calculations, I assume a constant gas-to-dust mass ratio $R$~=~100 and compute the dust density $\rho_\mathrm{i}^\mathrm{dust}$~=~$\rho_\mathrm{i} / R$.
The quantity $\vec{r}_0$ represents the position vector of the central cell of the sphere.
Note that the sphere component only contributes within its radius $r_{\mathrm{s}} \geq |\vec{r}-\vec{r}_0|$.

I generate model cylinders with $n_{\mathrm{x}}$, $n_{\mathrm{y}}$, and $n_{\mathrm{z}}$ grid elements, each having the edge length $l_\mathrm{cell}$, in $x$-, $y$-, and $z$-direction, respectively.
Table~\ref{tab:viewing_cyl_details} provides the parameters I use for the calculations.
These parameters are consistent with those \citet{Arzoumanian2011} find on average for nearby filaments.

\begin{table}
	\centering
	\begin{tabular}{l|c|c|c}
          Parameter	& Hom.\ cylinder	& Cylinder		& Cylinder \& sphere \\ \hline
			Model ID 					& homcyl		& cyl 		& cylsph \\ \hline
			$n_{\mathrm{x}}$ / $n_{\mathrm{y}}$ / $n_{\mathrm{z}}$		& 200 / 100 / 100 	& 500 / 50 / 50	& 500 / 50 / 50 \\
			$l_{\mathrm{cell}}$ [pc]						& 0.05		& 0.05		& 0.05 \\
			$R_{\mathrm{flat}}$ [pc]						& 0.05		& 0.03 		& 0.03 \\
			$A_{\mathrm{V}}^{\mathrm{mean}}$ [mag]			& 1948		& 10		& 12 \\
			$A_{\mathrm{V}}^{\mathrm{max}}$ [mag]			& 2432		& 64		& 2296 \\
			$A_{\mathrm{250 \mu m}}^{\mathrm{mean}}$ [mag]	& 0.840		& 0.004		& 0.005 \\
			$A_{\mathrm{250 \mu m}}^{\mathrm{max}}$ [mag]	& 1.05		& 0.03		& 0.99 \\ \hline
			\textit{cylinder}								& 			&	 	& \\ \hline
			$l_c$ [pc]										& 10 			& 25		& 25 \\
			$r_c$ [pc]										& 2.5 			& 1.25		& 1.25 \\
			$\rho_c$ [g cm$^{-3}$]							& 10$^{-20}$ 		& 10$^{-20}$	& 10$^{-20}$ \\
			$p_c$ 											& 0.0 			& 1.6		& 1.6 \\ \hline
			\textit{sphere}									& 			& 		& \\ \hline
			$\vec{r}_0$ [grid units]						& ---			& ---		& (0.15 $n_{\mathrm{x}}$ , $\frac{n_{\mathrm{y}}}{2}$ , $\frac{n_{\mathrm{z}}}{2}$) \\
			$r_s$ [pc]										& --- 			& --- 		& 0.2 \\
			$\rho_s$ [g cm$^{-3}$]							& --- 			& --- 		& 10$^{-19}$ \\
			$p_s$											& --- 			& --- 		& 2.0 \\
	\end{tabular}
    \caption[Input parameters used for cylinders]{Input parameters used for cylinders.
    	Note that the numbers for mean and maximum extinction at 0.55$\mu$m, $A_{\mathrm{V}}$, and 250$\mu$m, $A_{\mathrm{250 \mu m}}$ are valid for ($\theta$,$\varphi$) = (0$^\circ$,0$^\circ$) and may vary for other sight lines.
		}
	\label{tab:viewing_cyl_details}
\end{table}

The dust within the cylinders is isothermal, but for different runs I vary the dust temperature, $T_\mathrm{iso}$, between 10~K and 20~K.
I rotate the models by 0$^\circ$, 45$^\circ$, and 90$^\circ$ in inclination, $\theta$, and position angle, $\varphi$, to look at the structures from different LoSs.

At each LoS and for each temperature, I produce flux density maps at $\lambda$~=~250~$\mu$m analytically using a modified black body function, given by:
\begin{equation}
	F_{\lambda}~=~B_{\lambda}(T_\mathrm{d}) \cdot \left( 1 - e^{-\kappa_{\lambda} \, \mu_{H_2} \, m_p  \, N_\mathrm{tot}} \right) \, ,
	\label{equ:viewing_cyl_def_mbnu}
\end{equation}
where $B_{\lambda}$ is the Planck function at the wavelength 
$\lambda$, $T_\mathrm{d}$ is the dust temperature, $\kappa_{\lambda}$ the dust opacity at $\lambda$, $\mu_{H_2}$~=~2.33 the mean molecular weight per hydrogen molecule, $m_p$ the proton mass, and $N_\mathrm{tot}$ the total column density along the LoS. 

Note that I choose values for $R_\mathrm{flat}$ that are equal or smaller than $l_\mathrm{cell}$.
This means that the flattened region of the cylinders cannot be resolved.
I have tested how sensitive the results are to the resolution by varying the ratio of $R_\mathrm{flat}$ / $l_\mathrm{cell}$ and found no confusion.

\subsection{Flux density spectral energy distribution}\label{viewing:cyl_analysis}

\begin{figure}
	\centering
    \includegraphics[width=\textwidth]{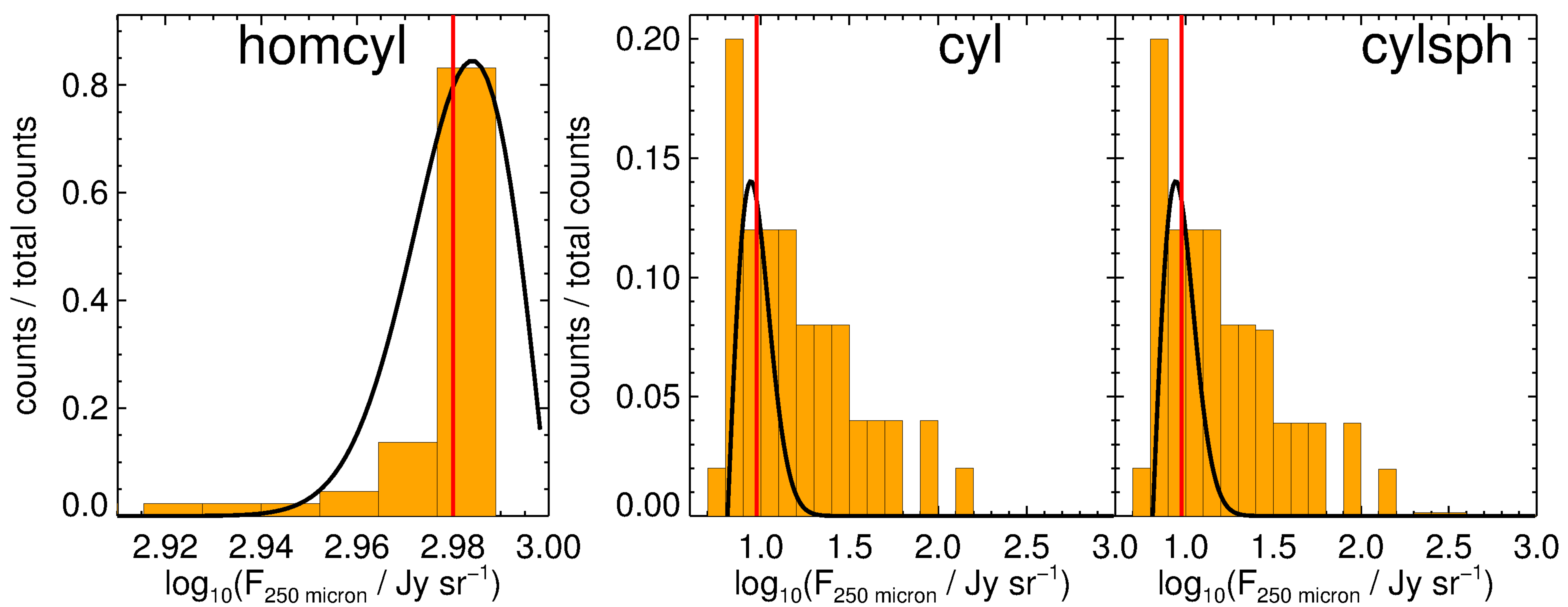}
    \caption[250~$\mu$m flux density PDFs]{Probability density functions (PDFs) showing the 250~$\mu$m flux density, $F_\mathrm{250 \mu m}$. 
    	From left to right, the plots show the PDFs of the \textit{homcyl}, \textit{cyl}, and \textit{cylsph} models, all at ($\theta$,$\varphi$)~=~(0$^\circ$,0$^\circ$) and $T_\mathrm{iso}$~=~10~K. 
		The black lines show the Rayleigh distributions fitted to the PDFs.
		The fitted mean values are marked with red lines.}
    \label{pic:viewing_cylinders_pdf}
    \vspace*{-\baselineskip}
\end{figure}

Fig.~\ref{pic:viewing_cylinders_pdf} shows the flux density probability density functions (PDFs) of \textit{homcyl}, \textit{cyl}, and \textit{cylsph} at $T_\mathrm{iso}$~=~10~K and ($\theta$,$\varphi$)~=~(0$^\circ$,0$^\circ$). 
In the case of \textit{homcyl}, $\sim$90\% of all pixels have values close to the peak flux density (see Fig.~\ref{pic:viewing_cylinders_pdf} \textit{left}) due the homogeneous density distribution.
In \textit{cyl} and \textit{cylsph}, only a small fraction of the models contain dust at higher density owing to a steep density gradient. 
This means that the flux density PDFs are dominated by the low-density regime (see Fig.~\ref{pic:viewing_cylinders_pdf} \textit{middle} and \textit{right}).
There is a small contribution (within 158--631~Jy~sr$^{-1}$) by the sphere within \textit{cylsph}, but due to its small size compared to the total volume this contribution is not significant.

\begin{figure}
	\centering
    \begin{subfigure}[c]{\textwidth}
    	\includegraphics[width=\textwidth]{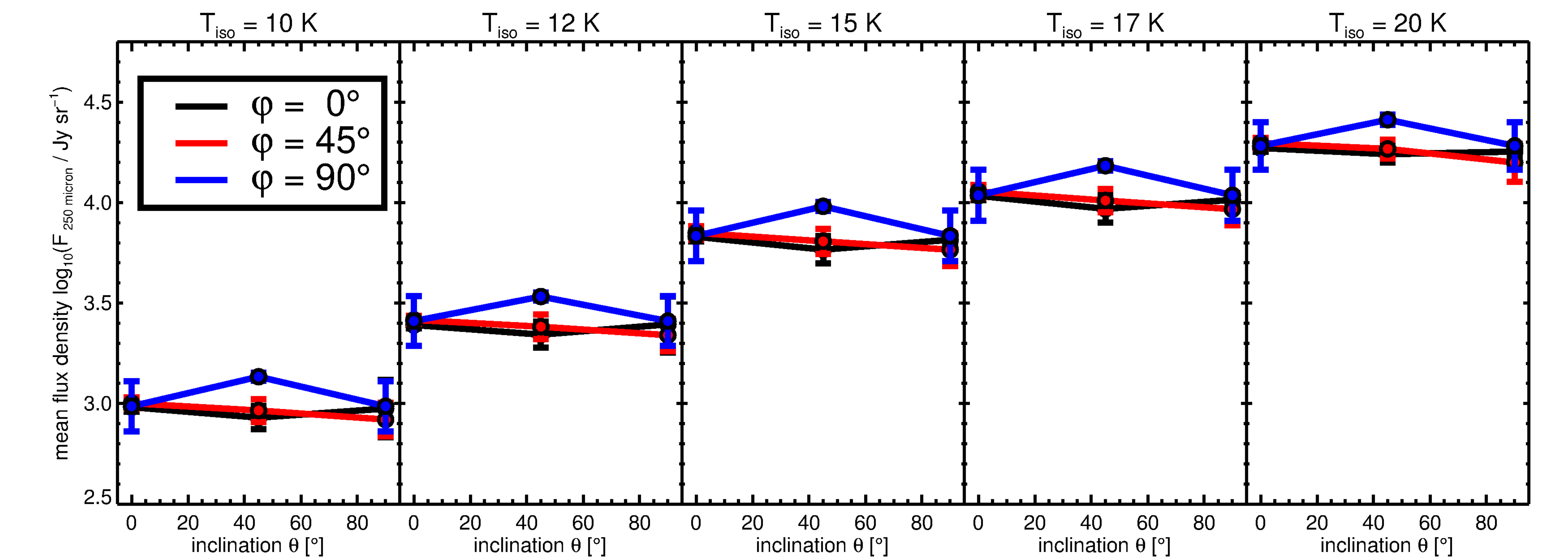}
    	\caption{homcyl}
    	\label{pic:viewing_ana_stat_homcyl}
	\end{subfigure}
		
	\begin{subfigure}[c]{\textwidth}
    	\includegraphics[width=\textwidth]{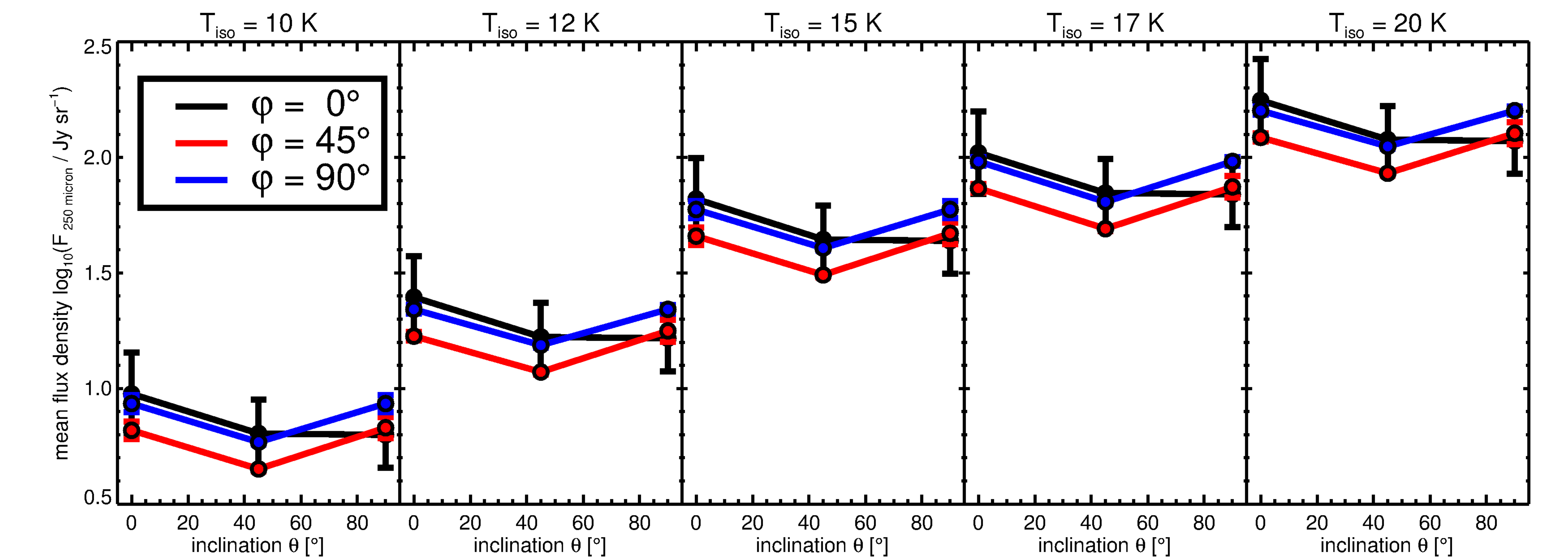}
        \caption{cyl}
        \label{pic:viewing_ana_stat_cyl}
	\end{subfigure}
		
	\begin{subfigure}[c]{\textwidth}
    	\includegraphics[width=\textwidth]{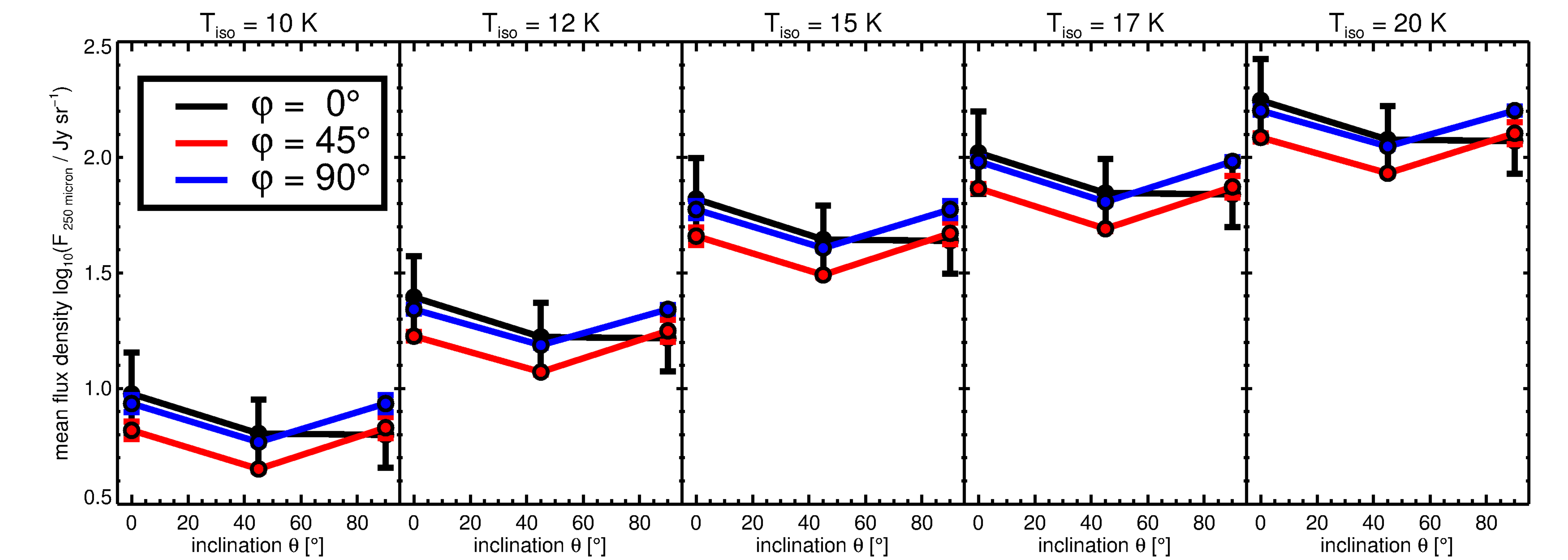}
        \caption{cylsph}
        \label{pic:viewing_ana_stat_cylsph}
	\end{subfigure}
		
	\caption[Mean flux density as function of temperature, inclination, and position angle]{Mean flux density, $\log_{\mathrm{10}}$($\xbar{F}_\mathrm{250 \mu m}$), of \textit{homcyl}, \textit{cyl}, and \textit{cylsph} as function of temperature, inclination, and position angle. The error bars illustrate the errors of the mean value. I present the results for $\theta$ and $\varphi$ at 0$^\circ$, 45$^\circ$ and 90$^\circ$. The lines are interpolations that represent well the true values.
		}
		\label{pic:viewing_ana_stat}
\end{figure}

To quantify the flux density PDFs better, I fit them with Rayleigh distributions, given by
\begin{equation}
	f(x)~=~A \cdot \frac{x - x_\mathrm{0}}{\sigma^2} \cdot e^{- \frac{\left(x - x_\mathrm{0}\right)^2}{2 \sigma^2}},
	\label{equ:viewing_cyl_def_rayleigh}
\end{equation}
where A is a scaling factor for the amplitude, $x_\mathrm{0}$ is a parameter shifting the distribution along x and $\sigma$ is the standard deviation.
The mean value $\xbar{x}$, which is equivalent to the position of the distribution peak, is given by $x_\mathrm{0} + \mathrm{sgn}(A) \sqrt{\frac{\pi}{2}} \, \sigma$.
In Fig.~\ref{pic:viewing_ana_stat} I show the mean values of the flux density PDFs, $\xbar{F}_\mathrm{250 \mu m}$, as function of dust temperature, inclination, and position angle.

Note that from now on I discuss only the behaviour of these mean values.
The mean value is the most favourable statistical parameter to compare with observational studies since it is commonly used, reflects the overall characteristics of the object, and is less affected by resolution effects.
Alternatively, one can use the maximal flux density of the distribution or its standard deviation.
The advantage of looking for the areas of maximal flux density is that it is the most straightforward way to identify the densest and/or warmest regions within a filament.
I.e. the maximal flux density would be the best parameter in differentiating between \textit{cyl} and \textit{cylsph}.
However, the maximal flux density is not able to trace the over-all properties of a given object.

The standard deviation, $\sigma$, on the contrary, is more promising.
I find that there are variations in $\sigma$ for LoSs along $\varphi$~=~0$^\circ$ and $\varphi >$~0$^\circ$ (meaning both $\varphi$~=~45$^\circ$ and 90$^\circ$).
These variations are less revealing since they are independent of the dust temperature.
Thus, compared to the mean value, one would loss information about the evolutionary stage of the filament.
Furthermore, the variations are not uniform.
While $\sigma$ increases from 0.01~Jy~sr$^{-1}$ along $\varphi$~=~0$^\circ$ to about 0.13~Jy~sr$^{-1}$ along $\varphi >$ 0$^\circ$ for \textit{homcyl}, it decreases for \textit{cyl} and \textit{cylsph} from 0.13~Jy~sr$^{-1}$ along $\varphi$~=~0$^\circ$ to about 0.05~Jy~sr$^{-1}$ along $\varphi >$ 0$^\circ$.
This means that the changes are indeed significant in numbers, but they also depend on how the matter is distributed.
Additionally, a sampling of $\sigma$ requires the filament to be well-resolved, which might become a problem the fainter and more distant the filament is located from us.

For the test cylinders, one sees that the mean flux increases with increasing temperature.
For a given temperature, the average variations are on the order of 0.2~dex, which corresponds to a factor of about 1.58.
This is in agreement with \citet{Arzoumanian2011} who predicted that their observed column densities are on average overestimated by a factor of $<\cos(\theta)^{-1}> \,~=~\pi/2 \approx 1.57$ due to the unknown inclinations.

In \textit{homcyl} (Fig.~\ref{pic:viewing_ana_stat_homcyl}), one sees that the mean flux density is mostly constant, with small variations within 0.2~dex.
In the case of \textit{cyl} and \textit{cylsph} (Figs.~\ref{pic:viewing_ana_stat_cyl} and \ref{pic:viewing_ana_stat_cylsph}), the variations are slightly larger, but within 0.3~dex (corresponding to a factor of 2.0).
In general, the mean flux density decreases with increasing inclination.
This is because the area within the flux density is amplified, due to larger amounts of dust along the LoS, becomes smaller and statistically less significant.
Thus, the PDFs are even more dominated by the more diffuse dust, decreasing the value of $\xbar{F}_\mathrm{250 \mu m}$.

When comparing the results of \textit{cyl} and \textit{cylsph}, one sees no contribution of the sphere to $\xbar{F}_\mathrm{250 \mu m}$. 
This means that the sphere is too small compared to the whole cylinder and does not lead to an increase of the mean flux density.
I have performed additional tests and found that increasing the density by several orders of magnitude of the sphere alone does not have any impact.
When I insert more spheres such that the spheres contain at least 10\% of the total mass of the models, one sees notable fluctuations in the mean flux density.
This means that, as long as a core is small in size and mass compared to the filament, it can only be studied directly by removing the filamentary background. 
For example, \citet{Montier2010} filtered the local background from \textit{Planck}-HFI maps to reduce the influence of large-scale structures on the source detection and build up their \textit{Planck} catalogue of cold cores.
This is comparable to using the maximum flux density to identify objects of interest instead of the mean flux density.

In summary, I predict little to no variations in the mean flux density PDFs for elongated cylindrical structures based on their geometry.
They are not significant enough to allow conclusions on viewing angles based on dust observations alone.
I do not see significant changes in the PDF distributions when inserting a core-like sphere into the cylinder. 
This implies that as long as a single core does not contain a significant fraction of the mass compared to its surrounding filament it does not influence the average properties of the whole filament.

\section{Observation motivated models}\label{viewing:fila}

In this section, I study the LoS effects on models that are more complex in both their density and temperature structures and better represent real filaments.
I discuss two main questions:
Firstly, how would one observe filamentary structures at different viewing angles?
Secondly, are there observational criteria, such as the dust temperature and total column density, which enable us to reconstruct elongated structures along the LoS by dust observations alone?

\begin{figure}
	\begin{subfigure}[l]{0.485\textwidth}
		\includegraphics[width=\textwidth]{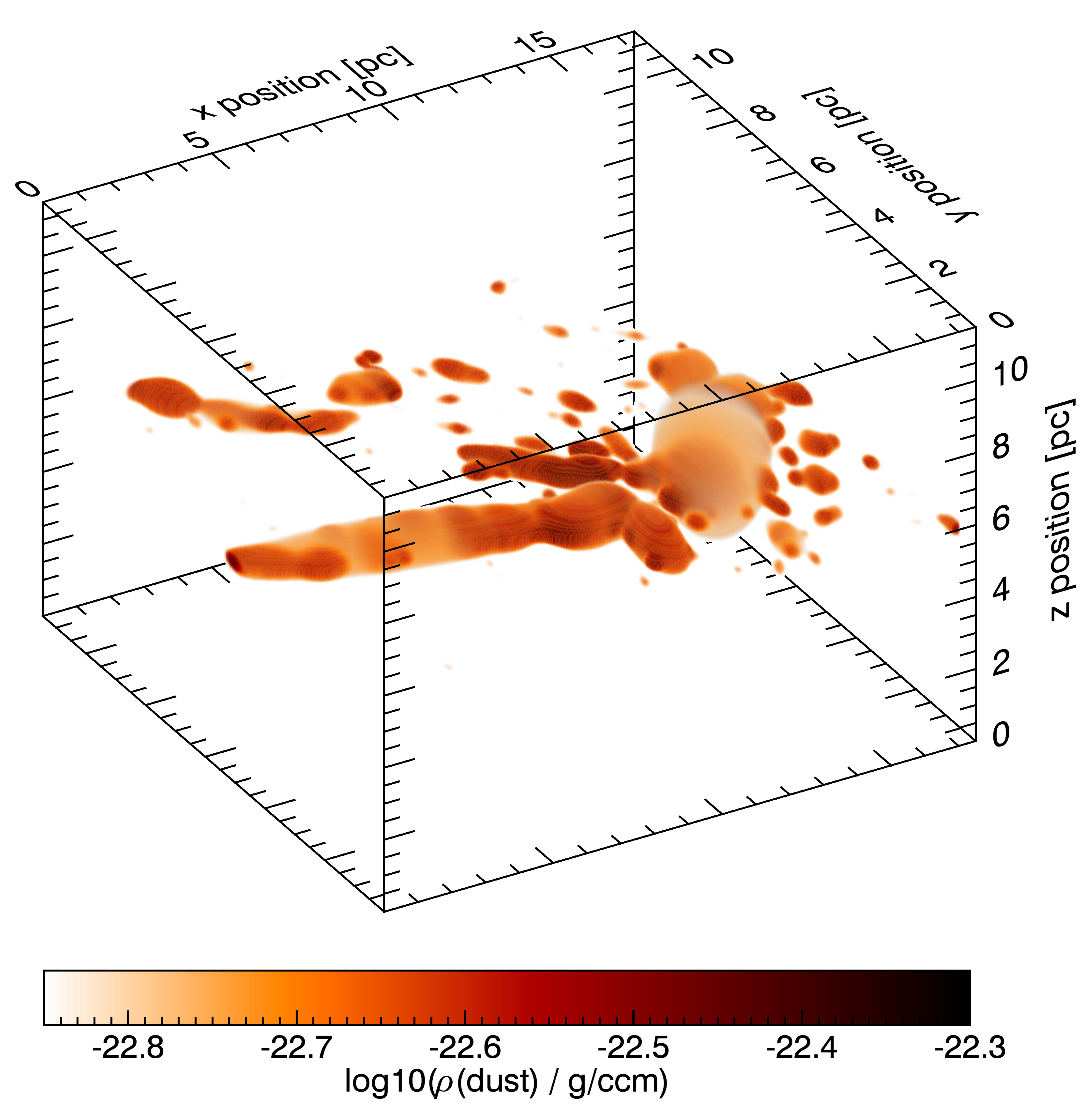}
		\subcaption{$\rho$~Ophiuchi cloud. }
		\label{pic:viewing_fila_appli_rhooph}
	\end{subfigure} \hfill
	\begin{subfigure}[r]{0.485\textwidth}
		\includegraphics[width=\textwidth]{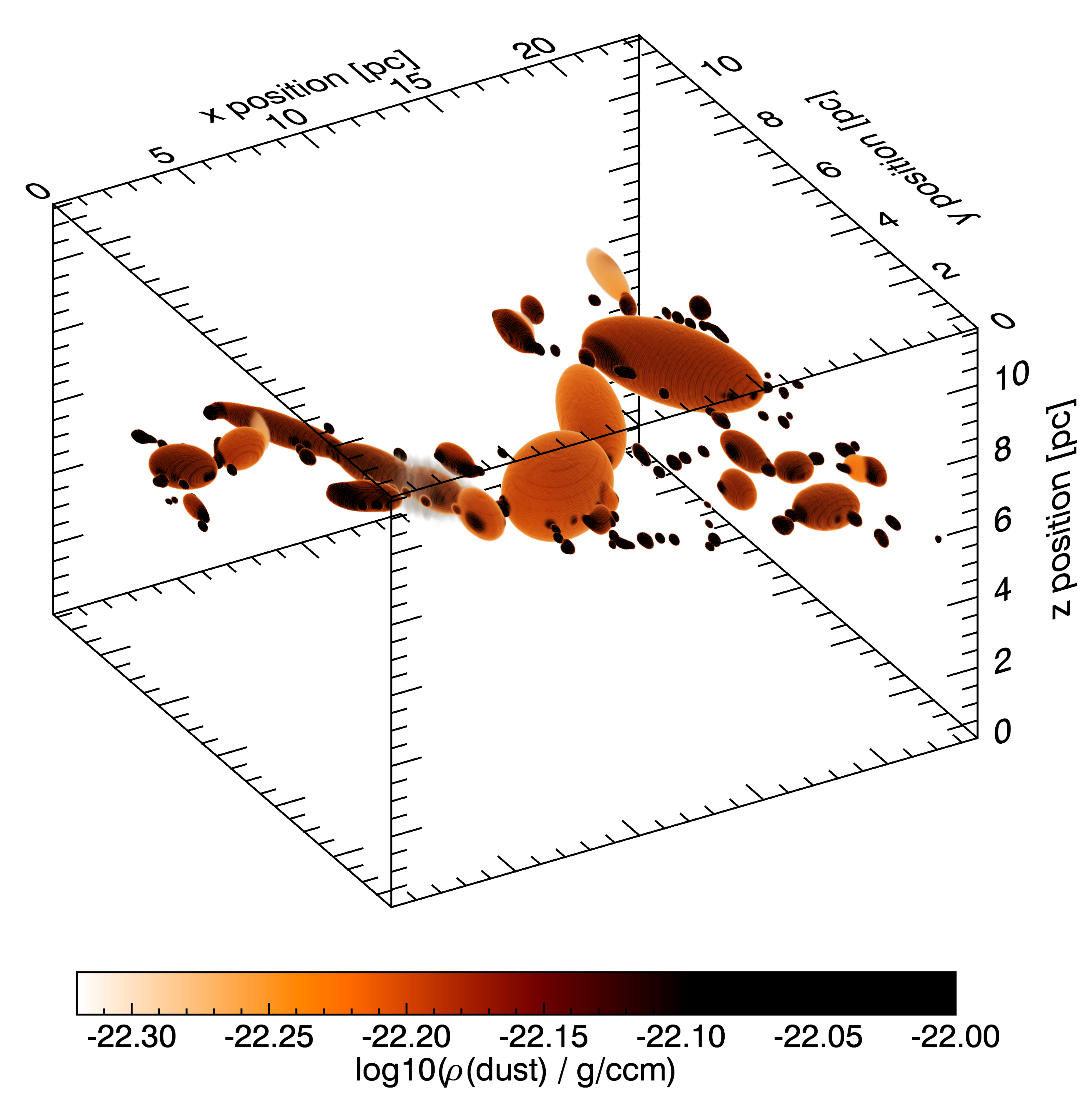}
        \subcaption{G11.11 filament, alias \textit{Snake}. }
        \label{pic:viewing_fila_appli_g11}
	\end{subfigure}
    
	\caption[3D images of the dust density distribution of the $\rho$~Ophiuchi and G11.11 models]{3D images of the dust density distribution of the $\rho$~Ophiuchi \citep[based on][]{Kainulainen2009} and G11.11 models \citep[based on][]{Kainulainen2013b}. 
      The models consist of a number of small, dense clumps that are surrounded by diffuse envelopes. 
      Note that I only plot the most dense parts of the models and not the more diffuse envelopes since those would obscure the central clumps and cores.
      The real range of total column densities and dust temperatures are summarised in Table~\ref{tab:viewing_fila_models_derived}.
	}
	\label{pic:viewing_fila_appli}
\end{figure}

Therefore, I use \texttt{MC} to derive the dust temperature and emission in the FIR and sub-mm self-consistently (contrary to the isothermal models in Sect.~\ref{viewing:cyl}).
The two 3D molecular cloud models are constructed by fitting superpositions of spheroids on the observed column density maps of those clouds.
Fig.~\ref{pic:viewing_fila_appli} shows 3D images of the models, one based on the $\rho$~Ophiuchi cloud \citep[Fig.~\ref{pic:viewing_fila_appli_rhooph},][]{Kainulainen2009,Kainulainen2014} and on the G11.11 \textit{Snake} filament \citep[Fig.~\ref{pic:viewing_fila_appli_g11},][]{Kainulainen2013b,Kainulainen2014}.
In both models, the x-y-plane corresponds to the plane-of-the-sky while the z-axis represents the LoS.
Table~\ref{tab:viewing_fila_models_details} lists the basic parameters I used for the simulations and Table~\ref{tab:viewing_fila_models_derived} provides a summary of the volume and column density, dust temperature, and mean flux density statistics of both models.

\begin{table}
	\renewcommand{\arraystretch}{1.5}
	\centering
	\begin{tabular}{l|c|c}
		Parameter & $\rho$ Ophiuchi & G11.11 \textit{Snake} \\ \hline
		Cube size ($n_\mathrm{x}, n_{\mathrm{y}}, n_{\mathrm{z}}$)	& (370, 242, 243)	& (1181, 581, 581) \\
		Edge length $l_\mathrm{cell}$ [pc]				& 0.05	& 0.02 \\
		ISRF strength $w$ 								& 1.0	& 1.0 \\
		Total gas mass [$M_\odot$]						& 8100	& 49745 \\
		Maximal $A_\mathrm{V}^\mathrm{z-axis}$ [mag]	& 40	& 70
	\end{tabular}
	\caption[Input parameters used for filamentary models]{Input parameters used for filamentary models.}
	\label{tab:viewing_fila_models_details}
	\renewcommand{\arraystretch}{0.66}
\end{table}

\begin{table}[h!t]
	\renewcommand{\arraystretch}{1.3}
	\begin{center}
	\begin{tabular}{l|r|r|r|r}
		Quantity & Mean & Minimum & Median & Maximum \\ \hline
		\multicolumn{5}{l}{$\rho$~Ophiuchi} \\ \hline
		$\rho_\mathrm{gas}$ [10$^{-22}$ g cm$^{-3}$]	& 2.7	& 0.05	& 1.4	& 835.7 \\
		$N_\mathrm{tot}^\mathrm{real}$ [10$^{21}$ cm$^{-2}$]	& 5.2	& 0.001	& 3.4	& 220.0 \\
		$N_\mathrm{tot}^\mathrm{eff}$ [10$^{21}$ cm$^{-2}$]	& 13.0	& 0.001	& 6.0	& 610.0 \\
		$T_\mathrm{d}^\mathrm{eff}$ [K]			& 12.9	& 9.7	& 12.6	& 20.0 \\
		$\xbar{F}_\mathrm{250 \mu m}$ [Jy sr$^{-1}$] 	& 12.3	& 6.7	& 13.5	& 19.9 \\ \hline
		\multicolumn{5}{l}{G11.11 \textit{Snake}} \\ \hline
		$\rho_\mathrm{gas}$ [10$^{-22}$ g cm$^{-3}$]	& 11.0	& 0.07	& 5.3	& 1858.5 \\
		$N_\mathrm{tot}^\mathrm{real}$ [10$^{21}$ cm$^{-2}$] 	& 20.0	& 0.001	& 14.0	& 690.0 \\
		$N_\mathrm{tot}^\mathrm{eff}$ [10$^{21}$ cm$^{-2}$] 	& 6.9	& 0.001	& 3.3	& 500.0 \\
		$T_\mathrm{d}^\mathrm{eff}$ [K] 			& 13.1	& 8.6	& 12.6	& 23.9 \\
		$\xbar{F}_\mathrm{250 \mu m}$ [Jy sr$^{-1}$] 	& 6.6	& 2.4	& 7.6	& 10.9 
	\end{tabular}
	\renewcommand{\arraystretch}{1.0/0.75}
	\end{center}
	\caption[Statistical summary of derived parameters based on the $\rho$~Ophiuchi and the \textit{Snake} models]{Statistical summary of derived parameters based on the $\rho$~Ophiuchi and the \textit{Snake} models.
        Shown are the input total column density, $N_\mathrm{tot}^\mathrm{real}$, as derived directly from the input volume density models, the effective total column density, $N_\mathrm{tot}^\mathrm{eff}$, representing the column density and the effective dust temperature, $T_\mathrm{d}^\mathrm{eff}$, both derived from SED fitting, and the mean flux density at 250 $\mu$m based on our synthetic images.
      The given numbers are based on the values at all viewing angles.
      Note that I show the arithmetic mean values here, which do not necessarily need to match the mean values of the Rayleigh distributions.
      Note that I give values for the total density here. 
      The code, however, requires the dust density as input that I compute by assuming a gas-to-dust mass ratio of $R$ = 100.
}
	\label{tab:viewing_fila_models_derived}

\vspace*{-\baselineskip}
\end{table}

For the dust temperatures I assume that the dust is only heated by an external, isotropic, heating radiation field.
I use the ISRF model by \citet{Mathis1983} which consists of a combination of three diluted black-body spectra, with effective temperatures of 7500~K, 4000~K, and 3000~K, and an additional UV excess \citep{Mathis1983}. 
I produce images by ray-tracing \citep[see ][]{Heymann2012} at 881 wavelengths within a range from 0.43~\AA{} to 1.2~cm.
To reduce the Monte Carlo noise I use 5~$\times$~10$^4$ and 10$^6$ photon packages per frequency grid point for the $\rho$~Ophiuchi and \textit{Snake} model, respectively. 
Previous tests have demonstrated that this is sufficient to sample the temperature within all cells continuously, which is reflected by a noise-less spectrum.
\pagebreak

\noindent Using the ISRF as the heating source is a simplification.
The $\rho$~Ophiuchi cloud, e.g., is a site of active star formation \citep{Wilking2008} that heats the cloud internally, which leads to significant temperature gradients and a caveat of my model.

For the dust I use a mixture of amorphous carbon (aC) and silicate (Si) dust grains.
The grain sizes range between 16 to 128 nm and 32 to 256 nm for aC and Si grains, respectively, and follow the size distribution by \citet{MRN1977} with number density $n \, \propto \, a^{-3.5}$ d$a$.
The cross-sections and opacities are based on the dust models of \citet{Siebenmorgen2014}.
I choose spherical grains and abundances relative to H$_2$ of $\chi_{\mathrm{aC}}$~=~2.5~$\times$~10$^{-3}$ and $\chi_{\mathrm{Si}}$~=~4.8~$\times$~10$^{-3}$, respectively.
This ratio is consistent with typical values of dust in the ISM in the solar neighbourhood \citep{Dwek2005}.
\pagebreak

\noindent I am aware that the dust grain model is also a simplification.
Studies by, e.g., \citet{Ossenkopf1994}, \citet{Stepnik2003}, \citet{Steinacker2015}, and \citet{Lefevre2016} indicate that dust grains in dense parts of the ISM grow to larger and more irregularly shaped aggregates.
In the future, a similar study needs to be conduced with an extended parameter space to study the influence of grain properties in more details.
For now, I keep the simplified set-up and focus on analysing filaments at different viewing angles.

I consider 40 different directions by varying the inclination, $\theta$, between 0$^\circ$ and 315$^\circ$ and position angle, $\varphi$, between 0$^\circ$ and 180$^\circ$, both in steps of 45$^\circ$.
I focus on the analysis on the FIR and sub-mm part of the spectrum since this part is dominated by the thermal emission of dust.
To mimic \textit{Herschel} SPIRE observations (at 250, 350, and 500~$\mu$m) I smooth the flux density maps to the resolution of the SPIRE 500~$\mu$m data using Gaussian point-spread functions.
For each pixel of the synthetic intensity maps, I fit a modified black body to the flux values at $\lambda$~=~250, 350 and 500~$\mu$m, following the descriptions in \citet[][see also Eq.\ (\ref{equ:viewing_cyl_def_mbnu})]{Koenyves2010}.
I assume a mass absorption coefficient $\kappa_{\nu}$~=~0.1 cm$^2$~g$^{-1} \left( \frac{\nu}{\mbox{1000~GHz}} \right)^\beta$.
For the dust emissivity index, $\beta$, I use 1.75 \citep{Koenyves2010,Wang2015}.
I have performed the same analysis with $\beta$~=~2.0 and recover similar results within the typical observational uncertainties.

I use the \texttt{idl} routine \texttt{mpfitfun} \citep{Markwardt2009} to perform the fittings, with the effective dust temperature, $T_\mathrm{d}^\mathrm{eff}$, and total column density as free parameters, $N_\mathrm{tot}^\mathrm{eff}$.
Fig.~\ref{pic:viewing_rhooph_example} presents results based on the $\rho$~Ophiuchi model, including flux density, effective column density, and effective dust temperature maps at ($\theta$,$\varphi$)~=~(0$^\circ$,0$^\circ$) and (45$^\circ$,45$^\circ$).

I again emphasise that $T_\mathrm{d}^\mathrm{eff}$ and $N_\mathrm{tot}^\mathrm{eff}$ are not direct results of the RT simulations, but derived by SED fitting of three data points and assuming a power-law opacity relation ($\kappa \, \propto \, \nu^\beta$).
This method introduces some degree of degeneracy and uncertainties.
For this reason, both quantities are less accurate than the values resulting from the RT procedure.
I find a correlation between the real column density, which I computed directly from the input volume density distribution, and the effective column density derived by SED fitting.
This correlation is lost by some scatter owing to the inaccuracies of fitting an entire SED to three points with only one single temperature along the LoS.
However, I proceed with the derived $T_\mathrm{d}^\mathrm{eff}$ and $N_\mathrm{tot}^\mathrm{eff}$ since they are equivalent to dust temperatures and column densities obtained from observed fluxes.

\subsection{$\rho$~Ophiuchi model}\label{viewing:rhooph}

\begin{figure}
  \centering
  \includegraphics[width=0.98\textwidth]{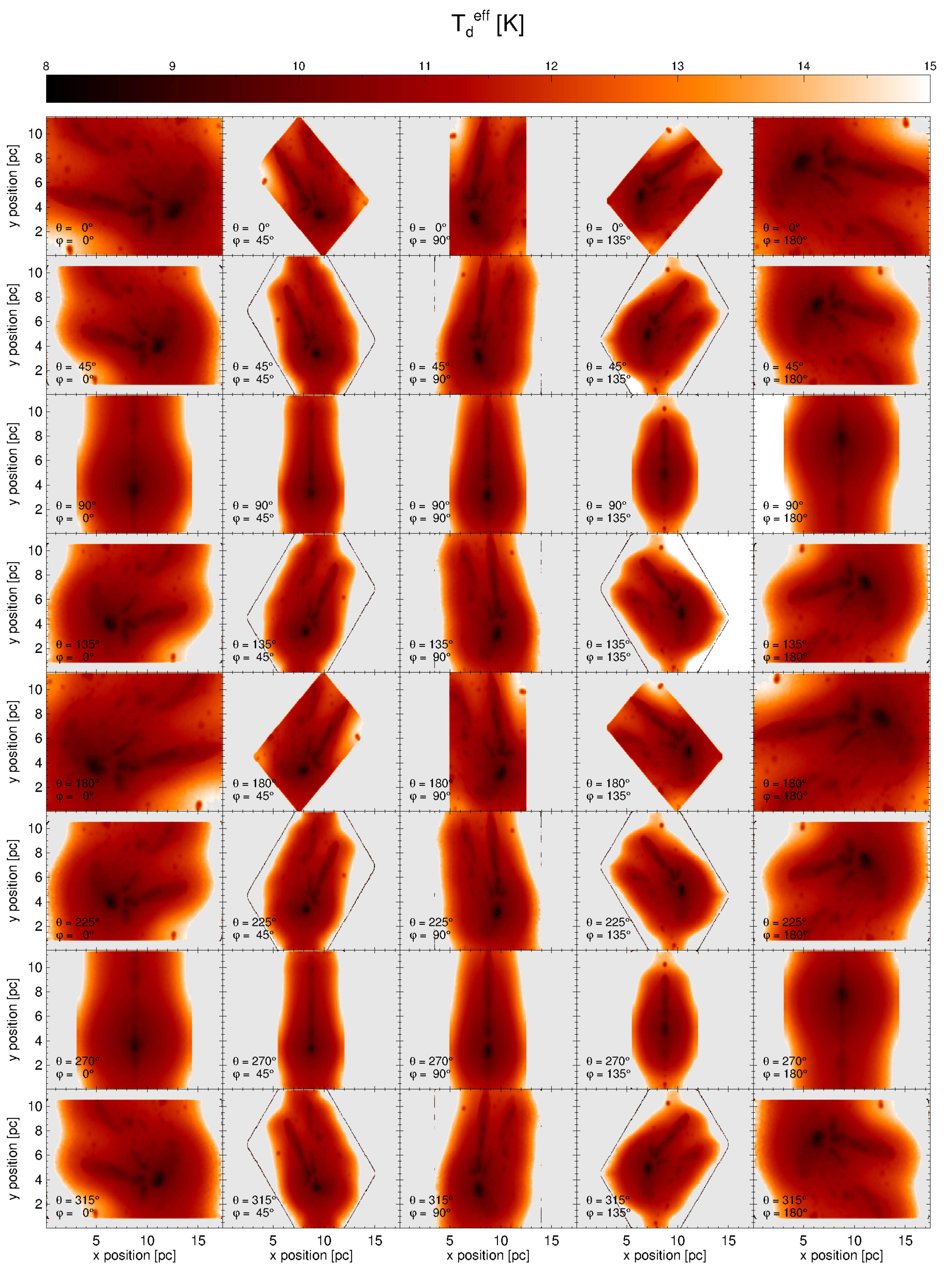}
  \caption[$T_\mathrm{d}^\mathrm{eff}$ maps of $\rho$~Ophiuchi cloud model]{$\rho$~Ophiuchi cloud model. Maps of effective dust temperature along the LoS, $T_\mathrm{d}^\mathrm{eff}$, that I derive by pixel-by-pixel SED fitting. }
  \label{pic:viewing_rhooph_tmap}
\end{figure}

\begin{figure}
  \centering
  \includegraphics[width=0.98\textwidth]{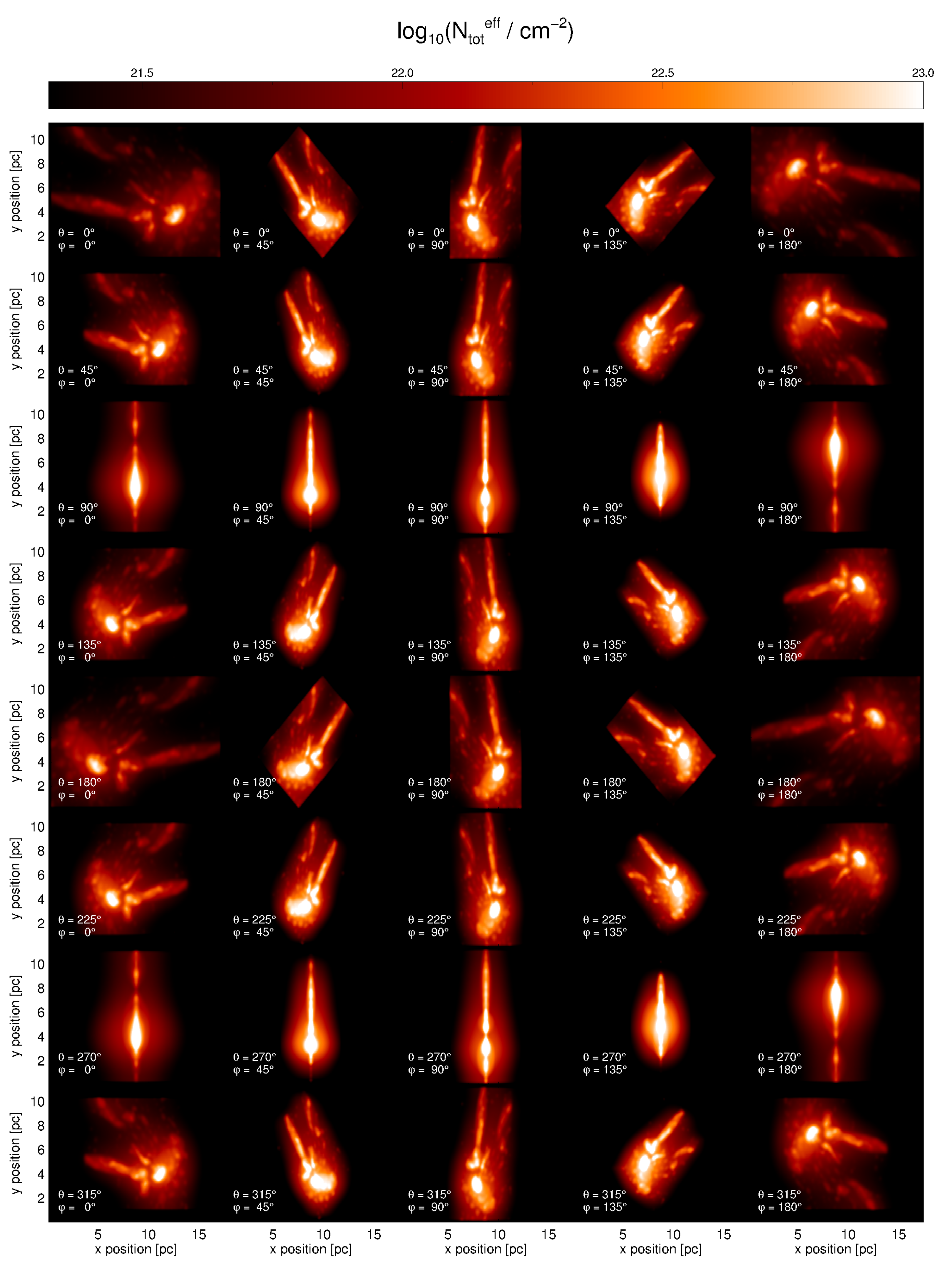}
  \caption[$N_\mathrm{tot}^\mathrm{eff}$ maps of $\rho$~Ophiuchi cloud model]{$\rho$~Ophiuchi cloud model. As Fig.~\ref{pic:viewing_rhooph_tmap} showing the effective total column density along the LoS, $N_\mathrm{tot}^\mathrm{eff}$.}
  \label{pic:viewing_rhooph_nmap}
\end{figure}

Figs.~\ref{pic:viewing_rhooph_tmap} and \ref{pic:viewing_rhooph_nmap} show maps of effective dust temperature and total column density based on the $\rho$~Ophiuchi model, respectively.
One sees that the images are symmetric in the sense that for a given LoS ($\theta$,$\varphi$) the image is the same as the image along ($\theta$+180$^\circ$,$\varphi$), just mirrored.
This is because as long as the dust is optically thin the distribution along the LoS dimension is basically erased and one sees the same profiles both ways.
For molecular lines, however, the LoS will make a significant difference \citep{Chira2014}.

Looking at the model along different LoSs, one sees significant changes in the morphology. 
The more I incline the object the clumpier and more elliptic the morphology becomes.
This is expected based on the structure of the input model.
One also observes that the effective total column density of the central region becomes higher while the gradient between central and outer regions becomes steeper the more I incline the model.
The central effective dust temperatures, however, do not change significantly, whereas the profile becomes flatter.
Fig.~\ref{pic:viewing_rhooph_example} shows a detailed example by offering maps of synthetic 250~$\mu$m flux density, $F_\mathrm{250 \mu m}$, the effective total column density, $N_\mathrm{tot}^\mathrm{eff}$, and effective dust temperature, $T_\mathrm{d}^\mathrm{eff}$, as seen at ($\theta$,$\varphi$)~=~(0$^\circ$,0$^\circ$) and ($\theta$,$\varphi$)~=~(45$^\circ$,45$^\circ$), as well as the respective PDFs.
Note that since the dust is optically thin in the FIR the flux and column densities behave in the same way.
This is why I focus the discussion on $N_\mathrm{tot}^\mathrm{eff}$.

\begin{mysidewaysfigure}
	\includegraphics[width=\textwidth]{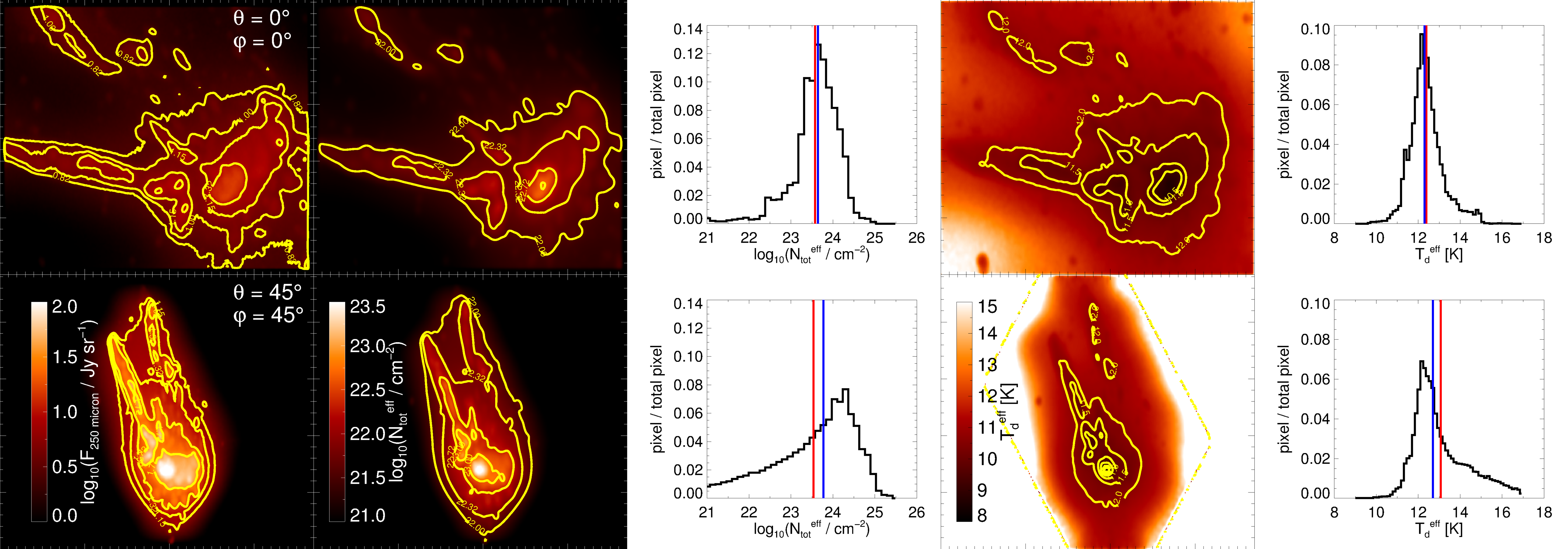}
 	\caption[Example of results based on $\rho$~Ophiuchi]{Results based on the $\rho$~Ophiuchi cloud model at ($\theta$,$\varphi$)~=~(0$^\circ$,0$^\circ$) (\textit{top}) and ($\theta$,$\varphi$)~=~(45$^\circ$,45$^\circ$) (\textit{bottom}).
  		From \textit{left} to \textit{right} the figures show the maps of synthetic intensity at 250~$\mu$m, maps of effective total column density with corresponding PDFs, and the maps of effective dust temperature with corresponding PDF.
  		The contours in the maps are at $\log_{\mathrm{10}}\left(F_\mathrm{250 \mu m} \, / \, \mathrm{Jy \, beam}^{-1} \right)~=~$ \{0.82,1.0,1.15,1.32,1.5\}, $\log_{\mathrm{10}} \left(N_\mathrm{tot}^\mathrm{eff} \, / \, \mathrm{cm}^{-2} \right)~=~$ \{22.0,22.32,22.72,23.0\}, and $T_\mathrm{d}^\mathrm{eff}~=~$ \{10.5,11.0,11.5,12.0\} K.
  		The red lines in the PDFs indicate the arithmetic mean values of the distributions, the blue lines the median values.
 	}
	\label{pic:viewing_rhooph_example}
\end{mysidewaysfigure}

\noindent The effective column density is concentrated towards the longest axis of the model at \linebreak ($\theta$,$\varphi$)~=~(45$^\circ$,45$^\circ$) compared to ($\theta$,$\varphi$)~=~(0$^\circ$,0$^\circ$), leading to higher intensities at similar effective dust temperatures.
If one compares the regions enclosed by the contours at both viewing angles one can identify the same structures at ($\theta$,$\varphi$)~=~(45$^\circ$,45$^\circ$) as in ($\theta$,$\varphi$)~=~(0$^\circ$,0$^\circ$) (considering the influence of rotation), but at higher values in the case of the intensity and effective column density.
In the case of effective dust temperature, the respective regions are indicated by the contours of the same values. 

At ($\theta$,$\varphi$)~=~(0$^\circ$,0$^\circ$), the PDFs for both the effective total column density and dust temperature are log-normal with a peak at $\xbar{N}_\mathrm{tot}^\mathrm{eff}$~=~4.4~$\times$~10$^{21}$~cm$^{-2}$ and $\xbar{T}_\mathrm{d}^\mathrm{eff}$~=~12.2 K.
The PDFs of ($\theta$,$\varphi$)~=~(45$^\circ$,45$^\circ$), however, have the shape of a log-normal distribution with a power-law tail towards lower column densities and higher dust temperatures, peaking at $\xbar{N}_\mathrm{tot}^\mathrm{eff}$~=~2.0~$\times$~10$^{22}$~cm$^{-2}$ and $\xbar{T}_\mathrm{d}^\mathrm{eff}$~=~12.2 K.

The main point that can be deduced from this result is that observing filaments along different LoSs do not fake typical signatures of star formation in column density PDFs \citep[e.g.][]{Schneider2013,Stutz2015}.
The tails towards lower column densities in the PDFs are normally within the noise range and cut during the data reduction process.
I discuss this in Sect.~\ref{viewing:g11} in more detail.

\begin{figure}
	\centering
    
    \begin{subfigure}{\textwidth}
        \centering
        \includegraphics[height=0.42\textheight]{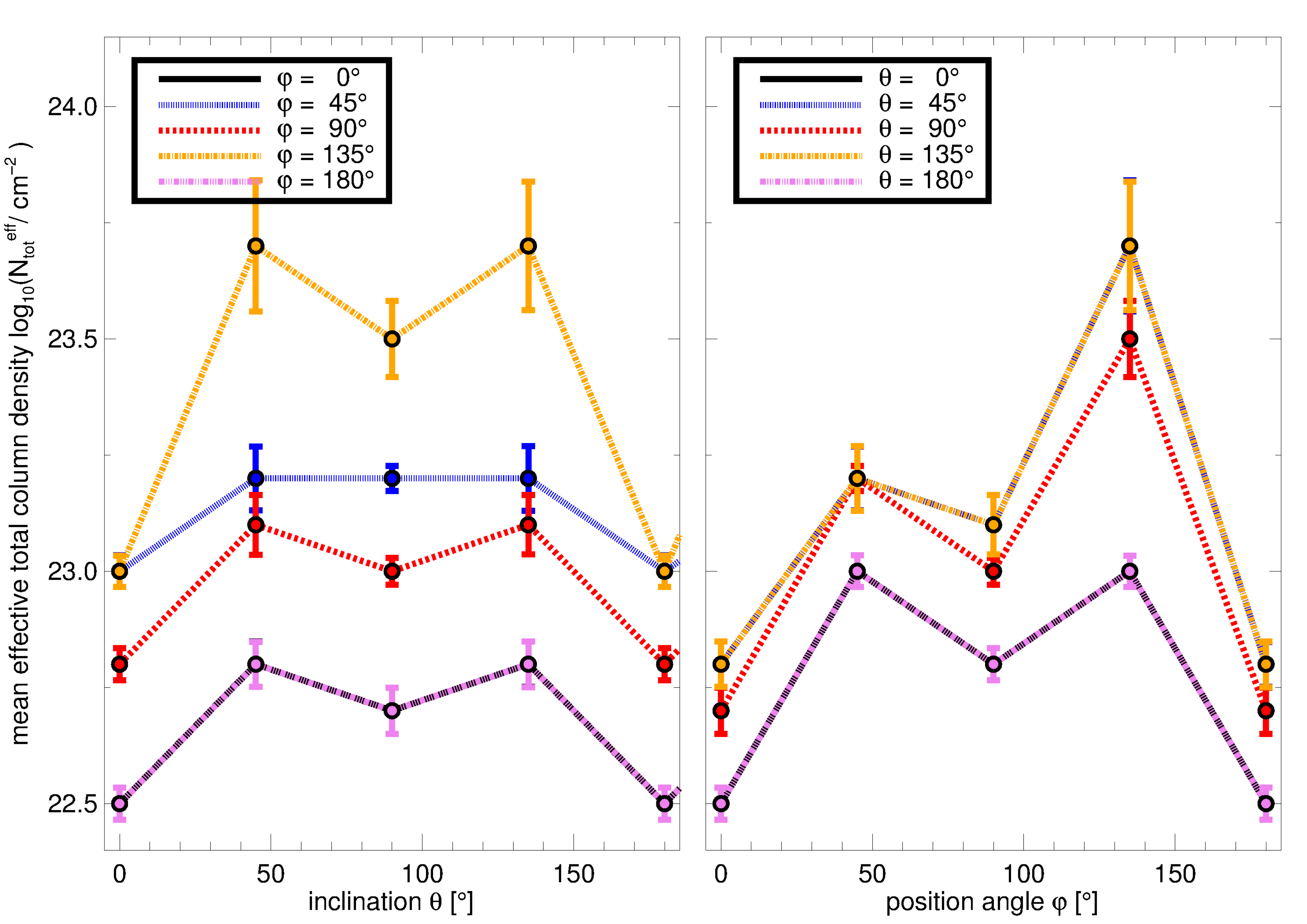}
        \caption{mean effective column density}
        \label{pic:viewing_rhooph_histostat_nmap}
    \end{subfigure}
    
    \begin{subfigure}{\textwidth}
        \centering
        \includegraphics[height=0.42\textheight]{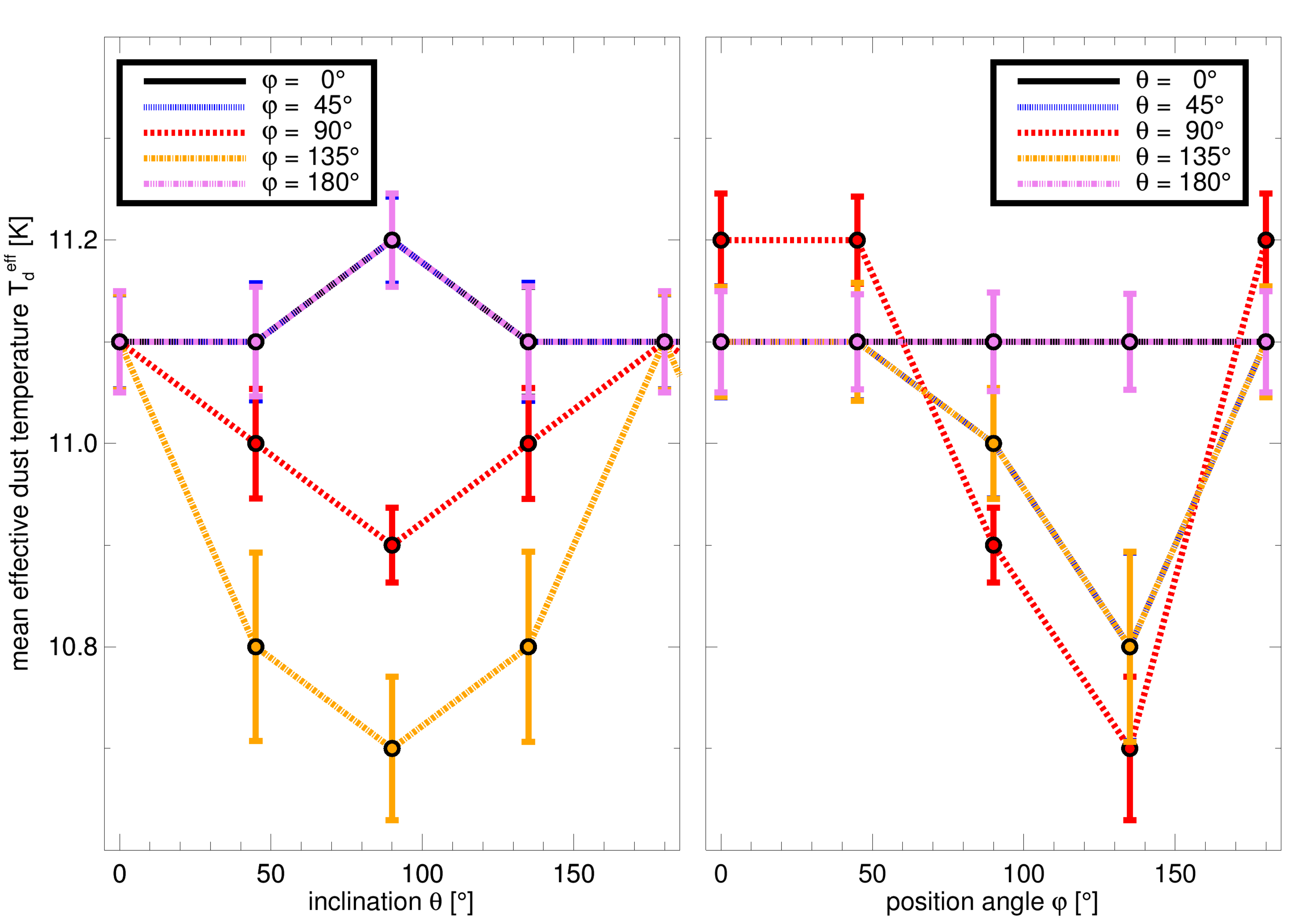}
        \caption{mean effective dust temperature}
        \label{pic:viewing_rhooph_histostat_tmap}
    \end{subfigure}
    
    \caption[Mean column density and dust temperature as function of observational angle for $\rho$~Ophiuchi cloud model]{$\rho$~Ophiuchi cloud model. The plots show the mean values of the effective total column density PDFs, $\xbar{N}_\mathrm{tot}^\mathrm{eff}$ (\textit{top}), and effect dust temperature, $\xbar{T}_\mathrm{d}^\mathrm{eff}$ (\textit{bottom}), as a function of the inclination, $\theta$, and position angle, $\varphi$. The error bars illustrate 1$\sigma$ errors. Note that I only show the values for inclinations ${\theta \, \leq\mathrm{ 180}^\circ}$ because the values for $\theta$ and $\theta$+180$^\circ$ are identical.
     }
    \label{pic:viewing_rhooph_histostat}
\end{figure}

Fig.~\ref{pic:viewing_rhooph_histostat} plots the values of $\xbar{N}_\mathrm{tot}^\mathrm{eff}$ and $\xbar{T}_\mathrm{d}^\mathrm{eff}$ from fitting Rayleigh distributions to the PDFs (see Sect.~\ref{viewing:cyl}) as a function of viewing angle.
I find large variations in $\xbar{N}_\mathrm{tot}^\mathrm{eff}$ compared to the non-inclined LoS ($\theta$,$\varphi$)~=~(0$^\circ$,0$^\circ$).
The maximal $\xbar{N}_\mathrm{tot}^\mathrm{eff}$ is detected at ($\theta$,$\varphi$)~=~(45$^\circ$,135$^\circ$) and (135$^\circ$,135$^\circ$) at a factor of 50 higher than the ($\theta$,$\varphi$)~=~(0$^\circ$,0$^\circ$) case.
These average column densities are clearly above the average range of mean total column density in the literature \citep[5~$\times$~10$^{20}$ -- 3.16~$\times$~10$^{22}$~cm$^{-2}$,][]{Koenyves2010,Juvela2012b,Palmeirim2013,Roy2014}.

In the contrary, the variations of $\xbar{T}_\mathrm{d}^\mathrm{eff}$ are below 0.5~K.
These differences are on the same order as typical noise levels and are not detectable.
For a fixed $\theta$~=~0$^\circ$, the $\xbar{T}_\mathrm{d}^\mathrm{eff}$ remains constant.
All values for $\xbar{T}_\mathrm{d}^\mathrm{eff}$ here are within a range that agrees with observational findings of pre-stellar cores and filaments \citep[10 -- 15~K, e.g.~][]{Juvela2012b}.

Comparing the results of the $\rho$~Ophiuchi model with those in Sect.~\ref{viewing:cyl}, the closest model with respect to the density profiles would be \textit{cylsph}.
I argue that the differences in behaviour result from the fact that the $\rho$~Ophiuchi model is dominated by its irregular, clumpy structure which makes it more sensitive to changes in the viewing angles than a smooth, regular structure.
Furthermore, the $\rho$~Ophiuchi model is not isothermal, unlike the individual spheroids the model is made of.
Since the individual spheroids are approximately homogeneous, the whole $\rho$~Ophiuchi model can be described as a superposition of \textit{homcyl}-like elements which, however, has not shown significant variations along different LoSs as one observes them here.

Thus, more complicated structures are significantly more affected by changes in orientation and geometry.
However, the column density alone is not enough to trace back the inclination of an observed filament.
Only a significant increase in column density at a constant average dust temperature can give hints on how the matter is distributed along a given LoS.

Just inclining the model does not change the thermal processes (heating, cooling) within the object. 
If the density within the object increases (for example, by collapsing material) the dust cools more efficiently and $T_\mathrm{d}^\mathrm{eff}$ decreases before the first protostar forms.
Thus, if $N_\mathrm{tot}^\mathrm{eff}$ and $T_\mathrm{d}^\mathrm{eff}$ do not behave anti-proportionally, this indicates that the object is elongated along the LoS.
In this case, it is necessary to re-construct the density distribution along the LoS by line observations.

\subsection{G11.11 \textit{Snake} model}\label{viewing:g11}

I repeat the analysis for the \textit{Snake} model and show the maps of effective dust temperature and total column density in Figs.~\ref{pic:viewing_g11_tmap} and \ref{pic:viewing_g11_nmap}, respectively.
Note that I only print the maps within (0$^\circ~\leq~\theta~\leq$~135$^\circ$,~0$^\circ~\leq~\varphi~\leq$~180$^\circ$) due to the symmetries I have already discussed in Sect.~\ref{viewing:rhooph}.

\begin{mysidewaysfigure}
  \includegraphics[width=0.98\textwidth]{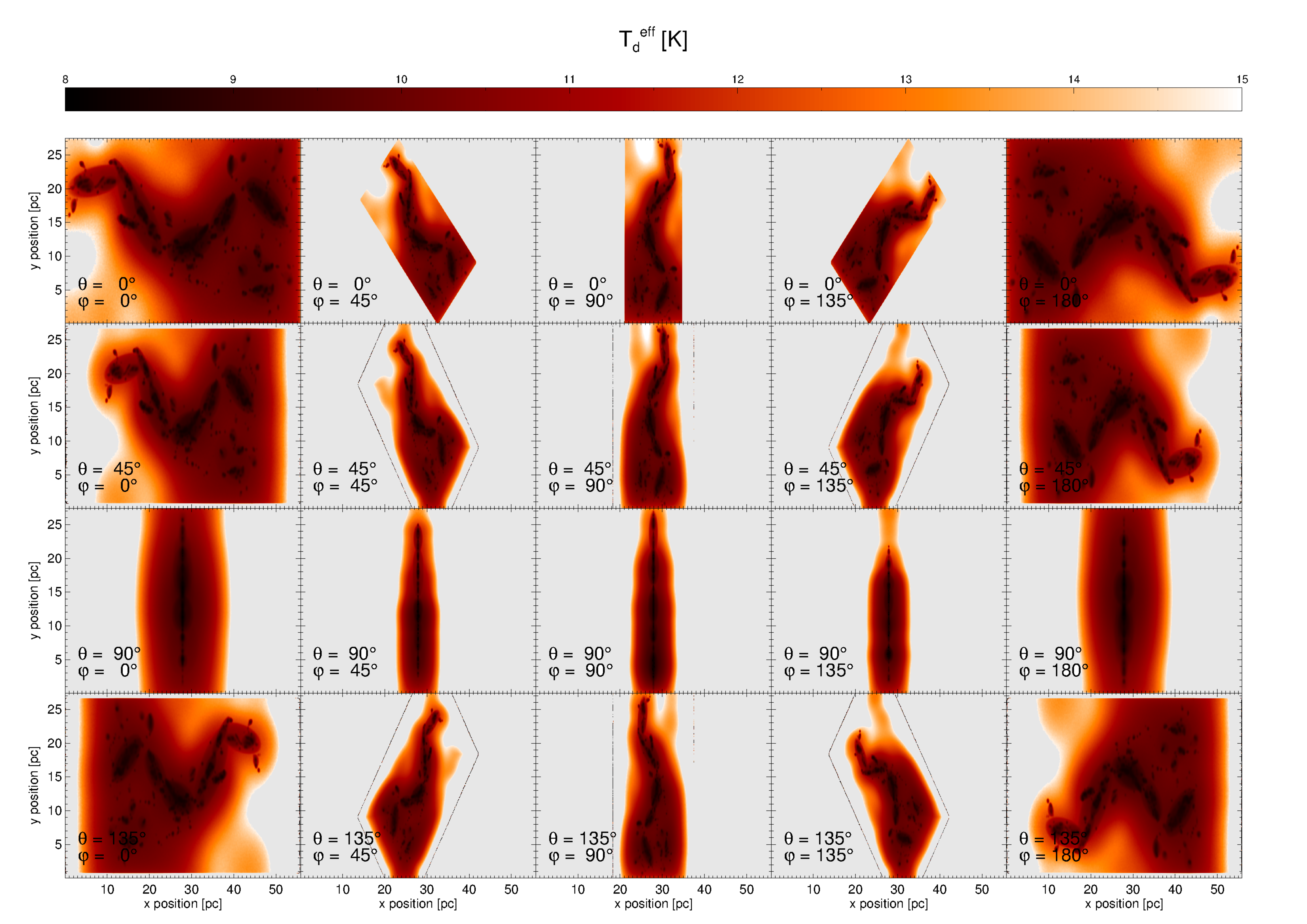}
  \caption[$T_\mathrm{d}^\mathrm{eff}$ maps of G11.11 model]{G11.11 \textit{Snake} model. Maps of effective dust temperature along the LoS, $T_\mathrm{d}^\mathrm{eff}$, that I derive by pixel-by-pixel SED fitting.
  }
  \label{pic:viewing_g11_tmap}
\end{mysidewaysfigure}

\begin{mysidewaysfigure}
  \includegraphics[width=0.98\textwidth]{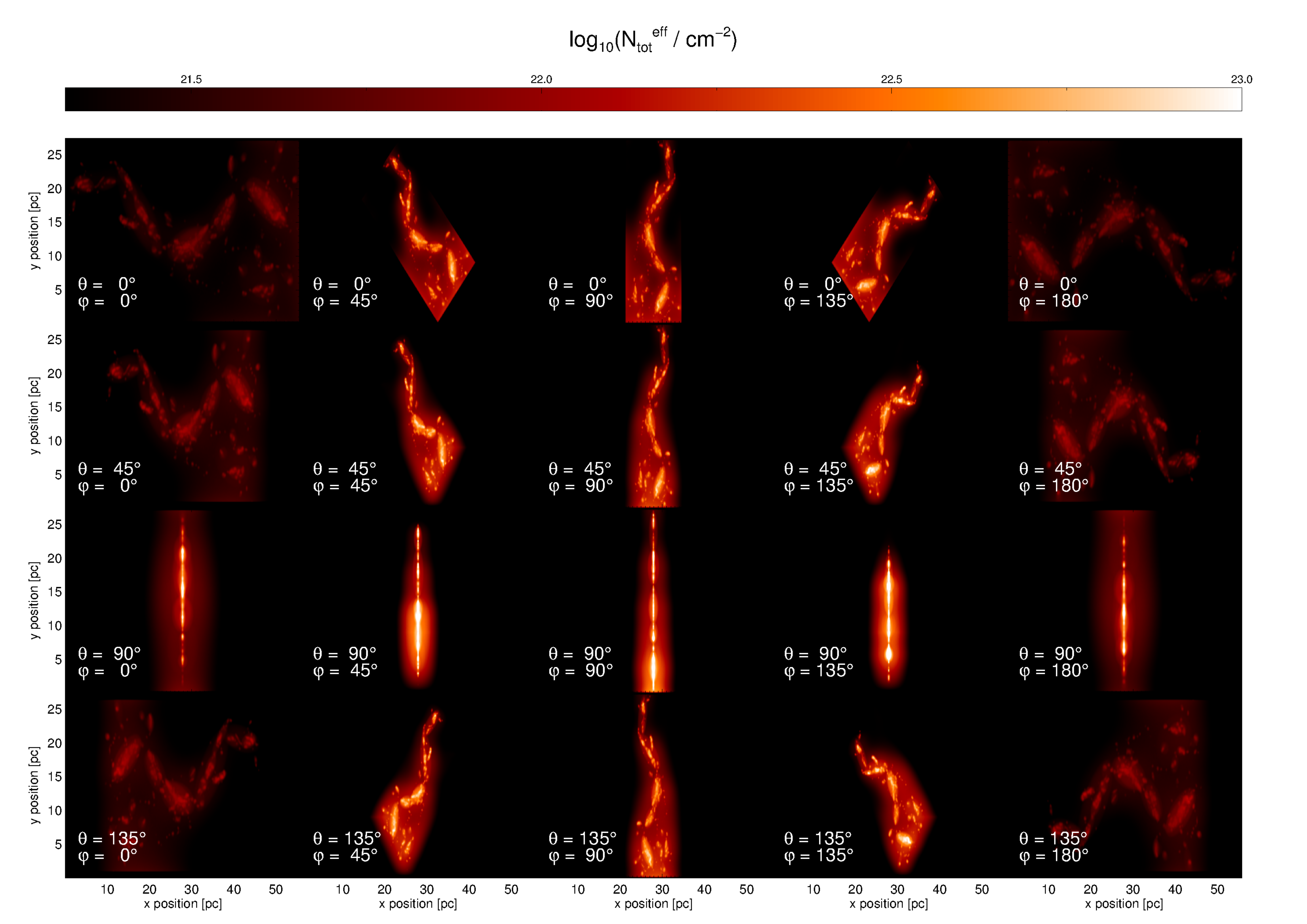}
  \caption[$N_\mathrm{tot}^\mathrm{eff}$ maps of G11.11 model]{G11.11 \textit{Snake} model. As Fig.~\ref{pic:viewing_g11_tmap} showing the effective total column density along the LoS, $N_\mathrm{tot}^\mathrm{eff}$.}
  \label{pic:viewing_g11_nmap}
\end{mysidewaysfigure}

\begin{figure}
	\centering
    
    \begin{subfigure}{\textwidth}
        \centering
        \includegraphics[height=0.42\textheight]{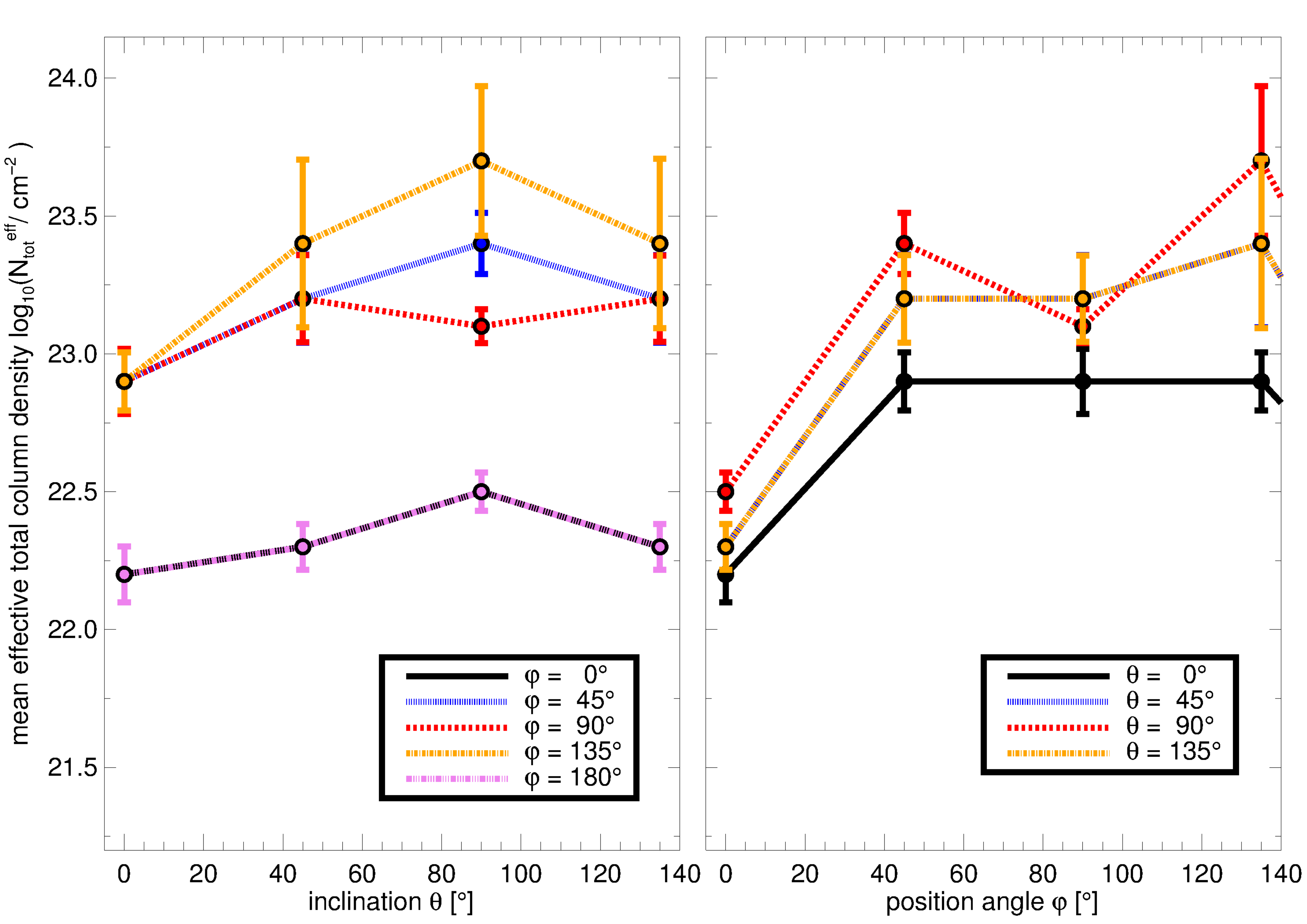}
        \caption{mean effective column density}
        \label{pic:viewing_g11_histostat_nmap}
    \end{subfigure}
    
    \begin{subfigure}{\textwidth}
        \centering
        \includegraphics[height=0.42\textheight]{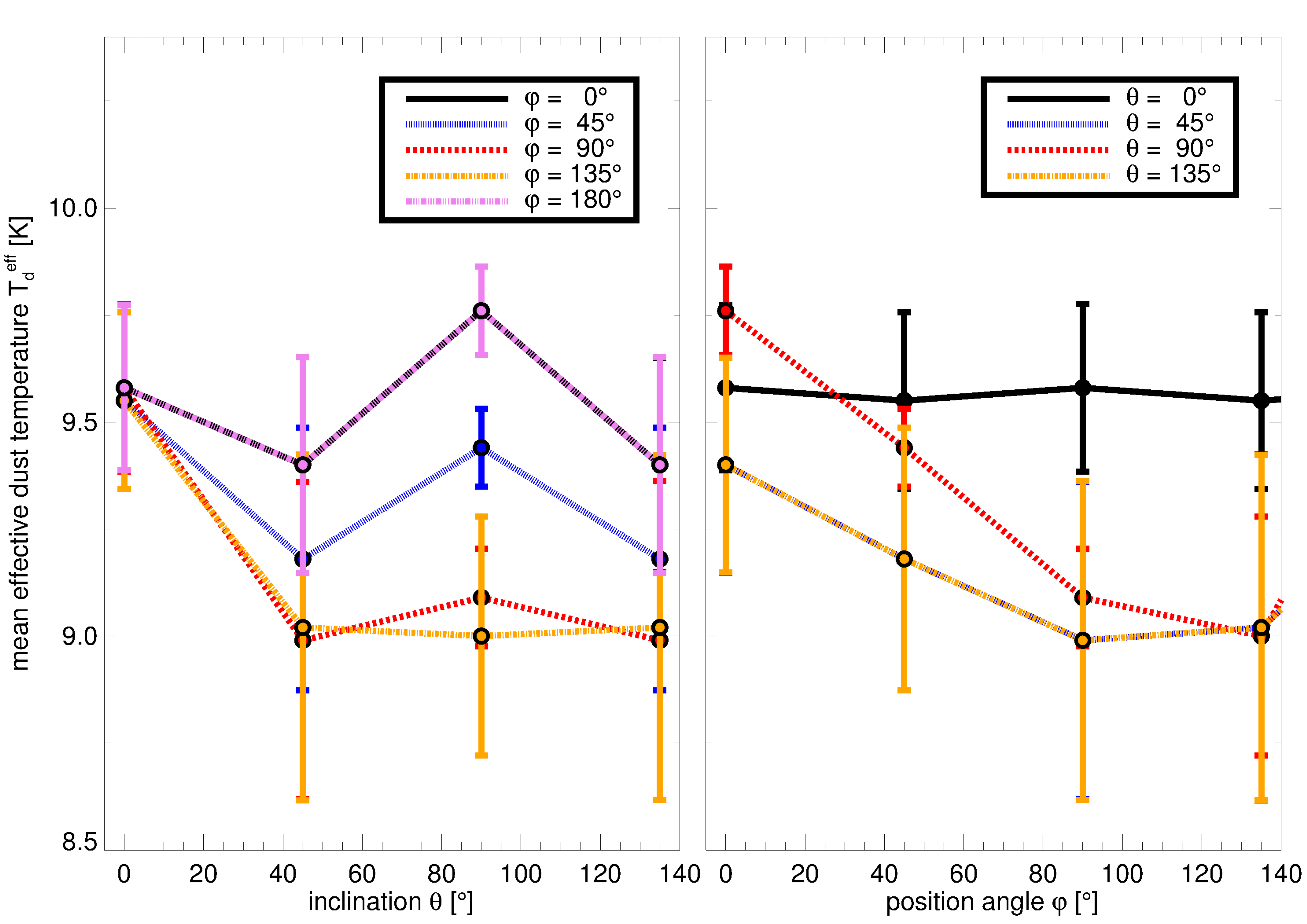}
        \caption{mean effective dust temperature}
        \label{pic:viewing_g11_histostat_tmap}
    \end{subfigure}
    
    \caption[Mean column density and dust temperature as function of observational angle for G11.11]{As Fig.~\ref{pic:viewing_rhooph_histostat}, but for G11.11 \textit{Snake}. }
    \label{pic:viewing_g11_histostat}
\end{figure}

One observes that rotations do not skew the projected morphology of \textit{Snake} as much as it has done for $\rho$~Ophiuchi.
At all viewing angles, except $\theta$~=~90$^\circ$, one can see the sinusoidal structure that is typical for the \textit{Snake}, although its amplitude slightly changes.
The distribution of effective dust temperature, however, remains approximately constant, as it has been the case for $\rho$~Ophiuchi.
The values of the average effective dust temperature $\xbar{T}_\mathrm{d}^\mathrm{eff}$ (see Fig.~\ref{pic:viewing_g11_histostat_tmap}) vary within 0.8 K, which is insignificant in terms of observational uncertainties \citep[e.g.,][]{Juvela2012b}.

As with the $\rho$~Ophiuchi model, one sees significant variations in the effective total column density with the viewing angle.
As plotted in Fig.~\ref{pic:viewing_g11_histostat_nmap}, the values of $\xbar{N}_\mathrm{tot}^\mathrm{eff}$ at ($\theta$,$\varphi$)~=~(0$^\circ$,0$^\circ$) and ($\theta$,$\varphi$)~=~(0$^\circ$,45$^\circ$) change by almost two orders of magnitude.
In general, one observes an increase in $\xbar{N}_\mathrm{tot}^\mathrm{eff}$ by factors between 16 and 320 compared to ($\theta$,$\varphi$)~=~(0$^\circ$,0$^\circ$).
The steepest rise is found when rotating the model between ($\theta$,$\varphi$)~=~(0$^\circ$,0$^\circ$) to positions with $\theta$~=~45$^\circ$ or $\varphi$~=~45$^\circ$.
For steeper angles, the column densities do not change as significantly.

Thus, the distributions of effective total column density and dust temperature of the \textit{Snake} model behave similarly to the $\rho$~Ophiuchi model.
The elongated structure of the model amplifies the variations of the small components in two ways;
firstly, the $N_\mathrm{tot}^\mathrm{eff}$ rises higher the more its long axis is aligned with the LoS;
secondly, the radial temperature gradient is larger since the filament is not dense enough to be isothermal.

\begin{figure}[h!t]
    \centering
    \includegraphics[width=\textwidth]{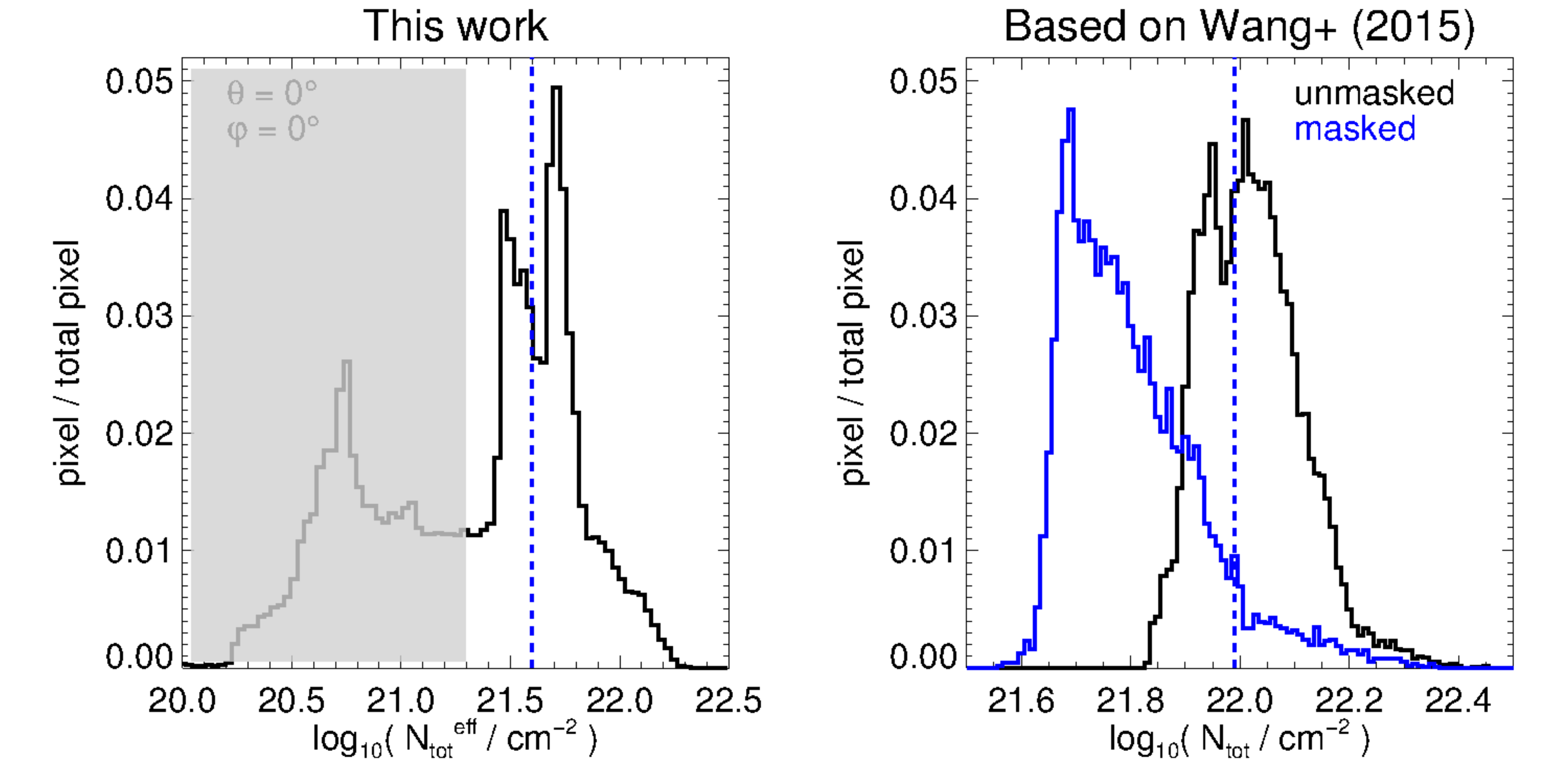}
    \caption[Comparision of reconstructed and observed column density PDFs]{Comparison of (effective) total column density PDFs of \textit{Snake} based on data of this work (\textit{left}) and \citet[\textit{right}]{Wang2015}.
      The black lines show the original data of the simulation and observation, respectively. 
      The grey-shaded area on the left plot indicates the range that would be below the detection limit of \textit{Herschel}.
      The blue-dotted line in both images indicates the cut limit which is used for noise reduction.
      The solid blue line in the right plot shows the total column density PDF based on the masked intensity maps.
    }
    \label{pic:viewing_g11_comp_wang}
\end{figure}

Note that the errors of the mean values of the distributions are larger than those in Sect.~\ref{viewing:rhooph}.
This is due to the more complex structure of the PDFs.
An example is shown in Fig.~\ref{pic:viewing_g11_comp_wang}.
The left panel plots the effective total column density PDF (black line) of the \textit{Snake} model as observed at ($\theta$,$\varphi$)~=~(0$^\circ$,0$^\circ$).
One sees that the PDF splits into three peaks in total; two at higher column densities ($\sim$3.2--5~$\times$~10$^{21}$~cm$^{-2}$), which are relatively close to each other and a third peak at low column densities ($\sim$5~$\times$~10$^{20}$~cm$^{-2}$).
A similar profile is seen with the effective dust temperature PDF. 
Since I only fit a Rayleigh distribution with one peak to the PDF, structures like this are a major source for uncertainties. 
Of course, I could reduce the errors by fitting a super-position of Rayleigh distributions for each component.
However, in terms of physical interpretation this would mean that I assume that the model consists of at least three independent components.
Yet, finding these peaks has already been unexpected since there is no hint in the literature that this structure has been observed for the real \textit{Snake} cloud.

I conclude that this is due to observational limitations and reduction processes (Ke Wang, private communication).
The right side of Fig.~\ref{pic:viewing_g11_comp_wang} illustrates the total column density PDF derived by \citet{Wang2015}.
The black solid line shows the column density PDF derived from the unmasked (calibrated, but not noise-reduced) intensity map, and with the blue solid line the PDF for the masked map.
The blue dashed line indicates the level at which the originally observed intensity maps have been cut in the masking process.

For \citet{Wang2015} fluxes corresponding to total column densities below 2.5~$\times$~10$^{21}$~cm$^{-2}$ are below the sensitivity limit of \textit{Herschel} and would not be observed.
I indicate this part with a grey box in the PDF in the left panel.
This means that I could detect the two peaks at higher column densities with observations, but not the third one at low column densities.
This is consistent with the fitting routine of \citet{Kainulainen2014}, which weights the highest column density regimes more than the low column density parts.
Thus, the lower peak belongs to the parts of the observations that lie within the noise and is cut-off during the noise correction process.
Applying all these factors, a single-peak PDF is left that agrees with the observed findings.

\noindent Note that the peak positions in the PDFs of \citet{Wang2015} and this work are not the same.
This has several reasons.
First, the model filaments were derived from extinction maps in the NIR, whereas \citet{Wang2015} used emission maps in the sub-mm.
Even though these techniques result in column densities that agree on each other on average \citep[e.g.,][]{Kainulainen2013b}, regional differences may exist.
While extinction maps are limited at high column densities, emission maps are most sensitive there.
Emission maps, though, are more temperature-dependent and their interpretation require more knowledge about the dust grain properties than it is the case for extinction maps.
Second, the models in my study are only heated externally.
Both \citet{Henning2010} and \citet{Wang2015} show that this does not entirely match reality.
The \textit{Snake} filament contains cores that are associated with protostars and ongoing star formation processes.
While the results of \textit{cylsph} in Sect.~\ref{viewing:cyl} suggest that missing individual cores do not affect the column density statistics of the filament, a larger population of cores would contribute to the incoming radiation field if they contain embedded sources and thus change the outputs of the RT code.
Moreover, I do not consider the temperature gradient in \textit{cylsph}, which is expected for the \textit{Snake} filament.
The influence of different external heating fields and/or internal heating sources needs to be investigated in more detail in future studies.

\begin{figure}
	\centering
    
    \begin{subfigure}{\textwidth}
        \centering
        \includegraphics[height=0.42\textheight]{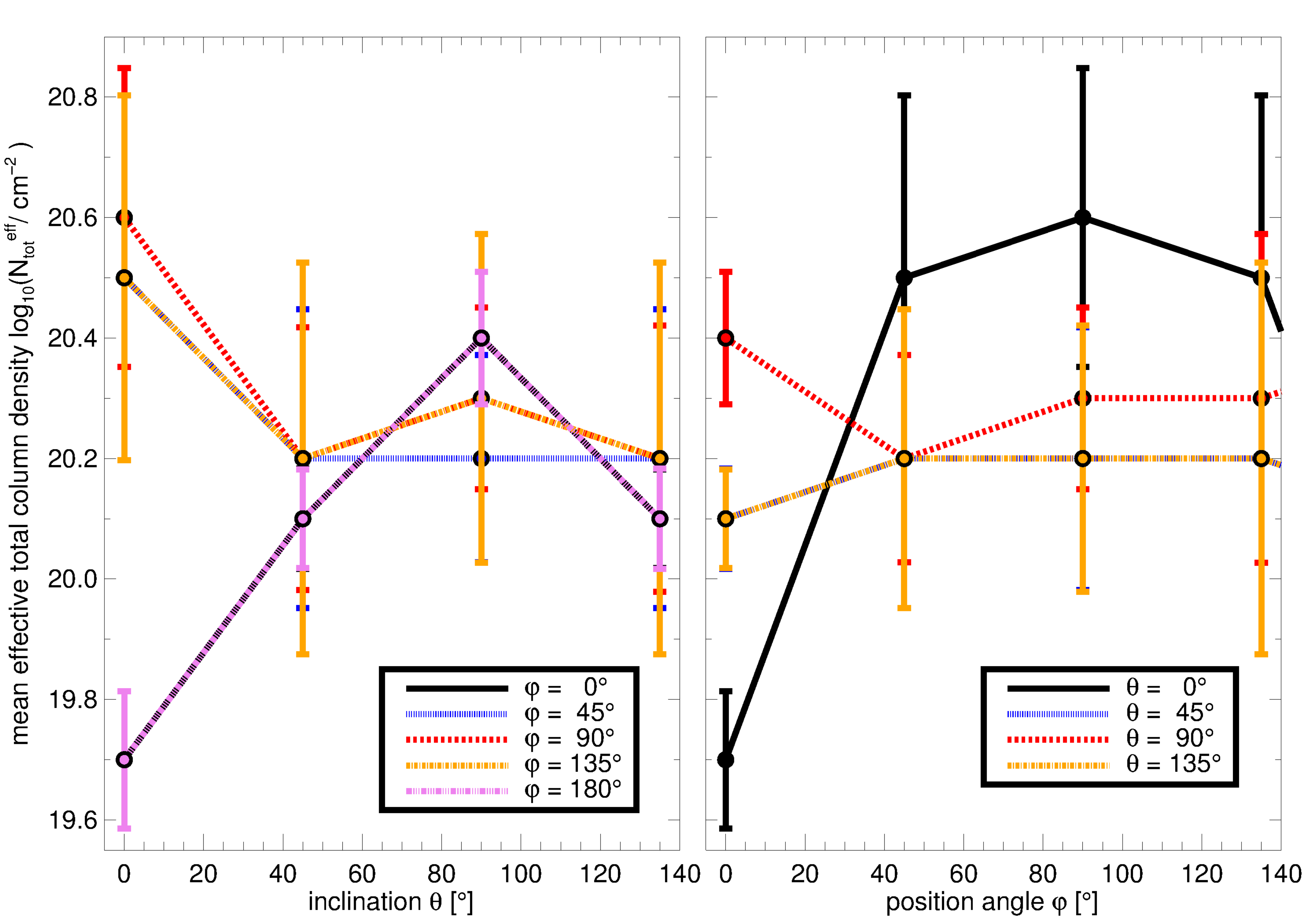}
        \caption{mean effective column density}
        \label{pic:viewing_g11_intcut_histostat_nmap}
    \end{subfigure}
    
    \begin{subfigure}{\textwidth}
        \centering
        \includegraphics[height=0.42\textheight]{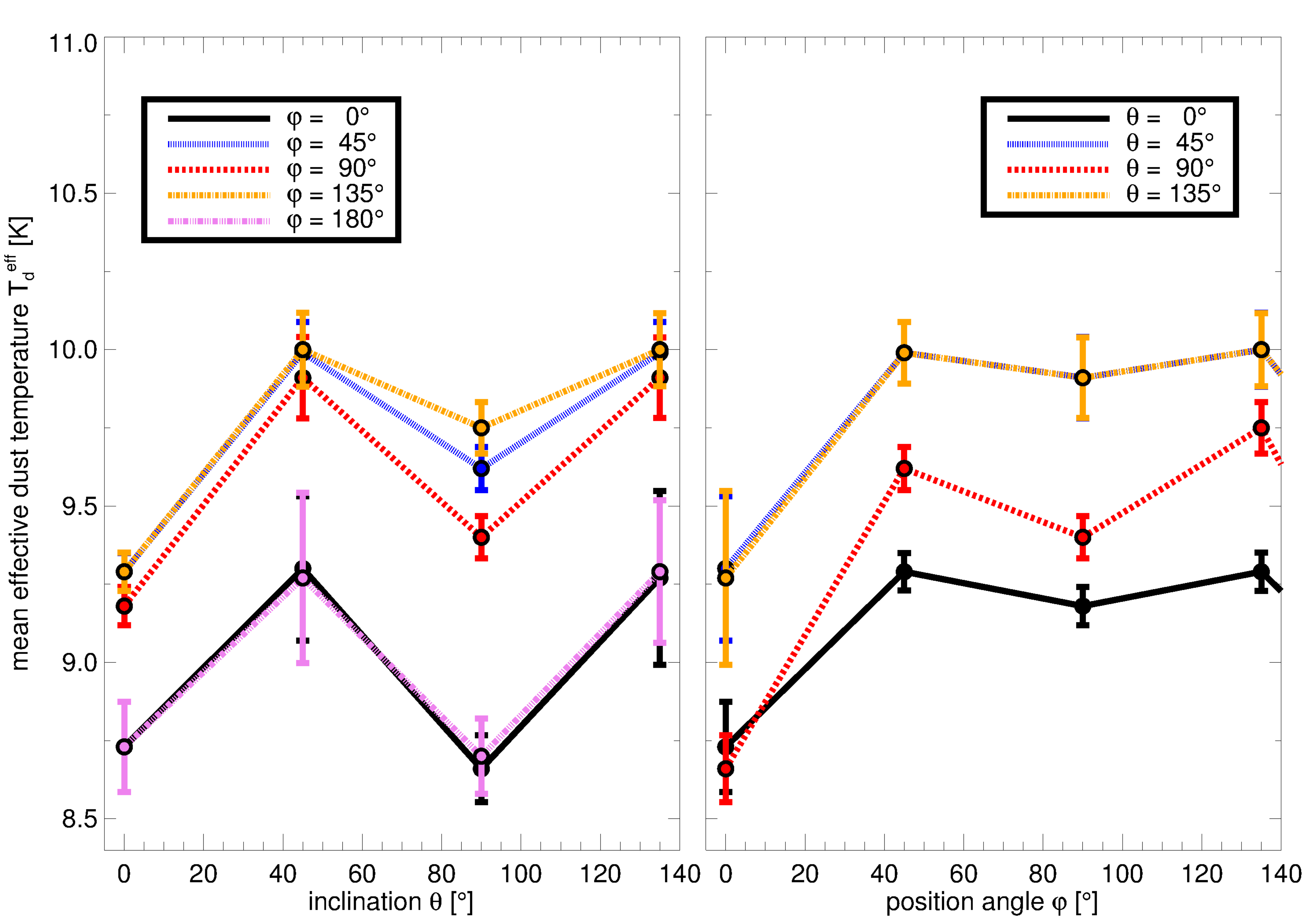}
        \caption{mean effective dust temperature}
        \label{pic:viewing_g11_intcut_histostat_tmap}
    \end{subfigure}
    
    \caption[Mean column density and dust temperature as function of observational angle for masked G11.11]{As Fig.~\ref{pic:viewing_g11_histostat} but based on the masked maps of the G11.11 \textit{Snake} model. }
    \label{pic:viewing_g11_intcut_histostat}
\end{figure}

I test the effect of introducing column density cuts, in order to focus on significant signals above a certain noise level.
The results for the $\xbar{N}_\mathrm{tot}^\mathrm{eff}$ and $\xbar{T}_\mathrm{d}^\mathrm{eff}$ are plotted in Fig.~\ref{pic:viewing_g11_intcut_histostat}.
I find that the effective total column density are lower than before by about one order of magnitude, whereas the effective dust temperatures keep relatively constant although the variations become larger.

\pagebreak

\section{Summary}\label{viewing:conclusion}

In this chapter, I investigate the dependence of effective total column density and dust temperature of filaments on the viewing angle and found the following.
\medskip

\begin{itemize}
	\item One sees that the mean flux density of simple cylinders rises with increasing dust temperature, but it is not significantly influenced by changing the viewing angle.
	\item For the 3D models of the $\rho$~Ophiuchi cloud and the G11.11 \textit{Snake}, one observes that the mean effective dust temperature is approximately constant, whereas mean effective dust number density changes significantly, especially when I rotate the models into the direction where their long axis is parallel to the LoS. Since the dust emission is optically thin, the column densities strongly depend on the viewing angle. The dust temperature is determined by the local heating and cooling processes, which is unchanged when I rotate the models. 
    \item I investigate how sensitivity limits and noise correction procedures influence the results. I find that common data reduction processes reduce the level of variation in column density. The variations in effective dust temperature have increased, but have been still insignificant enough in the observational context.
\end{itemize}

%\chapterimage{head02.pdf}

\chapter[Project II: Fragmenting Filaments in Simulations]{Fragmenting Filaments in Simulations\footnotemark[6]}\label{frag}

\footnotetext[6]{The content and results presented in this chapter are published in \citet{Chira2017}.}

\vspace*{-\baselineskip}
In this chapter, I shift the focus from the detection of filaments in molecular clouds to how they are connected to the formation of pre-stellar cores, and thus to the onset of star formation.

As mention in Chapter~\ref{intro}, filaments are supposed to represent an essential phase in the earliest stages of star formation.
This idea is motivated by observations that show that pre-stellar cores and young stellar clusters are preferentially located within filaments or at intersections between them \citep{Schmalzl2010,Myers2011,Schneider2012,Koenyves2015}.
Furthermore, there are studies that suggest that filaments not only provide the sites for the formation of star-forming cores, but are actively controlling star formation activities \citep[e.g.,][]{Zhang2009,Pineda2015,Henshaw2016b}.

Thus, the questions I address in this chapter are:
How do filaments evolve and do cores form by the fragmentation of those filaments in numerical simulations? 
Is the fragmentation picture seen in simulations in agreement with the quasi-static analytic framework of gravitational fragmentation?

For the latter, I focus on comparing the filaments in the simulation with models describing hydrostatic cylinders that fragment due to linear perturbations \citep[Sect.~\ref{frag:theory}]{Ostriker1964a,Larson1985,Padoan1999}.
Note that some studies \citep[e.g., by][]{Lee1999,Hartmann2002,Henshaw2016b,Clarke2017,Gritschneder2017} demonstrate that other environmental conditions, such as turbulence, accretion, or magnetic fields, can introduce additional fragmentation modes.
Therefore, another key question is whether a quasi-static description is justified for characterising and predicting the evolution of filaments.

One way to test this is to use numerical simulations that form filaments self-consistently and compare the properties of the resulting structures with those predicted by the simple model.
In doing so, one can examine how each process, and combination of different forces, affects the evolution of filaments, and whether a quasi-static model can capture the essential physics involved in the process of fragmentation.

Since the forces acting on filaments are not directly observable, one needs to analyse the properties of the filaments and fragments as observational diagnostics.
For the analysis here, I choose the mass per unit length, or line mass, to characterise the filaments.
The line mass is argued to determine the filament's stability in the quasi-static model \citep{Ostriker1964a,Ostriker1964b,Nagasawa1987,Inutsuka1992,Fiege2000a,Fischera2012a,Myers2017}.
Similar to the Jeans analysis for spheres \citep{Jeans1902}, there is a critical line mass that marks the transition between states of equilibrium and gravitational collapse. 
In principle, the line mass is an easy to measure quantity since it only requires the length and the enclosed mass of the respective filament.
However, there are many effects that need to be considered when measuring the line mass from observational data, such as inclination, optical depth, uncertainties in distance estimations, and overlap effects \citep{Ballesteros2002,ZamoraAviles2017}.

I analyse filaments that have formed self-consistently in molecular cloud simulations and compare their fragmentation behaviour with predictions of analytic cylindrical models (see Sect.~\ref{frag:theory}).
In Sect.~\ref{frag:clouds}, I present the 3D FLASH adaptive mesh refinement~(AMR) simulations and the three model clouds, and detail the methods I use for the analysis in Sect.~\ref{frag:prop}.
That includes a performance comparison of commonly used filament finder codes in Sect.~\ref{frag:prop_filfinder}.
Although the data does not suffer from observational biases and uncertainties, the comparison clearly demonstrates how the different approaches of those codes can influence the results, and consequently also the conclusions drawn based on the results, even in an idealised environment.
The results of the analysis are outlined and discussed in Sect.~\ref{frag:thermsupport}.

\section{Analytic Models of Filaments}\label{frag:theory}

The usual analytical model for filaments describes them as hydrostatic cylinders that fragment due to linear perturbations \citep{Ostriker1964a,Ostriker1964b,Larson1985,Padoan1999}.
According to this model, the cylinder is infinitely long and filled with self-gravitating, polytropic gas. 
In this case, the internal pressure of the gas, $P$, and its volume mass density, $\rho$ are proportional to each other,
\begin{equation}
	P = K_n \rho^{1 + \frac{1}{n}} ,
	\label{equ:intro_theory_polytropic_def}
\end{equation}
with $K_n$ being the proportionality constant and $n$ a parameter describing the state of the contained gas.
For an isothermal ideal gas, as one can assume it for the ISM, $n~=~\infty$.
This means that Eq.~(\ref{equ:intro_theory_polytropic_def}) transforms to
\begin{equation}
	P = K_\infty \rho = \frac{k_B T}{\mu m_p} \rho ,
	\label{equ:intro_theory_polytropic_ideal}
\end{equation}
where $k_B$ is the Boltzmann constant, $T$ the temperature of the gas, $\mu$ its molecular weight, and $m_p$ the mass of a proton.
Assuming that the gas is in equilibrium, one can insert Eq.~(\ref{equ:intro_theory_polytropic_ideal}) into the equation of hydrostatic equilibrium,
\begin{equation}
	\nabla \Phi = \frac{1}{\rho} \nabla P ,
	\label{equ:intro_theory_hydrostaticequ_def}
\end{equation}
with $\Phi$ being the gravitational potential of the gas, and combine it with Poisson's equation,
\begin{equation}
	\Delta \Phi = - 4 \pi G \rho , 
	\label{equ:intro_theory_poisson_def}
\end{equation}
where $G$ is the Gravitational constant.
Following the descriptions in \citet{Ostriker1964b}, this returns in the following radial density distribution:
\begin{equation}
	\rho (\xi) = \rho_0 \frac{1}{\left( 1 + \frac{1}{8} \xi^2  \right)^2}.
	\label{equ:intro_theory_ostriker_rho}
\end{equation}
Thereby, the radius, $r$, of the cylinder is substituted by its length, $\xi$, via $r = a_i \xi$.

\noindent The motivation of this parameterisation becomes more obvious when one computes the mass per unit length, or line mass hereafter, of the cylinder.
The line mass, $M_\mathrm{lin}$, is generally given by the ratio of the cylinder's mass, $M_\mathrm{tot}$, and length, $\ell$,
\begin{equation}
	M_\mathrm{lin} = \frac{M_\mathrm{tot}}{\ell} .
	\label{equ:intro_theory_mlin_def}
\end{equation}
The line mass is an interesting parameters since it is supposed to determine a filament's stability state as the Jeans mass does so for spheres \citep{Ostriker1964b,Nagasawa1987,Inutsuka1992,Fiege2000a,Fischera2012a}.
This means that a cylindrical filament is only then thermally supported against collapse if its line mass remains below a critical value. 
Otherwise, initially small perturbations within the filaments can grow under the influence of self-gravity.
This then leads to radial collapse and fragmentation. 

\noindent Using Eq.~(\ref{equ:intro_theory_ostriker_rho}), the critical line mass of an isothermal cylinder is given by,
\begin{equation}
	M_\mathrm{lin}^\mathrm{crit} = \frac{2 k_B T}{\mu m_p G} \frac{1}{1 + \frac{8}{\xi^2}} .
	\label{equ:intro_theory_mlin_finite}
\end{equation}
As the length of the cylinder approaches infinity, $\xi \rightarrow \infty$, the line mass of the cylinder converts towards
\begin{equation}
	M_\mathrm{lin}^\mathrm{crit} = \frac{2 k_B T}{\mu m_p G} = 2 \frac{c_s^2}{G} \approx 16.7 \left( \frac{T}{\mathrm{10~K}} \right) \frac{M_\odot}{\mathrm{pc}} , 
	\label{equ:intro_theory_mlin_infinite}
\end{equation}
with $c_s$ being the thermal sound speed.

The equations summarised above offer some very important insights as some of the diagnostics can be directly compared to observations. 
For example, Eq.~(\ref{equ:intro_theory_ostriker_rho}) implies that the density distribution of a thermally supported isothermal cylinder of self-gravitating ideal gas should fall with $r^{-4}$.
This is much steeper than the density distribution slopes that are normally assumed for isothermal spheres, that are used to model molecular clouds, clumps, and pre-stellar cores \citep[$\rho \propto r^{-1} / r^{-2}$, e.g.,][]{MacLaren1988}.
It is also much steeper than the slopes on the order $\rho \propto r^{-1.3-2.4}$ that are found for filaments in near-by star-forming regions \citep{Arzoumanian2011}.
These differences are significant and suggest that the assumptions above may not be enough to describe real filaments.

Additionally, some studies \citep[e.g.,][]	{Lee1999,Hartmann2002,Henshaw2016b,Clarke2017,Gritschneder2017} demonstrate that other environmental conditions, such as (supersonic) turbulence, accretion, or magnetic fields, can introduce additional fragmentation models.
This means that the filaments are not only supported by thermal pressure, which is reflected by the sound speed in Eq.~(\ref{equ:intro_theory_mlin_infinite}), but also by turbulence:
\begin{equation}	
	M_\mathrm{lin}^\mathrm{turb} = 2 \frac{\sigma_\mathrm{tot}^2}{G} .
	\label{equ:intro_theory_mlin_turb}
\end{equation}
Thereby, the total velocity dispersion, $\sigma_\mathrm{tot}$, is given by the sum of thermal, $\sigma_\mathrm{s}$, and non-thermal velocity dispersion, $\sigma_\mathrm{nt}$:
\begin{equation}
	\sigma_\mathrm{tot}^2 = \sigma_\mathrm{s}^2 + \sigma_\mathrm{nt}^2 .
	\label{equ:intro_theory_sigmatot}
\end{equation}
\pagebreak

\noindent In the case of magnetic fields, \citet{Fiege2000a} show that the critical line mass of a cylindrical filament is increased by,
\begin{equation}
	M_\mathrm{lin}^\mathrm{mag} = M_\mathrm{lin}^\mathrm{crit} \cdot \left( 1 - \frac{\mathcal{M}}{|\mathcal{W}~|} \right)
	\label{equ:intro_theory_mlin_mag}
\end{equation}
where $\mathcal{M}$ and $\mathcal{W}$ are the magnetic and potential energy density, respectively.
However, studies by, for example, \citet{Seifried2015} demonstrate that not only the strength of the magnetic field is relevant for equilibrium state of the cylinder, but also its orientation relative to its elongated axis.

Furthermore, in reality filaments are generally short-lived substructures within globally fragmenting molecular clouds and normally neither quiescent nor isolated.
External gravitational potentials or pressures can introduce additional perturbations that either support or compress the filaments. 
Therefore, the question is whether a quasi-static description of the evolution of filaments is justified.
One way to test this is to use numerical simulations that form filaments self-consistently and compare the properties of the resulting structures with those predicted by the simple model.
In doing so, one can examine how each force, as well as a combination of different forces, affects the evolution of the filaments, and whether a quasi-static model can capture the essential physics involved in the process of fragmentation.

One of such studies is presented by \citet{Federrath2016} has performed AMR simulations of a 3D periodic box representing a single molecular cloud.
In his simulations, \citet{Federrath2016} included different combination of forces (gravity, magnetic fields, turbulence, jet feedback) interacting with each other.
He finds that the filaments that have formed in his simulations have similar properties as observed filaments.
This means that the radial profiles of the filaments are much shallower than predicted by \citeauthor{Ostriker1964b}'s cylinder ($\rho \propto r^{-2.0}$), but also that it has a finite radius, which is not the case in the isothermal cylinder according to \citet{Ostriker1964b}.
However, \citet{Federrath2016} also demonstrate that, as long as gravity does not act alone, both the widths and the central column densities of the filaments are degenerate from the physical processes that have created them, as well as the global star formation efficiency of the parental cloud.
Consequently, those parameters cannot be suitable tracers for the formation and evolution of filaments.

In this chapter, I examine the predictability of the simple cylindrical model on filaments that have formed self-consistently in the Galactic box simulations of \citet[Sect.~\ref{frag:clouds}]{IbanezMejia2016}.
Thereby, I focus on the line masses of filaments and investigate to which extent it reliably reflects the stability state of the filaments.

\section{Model Clouds}\label{frag:clouds}

\begin{figure}[h!t]
	\centering
	\begin{subfigure}{0.32\textwidth}
		\includegraphics[width=\textwidth]{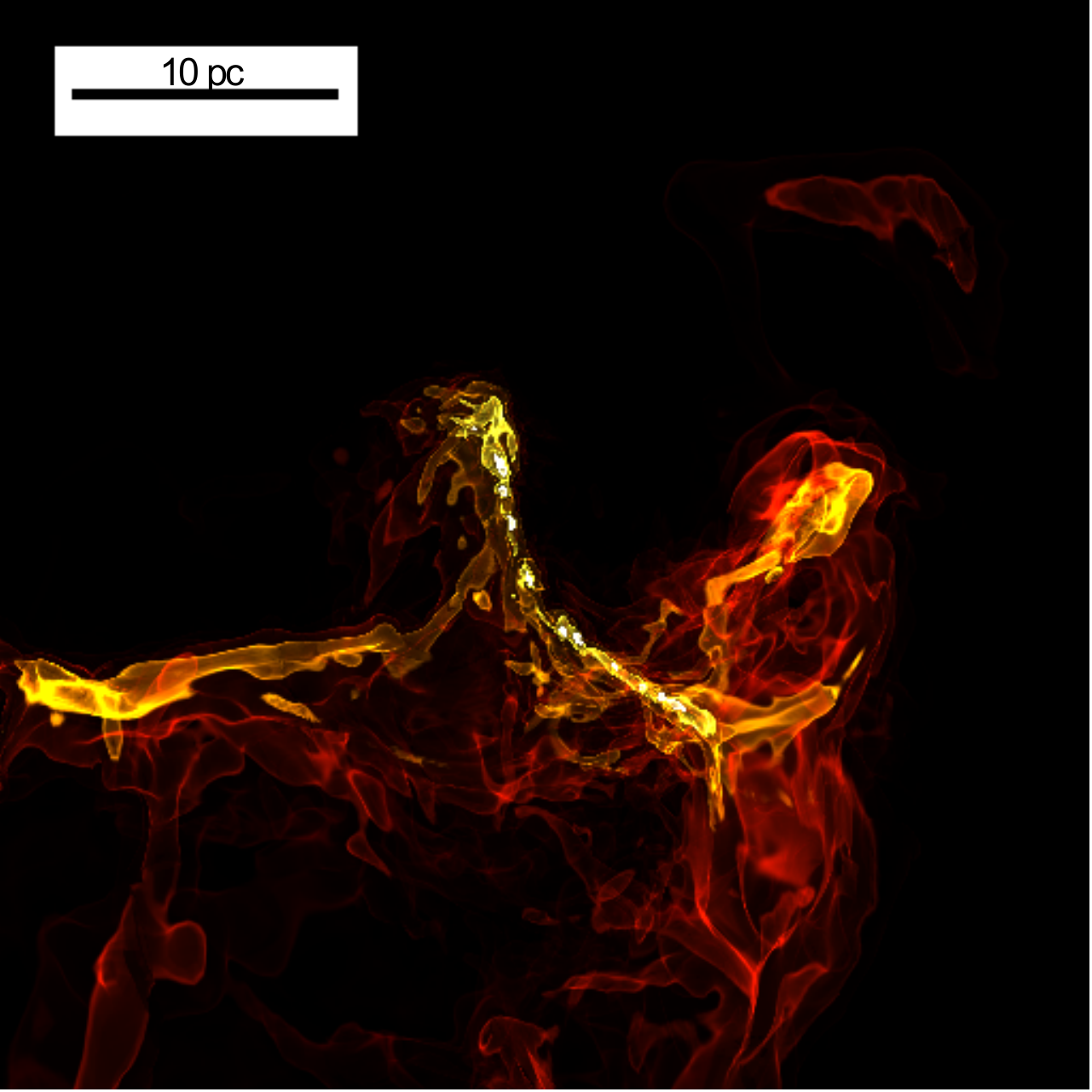}
		\caption{\texttt{M3}, contour colour table}
		\label{pic:fragment_methods_m3e3_0040_3d_v1}
	\end{subfigure}
	\begin{subfigure}{0.32\textwidth}
		\includegraphics[width=\textwidth]{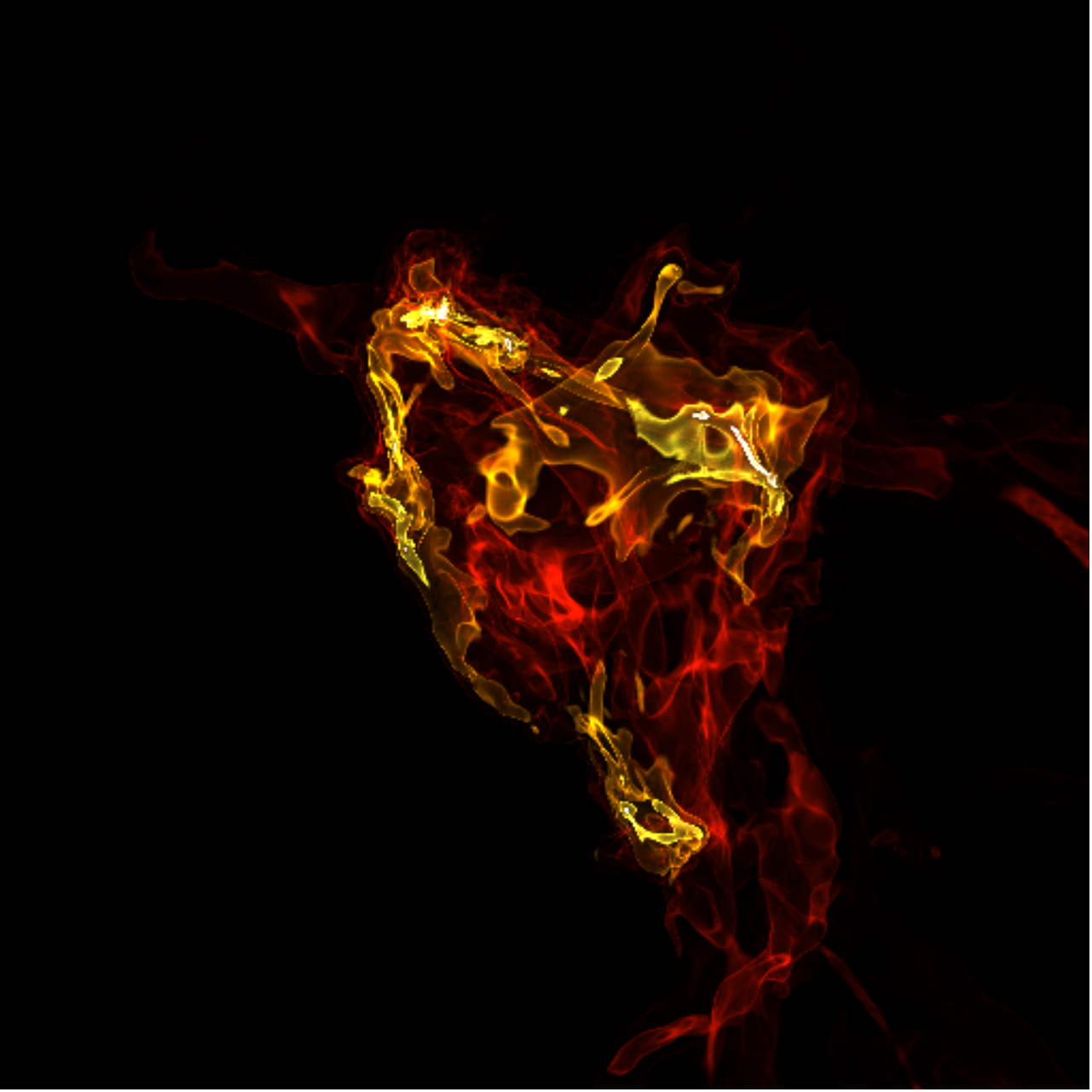}
		\caption{\texttt{M4}, contour colour table}
		\label{pic:fragment_methods_m4e3_0040_3d_v1}
	\end{subfigure}
	\begin{subfigure}{0.32\textwidth}
		\includegraphics[width=\textwidth]{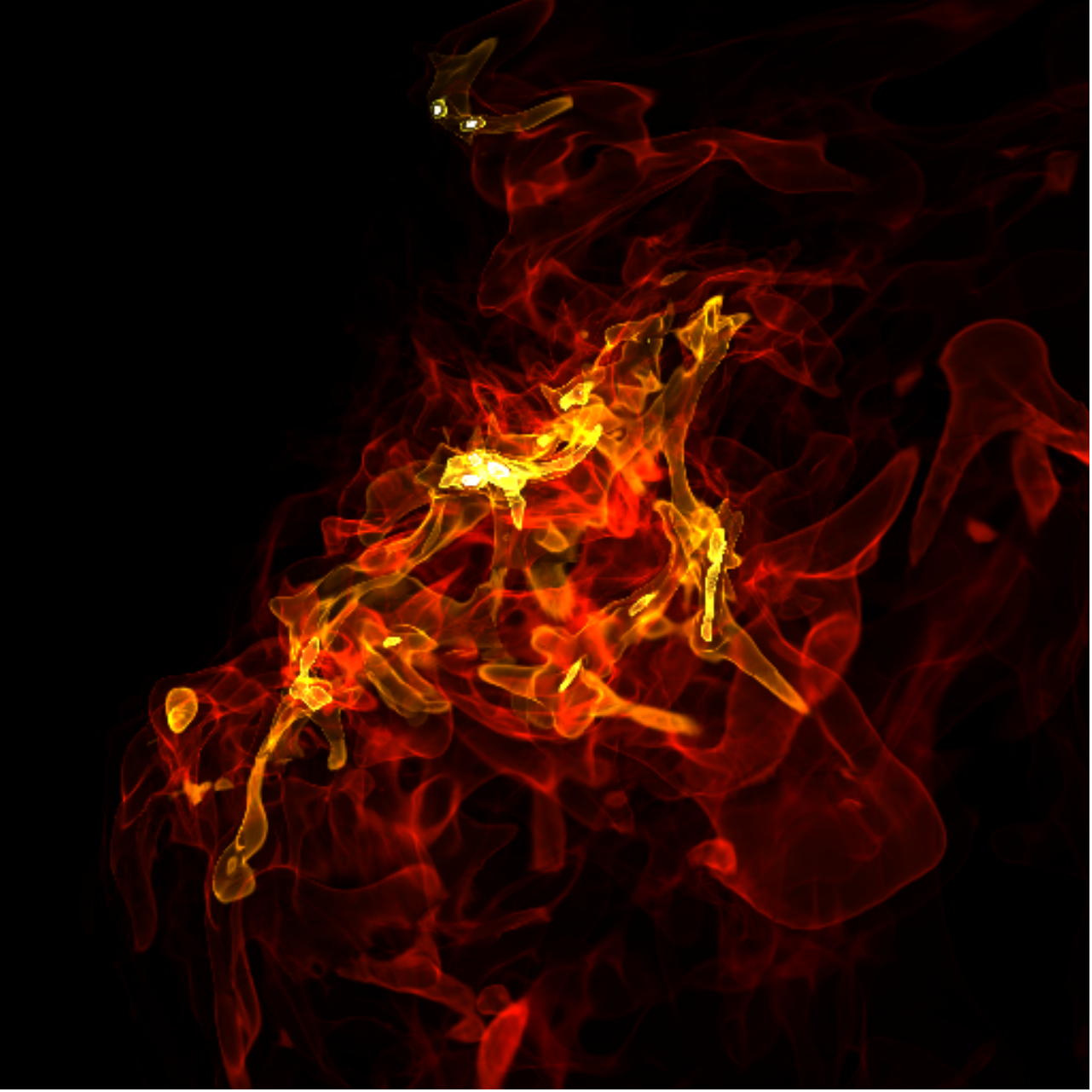}
		\caption{\texttt{M8}, contour colour table}
		\label{pic:fragment_methods_m8e3_0040_3d_v1}
	\end{subfigure}
	
	\begin{subfigure}{0.32\textwidth}
		\includegraphics[width=\textwidth]{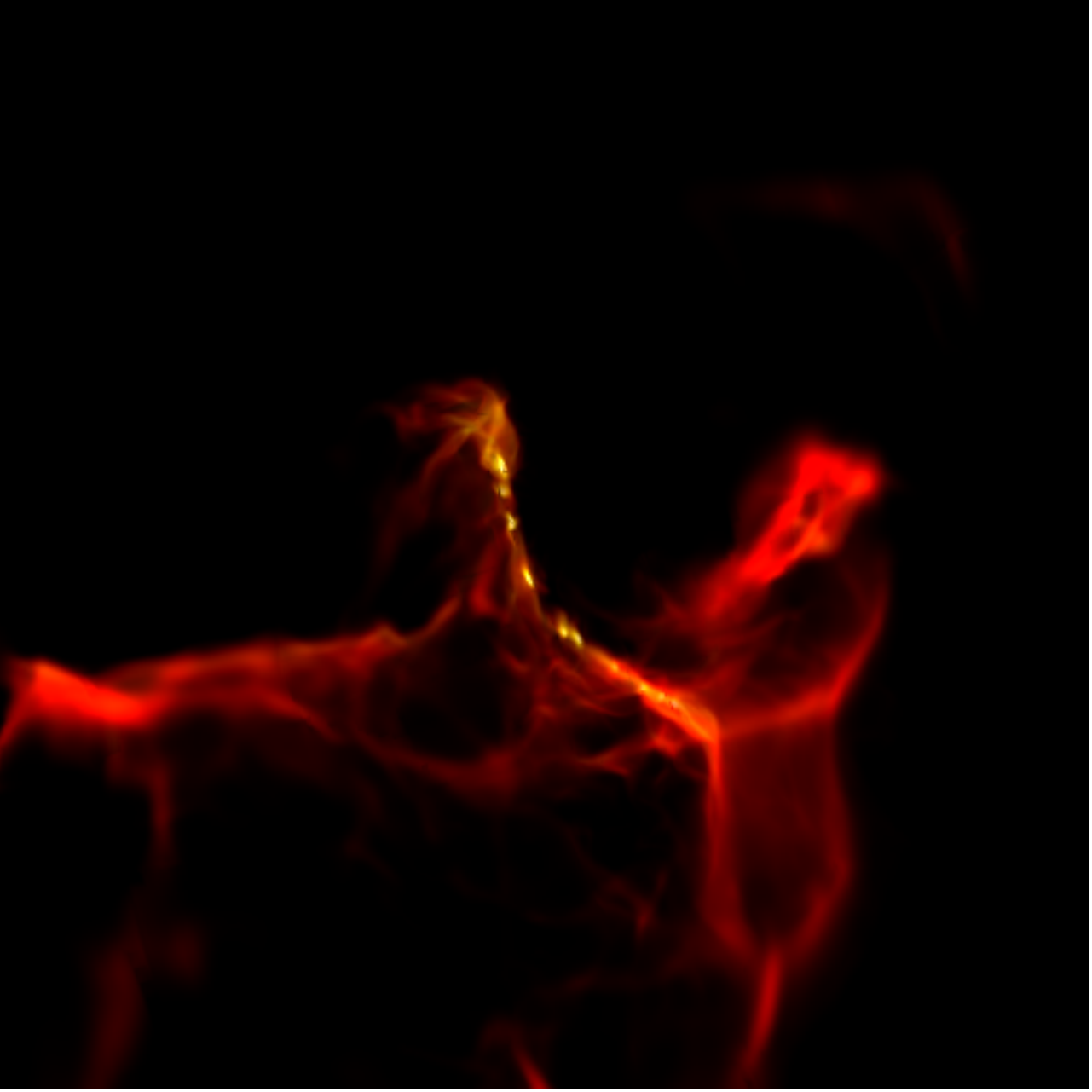}
		\caption{\texttt{M3}, continuous colour table}
		\label{pic:fragment_methods_m3e3_0040_3d_v2}
	\end{subfigure}
	\begin{subfigure}{0.32\textwidth}
		\includegraphics[width=\textwidth]{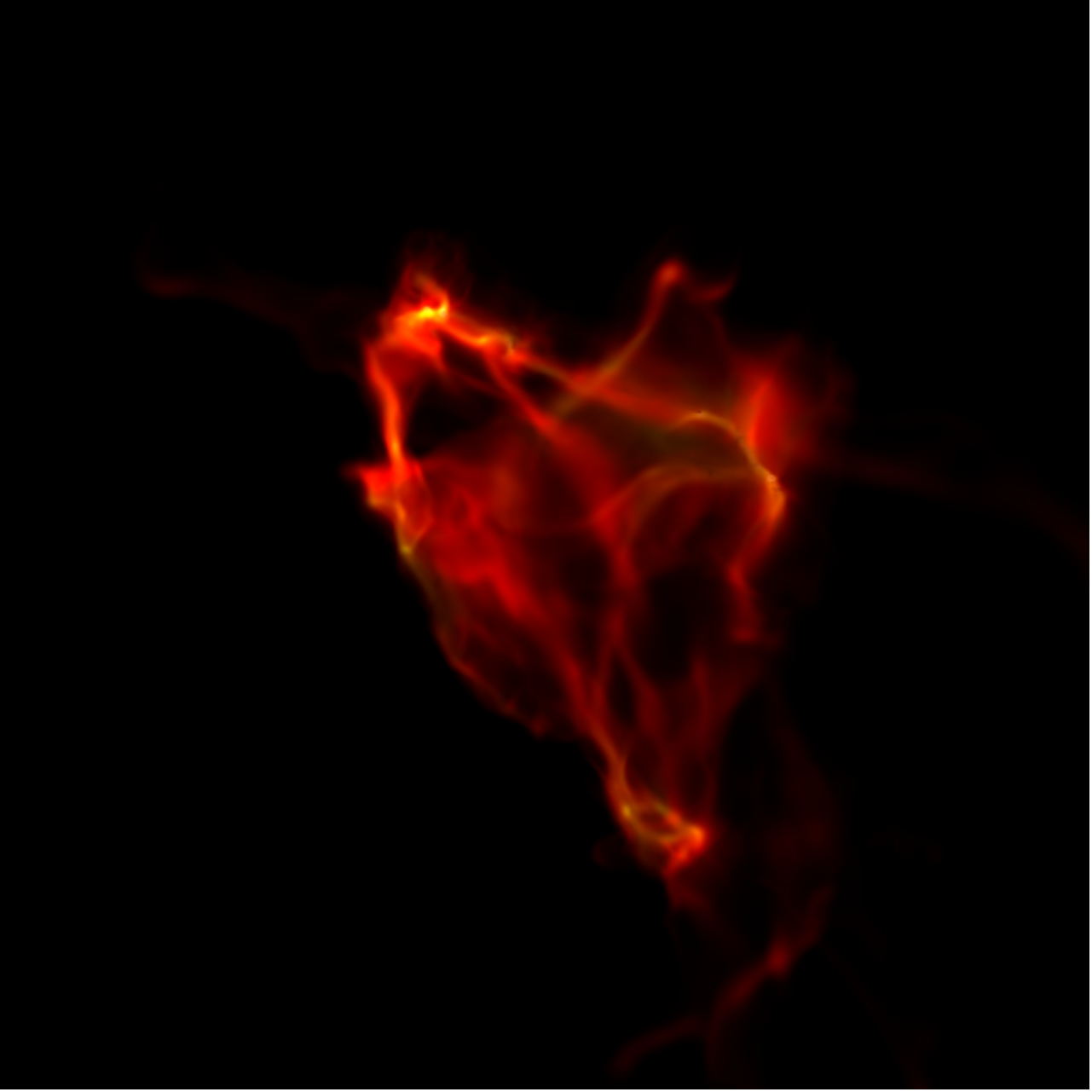}
		\caption{\texttt{M4}, continuous colour table}
		\label{pic:fragment_methods_m4e3_0040_3d_v2}
	\end{subfigure}
	\begin{subfigure}{0.32\textwidth}
		\includegraphics[width=\textwidth]{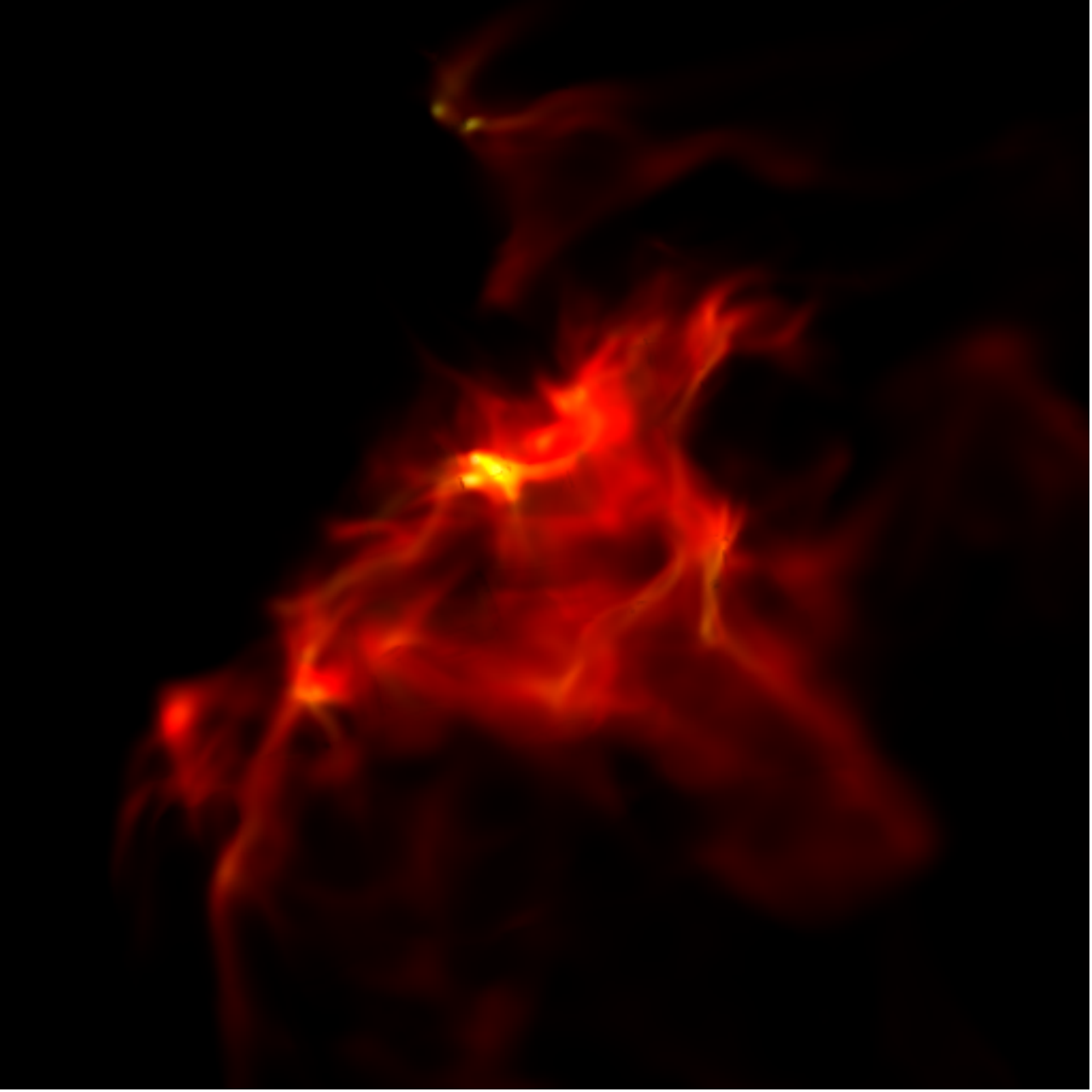}
		\caption{\texttt{M8}, continuous colour table}
		\label{pic:fragment_methods_m8e3_0040_3d_v2}
	\end{subfigure}
	
	\caption[Volume rendered examples of model clouds \texttt{M3}, \texttt{M4}, and \texttt{M8}]{Example volume rendered plots showing \texttt{M3} (\textit{left}), \texttt{M4} (\textit{middle}), and \texttt{M8} (\textit{right}) at $t$~=~4.0~Myr. 
	The colours in the upper panel represent contours at volume densities of 10$^{-23}$~(dark red), 10$^{-22}$ (light red), 10$^{-21}$ (orange), 10$^{-20}$ (yellow), and 10$^{-19}$~g~cm$^{-3}$ (white), while the colour table in the plots in the lower panel is continuous over the same range of volume density.
	}
	\label{pic:fragment_methods_volrender_3d}
\end{figure}

The filamentary structures I analyse in this chapter come from numerical simulations of dense cloud formation by \citet{IbanezMejia2016}.
The simulations have been performed with the 3D magnetohydrodynamics~(MHD), AMR FLASH code \citep{Fryxell2000}, and are of a $1~\times~1~\times~40$~kpc$^3$ vertical box of the ISM in a disk galaxy, with the galactic midplane located in the middle of the box.
Turbulence is driven by the injection of discrete, thermal supernova explosions. 
SN rates are normalized to the galactic SN rates \citep{Tammann1994TheRate}, with 6.58 and 27.4~Myr$^{-1}$~kpc$^{-2}$ for Type~Ia and core-collapse SNe, respectively.
The positioning of SN explosions is random in the horizontal direction and exponentially decaying in the vertical direction with scale heights of 90 and 325~pc for type~Ia and core-collapse SNe.
SN clustering is also taken into account by assuming 3/5 of the core-collapse population are correlated in space and time.
Magnetic fields are included in the simulation, with an initial field strength of 5~$\mu$G at the midplane that exponentially decays in the vertical direction. 
The fields are initially uniform and oriented in the horizontal direction. 
They are allowed to evolve self-consistently with the simulation.
The simulations also include distributed photoelectric heating and radiative cooling.
The photoelectric heating of dust grains \citep{Bakes1994TheHydrocarbons} is implemented with a background FUV intensity of $G_{0}=1.7$ with a vertical scale height of 300~pc.
Radiative cooling rates are appropriate for a solar metallicity, optically thin gas, with a constant ionization fraction of $\chi_{e}=0.01$ for temperatures below $2\times~10^{4}$~K \citep{Dalgarno1972HeatingRegions}, and resonant line cooling for collisionally ionized gas at higher temperatures \citep{Sutherland1993CoolingPlasmas}.
The simulation initially includes only a static galactic gravitational potential, accounting for a stellar disk and a spherical dark matter halo.
The potential is uniform in the horizontal direction; in the vertical direction, for altitudes $z \leq 8$~kpc, it follows a parametrized model of the mass distribution of the Milky Way \citep{Dehnen1998MassWay}, and at higher altitudes, it smoothly transforms into the outer halo profile described by \citet{Navarro1996TheHalos} with a scale length of 20~kpc \citep{Hill2012}.
The simulation runs for $\sim$250~Myr without gas self-gravity in order to develop a multiphase, turbulent ISM, where dense structures form self-consistently in turbulent, convergent flows.
Self-gravity is then switched on to follow the formation and evolution of clouds over the next 6~Myr.

Three dense clouds from the cloud population found in the simulation 
were selected for high-resolution re-simulation by \citet{Ibanez-Mejia2017}. 
Once one of these clouds was identified in the cloud catalogue, a higher-resolution refinement region was defined around the region where the cloud would form in a checkpoint prior to the onset of self-gravity. 
Gas self-gravity was then turned on, and the evolution and collapse of the cloud were followed.
For the investigations, I focus on these three clouds and map a (40~pc)$^{3}$ volume enclosing each cloud with $400~\times~400~\times~400$ grid cells, with an effective spatial resolution of $\Delta x_\mathrm{min}=0.1$~pc.
I consider objects to be fully resolved if their local Jeans length $\lambda_J > 4~\Delta x_\mathrm{min}$, corresponding to a maximum resolved density at 10~K of $8~\times~10^3$~cm$^{-3}$ \citep[e.g.][Eq.~(15)]{Ibanez-Mejia2017}.
This means that I can trace fragmentation down to 0.4~pc, but cannot fully resolve objects that form at smaller scales.
The clouds have total masses on the order of $3~\times~10^3$, $4~\times~10^3$, and $8~\times~10^3$~M$_{\odot}$ (hereafter denoted models \texttt{M3}, \texttt{M4}, and \texttt{M8}).
Examples are shown in Fig.~\ref{pic:fragment_methods_volrender_3d}.

\section{Identification and Characterisation of Filaments and Fragments}\label{frag:prop}

In this section, I provide details of how I identify the filaments and fragments in the model clouds and derive their properties. 
I start by introducing the approaches that are generally used in observational studies (Sect.~\ref{frag:prop_obsfil}).
In Sect.~\ref{frag:prop_filfinder} I present a sample of available filament finder codes and compare their performance to motivate why I use \texttt{DisPerSe} for my further investigations.
Sect.~\ref{frag:prop_mlin} and Sect.~\ref{frag:prop_frag} describe how I derive the properties of the filaments and fragments, respectively.

\subsection{Observational Approaches}\label{frag:prop_obsfil}

Despite the variety of filamentary studies in observations (see Chapter~\ref{intro}), there is no unique definition of what filaments exactly are. 
This leads to a large variety of subsamples that needs to be clearly characterised in every study on filaments and often has no overlap with other samples. 
However, the inaccuracy provides the possibility to detect filaments with a variety of methods, of which the most prominent methods are:
\begin{enumerate}[label=(\Alph*)]
\item Filaments as elongated intensity (or column density) enhanced objects in emission \citep{Wang2015,Li2016}: \label{frag:prop_obsfil_emission} \newline 
	This method takes advantage that filaments consists of dense gas at low temperatures. 
	These are the ideal conditions for the existence of collisionally excited molecules and larger thermally emitting dust grains.
	The emission can then be observed at FIR, sub-mm, and radio wavelengths.
	Emission maps reveal the structure of the filaments and, in the case of molecular emission, the kinematics of the contained gas.
	In principle, this method can be applied for both isolated and in molecular clouds embedded filaments, as well as simulated filaments. 
	In the case of observations, though, observers need to be aware of many uncertainties introduced by, i.e.~the optical depths, the degenerate emission behaviour of dust grains, and the strong correlation between measured intensities and local temperatures.  
\item Filaments as elongated objects that appear obscured relative to a bright background due to absorption \citep{Kainulainen2013a,Ragan2014} \label{frag:prop_obsfil_extinction} \newline
	This method is particularly useful for isolated filaments as they can be easily detected as dark structures against the bright Galactic background.
	Embedded filaments can be detected this way, as well, if they are located between the observer and a recently formed, bright cluster of stars.
	The advantage of this method is that dust absorption does not depend on the local thermal and kinetic conditions that control emission rates. 
	However, the method strongly depends on the properties of dust grains. 
	Although these have been part of many investigations little is known about the exact composition of interstellar dust and cloud-to-cloud variations in abundances of individual dust particle groups. \pagebreak
\item Filaments as connecting bridges of matter between recently formed pre-stellar and protostellar cores \citep{Lenfestey2013,Wang2016} \label{frag:prop_obsfil_mst} \newline
	This method is only applicable to slightly more evolved filaments that already contain pre-stellar and protostellar cores. 
	Assuming that they have formed though the fragmentation of their parental filaments, these cores are supposed to be close to the skeleton of the filament (meaning the main, long axis defining the filament).
	Thus, connecting the cores would resemble the structure of the initial filament.
	This approach clearly has the advantage that it only requires a core catalogue for identifying potential filament candidates by looking for abundance enhancements of cores that have an elongated geometry.
	As the filaments evolve and the cores dissolve the actual skeletons and core-resembled axes diverge from each other, though. 
	Therefore, this method requires a good knowledge about the evolutionary state of the filament and the origin of the contained cores.
\end{enumerate} \smallskip

\noindent Preferentially, at least two of these methods should be applied to characterise and examine filaments. 
Especially the combination of gas emission and dust absorption seems promising.
While dust traces a wide range of density and is less affected by LoS effects, line observations provide insights on the kinematics within the filament, proving that the filament is indeed an coherent object.
For most of the Galactic plane dust emission surveys exist that cover most of the spectrum from NIR to radio wavelengths \citep[e.g., by \textit{Spitzer}, ATLASGAL, or \textit{Herschel};][]{spitzer_Werner2004,atlasgal_Schuller2009,herschel_Pilbratt2010}. 
For these reasons, investigations of dust observations have become popular over the past years.
However, for most of these observations there are no molecular counterpart observations yet.

In this chapter, I mainly detect and analyse the filaments using method \ref{frag:prop_obsfil_emission}.
This means that I study the filaments based on simulated density distribution that are, in projection, comparable to column density maps obtained by dust emission.
In Sect.~\ref{frag:prop_filfinder}, I present different commonly used, publicly available filament finder codes.
Most of them identify filaments according to method \ref{frag:prop_obsfil_emission} or \ref{frag:prop_obsfil_extinction}.
The so-called minimal spanning tree (\texttt{MST}, see Sect.~\ref{frag:prop_filfinder}) is based on method \ref{frag:prop_obsfil_mst}. 
I evaluate the performance of the different codes, and thus of the different approaches, by comparing the structures and mean properties of filaments the individual codes return.

\subsection{Filament Finder Codes}\label{frag:prop_filfinder}

In this part, I introduce a sample of available filament finder codes and compare their performance.
This comparison is necessary because all codes are based on different assumptions.
Thus, it is not clear that the resulting structures are indeed similar and comparable.
Furthermore, there is no standard code that is used by the majority of studies or for creating filament catalogues.
Thus, if one wants to compare the properties of filaments examined by different studies, it is crucial to understand how the properties of those filaments depend on the finder code.

There are a variety of algorithms available for identifying filamentary structures in molecular clouds.
Naturally, there are more filament finders that work with 2D data than with 3D data, since 2D algorithms can directly be applied to observed data like dust intensity or column density maps, while 3D filament finders require data with a resolved third dimension, like the local standard of rest velocity observed spectroscopically.
The latter is not only observationally demanding, but computationally expensive to analyse.
\pagebreak

\noindent Here I offer a summary of considered filament finder codes:
\begin{itemize}
\item The \texttt{Discrete Persistent Structures Extractor} \citep[\texttt{DisPerSe}][]{Sousbie2011a} extracts coherent structures by evaluating the gradients between individual grid cells and the robustness of the topological features found. It was originally written for finding structures (both over- and under-densities) in cosmological data, but can be applied to other applications. The advantage of \texttt{DisPerSe} is that it is independent of the content and dimension of data it receives. Therefore, it can be directly applied to volume density cubes as well as column density maps and returns filamentary structures in both based on the same algorithms.
\item \texttt{FilFinder} \citep{FilFinder_Koch2015} was written to extract filamentary structures in molecular clouds observed by the \textit{Herschel} Gould Belt Survey \citep{Andre2010}. It does this by reducing the areas of interest (parts of molecular clouds with intensities above a specified threshold) to topological skeletons. Therefore, each element of the skeletons represents the medial position of the areas of interest within the boundaries. Unfortunately, \texttt{FilFinder} can only be applied to 2D maps.
\item \texttt{astrodendro}\footnote[7]{\url{http://dendrograms.readthedocs.io}} creates dendrogram trees representing the hierarchical structure of the underlying data. That means that the code reveals how individual regions are connected with each other. Those regions are then classified into trunks, branches, and leaves that represent molecular clouds, clumps and cores in this context. Thus, \texttt{astrodendro} does not identify filamentary structures, but it is well suited for tracing fragments within the filaments or for preparing large data sets, so the actual filament finder can focus on the regions of interest.
\item \texttt{SCIMES} \citep{SCIMES_Colombo2015} is based on dendrograms, as well. However, while \texttt{astrodendro} just detects structures, \texttt{SCIMES} weights the branches and leaves according to user-defined affinities (for example, minimal size, or maximal separation along position or velocity axes) and organises the dendrogram tree accordingly. It returns weighted branches as clusters that, in this context, represent individual filaments. Note that \texttt{SCIMES} itself is not a filament finder, but just returns regions that are likely to contain compact substructures, such as filaments. In order to obtain the filaments one additionally needs a filament finder. \\
	There are some advantages to combining \texttt{SCIMES} and \texttt{DisPerSe}. On the one hand, it is computationally more efficient to first create masks of the relevant regions before applying \texttt{DisPerSe} on those masks. On the other hand, most filament finders return their structures without any weighting. Hence, the users would need to distinguish the filaments from each other by hand, which is impractical for larger datasets like ours. This step can be transferred to \texttt{SCIMES} as well, by applying the masks on the filament finder outputs.
\item The \texttt{Minimal Spanning Trees} \citep[MST, review by][]{Nesetril2001} algorithm is used for optimising costs by minimising the lengths of grids and efficiency of networks. It can also be used to find coherent, filamentary structures as, for example, \citet{Wang2016} demonstrated. With the MST one can define filaments by connecting the leaves found by \texttt{astrodendro} representing pre-stellar cores in molecular clouds according to the requirements. I use this method as an alternative method for identifying filaments in 3D. \texttt{DisPerSe} also uses the MST algorithm to connect individual skeleton segments with each other. The difference here is that I use MSTs for connecting the fragments I identified with \texttt{astrodendro} and define the straight lines between them as filaments. Naturally, the separation between the fragments is on average much larger than those between \texttt{DisPerSe}'s segments, leading to less accurate curvatures.
\end{itemize}

\subsubsection{Comparison of Filaments Identified by Different Filament Finder Algorithms}\label{frag:prop_filfinder_comp}

I compare the structures identified by \texttt{DisPerSe} with those detected by other codes. 
All of them return similar structures in the densest regions of the clouds but show pronounced differences in the more diffuse envelopes.
This behaviour has a significant impact on the measured properties of the filaments. 
The consequence is that studies that do not use the same filament finding techniques are not directly comparable unless it has been proven that the identified structures are indeed similar.

\begin{figure}
	\centering
	\includegraphics[width=\textwidth]{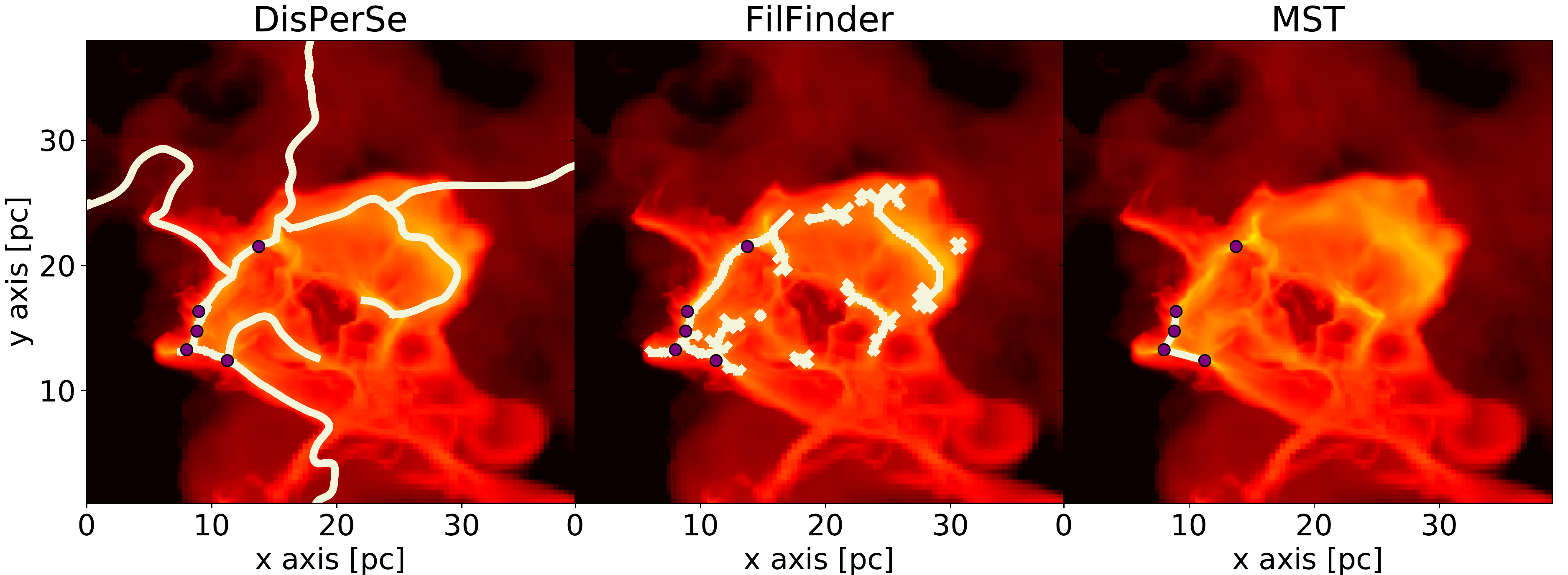}
	\caption[Example for structures identified by \texttt{DisPerSe}, \texttt{FilFinder}, and \texttt{MST}]{Example for structures identified by \texttt{DisPerSe} (\textit{left}), \texttt{FilFinder} (\textit{middle}), and \texttt{MST} applied to fragments from \texttt{astrodendro}  (\textit{right}).
		The background shows the column density map of \texttt{M4} at $t$ = 2.7 Myr along the z axis, ranging between 10$^{18}$ to 10$^{24}$ cm$^{-2}$.
		The purple dots represent the position of cores identified by \texttt{astrodendro}.
		The white lines illustrate filamentary structures found by the three filament finders, applying a column density cut at N$_\mathrm{th} = 1.5 \times 10^{21}$~cm$^{-2}$. 
	}
	\label{pic:frag_filfinder_example_filfinders}
\end{figure}

\begin{figure}
	\centering
	\includegraphics[width=\textwidth]{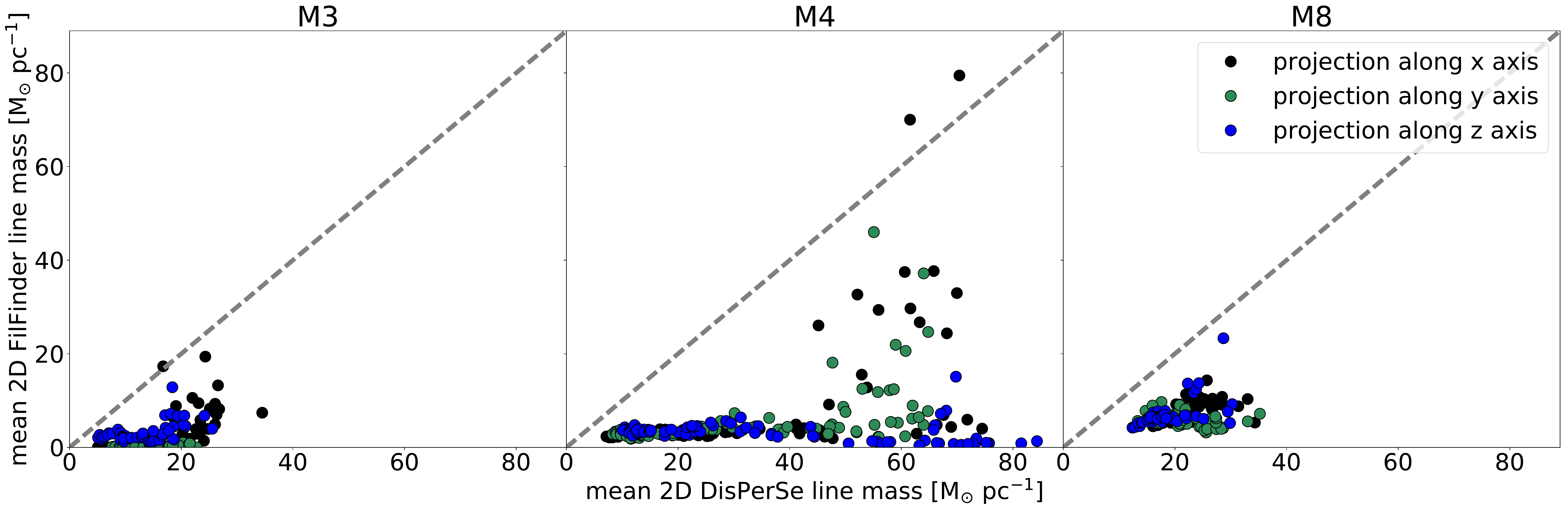}
	\caption[Comparison of mean 2D line masses, $\langle M_{lin,2D}\rangle$ with \texttt{DisPerSe} and \texttt{FilFinder}]{Comparison of mean 2D line masses, $\langle M_{lin,2D}\rangle$, derived with all filaments at the time steps of the simulations based on column density maps.
		The maps have been produced by projecting the volume density cubes along the $x$-axis (black dots), $y$-axis (green dots), and $z$-axis (blue dots).
		The values plotted correspond to the filaments identified by \texttt{DisPerSe} (ordinate) and \texttt{FilFinder} (abscissa).
	}
	\label{pic:frag_filfinder_compml2d_disp_filf}
% \end{figure}
% 
~\vfill

% \begin{figure}
% 	\centering
	\includegraphics[width=\textwidth]{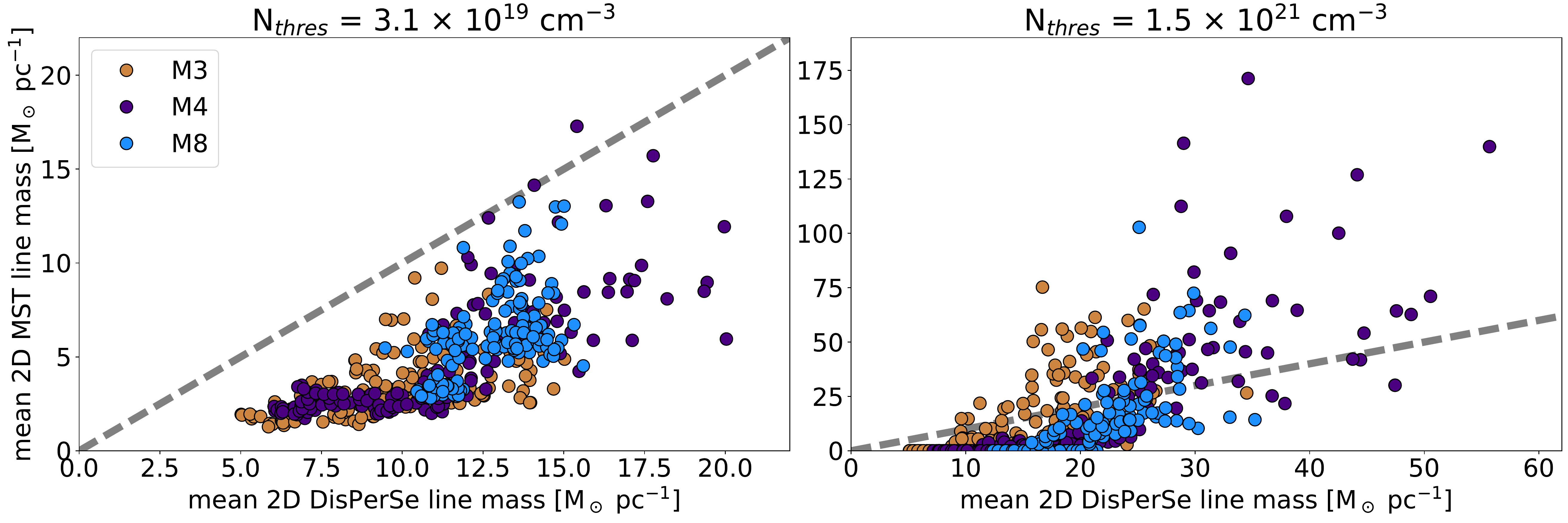}
	\caption[Comparison of $\langle M_{lin,2D}\rangle$ with \texttt{DisPerSe} and \texttt{MST}]{Comparison of $\langle M_{lin,2D}\rangle$ derived with all filaments derived using 2D column density maps at the time steps of the simulations.
		The values plotted correspond to the filaments identified by \texttt{DisPerSe} (ordinate) and by connecting \texttt{astrodendro} cores with \texttt{MST} (abscissa)
	}
	\label{pic:frag_filfinder_compml2d_disp_mst}
% \end{figure}
~\vfill

% 
% \begin{figure}
% 	\centering
	\includegraphics[width=\textwidth]{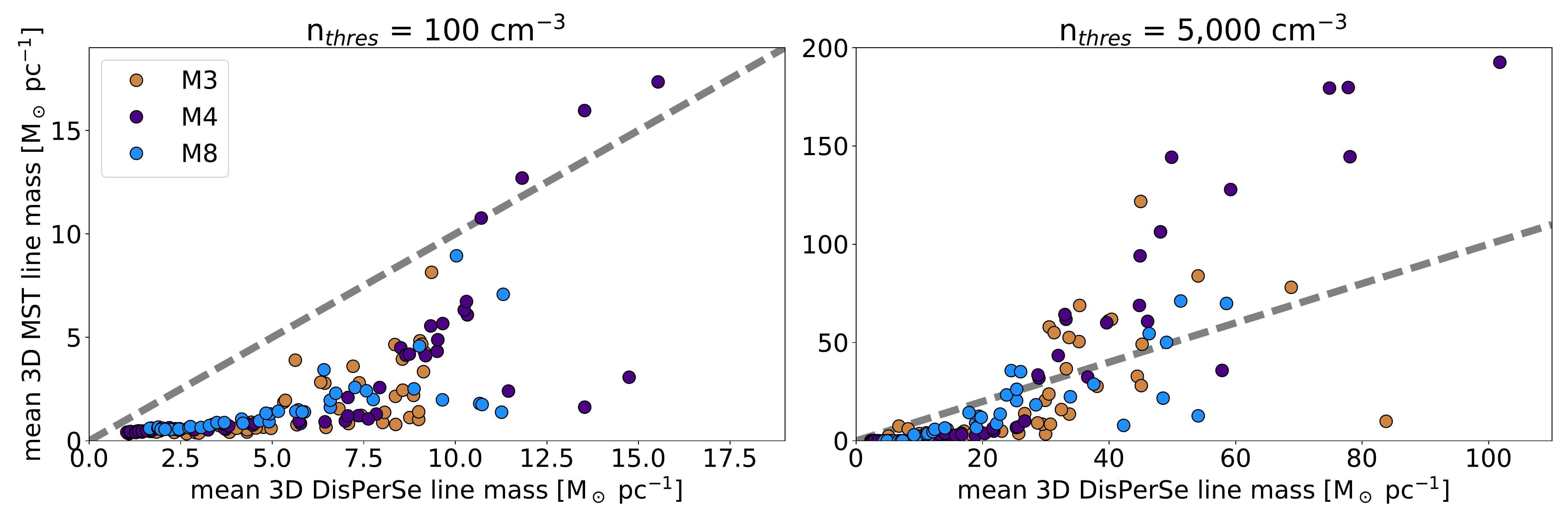}
	\caption[Comparison of $\langle M_{lin,3D}\rangle$ with \texttt{DisPerSe} and \texttt{MST}]{Comparison of $\langle M_{lin,3D}\rangle$ with all filaments derived using volume density cubes at the time steps of the simulations.
		The values plotted correspond to the filaments identified by \texttt{DisPerSe} (ordinate) and by connecting \texttt{astrodendro} cores with \texttt{MST} (abscissa).
	}
	\label{pic:frag_filfinder_compml3d_disp_mst}
\end{figure}

For evaluating and interpreting the results (see Sect.~\ref{frag:thermsupport}), it is essential to understand and compare the performance of the underlying filament finder algorithms. 
Fig.~\ref{pic:frag_filfinder_example_filfinders} gives an example of this.
The background of each panel shows the column density map of the \texttt{M4} model projected along the $z$-axis.
The purple dots represent the position of the fragments identified by \texttt{astrodendro}.
The white lines illustrate the filaments found by \texttt{DisPerSe} (\textit{left}), \texttt{FilFinder} (\textit{center}), and \texttt{MST} (\textit{right}) applied to fragments from \texttt{astrodendro}, when using a column density threshold of N$_\mathrm{th}$~=~1.5~$\times$~10$^{21}$~cm$^{-2}$.
One sees that the codes return widely varying structures, although they agree well where the column density is highest.
Only \texttt{DisPerSe} follows the filaments that connect the cloud to the ISM.
This has a huge impact on the physical properties such as total length, enclosed mass, and the field of interest in general.

Since the different methods do not identify the same structures I cannot compare filaments individually. 
For evaluating the influence the underlying algorithms have on the structures they return, I measure the average line masses of all filaments detected within the respective cloud at a given time (analogously to Sect.~\ref{frag:thermsupport_mean}).
In Figs.~\ref{pic:frag_filfinder_compml2d_disp_filf}, \ref{pic:frag_filfinder_compml2d_disp_mst}, and \ref{pic:frag_filfinder_compml3d_disp_mst} I compare those average line masses based on the filaments returned by the individual codes in 2D and 3D respectively.
one sees that not only does the morphology of the filaments differ significantly using different codes, but also the properties of the structures.

I focus the further analysis on the filamentary skeletons identified by \texttt{DisPerSe} because one can automatically run the code on both 2D and 3D data and work with structures based on the same algorithm and parameter dependence.
Furthermore, \texttt{DisPerSe} is the code that is least sensitive to the input parameters since it is the only code that considers gradients in the matter distribution automatically, giving the skeletons a physical meaning.

\subsection{Filament Properties}\label{frag:prop_mlin}

In order to obtain results that can be qualitatively compared to observations I use \texttt{DisPerSe} to identify the filaments.
I follow the approximation of \citet{IbanezMejia2016} and use two identification thresholds:
(a) a low-density threshold at n$_\mathrm{th}~=~$~100~cm$^{-3}$ which defines the volume of the cloud, and corresponds very roughly to the density at which CO emission can first be detected; and
(b) a high-density threshold at n$_\mathrm{th}~=~$~5,000~cm$^{-3}$ that points to the denser clumps within the cloud that have a high potential for forming stars \citep[see, e.g.,][]{Kainulainen2014}.
Note that I approach the number density resolution (see Sect.~\ref{frag:clouds}) when using this high-density threshold. 
However, I have found that using a lower threshold that is still of the same order of magnitude has no qualitative effect on the structures and properties of the filaments, or their time evolution in general (see also the analysis of the dense gas mass fraction in Sect.~\ref{frag:thermsupport_mean} and Fig.~\ref{pic:frag_fil3d_dgmf}).

\begin{figure}
	\centering
	\begin{minipage}{0.495\textwidth}
			\centering
		  	\includegraphics[width=0.98\textwidth]{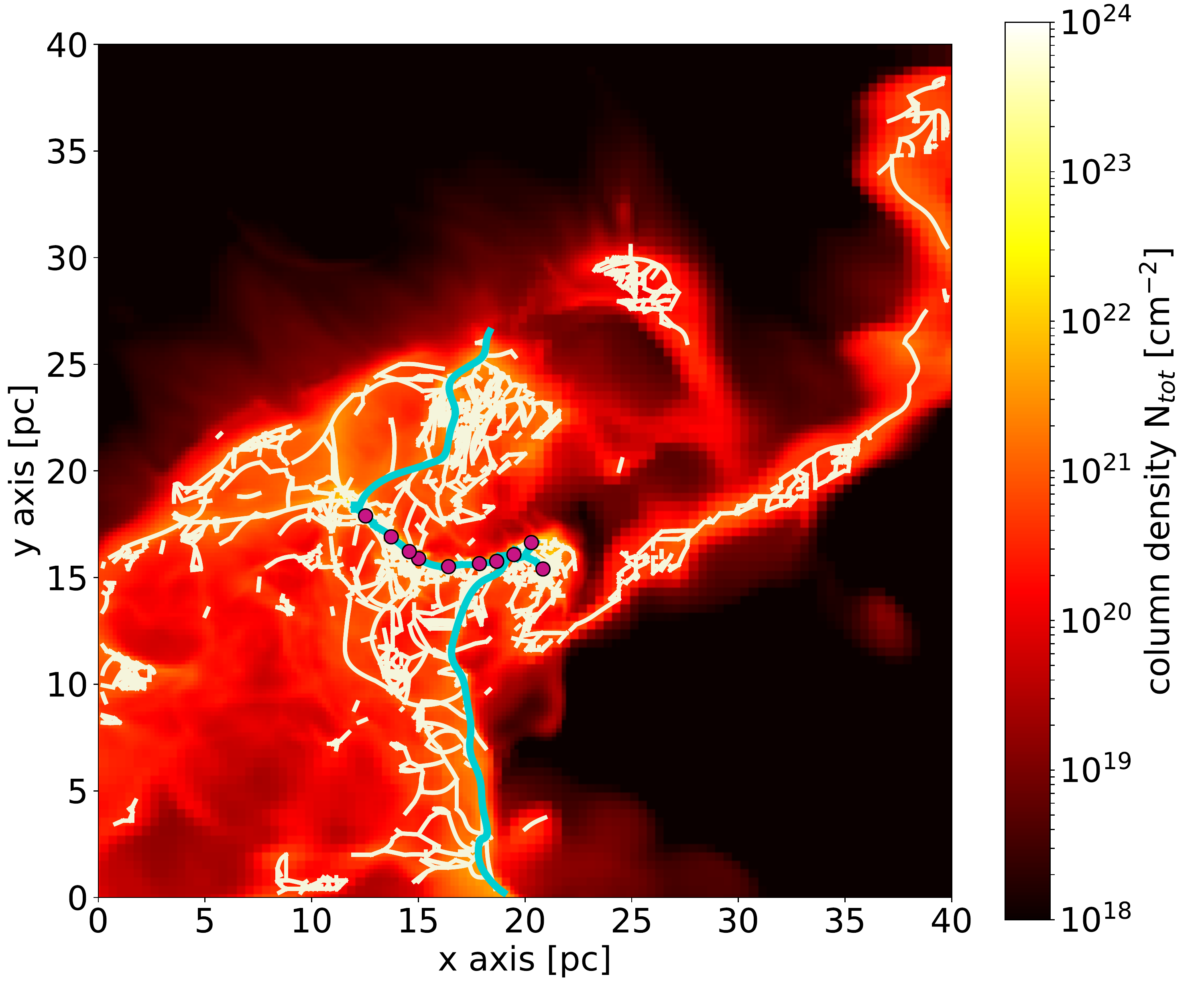}
			\caption[Example showing the effect of density threshold when identifying filaments]{Column density map of \texttt{M3} at $t$~=~3.8~Myr when projected along the z axis with filaments identified using the 3D low density threshold of 100~cm$^{-3}$ (projected, \textit{white lines}), and the 3D high density threshold at 5,000~cm$^{-3}$ (projected, \textit{light blue lines}). The projected positions of identified density fragments are shown with purple dots. }
			\label{pic:frag_example_ntot_lowhigh}
	\end{minipage}
	\begin{minipage}{0.495\textwidth}
			\centering
			\includegraphics[width=0.98\textwidth]{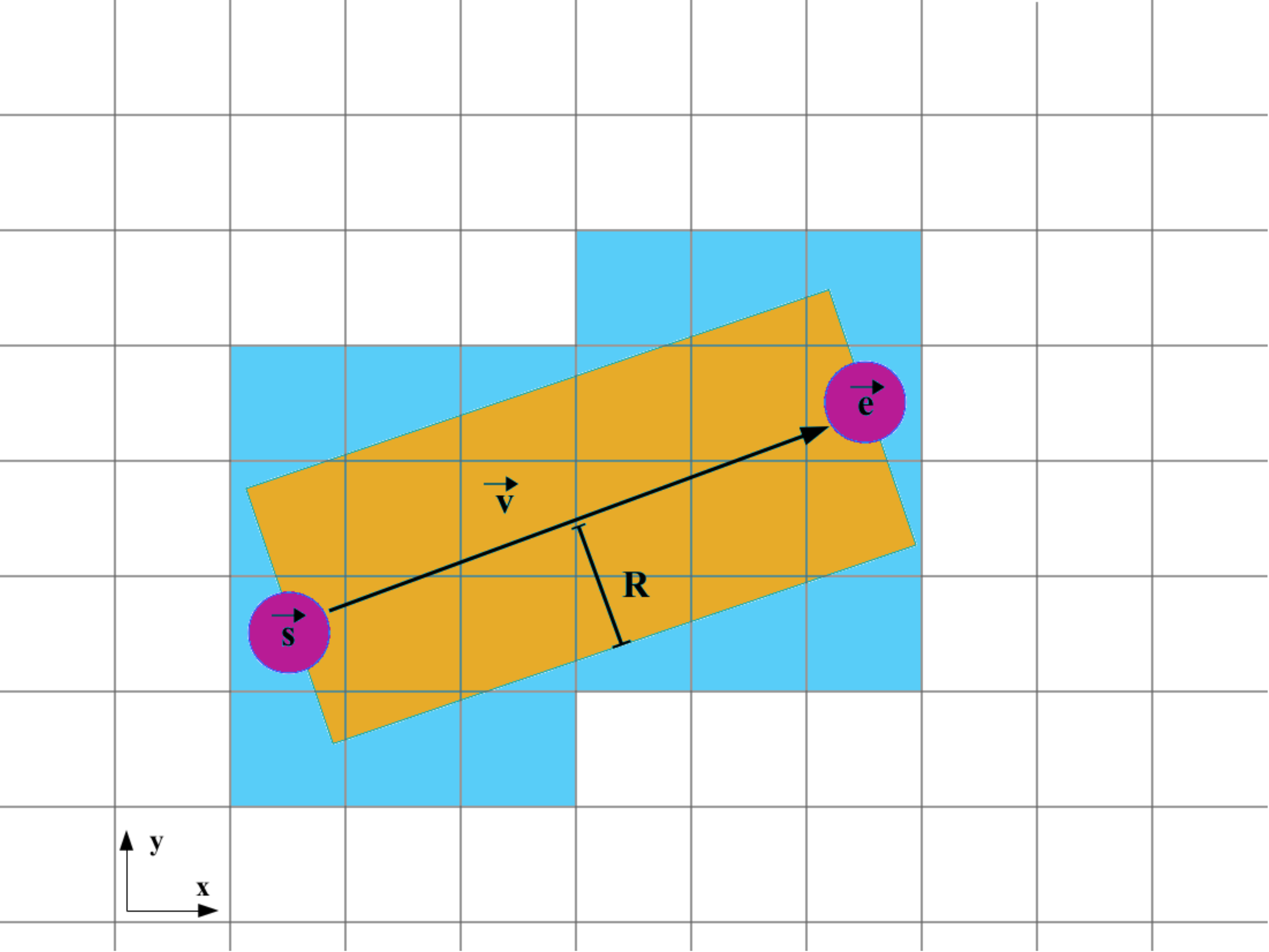}
			\caption[Sketch demonstrating how the properties of filaments are derived]{2D sketch describing how filament volume is defined. The purple circles represent the positions of the starting point $\protect\mathbf{s}$, and end point $\protect\mathbf{e}$ on the data grid. The filament is defined as the orange cylinder with axis $\protect\vec{v}$ and radius $R$. The blue area shows the  grid cells actually used to compute the mass, volume, and other quantities.
			}
			\label{pic:fragment_methods_filident_sketch}
	\end{minipage}
\end{figure}

Fig.~\ref{pic:frag_example_ntot_lowhigh} illustrates with an example how strongly the structures of the identified skeletons depend on the considered threshold.
The map shows a column density map of \texttt{M3} at $t$=~3.8~Myr that has been produced by projecting the volume density cube along the z-axis.
The map illustrates a projection of the filament spines identified by \texttt{DisPerSe} in the 3D position-position- position space. 
Note that I only include the spines of the filaments with densities exceeding the density threshold used for their identification, i.e. 100 or 5000~cm$^{-3}$.
This restriction is necessary because of the way \texttt{DisPerSe} identifies filaments. 
\texttt{DisPerSe} only needs a seed point with a (column) density above the specified threshold, and then connects local maxima through tangent lines to the (column) density gradients. 
If no restriction to the (column) density gradients are set, \texttt{DisPerSe} will eventually connect all local maxima, even if they are below the specified (column) density threshold. 
In this work, I set no restriction to the (column) density gradient for the filament identification. 
However, once the filamentary network was identified, I kept only those filaments whose spines are above our desired (column) density threshold.

\pagebreak

\noindent \texttt{DisPerSe} derives the skeletons of filaments based on the local gradients and user-set thresholds.
It then uses the MST method to find the nearest neighbour for each element of the skeletons.
Most other filament finders neglect this step and leave it to the user to assemble the individual points to filaments, which may not be as consistent as the algorithm \texttt{DisPerSe} offers. 

In the end, \texttt{DisPerSe} returns the starting points $\vec{s}_i$ and end points $\vec{e}_i$ of each segment $i$ of an identified skeleton.
With these points, I calculate the length of the filament,
\begin{equation}
	\ell_\mathrm{tot}~=~\sum_i | \vec{v}_i |~=~\sum_i | \vec{e}_i - \vec{s}_i | \, ,
	\label{equ:methods_ltot_def}
\end{equation}
where $\vec{v}_i$ is the direction vector pointing from $\vec{s}_i$ to $\vec{e}_i$.

I define the volume of the filaments as the set of grid cells $\vec{x}_i$ that are located along the direction vector $\vec{v}_i$ between $\vec{s}_i$ and $\vec{e}_i$, or have any part that lies within a distance $R$ to these cells.
Fig.~\ref{pic:fragment_methods_filident_sketch} illustrates this in a 2D example. 
The orange area represents the analytic volume of the filament while the blue area marks the grid cells that are actually used.

The range of filament widths that I find is wide, partly due to the complexity of profiles (e.g., from crossing filaments) that makes it difficult to measure relevant quantities like the full-width half maximum (FWHM). 
For simplicity, I assume that the radius of the filaments is $R~=~0.3$~pc everywhere.
This value agrees with average filament widths I find when fitting Gaussians to the line profiles of the filaments.
Yet, numerical resolution limits prevent filaments in the model from collapsing to smaller extensions.	
The enclosed mass of a filament is given by,
\begin{equation}
	M_\mathrm{tot}~=~\sum_{\vec{x}_i \, \in \, \mathrm{filament}} \rho(\vec{x}_i) \cdot \Delta x_\mathrm{min}^3 \, ,
	\label{equ:methods_mtot_def}
\end{equation}
where $\rho$ is the mass density within the grid cell $\vec{x}_i$; and the line mass by,
\begin{equation}
	M_\mathrm{lin}~=~\frac{M_\mathrm{tot}}{\ell_\mathrm{tot}} .
	\label{equ:fragment_prop_mlin}
\end{equation}

\subsection{Fragment Identification and Properties}\label{frag:prop_frag}

To identify fragments, I use \texttt{astrodendro} (see Sect.~\ref{frag:prop_filfinder}).
This method computes dendrogram trees that represent the hierarchical structure of the underlying matter distribution in order to unravel how sub-structures relate to each other.
This way one can easily identify individual clumps, filaments, and fragments.

I identify fragments as highest level leaves in the dendrograms derived with a minimal density threshold at \texttt{min\_value}~=~5,000~cm$^{-3}$ and a minimum of \texttt{min\_npix}~=~20~cells.
I define the volume of the fragment as a sphere with a radius of $R_\mathrm{f}$~=~0.3~pc around the central position of the respective dendrogram leaf.

In order to find evidence for the fragments in the simulation forming by gravitational fragmentation following the cylindrical fragmentation criterion, I make a first estimate of how gravitationally bound the fragments are.
For this, I compute their virial parameters, given by \citep{Bertoldi1992}:
\begin{equation}
	\alpha~=~\frac{\mathrm{2} E_\mathrm{kin}}{E_\mathrm{pot}}~=~\frac{\mathrm{5} \sigma R_\mathrm{f}}{G M_\mathrm{f}}
	\label{equ:frag_virial_parameter}
\end{equation}
for each of the fragments at any time step.
Here, $G$ is the gravitational constant, $M_\mathrm{f}$ the enclosed mass of the fragment, $R_\mathrm{f}$~=~0.3~pc its radius and $\sigma$ its velocity dispersion, which I calculate following Eqs.~(9)~\&~(10) in \citet{IbanezMejia2016}:
\begin{equation}
	\sigma^\mathrm{2}~=~\frac{1}{3} \, \frac{\sum_{|\vec{x}-\vec{x}_0| < \mathrm{R}_f} \, \rho(\vec{x}) \, (\vec{v} - \bar{\vec{v}})^2}{\sum \rho(\vec{x})} + \bar{c}_\mathrm{s}^2 ,
\end{equation}
with $\vec{x}_0$ being the central position of the fragment, $\bar{\vec{v}}$ the average gas velocity vector and $\bar{c}_\mathrm{s}$ the average sound speed.
Note that this estimate only takes the ratio between volumetric kinetic and gravitational energy into account.
Other terms, such as magnetic fields or thermal or kinetic surface pressure \citep{McKee1992}, are neglected. 
This is a common approximation used in both observational and theoretical studies, although \citet{Ballesteros-Paredes2006} suggest that these additional terms can be as important for objects embedded within molecular clouds as the balance of volumetric kinetic and gravitational energy.

\section{Thermal Support and Fragmentation of Filaments}\label{frag:thermsupport}

I apply \texttt{DisPerSe} on the 3D volume density cubes extracted from the simulations of all three model clouds for all simulated time steps.
In the following discussion, I investigate the fragmentation behaviour of those filaments based on the 3D data, looking on both, the average of all filaments (Sect.~\ref{frag:thermsupport_mean}) and the behaviour of individual filaments (Sect.~\ref{frag:thermsupport_indv}).  
Sect.~\ref{frag:thermsupport_frags} provides an analysis of the fragments' properties and their spacing manner.
In Sect.~\ref{frag:thermsupport_comp3d2d}, I target the question how our previous results based on the full 3D data relate to what would be observed based on 2D projections. 

\subsection{Mean Properties and Evolution of 3D Filaments}\label{frag:thermsupport_mean}

I begin the analysis by studying the average properties and time evolution of the filaments I find in the model clouds in 3D, and how those properties relate to the global characteristics of the surrounding clouds.

\begin{figure}
	\centering
	\includegraphics[width=\textwidth]{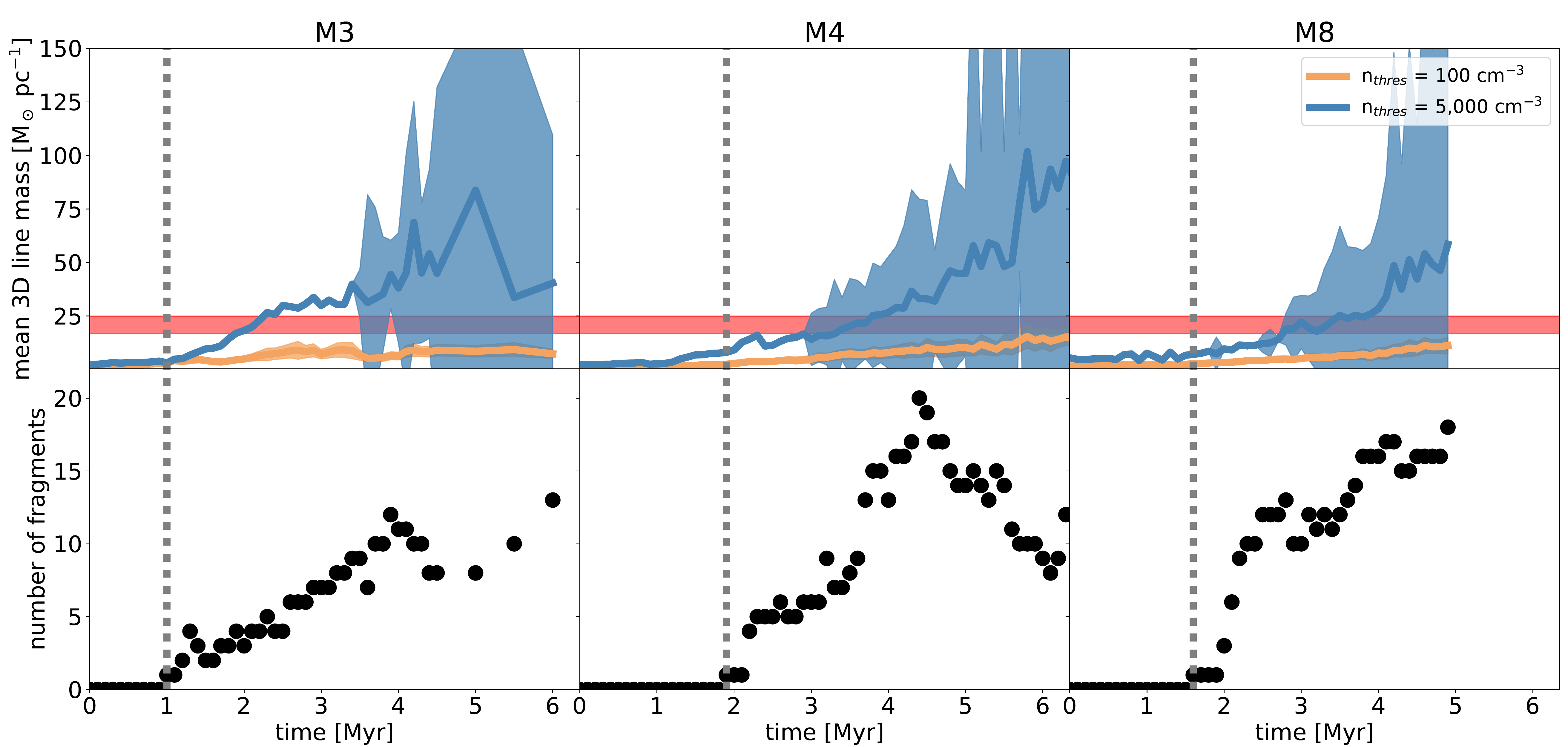}	
	\caption[Average line masses of 3D filaments]{Average line mass of all filaments $\langle M_\mathrm{lin,3D} \rangle$ (\textit{top}) and number of fragments (\textit{bottom}) found in models \texttt{M3} (\textit{left}), \texttt{M4} (\textit{middle}), and \texttt{M8} (\textit{right}) as functions of time.
		The coloured areas illustrate the 1$\sigma$ range.
		The red area indicates the critical value of $M_\mathrm{lin,3D}$ for an isothermal cylinder with temperatures between 10--15~K.
		The grey dashed lines mark the times when the first fragments in each simulation form.
	}
	\label{pic:frag_numfrags}
\end{figure}

In the top panel of Fig.~\ref{pic:frag_numfrags} I show the average line mass $\langle M_\mathrm{lin,3D} \rangle$ of all the filaments identified in each of the three clouds as function of time $t$ after self-gravity has been activated in the simulations.
one sees that the average line masses within all clouds increase with time and with increasing identification threshold $n_\mathrm{th}$.
This is expected considering the structures I obtain when using different thresholds. 
When I apply a high threshold I obtain compact structures connecting the sites of first fragment formation, where the gas is more concentrated.
In contrast, with a lower threshold I identify a larger number of low mass structures in more diffuse regions of the clouds (see Fig.~\ref{pic:frag_example_ntot_lowhigh}).
As a result, the average line mass of low density filaments is always lower compared to using a higher threshold.

\begin{figure}
	\centering
	\includegraphics[width=\textwidth]{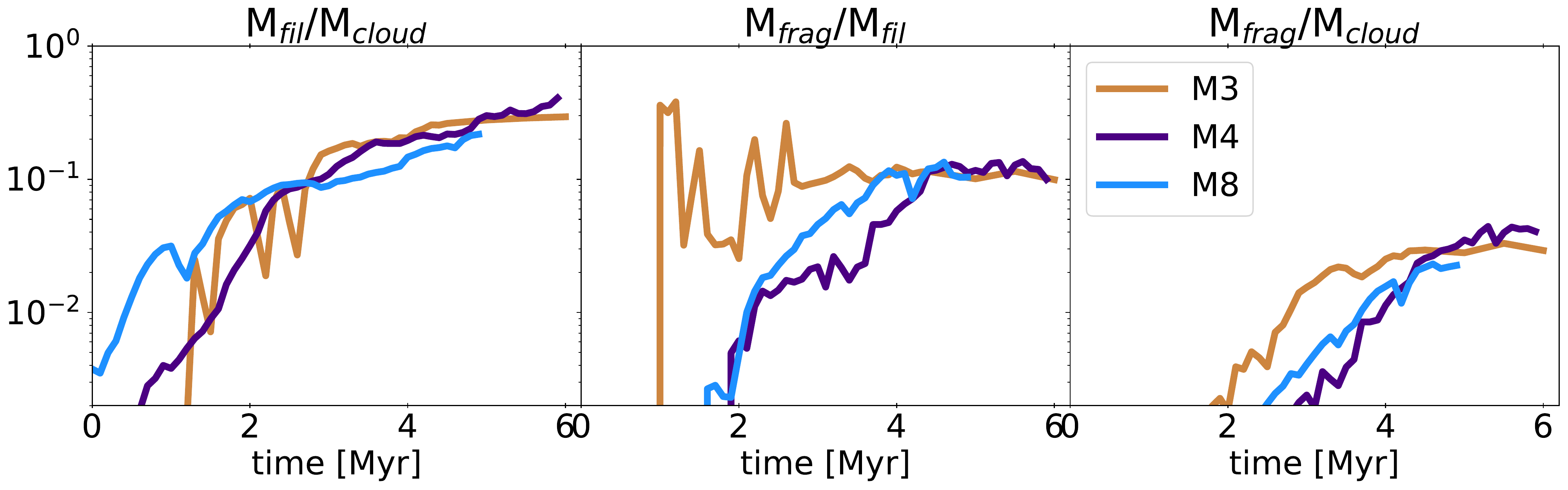}
	\caption[Time evolution of mass fractions of clouds, filaments and fragments]{Time evolution of ratios of filament and fragment masses $M_\mathrm{fil}$ and $M_\mathrm{frag}$ with respect to cloud masses  $M_\mathrm{cloud}$ and each other, shown for the objects within \texttt{M3}~(\textit{brown}), \texttt{M4}~(\textit{purple}), and \texttt{M8}~(\textit{blue}).
		The filaments considered have been identified with $n_\mathrm{th}=100$~cm$^{-3}$. 
	}
	\label{pic:frag_massfrac}
	\vspace{\baselineskip}
	
	\vfill

	\includegraphics[width=\textwidth]{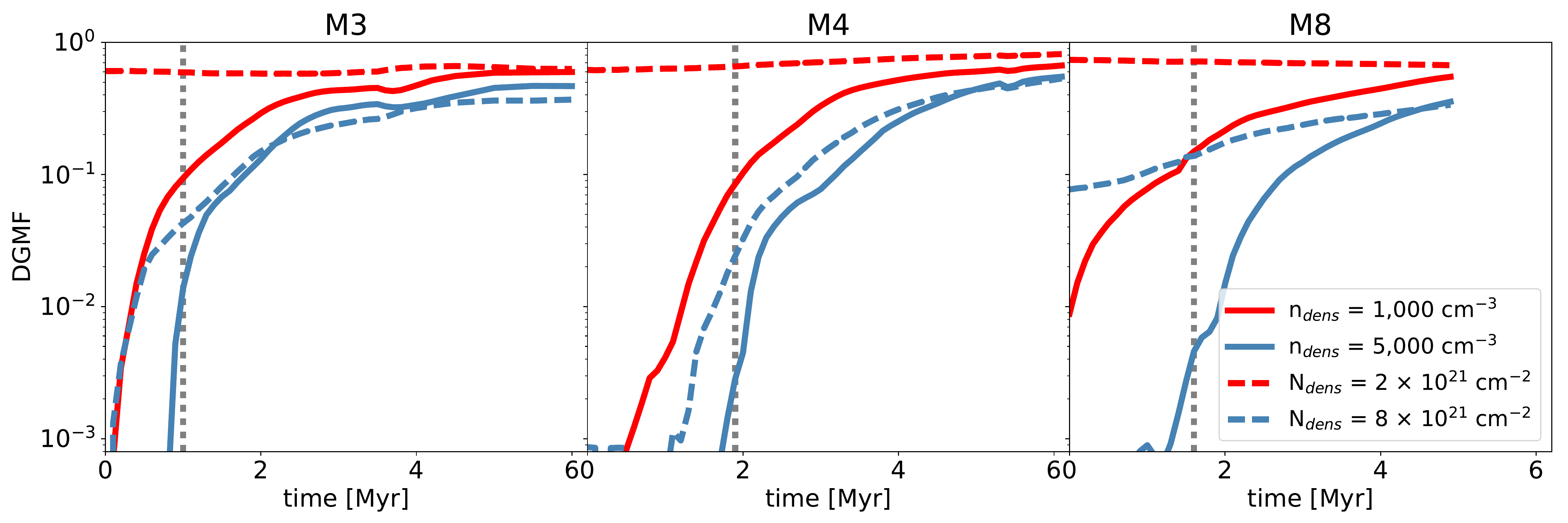}
	\caption[Time evolution dense gas mass fraction]{Evolution of DGMF in the model clouds as a function of time. Solid lines give DGMFs measured from 3D volume density cubes, while dashed lines are from 2D column density maps. 
Red lines correspond to dense gas above number and column density thresholds of 1000~cm$^{-3}$ and $2~\times~10^{21}$~cm$^{-2}$, and blue lines correspond to dense gas above 5000~cm$^{-3}$ and $8~\times~10^{21}$~cm$^{-2}$, respectively.
		The grey dashed lines mark the times when the first fragments in each simulation form.
	}
	\label{pic:frag_fil3d_dgmf}
	\vspace{\baselineskip}

	\vfill

	\includegraphics[width=\textwidth]{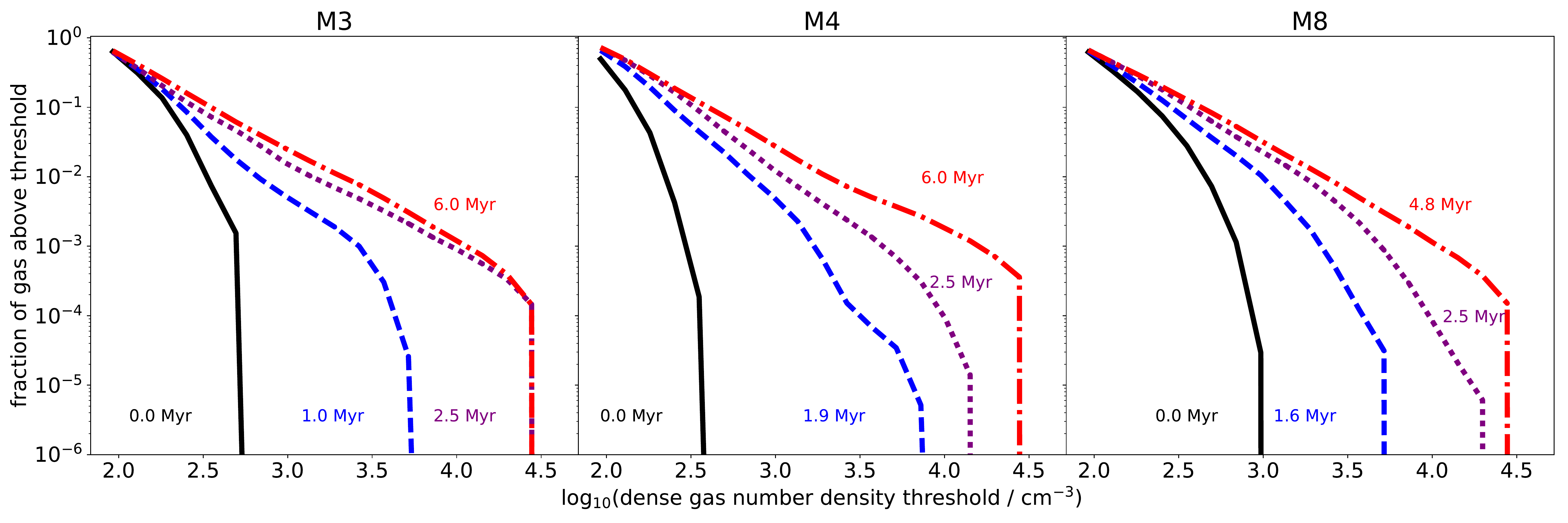}
	\caption[Time evolution of fraction of gas above given number density threshold]{Fraction of gas above the given gas number density threshold (ordinate), $n_\mathrm{dens}$, within \texttt{M3}~(\textit{left}), \texttt{M4}~(\textit{middle}), and \texttt{M8}~(\textit{right}). 
		The lines show the distribution at the time self-gravity has been activated, $t=$0~Myr (black solid line), the time the first fragments are detected (blue dashed line), $t=$2.5~Myr (violet dotted line), and the final snapshot (red dashed-dotted line).
	}
	\label{pic:frag_rho_pdf}
\end{figure}

The increase of average line masses over time can be understood from the fact that the clouds formed by compression of gas, due to supernova shocks and global turbulent motions, in the time before the analysis. 
Thus, at $t=0$~Myr, when self-gravity is activated, the clouds are already highly self-gravitating and begin to collapse globally.
As a result, the filaments within the clouds gain more mass with time, which increases their line masses.
Only in the case of \texttt{M3} does the growth of average line masses of the low-density filaments stagnate after the first megayear. 
However, the average line mass of the high-density filaments  in all clouds steadily increases in time, suggesting that the filaments are collapsing gravitationally.
Fig.~\ref{pic:frag_massfrac} shows a clearer picture of how the mass within the cloud is distributed, and its time evolution.  
The top panel shows the fraction of mass contained in filaments, $M_\mathrm{fil}$, compared to the mass of the entire parental cloud, $M_\mathrm{cloud}$. 
In all clouds, the fraction rapidly and continuously increases in time.
Since the clouds accrete mass from their environments, the growth of the filament mass fraction means that filaments are forming and growing at rates not directly correlated to the evolution of their parental clouds.
A similar trend is observed in the mass fraction of the fragments with respect to the cloud mass for all three clouds (bottom panel of Fig.~\ref{pic:frag_massfrac}). 
However, the middle panel showing the mass ratio of filaments to cores shows a steep rise when cores are formed, but stagnates at later times suggesting that filaments and cores grow at the same rate.

In order to confirm this behaviour, I take a closer look at the dense gas mass fraction (DGMF) of the clouds.
I define the DGMF as the fraction of mass enclosed in cloud cells that contain gas above a given dense gas threshold, n$_\mathrm{dens}$, compared to the mass of the entire cloud:
\begin{equation}
	\mathrm{DGMF}(t,n_\mathrm{dens}) ~=~\frac{ \int_V n(t,\vec{x}) \, G(n_\mathrm{dens}) \,  dV}{\int_V n(t,\vec{x}) \, G(\mathrm{100~cm}^{-3}) \,  dV} \, ,
	\label{equ:frag_dgmf}
\end{equation}
with $n$ being the number density at the time $t$ and
\begin{equation}
	G(n_0)~=~\begin{cases}
		1 & n(t, \vec{x}) \geq n_0 \\
		0 & n(t, \vec{x}) < n_0
	\end{cases} .
	\label{equ:frag_g_n}
\end{equation}
Fig.~\ref{pic:frag_fil3d_dgmf} shows the evolution of the DGMFs of the three simulated clouds using two different dense gas thresholds, namely $n_\mathrm{dens}$~=~1,000~cm$^{-3}$ and $n_\mathrm{dens}$~=~5,000~cm$^{-3}$.

In all clouds one sees that the DGMF continues growing until the end of the simulation, with maximal values of 55--70\% for $n_\mathrm{dens}$~=~1,000~cm$^{-3}$ and 35--55\% for $n_\mathrm{dens}$~=~5,000~cm$^{-3}$. 
These clouds continue to accrete mass from their surroundings
\citep[see Figs.~3, 8, and 13 in][]{Ibanez-Mejia2017}, so I conclude that these clouds are collapsing faster than they are growing. 

One sees a similar behaviour in Fig.~\ref{pic:frag_rho_pdf} that shows the fraction of gas above a given number density threshold as a function of this number density threshold, $n_\mathrm{dens}$, for four individual time steps in the simulations.
In all clouds one sees that the maximum gas density grows with time until it reaches the limit of the numerical resolution.
Note that although I reach densities of order 30,000~cm$^{-3}$, I only resolve densities up to 8,000~cm$^{-3}$ reliably.
At the same time, the density distributions become flatter as the clouds evolve.
This means that the clouds collapse in their entirety and compress the enclosed mass into denser substructures such as fragments without necessarily losing their diffuse envelopes.
This likely occurs because of continuing accretion of diffuse gas from the surrounding ISM \citep{Ibanez-Mejia2017}.

In summary, one sees that the average evolution of the filaments is influenced by the global kinematics of their parental cloud. In particular, the way that the cloud transforms the mass it accretes from the ISM into dense substructures is related to the formation of fragments within the filaments. 
However, I also see that the properties of the dense gas strongly depend on the parameters used to define and identify it, such as the threshold density. 
This might become important for the key question of how well analytic models evaluate the stability of filaments and predict their fragmentation behaviour.

\begin{table}
	\centering
	\begin{tabular}{l|cccc}
		Quantity	& min	& mean	& median	& max \\ \hline
		\multicolumn{5}{l}{n$_\mathrm{th}$ = 100 cm$^{-3}$} \\ \hline
		Number density n [10$^{3}$ cm$^{-3}$]			&   0.1	&   1.7	&  0.5	& 2079	\\ 
		Gas temperature $T_\mathrm{gas}$ [K]			&  10.2	&  11.8	& 11.3	&   16.3	\\ 
		Total length $\ell_\mathrm{tot}$ [pc]			&   1.8	& 17.8	& 11.2	& 133.2		\\ 
		Total mass $M_\mathrm{tot}$ [M$_\odot$]			&   0.4	& 107.0	&  8.2	& 1138		\\ 
		Line mass $M_\mathrm{lin,3D}$ [M$_\odot$ pc$^{-1}$]	&   0.2	&   2.3	&  0.8	&   17.3	\\ \hline 
		\multicolumn{5}{l}{n$_\mathrm{th}$ = 5,000 cm$^{-3}$} \\ \hline
		Number density n [10$^{3}$ cm$^{-3}$]			&  2.5	&  26.0	&  13.5	& 2116		\\ 
		Gas temperature $T_\mathrm{gas}$ [K]			& 10.2	&  11.8	&  11.3	&   16.3	\\ 
		Total length $\ell_\mathrm{tot}$ [pc]			&  1.8	&   5.6	&   5.0	&   15.0	\\ 
		Total mass $M_\mathrm{tot}$ [M$_\odot$]			& 28.8	& 553.6	& 409.6	& 2553		\\ 
		Line mass $M_\mathrm{lin,3D}$ [M$_\odot$ pc$^{-1}$]		& 13.2	&  97.0	&  73.3	&  707.6	\\ \hline 
	\end{tabular}
	\caption{Summary of the properties of the 3D filaments.}
	\label{tab:fragment_results_prop3d}
\end{table}
%\vspace{-\baselineskip}

\subsection{Properties and Evolution of individual 3D Filaments}\label{frag:thermsupport_indv}

In this section, I investigate the evolution of individual filaments and compare the properties of fragmenting filaments with others. 
I confront these properties with the predictions of analytic models.
Table \ref{tab:fragment_results_prop3d} provides a summary of the properties of the examined filaments. 

\begin{figure}[h!t]
	\centering
	\begin{subfigure}{\textwidth}
		\includegraphics[width=\textwidth]{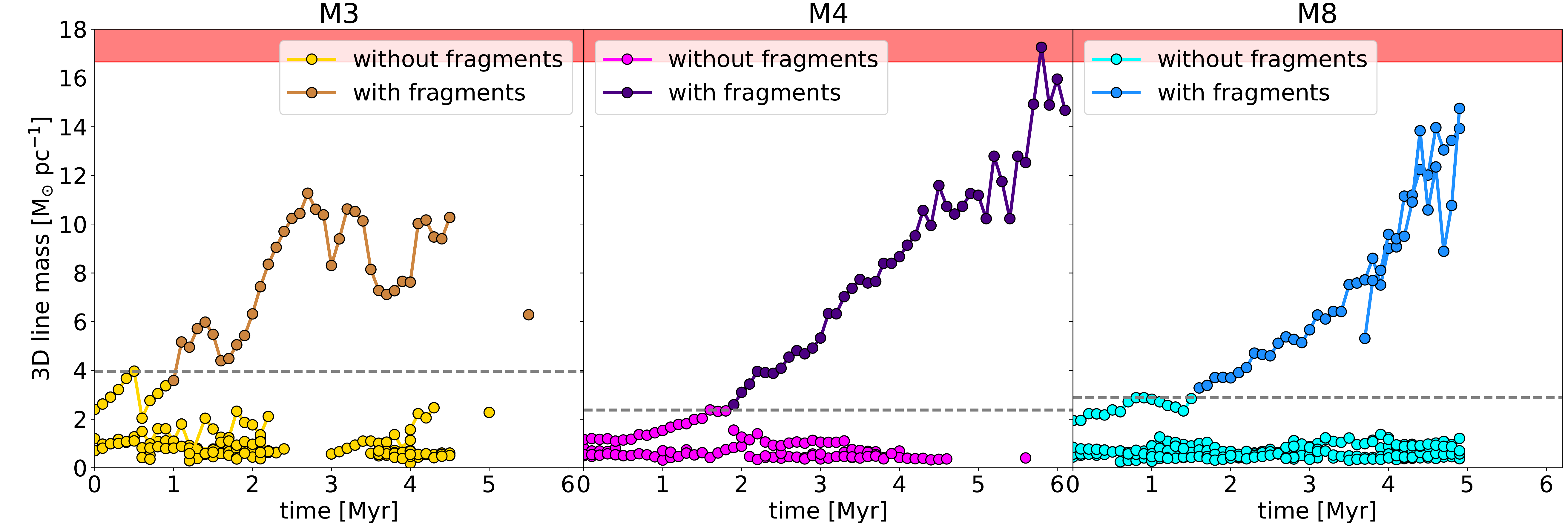}
		\caption{Filaments identified with threshold of 100~cm$^{-3}$.}
		\label{pic:frag_filsep_low}
	\end{subfigure}
	
	\vfill
	
	\begin{subfigure}{\textwidth}
		\includegraphics[width=\textwidth]{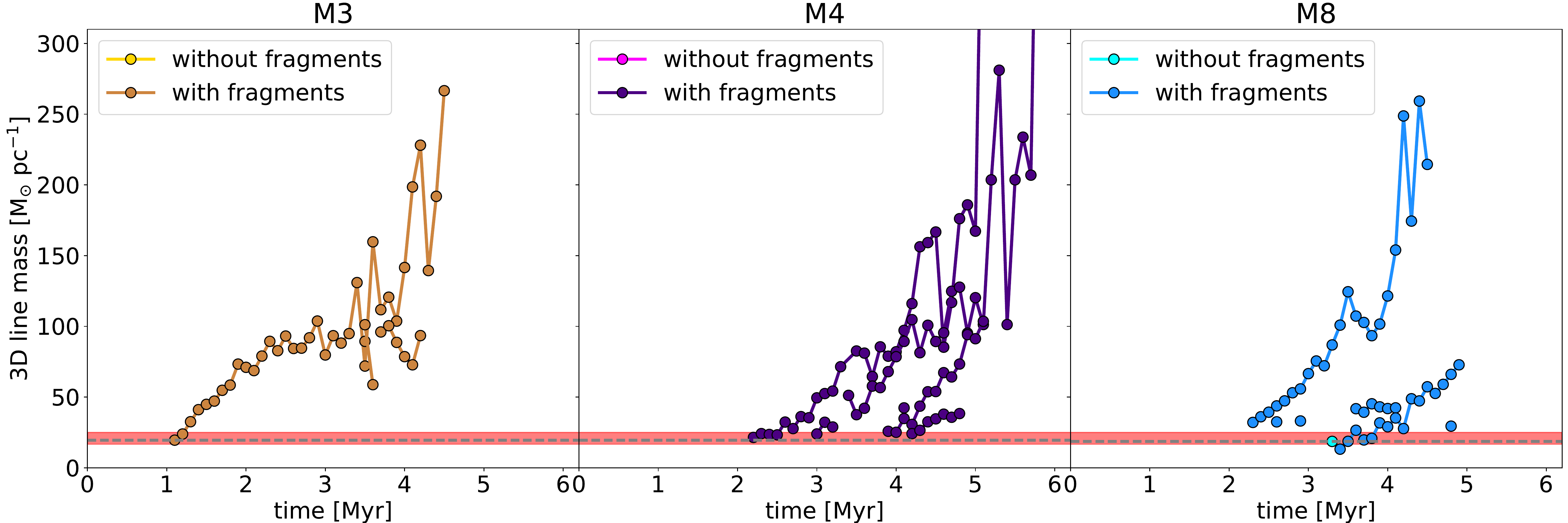}
		\caption{Filaments identified with threshold of 5,000~cm$^{-3}$.}
		\label{pic:frag_filsep_high}
	\end{subfigure}
	
	\caption[Evolution of line mass of individual 3D filaments]{Evolution of line mass of individual filaments.
		Each line represents a single filament, with dots showing its line mass at each output time step.
		The light dots illustrate the time steps when the respective filament does not contain any fragments, the darker dots show these time steps  when the filament contains at least one fragment.
		The red area illustrates the critical line mass at which isothermal filaments with temperatures between 10--15~K become unstable against collapse according to the \citet{Ostriker1964b} model.
		The grey dashed line shows the maximum line mass of filaments \textit{without} fragments in the simulation.
	}
	\label{pic:frag_filsep}
\end{figure}

Fig.~\ref{pic:frag_filsep} shows the time evolution of line masses of \textit{individual} filaments.
In the case of the low-density filaments (Fig.~\ref{pic:frag_filsep_low}), the transition from filaments without fragments to those with fragments occurs at very low line masses (4.0, 2.4, and 2.9~M$_\odot$~pc$^{-1}$ for \texttt{M3}, \texttt{M4}, and \texttt{M8}, respectively).

I compare the line masses at which fragmentation occurs in the model with the criterion for cylindrical filaments being in hydrostatic equilibrium.
In particular, I focus on the model by \citet{Ostriker1964b} which describes a filament as an infinitely long, isolated, isothermal cylinder filled with self-gravitating gas that is in balance between gravity and thermal pressure (see Sect.~\ref{frag:theory}). 
Such cylindrical models have become commonly used descriptions of filaments and set of initial conditions of its fragmentation as, according to \citeauthor{Ostriker1964b}, a cylindrical filament is only thermally supported against collapse if its line mass remains below this critical value.
The key question is whether or not the evolution of the filaments in the simulations follows this criterion.
Can the equilibrium configuration be regarded as a realistic initial condition for fragmentation?
Is the equilibrium configuration, in fact, ever reached?
\pagebreak

\noindent To address these questions, I mark the range of the analytic values for the critical line masses for typical gas temperatures between 10 and 15~K with red areas in Fig.~\ref{pic:frag_filsep}.
one sees that the low-density filaments in the samples start fragmenting at line masses far below the predicted critical values, and hardly reach such high line masses even at later stages in their evolution.

This suggests that the approximations of the cylindrical fragmentation model fail here.
one sees many differences between the filaments and those in the analytic model.
In particular, the filaments are neither isolated, nor in hydrostatic equilibrium, nor cylindrically symmetric.
Rather they are part of a hierarchically collapsing cloud, and interact with each other, e.~g.~by crossing each other or accreting gas.
Thus, the filaments both are subject to external pressure and may have large density perturbations that are outside the regime of linear growth.

Furthermore, studies \citep[e.g.,][]{Nagasawa1987,Inutsuka1992,Fiege2000b,Fischera2012a} have found that even filaments that are subcritical in terms of line mass can fragment.
They show that fragmentation itself does not show any clear imprints of the forces that originally formed the fragment (e.g., shock waves, or cloud-cloud collisions).
The conclusion is that there is no prediction for a threshold below which the fragmentation of filaments is prevented.
\pagebreak

\noindent The situation changes with the high-density filaments (Fig.~\ref{pic:frag_filsep_high}).
Here, the maximal line masses of filaments without fragments is between 18--20~M$_\odot$~pc$^{-1}$ in all clouds.
These transitional line masses appear to be in agreement with the critical line mass of the cylindrical fragmentation model.
The reason for this is that the properties of the filaments here are more likely comparable to those of the cylindrical fragmentation model since they are more likely isothermal, straighter, and closer to the sites where overdensities form compared to more diffuse filaments. 
Consequently, the objects I study here agree better with the properties of the analytic cylinders.
However, I detect only one dense filament without an embedded fragment for one single time step (in \texttt{M8}) and, thus, lack a statistically meaningful sample to draw final conclusions about the capability to predict the fragmentation behaviour of high-density filaments.
From the analysis of the low-density filaments, though, one sees that the configuration represented by the analytic model of cylindrical fragmentation is not universally part of the evolution of filaments in the simulations.

I conclude that a simple cylindrical model as described by \citet{Ostriker1964b} and the cylindrical fragmentation model does not represent the typical initial conditions for the fragmentation of a filament in the molecular cloud simulations.
Therefore, it is not a complete model for evaluating the stability of filaments.
To predict the fragmentation of a filament a more complex model is essential. 
One that not only considers the balance between internal self-gravity and thermal pressure, but also others, such as external hydrostatic, turbulent, or magnetic pressure. 
Furthermore, connecting such a model with observations needs to take into account that the properties of the filaments derived strongly depend on the parameters used to identify the filaments (as well as the filament finder code used, as discussed in Sect.~\ref{frag:prop_filfinder}).
This also means that whether or not observed filaments fragment as predicted by a cylindrical fragmentation model is strongly influenced by the identification parameters, as one can always choose parameters in a way that it perfectly suits the model. 
In our case, this is represented by the high-density filaments that only then contain fragments after exceeding the critical line mass. 
Since there is no unique, universal and physically motivated definition of what filaments the fragmentation models are not universal themselves.

\subsection{Properties and Evolution of Fragments}\label{frag:thermsupport_frags}

In this section, I discuss the properties of the fragments I have detected within the clouds using \texttt{astrodendro} (Sect.~\ref{frag:prop_frag}), including their time evolution, and connection to their parental filaments and to each other.
The main properties of the fragments are summarised in Table~\ref{tab:fragment_results_propfrag} showing that those fragments have similar properties as dense regions and condensations in star-forming regions \citep{Bergin2007}.

\begin{table}
	\centering
	\begin{tabular}{l|cccc}
		Quantity & min & mean & median & max \\ \hline
		Number density n [10$^{3}$ cm$^{-3}$]	&  4.1	& 27.1	& 27.1	& 2136		\\ 
		Total mass $M_\mathrm{tot}$ [M$_\odot$]	&  2.1	& 48.6	& 48.6	&  744		\\ 
		Gas temperature $T_\mathrm{gas}$ [K]	& 10.0	& 10.1	& 10.1	&   48.6	\\ 
		cl.~neighbour separation [pc]		&  0.4	&  1.2	&  1.0	&    4.0	\\
		virial parameter at first detection	&  0.5	&  3.5	&  3.5	&    7.7	\\  \hline
	\end{tabular}
    \caption{Summary of properties of fragments.}
    \label{tab:fragment_results_propfrag}
\end{table}

\begin{wrapfigure}{R}{0.45\textwidth}
	\vspace*{-1.25\baselineskip}
	\centering
	\includegraphics[width=0.45\textwidth]{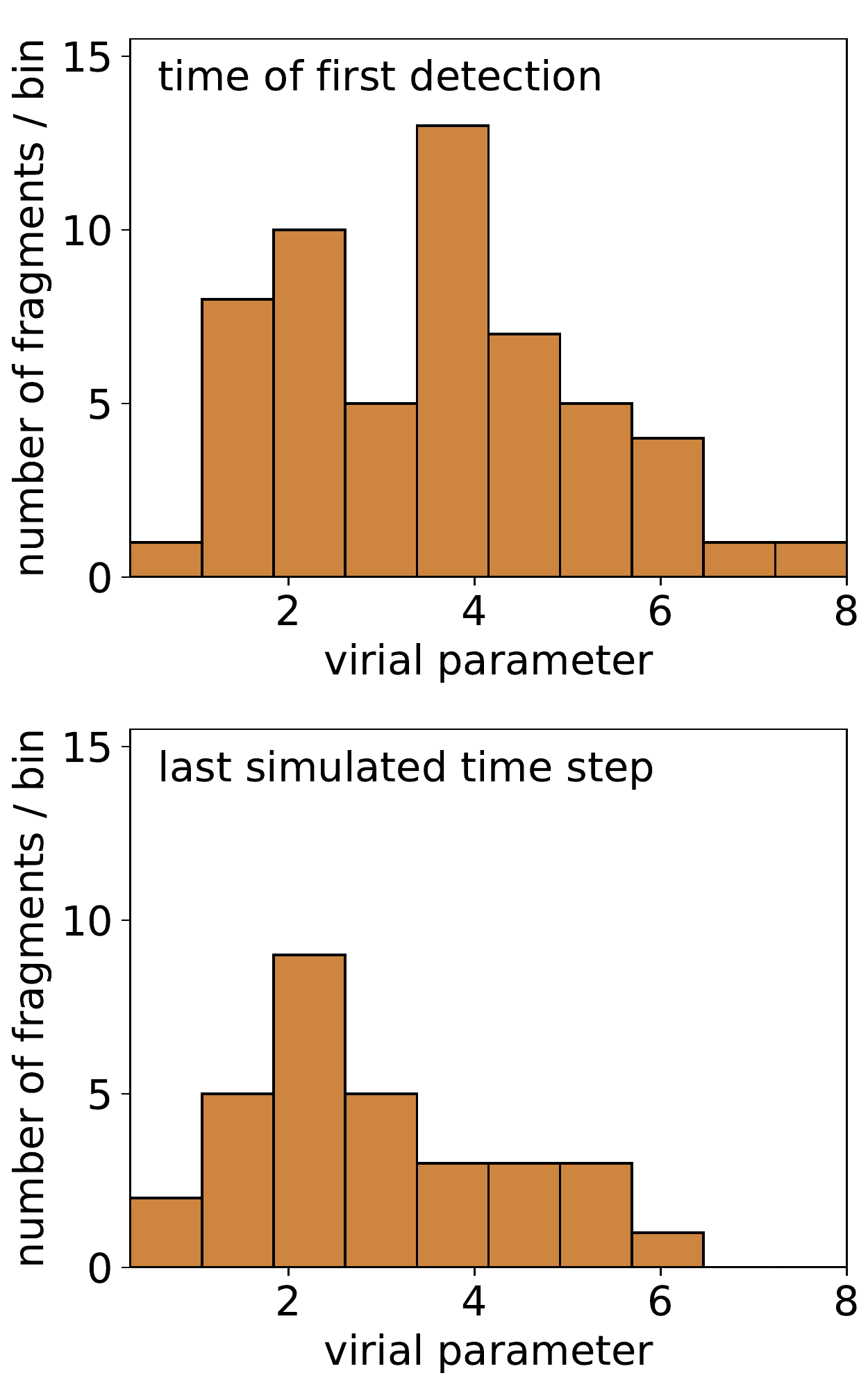}
	\caption[Histogram of volumetric virial parameters of fragments at the time when they have been initially detected]{Histogram of volumetric virial parameters of fragments (from all three clouds) at the time when they have been initially detected (\textit{top}) and in the last simulated time step (\textit{bottom}).
	}
	\label{pic:frag_dendro_alpha}
	\vspace*{-0.95\baselineskip}
\end{wrapfigure}

In the top row of Fig.~\ref{pic:frag_dendro_alpha} I show a histogram of virial parameters measured for the fragments at the time when they have been detected first.
According to Eq.~(\ref{equ:frag_virial_parameter}), the fragments are gravitationally bound if $\alpha < 2$. 
This criterion is fulfilled by only 20\% of the fragments.
The majority of fragments have virial parameters close to the mode of 3.5.
I argue that these fragments may nevertheless be bound objects because of the terms not accounted for by Eq.~(\ref{equ:frag_virial_parameter}), particularly the surface terms resulting from the collapse of the surrounding cloud.
Furthermore, one sees that the virial parameters decrease as the fragments evolve, which means that they become more bound.
The histogram in the bottom row of Fig.~\ref{pic:frag_dendro_alpha} confirms this as it shows that most of the fragments have virial parameters around 2 in the last simulated time steps. 

Note that, as explained by \citet{IbanezMejia2016,Ibanez-Mejia2017}, I can only give lower limits on the velocity dispersions, and thus the virial parameters, since I underresolve the turbulence on small scales in the simulations.
Resolving the subgrid scale turbulence may increase the energy by 33\% \citep{Ibanez-Mejia2017}, but likely won't prevent the fragments from beginning to collapse before the criteria for cylindrical fragmentation are fulfilled.
This consequently means that the fragmentation must be driven by neglected terms in the virial equation that can produce significant differences in the estimation of the fragments' boundness.

The bottom row of Fig.~\ref{pic:frag_numfrags} plots the number of identified fragments as function of time.
One sees that the number of fragments overall increases with time for each simulation.
However, there are cases when the number of fragments drops. 
The missing fragments are either disrupted, for example by shock waves or intracloud turbulence, or merge with each other, meaning that they approach each other too closely ($<$0.4~pc) to be distinguishable with the resolution. 

One sees that the first fragments form within the first 2~Myr, corresponding to 25--50\% of the clouds' free-fall times and the period when the growth of the DGMF is steepest (see Fig.~\ref{pic:frag_fil3d_dgmf}).
This indicates that the formation of fragments is primarily dominated by the compression of dense gas within the cloud.
This is verified by the ratio of mass contained within the fragments, $M_\mathrm{frag}$, relative to the mass of the entire cloud, $M_\mathrm{cloud}$, shown in the bottom panel of Fig.~\ref{pic:frag_massfrac}.
The $M_\mathrm{frag}/M_\mathrm{cloud}$ ratios evolve in a similar fashion as the DGMF.
Compared to the more constant growth of $M_\mathrm{cloud}$ or $M_\mathrm{fil}$, the ratio $M_\mathrm{frag}/M_\mathrm{cloud}$ increases rapidly before stagnating around 0.03.
Interestingly, if I approximate the star formation efficiency per free-fall time of the clouds with this ratio, the value agrees with typically observed efficiencies in molecular clouds \citep[e.g.,][]{Krumholz2007}.
This is consistent with the determination of star formation efficiency by the dynamics of gravitational collapse, though the relatively small sample and lack of feedback modelling does not allow definitive conclusions to be drawn.

Subsequently, I follow the evolution of the mass contained in the fragments relative to the mass contained in the filaments, $M_\mathrm{frag}/M_\mathrm{fil}$.
Fig.~\ref{pic:frag_massfrac} shows the ratios in the middle panel. 
In all clouds, one sees that the $M_\mathrm{frag}/M_\mathrm{fil}$ ratio increases rapidly within the first $\sim$2.5~Myr after the first fragments have formed, reaching maximal values of 15--40\%. 
This demonstrates that the fragments accrete mass from their parental filaments efficiently as long as there is a sufficient gas reservoir accessible, as is the case at the beginning of the simulations.
In doing so, they take up a significant fraction of the filaments' masses.

\noindent Another question that has been recently discussed in the literature is whether prestellar cores form in a regular pattern within filaments.
Observations suggest that cores condense at regular intervals along their parental filaments \citep[e.g.][]{Jackson2010,Hacar2011}.
The mean separations between the cores appear to correlate with the properties of the respective filament, ranging from a few tenths to several parsecs.
These observations seem consistent with theoretical models of periodic fragmentation \citep{Ostriker1964a,Nagasawa1987,Inutsuka1992,Fischera2012a}.
The instabilities causing fragmentation in these models have unique modes that depend on the initial conditions of the filament.
The wavelengths of these modes then define the mean separations between the forming cores.
However, other studies, both observational \citep[e.g.,][]{Enoch2006,Gutermuth2009} and theoretical \citep[e.g.,][]{Seifried2015,Clarke2017}, demonstrate that periodic fragmentation only occurs under special conditions (such as supersonic, purely compressive turbulent motions), if at all. 
In reality the conditions within the filaments are not as uniform as assumed by the models, so observed core patterns may also be the result of overlapping fragmentation modes.
According to those studies, filaments commonly fragment in a disordered, cluster-like fashion. 

\begin{figure}[h!t]
	\centering
	\includegraphics[width=\textwidth]{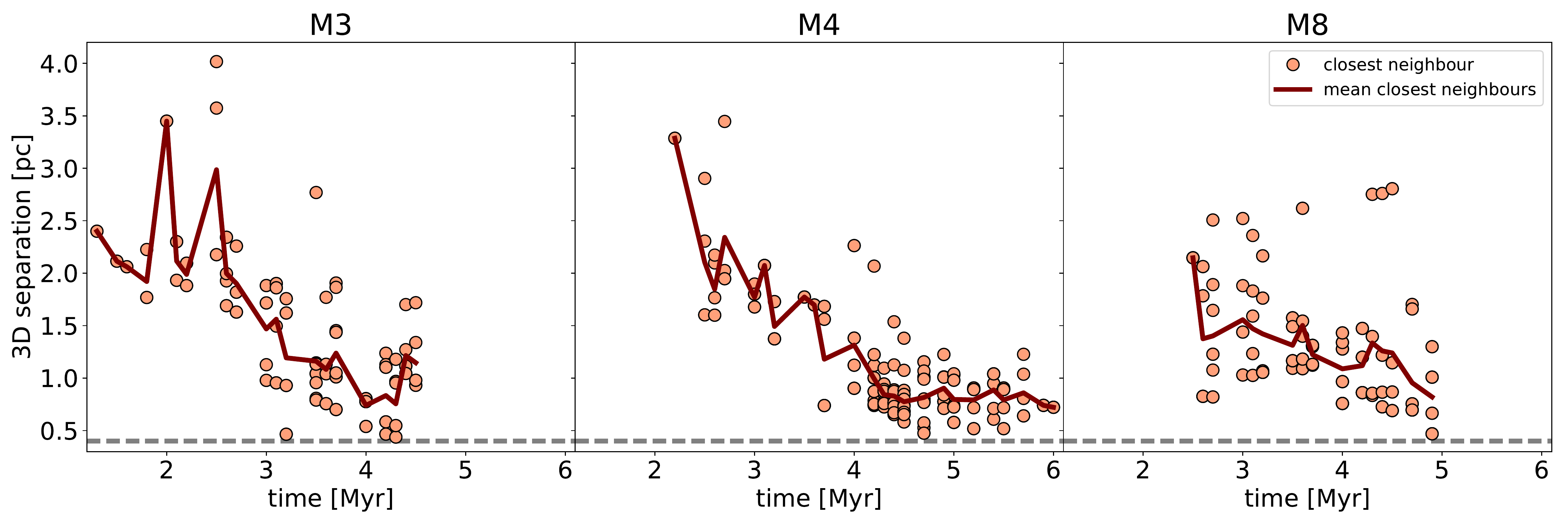}
	\caption[3D separations between fragments and their individual closest neighbours]{3D separations between fragments and their individual closest neighbours (orange dots) within the same filament within \texttt{M3} (\textit{left}), \texttt{M4} (\textit{middle}), and \texttt{M8} (\textit{right}) at each time step.
		The line illustrates the mean separation between closest neighbours over time.
		The grey dashed line represents the resolution limit at 0.4~pc below which I cannot distinguish individual cores from each other.
	}
	\label{pic:frag_fragments_meansep}

	\includegraphics[width=\textwidth]{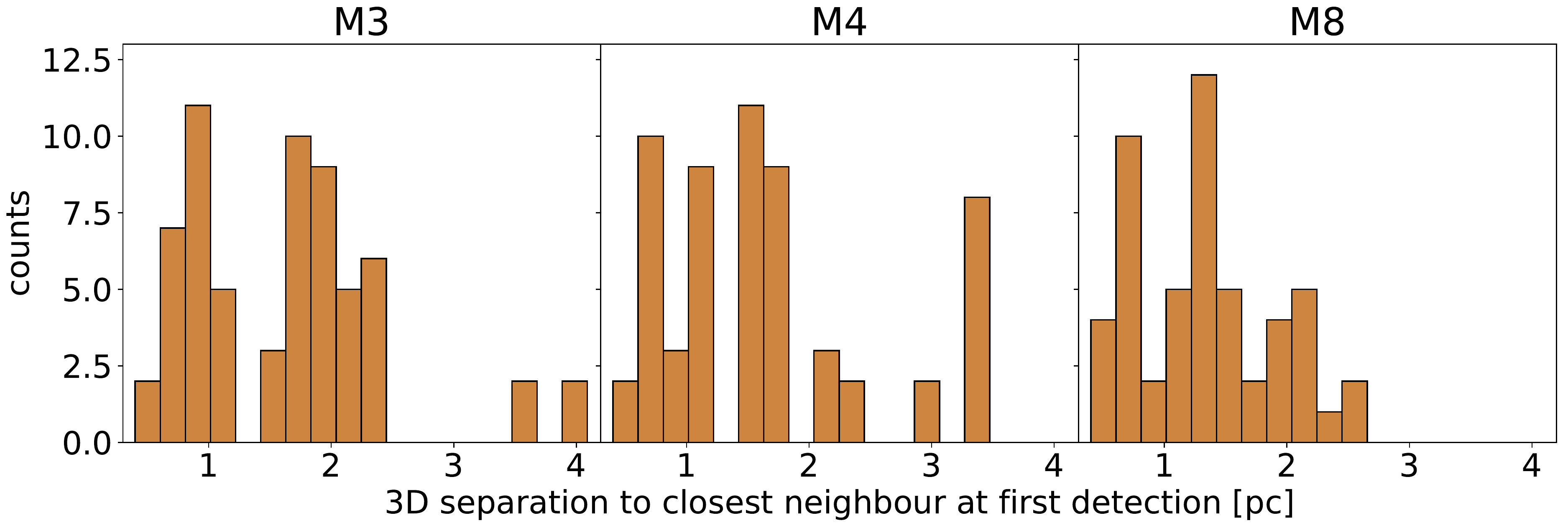}
	\caption[Histogram of separations of fragments to their closest neighbours]{Histogram of separations of fragments within \texttt{M3} (\textit{left}), \texttt{M4} (\textit{middle}), and \texttt{M8} (\textit{right}) to their closest neighbour at the time when the fragment was detected for the first time.
	}
	\label{pic:frag_hist_sep_first}
\end{figure}

Using the data, I test whether the fragments in the sample form with uniform separations.
Note that I can only detect separations that are larger than 0.4~pc due to the 0.1~pc resolution of the data grid and the 0.3~pc radius I assume for the fragments.
Fig.~\ref{pic:frag_fragments_meansep} shows the separations in 3D space of the individual fragments to their individual closest neighbour within the same filament as a function of time.
One sees that the fragments form, on average, at distances exceeding 2~pc from their closest neighbour at the beginning of the fragmentation process, but approach each other with time down to $<$1~pc or even merge (below 0.4~pc).

\begin{table}
	\centering
	\begin{tabular}{l|cccc}
		Quantity & min & mean & median & max \\ \hline
		\multicolumn{5}{l}{N$_\mathrm{th}$~=~3~$\times$~10$^{19}$ cm$^{-2}$} \\ \hline
		Col.~density N$_\mathrm{tot}$ [10$^{21}$ cm$^{-2}$]	&  0.7	&  8.4	&  4.2	&  313		\\ 
		Gas temperature T$_\mathrm{gas}$ [K]			& 11.1	& 56.7	& 56.4	&   97.8	\\ 
		Total length $\ell_\mathrm{tot}$ [pc]				&  1.8	&  4.8	& 56.4	&   58.5	\\ 
		Total mass M$_\mathrm{tot}$ [M$_\odot$]			&  0.7	& 46.0	& 17.0	& 1654		\\ 
		Line mass M$_\mathrm{lin,2D}$ [M$_\odot$ pc$^{-1}$]		&  0.4	&  9.6	&  5.0	&  321.1	\\ \hline 
		\multicolumn{5}{l}{N$_\mathrm{th}$~=~10$^{21}$ cm$^{-2}$} \\ \hline
		Col.~density N$_\mathrm{tot}$ [10$^{21}$ cm$^{-2}$]	&  0.6	& 13.4	&  6.0	&  295		\\ 
		Gas temperature T$_\mathrm{gas}$ [K]			& 13.2	& 51.8	& 51.1	&   97.8	\\ 
		Total length $\ell_\mathrm{tot}$ [pc]				&  1.8	&  4.3	&  2.5	&   35.3	\\ 
		Total mass M$_\mathrm{tot}$ [M$_\odot$]			&  1.3	& 71.8	& 24.1	& 1735		\\ 
		Line mass M$_\mathrm{lin,2D}$ [M$_\odot$ pc$^{-1}$]		&  0.6	& 16.5	&  7.4	&  371		\\ \hline 
	\end{tabular}
	\caption{Summary of properties of the 2D filaments.}
	\label{tab:fragment_results_prop2d}
\end{table}

To answer the question of whether there is a typical fragmentation scale one needs to consider the separations between closest neighbours at the moment the fragments form.
These are summarised in Fig.~\ref{pic:frag_hist_sep_first}. 
If there were a typical separation I should see a significant peak at that particular scale length, or a sequence of aliased peaks with equal separations. 
Looking at the histograms, one may argue that such sequences are seen in \texttt{M3} and \texttt{M8} with typical separations at 0.9 and 0.6~pc, respectively.
Both numbers exceed the local Jeans length by a substantial factor ($\sim$0.1~pc), but are only a factor of 2.3--1.5 larger than the resolution limit. 
Furthermore, in the case of \texttt{M4}, one does not see any significant peak in the distribution.
However, the number of fragments in the samples is too low to make a solid statement about whether there is a universal fragmentation scale length, and if it depends on the local physical conditions only.

\subsection{Properties and Evolution of Filaments in 2D}\label{frag:thermsupport_comp3d2d}

In the previous subsections, I have studied the properties and evolution of the filaments and fragments that I identified based on the full 3D simulation data.
In observations, however, such 3D data are not available \citep[see, however,][ for a method to reconstruct the volume density distribution]{Kainulainen2014}.
Instead, one observes the filaments projected onto the 2D plane of the sky.
This raises the questions of what one would observe if one applies the methods to the projected data and how these results compare to the results from 3D data.

In this section, I approach these questions by projecting the 3D volume density cubes onto 2D column density maps.
I project along the three major axes $x$,~$y$, and~$z$, in order to account for LoS specific variance. 
Note that these maps lack any additional observational effects, such as noise, beam, or optical depth effects, and are thus only idealised approximations to real observations.
However, such maps give the best case scenario to test the 3D-2D correspondence.
Since I neglect opacity, the column density maps are primarily comparable to optically thin (sub-)mm dust observations. 
To produce comparable synthetic spectral line emission maps, I would need to consider chemical abundances, as well as gas velocity, which is beyond the scope of this analysis.

\begin{wrapfigure}{R}{0.45\textwidth}
	\vspace*{-0.75\baselineskip}
	\centering
	\includegraphics[width=0.45\textwidth]{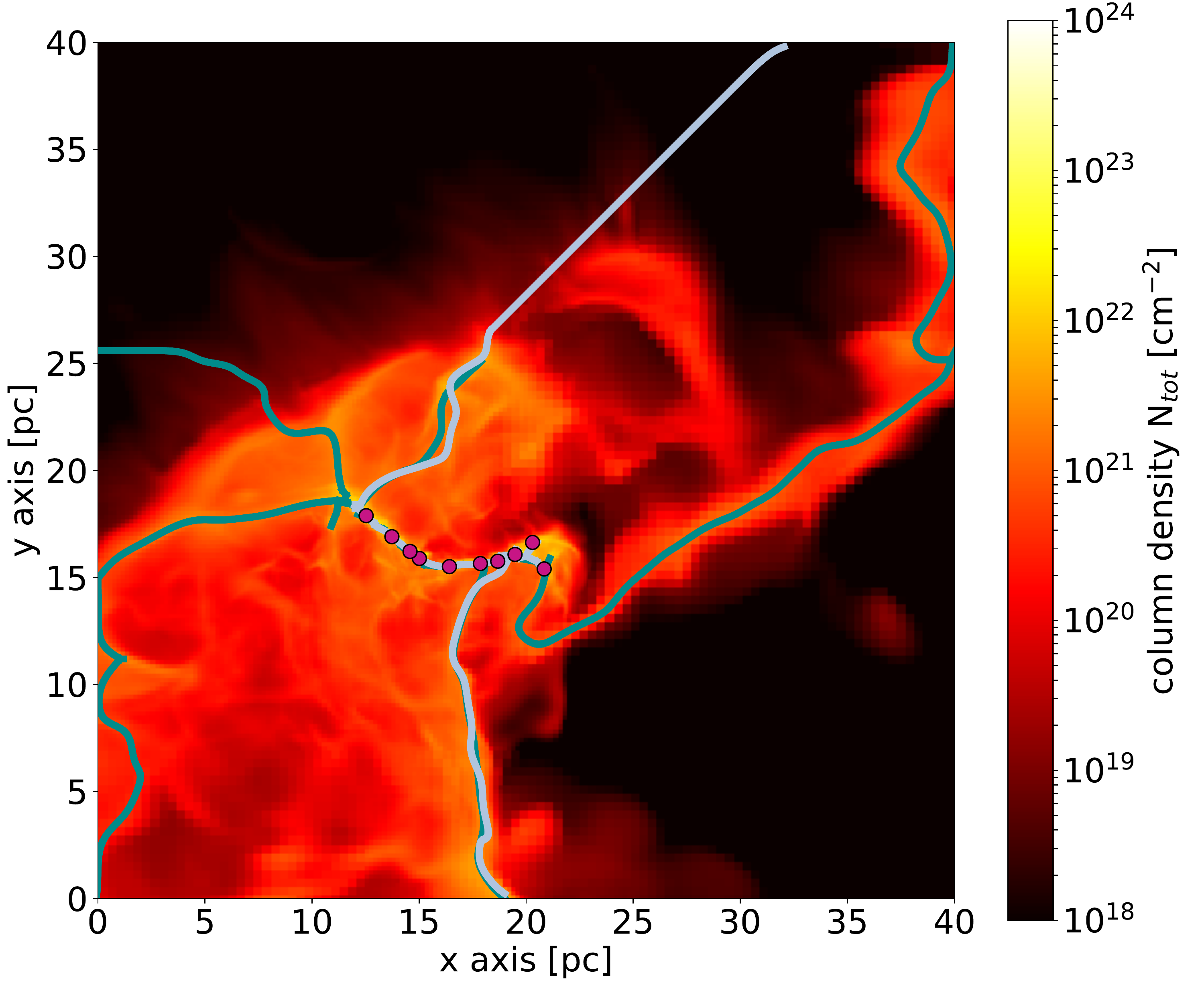}
	\caption[Example showing the effect of identifying filament based on 3D and 2D data]{Column density map of \texttt{M3} at $t$~=~3.8~Myr when projected along the z-axis. The over-plotted lines illustrate the high-density filaments found by DisPerSe based on the column density map (\textit{teal lines}) and the volume density cube (projected, \textit{light blue lines}, same as in Fig.~\ref{pic:frag_example_ntot_lowhigh}). Additionally, the projected position of the identified fragments are marked by with purple dots.
	}
	\label{pic:frag_example_fils_3d2d}
	\vspace*{-0.75\baselineskip}
\end{wrapfigure}

Analogously to Sect.~\ref{frag:thermsupport_mean}, I use \texttt{DisPerSe} to identify 2D filaments within the column density maps.
For this purpose, I convert the number density thresholds used in 3D into column density thresholds by assuming a path length of 0.1~pc, namely $n_\mathrm{th}~=~3~\times~10^{19}$~cm$^{-2}$ (corresponding to $n_\mathrm{th}~=~100$~cm$^{-3}$) and $n_\mathrm{th}=10^{21}$~cm$^{-2}$ (corresponding to $n_\mathrm{th}=5,000$~cm$^{-3}$).
I emphasise that the skeletons of the 2D filaments are independently identified based on the column density distributions and not the projections of the 3D filament skeletons.
Fig.~\ref{pic:frag_example_fils_3d2d} shows an example of the structures obtained, and Table~\ref{tab:fragment_results_prop2d} summarises their properties.
One sees that, similar to the structures detected in 3D, the 2D filaments' properties are influenced by the identification threshold, with the lengths of the filaments becoming shorter and line masses higher with higher thresholds.

Comparing the structures detected in 3D and 2D, however, reveals a more significant difference.
As shown in the example in Fig.~\ref{pic:frag_example_fils_3d2d}, I do find 2D counterparts for all 3D filaments, but there are 2D filaments that do not have matching 3D filaments identified with the same identification threshold.
This finding is a consequence of the projection:
structures with high volume densities typically have high column densities; conversely other structures with lower volume densities can appear denser in column densities, amplified by projection.

\begin{figure}
	\vspace{-0.65\baselineskip}
	\centering
	\includegraphics[width=\textwidth]{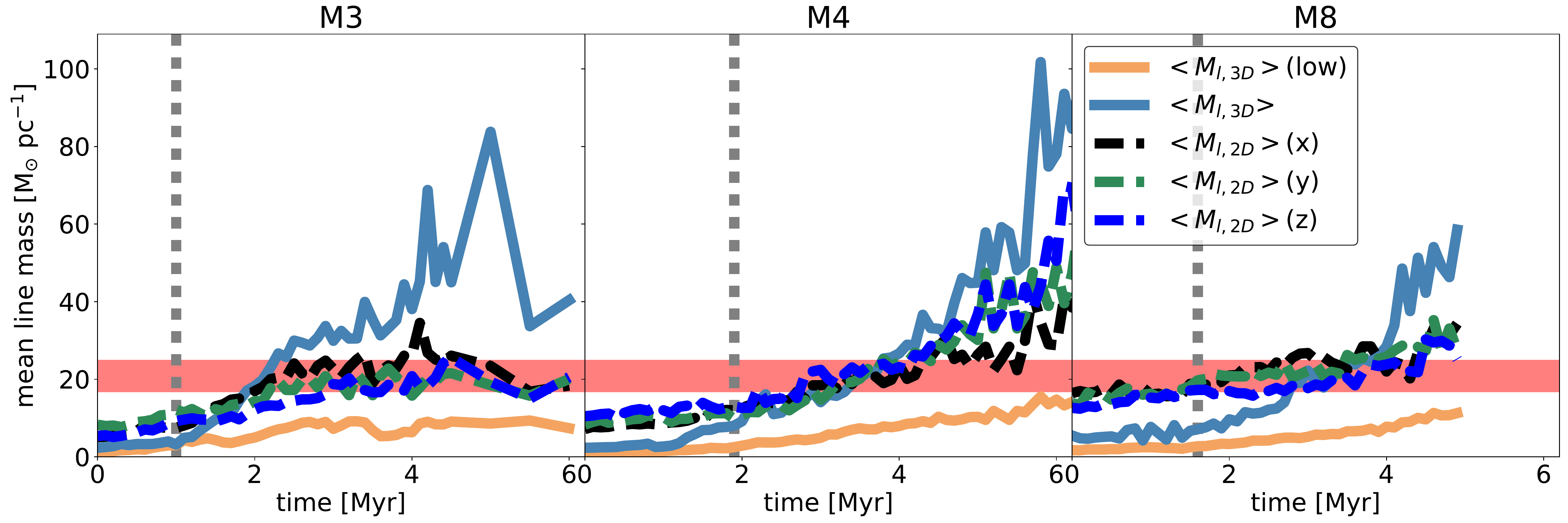}
	\caption[Evolution of $\langle M_\mathrm{lin,3D}\rangle$ and $\langle M_\mathrm{lin, 2D} \rangle$]{Evolution of $\langle M_\mathrm{lin,3D}\rangle$ (orange and blue lines) and $\langle M_\mathrm{lin, 2D}\rangle$ (black, green, and navy blue lines for projections along the $x$, $y$, and $z$ axes) for filaments that have been detected above $n_\mathrm{th}=5,000$~cm$^{-3}$ or $n_\mathrm{th} =10^{21}$~cm$^{-2}$, unless specified otherwise, in \texttt{M3} (\textit{left}), \texttt{M4} (\textit{middle}), and \texttt{M8} (\textit{right}).
	  The red area marks the theoretical critical line mass for isothermal gas at 10--15~K.
	  The grey dotted line shows the time step when the first fragment forms.
	}
	\label{pic:frag_ml3dall}
	\vspace*{-1.5\baselineskip}
\end{figure}

This also influences the measured properties of the 2D filaments, such as the 2D line masses, $M_\mathrm{lin,2D}$.
Analogously to Fig.~\ref{pic:frag_numfrags}, Fig.~\ref{pic:frag_ml3dall} shows the value of $\langle \mathrm{M}_\mathrm{lin,2D} \rangle$ for the dense 2D filaments as a function of time. 
One sees that the values lie between the values I measured based on the diffuse and dense 3D filaments.
If the dense 2D filaments exclusively represented the projection of dense 3D filaments, however, one would expect the trends in average line mass to evolve similarly.
This is not seen here.
On the contrary, the growth of $\langle \mathrm{M}_\mathrm{lin,2D} \rangle$ in the dense 2D filaments correlates with the growth of $\langle \mathrm{M}_\mathrm{lin,3D} \rangle$ of the diffuse 3D filaments.
As described before, this is because most of the dense 2D filaments are projections of diffuse 3D filaments, so the line mass of the 3D filaments is the main contributor of the average line mass of the dense 2D filaments.
Additional mass comes from material along the LoS, so $M_\mathrm{lin,2D} > M_\mathrm{lin,3D}$.

In summary, for a given identification threshold, all 3D filaments have counterparts in column density maps, but not necessarily vice-versa.
Projection not only maps dense 3D filaments onto the plane of the sky, but also merges less dense structures along the same LoS, producing structures that exceed the column density threshold.
Consequently, a comparison of the properties of 3D to 2D filaments is not directly possible if the corresponding identification thresholds are used for both samples.
However, if this is taken into account, the properties of 3D and 2D filaments evolve similarly, but with an offset due to the additional LoS mass projected onto the 2D filaments.

The difference in the density distributions measured in 3D and 2D is even more obvious in the overall properties of the clouds.
I measure the DGMF using column density by defining clouds as coherent volumes of gas with minimal number densities of 100~cm$^{-3}$. 
I choose a minimum path length of 0.6~pc based on the assumption that filaments and fragments have radii of 0.3~pc.
Note that this path length is not identical to the path length I used for computing the identification threshold of the column density filaments, $n_\mathrm{th}$, before.
However, by using a minimum path length of 0.6~pc ensures that the fraction of gas I hereafter consider mirrors gas of the real 3D cloud.
This way, I cannot only more reliably compare the DGMFs measured based on the 2D data with those based on the 3D data, but also with DGMFs in observed molecular clouds.
As a results I obtain a minimal column density of $N_\mathrm{min}=2~\times~10^{20}$~cm$^{-2}$. 
I then compute
\begin{equation}
	\mathrm{DGMF}(t,N_\mathrm{dens}) ~=~\frac{ \int_V N(t,\vec{x}) \, G(N_\mathrm{dens}) \,  dV}{\int_V N(t,\vec{x}) \, G(\mathrm{2~\times~10}^{20}\mathrm{~cm}^{-2}) \,  dV} \, ,
	\label{equ:frag_dgmf_2D}
\end{equation}
with $N$ being the column density at time $t$ and within the pixel $\vec{x}$, $N_\mathrm{dens}$ the column density above which the gas is defined as dense and
\begin{equation}
	G(N_0)~=~\begin{cases}
		1 & N(t, \vec{x}) \geq N_0 \\
		0 & N(t, \vec{x}) < N_0
	\end{cases} .
	\label{equ:frag_g_N}
\end{equation}
In Fig.~\ref{pic:frag_fil3d_dgmf}, the dashed lines show the evolution of the DGMF measured using this equation.

Analogous to Sect.~\ref{frag:thermsupport_mean}, I use two thresholds for tracing the evolution of dense gas within the clouds, namely $N_\mathrm{dens}$~=~2~$\times$~10$^{21}$~cm$^{-2}$ (corresponding to $n_\mathrm{dens}$~=~1,000~cm$^{-3}$) and \linebreak $N_\mathrm{dens}$~=~8~$\times$~10$^{21}$~cm$^{-2}$ (corresponding to $n_\mathrm{dens}$~=~5,000~cm$^{-3}$).
One sees that for most of the time the DGMF is about an order of magnitude higher than the corresponding DGMFs calculated from the volume density distributions, but maximal values (70--80\% for $n_\mathrm{dens}$~=~1,000~cm$^{-3}$ and 30--50\% for $n_\mathrm{dens}$~=~5,000~cm$^{-3}$) agree with each other.
In the case of a lower value of $n_\mathrm{dens}$, one sees that the 2D DGMFs are almost constant in time, or even slightly decreasing, in disagreement with
with the steady growth of the 3D DGMFs. 

In summary, one sees that the DGMF measured in 2D deviates from the true 3D value for most of the initial evolution of a particular cloud.
The DGMF measured in column density may show a completely different temporal behaviour, particularly when a low column density threshold is used to define dense gas.
However, one also sees that the individual filaments I identify in the 3D and 2D data and their properties agree decently with each other if the identification threshold in column density focuses on the range of volume density one wants to study and distinguishes unassociated gas along the LoS.

\pagebreak 	

\section{Summary}\label{frag:conclusions}

In this chapter, I analyse the properties and fragmentation of filaments forming within 3D AMR FLASH simulations of the self-gravitating, magnetised, supernova-driven ISM by \citet{IbanezMejia2016}. 
The main results are as follows.
\medskip

\begin{itemize}
	\item I find that the dense gas mass fraction (DGMF) steadily grows as a function of time. Although these clouds grow in mass as they accrete material from their environment, the DGMF continues growing in time, consistent with runaway gravitational collapse.
	\item I find that the average line masses of the filaments always increase in time, with significant differences depending on the volume (or column) density thresholds adopted for their identification. This and the continuously increasing filament-to-cloud mass ratio confirm that the gas of the parental clouds collapses into smaller scale structures, as the evolution of the DGMF has already indicated. 
	\item Filaments already start to fragment well before their line masses reach the critical mass for the collapse of uniform density, self-gravitating, hydrostatic cylinders \citep{Ostriker1964b}. This is true both for the line masses of individual filaments, as well as for the the average of all identified filaments. This implies that the filaments in the simulation never resemble the isolated, hydrostatic configuration of \citet{Ostriker1964b} that is commonly used as the initial condition in analytic filament evolution and fragmentation models. Instead, they are embedded in the hierarchical collapse of the larger cloud, and thus subject to substantial surface pressures.
	\item I compare the performance of different filament finder codes. I find that different codes clearly identify different structures, and further, that the filament properties derived depend strongly on the choice of input parameters.
	\item I compare the properties of the filaments identified in 3D density distributions from the models with those identified in projected 2D column density distributions. I find that, for a given identification threshold density, all 3D filaments have counterparts in 2D column density data, but not vice versa. This is because the 2D filaments may also be composed of the overlap of more diffuse structures along the given LoS that do not fulfil the identification criteria in 3D. As a consequence, the average properties of a sample of filaments and how they evolve in time are not well recovered from column density data. However, since all 3D filaments have counterparts in 2D, the correspondence is better in the case of individual isolated filaments.
\end{itemize}

%\chapterimage{head02.pdf} % Chapter heading image

\chapter[Project III: Turbulence in Simulated Clouds and Filaments]{Turbulence in Clouds \& Filaments}\label{turb}

In Chapter~\ref{frag}, I have emphasised the role and importance of fragmentation of filaments in the context of setting on the process of star formation.
I found that gravitational instability of thermally supported filaments alone is not sufficient to explain the formation of the first fragments, those that form within the first few megayears after self-gravity had been activated in the simulations.
Thus, one of the key questions is still not answered, namely: 
What causes the fragmentation of filaments?

As discussed before, the filaments in our model clouds are supposedly be stable against gravitational fragmentation due to thermal support for most of their evolution.
To trigger fragmentation, other forces are required that, at least locally, overcome the equilibrium state of the respective filament.
There are different mechanisms that are able to achieve this, for example colliding flows of gas within the filaments, infalling or crossing filaments, or supernova shock waves.
Although magnetic fields are included in the simulations and supposed to hold an important role in the evolution of molecular clouds and all their substructures, I focus on turbulent motions in this chapter.

In literature, turbulence has an ambiguous role in the context of star formation.
In most of the cases, turbulence is expected to stabilise molecular clouds on large scales \citep{Fleck1980,McKee1992,MacLow2003}, while feedback processes and shear motions heavily destabilise or even disrupt cloud-like structures, offering a formation scenario for large filaments \citep{Tan2013,Miyamoto2014}.
Yet, it is not entirely clear which mechanisms drive the turbulence within molecular clouds dominantly.
Thereby, all possible candidates are supposed to show different imprints in the observables. 
For example, turbulence that is driven by large-scale velocity dispersions during global collapse \citep{Ballesteros2011a,Ballesteros2011b,Hartmann2012} produces P-Cygni lines. 
These lines are, though, normally not observed, and this mechanism cannot explain the long lifetimes of giant molecular clouds.
The scenario in which internal feedback sources drive turbulence outwards \citep{Dekel2013,Krumholz2014} seems more promising. 
However, observations demonstrate that the required driving sources need to act on scales of entire clouds, which typical feedback processes cannot achieve \citep{Brunt2009,Brunt2013,Heyer2004}. 
\pagebreak

\noindent There have also been many theoretical studies examining the nature and origin of turbulence \citep[and references within]{MacLow2004}.
The most established work has been conducted by \citet{Kolmogorov1941} who investigated fully developed, incompressible turbulence.
The underlying assumptions, however, describe the very special scenario of a divergence-free velocity field within a cloud with homogeneous density.
Analytical studies without these assumptions are still rare, but exist.
\citet{She1994} and \citet{Boldyrev2002}, for example, generalise and extend the predicted scaling of the decay of turbulence to supersonic turbulence. 
\citet{Galtier2011} and \citet{Banerjee2013} provide an analytic description of the scaling of mass-weighted structure functions (see Sect.~\ref{turb:vsf}).

Thus, I extend the analysis of Chapter~\ref{frag} and focus now on the turbulent flows of gas within the model clouds and filaments.
The key questions I answer are the following:
What dominates the turbulence within the simulated molecular clouds?
Is there a method that can trace the dominant modes reliably?
And, do fragments form due to colliding flows within filaments?

In Sect.~\ref{turb:vsf}, I study the properties of turbulence on scales of the entire clouds by using velocity structure functions. 
I demonstrate that velocity structure functions are a useful tool to characterise the dominant driven source of turbulence in molecular clouds and can be applied on both simulated and observed data.
Sect.~\ref{turb:veldir} focusses on the filaments and examine the gas flows within them.

%%%%%%%%%%%%%%%%%%%%%%%%%%%%%%%%%%%%%%%%%%%%%%%%%%%%%%%%%%%%%%%%%%%%%%%%%%%%%%%%%%%%%%%%%%%%
\section{Characterising Turbulence in Molecular Clouds}\label{turb:vsf}

In this section, I examine the distribution of turbulent power throughout the entire modelled molecular clouds. 
Those clouds have been introduced in Sect.~\ref{frag:clouds}.
I use the so-called velocity structure functions that I define and describe in Sect.~\ref{turb:vsf_theory}.
In Sect.~\ref{turb:vsf_results} I discuss the results of the VSF analysis.

\subsection{Velocity Structure Functions}\label{turb:vsf_theory}

The velocity structure function (VSF) is a two-point correlation function that measures the mean velocity difference, $\delta{v}$, to the $p^\mathrm{th}$ order as function of lag distance, $\ell$, between the correlated points, or in my case cells.
Thereby, the VSF estimates the occurrence of symmetric motions (e.g., rotation, collapse, outflows), as well as rare events of random turbulent flows in velocity patterns that become more prominent the higher the order of the VSF, $p$, is \citep{Heyer2004}.
For an Eulerian grid the density weighted definition of the VSF, $\mathit{S}_p$, is given by
\begin{equation}
	\mathit{S}_p (\ell) = \frac{\langle \, \rho(\vec{x}) \rho(\vec{x}+\vec{\ell}) \, |\vec{v}(\vec{x}+\vec{\ell}) - \vec{v}(\vec{x})|^p  \, \rangle}{\langle  \, \rho(\vec{x}) \rho(\vec{x}+\vec{\ell}) \, \rangle} ,
    \label{equ:turb_vsf_def}
\end{equation}
\citep[and references within]{Padoan2016a} with $\vec{\ell}$ being the direction vector between the respective grid cells, $\ell = | \vec{\ell} |$ its length, $\vec{v}(\vec{x})$ the gas velocity vector and $\rho(\vec{x})$ the density within the grid cell $\vec{x}$, respectively.
One sees that each order has a physical meaning. 
For example, $\mathit{S}_1$ is correlated to the mean relative velocities between cells, reflecting the modes created by different gas flows.
$\mathit{S}_2$ is proportionally to the kinetic energy, making it a good probe of how the turbulent energy is transferred across different scales.

If the turbulence is fully developed the VSF is well-described by a power-law relation \citep{Kolmogorov1941,She1994,Boldyrev2002}:
\begin{equation}
	\mathit{S}_p (\ell) \propto \ell^{\zeta(p)} .
    \label{equ:turb_vsf_propto_zeta}
\end{equation}
\pagebreak

\noindent The scaling exponent of that power-law relation, $\zeta$, therefore, not only depends on the order of the VSF, but is also strongly influenced by the properties and composition of the underlying turbulence, like compressibility or Mach number.
Many studies on VSFs  distinguish between longitudinal and transverse velocity components, or compressible and solenoidal gas flow components since those are expected to behave differently, especially towards larger lag distances \citep{Gotoh2002,Schmidt2008,Benzi2010}.
However, these are mostly negligible.
This and the fact that those components are observationally very hard, if at all differentiable, are the reasons why I analyse all components in a common sample.

There are a few theoretical studies that predict values of $\zeta(p)$ depending on the status of turbulence.
For example, \citet{Kolmogorov1941} predicts the third-order exponent, $\zeta(3)$, to be exactly 1 for an incompressible, transonic flow.
This results in the commonly known prediction that the kinetic energy decays with $E_k(k) \propto k^{-\frac{5}{3}}$, with $k = \frac{2 \pi}{\ell}$ being the wavenumber of the turbulence mode.

For a supersonic flow, however, it is always greater or equal to unity.
Based on \citeauthor{Kolmogorov1941}'s work, \citet{She1994} and \citet{Boldyrev2002} have extended and generalised the analysis and predict the following.
For an incompressible filamentary flow \citet{She1994} predict that the VSFs scale with,
\begin{equation}
	\zeta_\mathrm{She}(p) = \frac{p}{9} + 2 \left[ 1 - \left( \frac{2}{3} \right)^{\frac{p}{3}} \right] = Z_\mathrm{She}(p) ,
    \label{equ:turb_vsf_she}
\end{equation}
while supersonic flows with sheet-like geometry are supposed to scale with \citep{Boldyrev2002},
\begin{equation}
	 \zeta_\mathrm{Boldyrev}(p) = \frac{p}{9} + 1 - \left( \frac{1}{3} \right)^{\frac{p}{3}} = Z_\mathrm{Boldyrev}(p) .
    \label{equ:turb_vsf_boldyrev}
\end{equation}
\citet{Benzi1993} have introduced the principle of "extended self-similarity" which propose that there is a fixed relation between the a VSF of $p^\mathrm{th}$ order and the 3$^\mathrm{rd}$ VSF, so that the ratio $Z(p) = \zeta(p) / \zeta(3)$ is constant over all lag scales.
Since the mentioned predictions of $\zeta(p)$ are normally normalised in a way that $\zeta(3)$~=~1 Eq.~(\ref{equ:turb_vsf_she}) and~(\ref{equ:turb_vsf_boldyrev}) also provide the predictions for $Z(p)$, respectively.

\noindent For the discussion below, I measure $\zeta$ by fitting a power-law, given by
\begin{equation}
	\ln\left[ S_p(\ell) \right] = \ln\left(A\right) + \zeta \ln(\ell) ,
    \label{equ:turb_vsf_fitting}
\end{equation}
with $A$ being the scaling factor of the power-law to the simulated measurements.
For the calculations, I only take those cells with a minimal number density of 100~cm$^{-3}$ into account as this threshold defines the volume of the clouds.

\subsection{Turbulence within Model Clouds}\label{turb:vsf_results}

In this section, I use the VSF to characterise the nature of turbulence in three model clouds.
I analyse the same clouds that are part of the simulations of \citet{IbanezMejia2016,Ibanez-Mejia2017} and that I have already introduced in Sect.~\ref{frag:clouds}.

\begin{figure}
	\centering
    \begin{subfigure}{0.7\textwidth}
		\includegraphics[width=\textwidth]{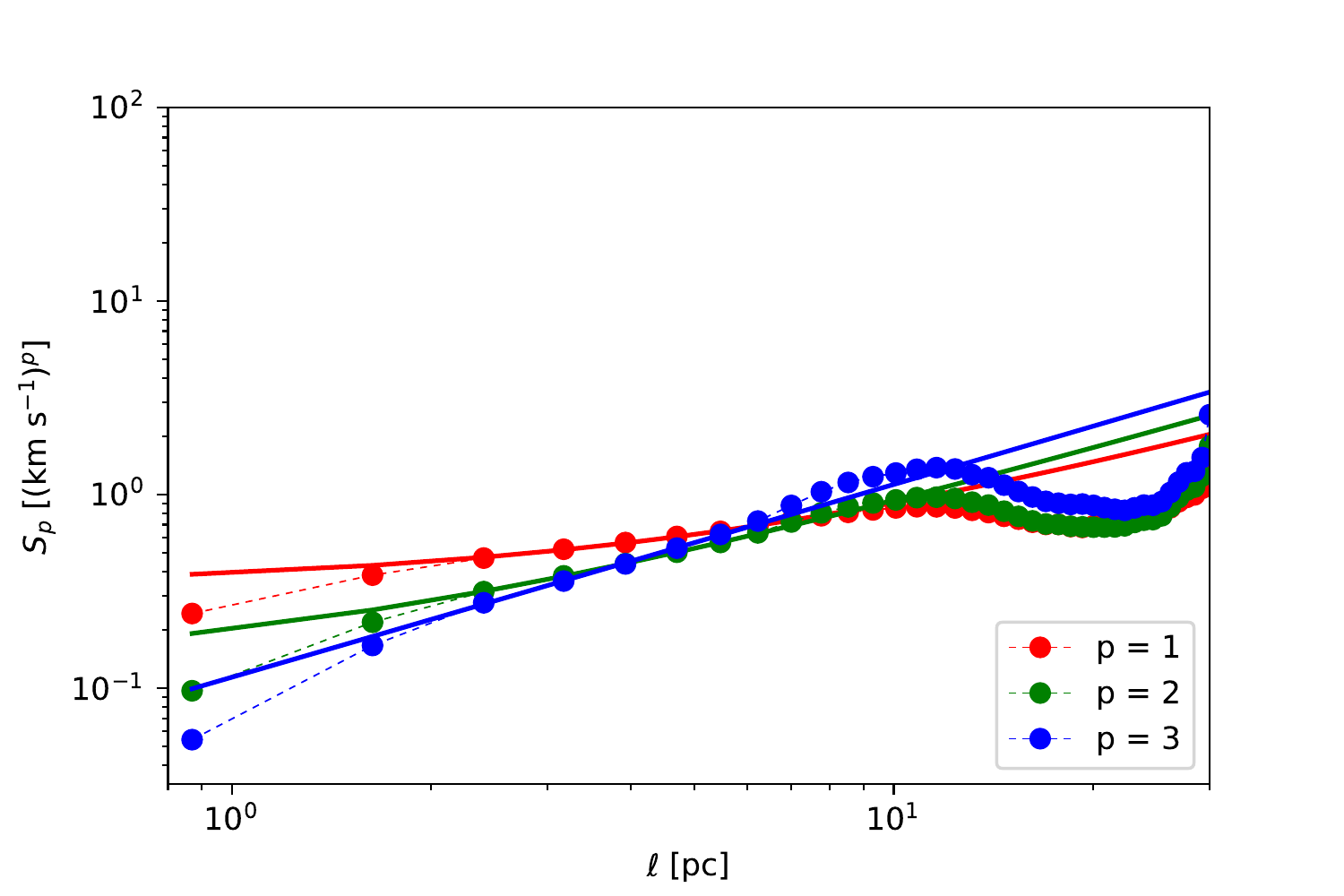}
		\caption{\texttt{M4} at $t$~=~1.2~Myr; decaying turbulence}
		\label{pic:turb_vsf_example_M4_0012}
    \end{subfigure}
    
    \begin{subfigure}{0.7\textwidth}
    	\includegraphics[width=\textwidth]{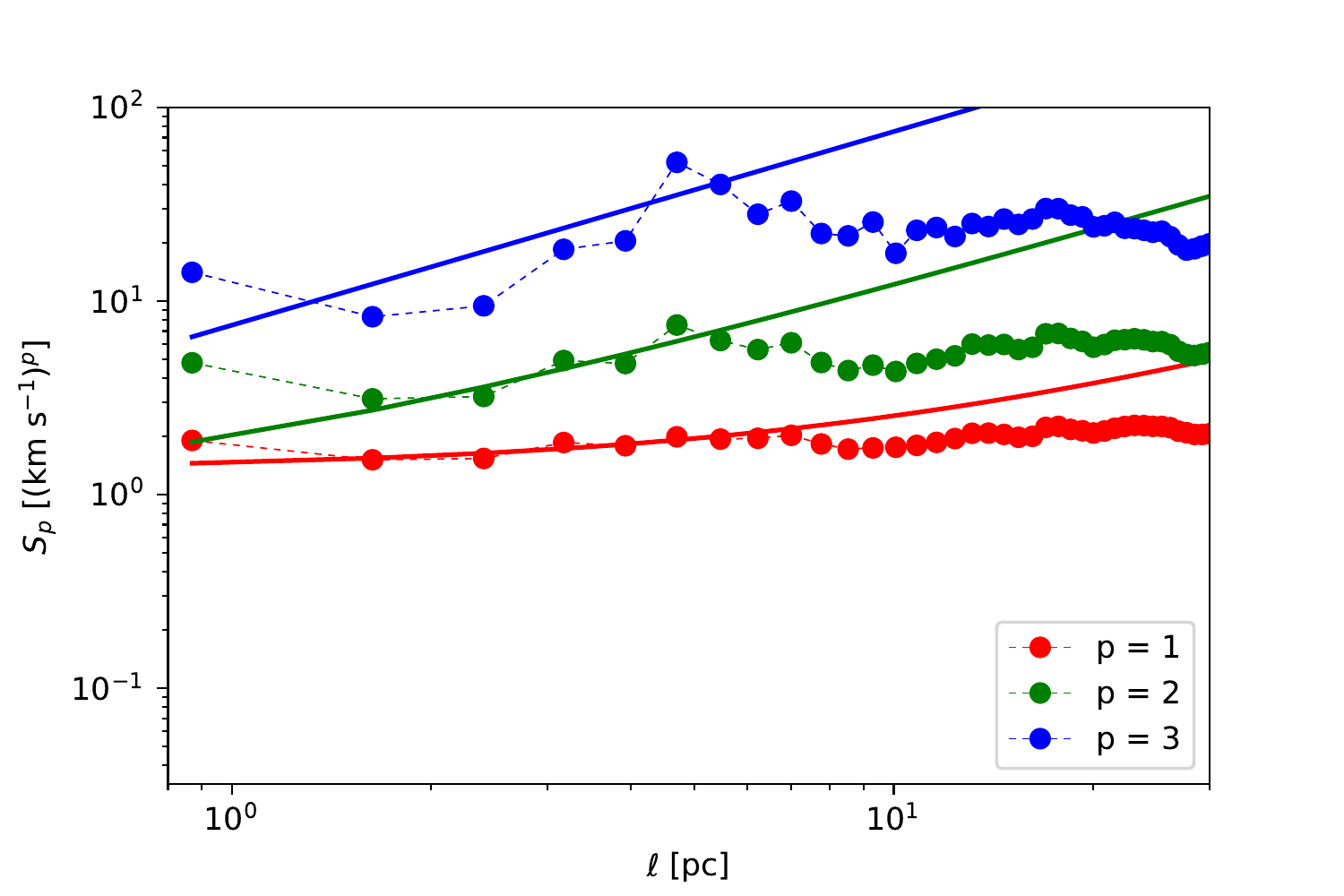}
		\caption{\texttt{M3} at $t$~=~3.5~Myr: supernova shocked}
		\label{pic:turb_vsf_example_M3_0035}
    \end{subfigure}
    
    \begin{subfigure}{0.7\textwidth}
    	\includegraphics[width=\textwidth]{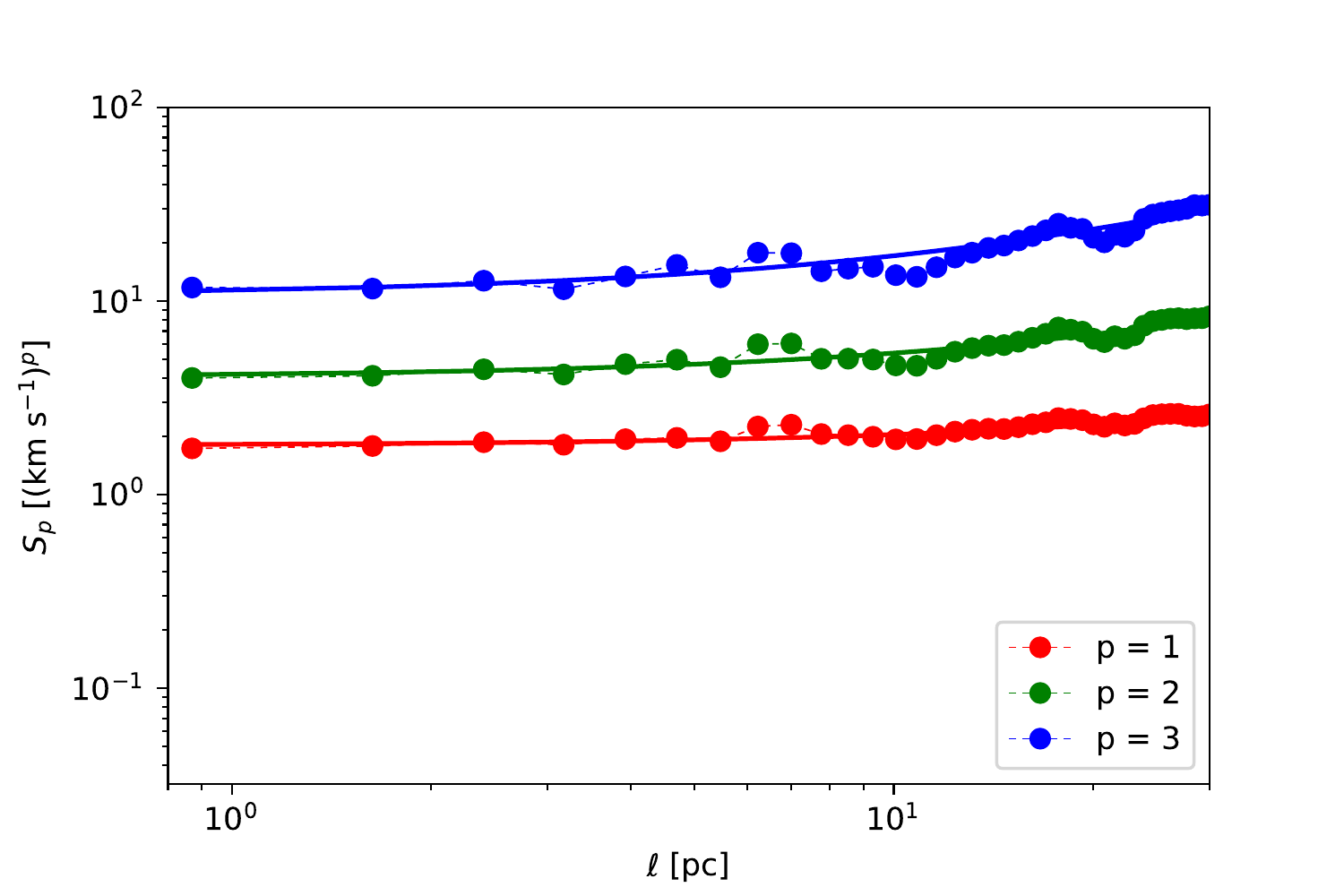}
		\caption{\texttt{M3} at $t$~=~4.0~Myr: gravitationally contracting}
		\label{pic:turb_vsf_example_M3_0040}
    \end{subfigure}
    \caption[Examples of velocity structure functions]{Example of velocity structure functions as function of the lag scale, $\ell$, and order, $p$.
		The dots (connected by dashed lines) illustrate the measured values based on the simulation data. 
		The solid lines represent the power-law relations fitted to the respective structure function.}
    \label{pic:turb_vsf_example}
\end{figure}

Fig.~\ref{pic:turb_vsf_example} shows three examples of how the VSFs look like, namely \texttt{M4} 1.2~Myr after self-gravity has been activated in the simulations  (Fig.~\ref{pic:turb_vsf_example_M4_0012}), \texttt{M3} at $t$~=~3.5~Myr (Fig.~\ref{pic:turb_vsf_example_M3_0035}) and \texttt{M3} at $t$~=~4.0~Myr (Fig.~\ref{pic:turb_vsf_example_M3_0040}).
All plots illustrate the VSFs of the orders $p$~=~1--3 that are computed based on the simulation data.

\pagebreak

\noindent The examples demonstrate that, in general, the measured VSFs cannot be described by a single power-law relation over the entire range of $\ell$.
Rather they are composed of three different regimes: 
one at small scales with $\ell \lesssim$~3~pc, a second one within 3~pc~$\lesssim \ell \lesssim$~10--15~pc, and the last one at large scales with $\ell >$~15~pc.
Therefore, only the small and intermediate ranges may be described by a common power-law relation.
On larger scales, one observes a local minimum before the VSFs either increase or saturate.
The location of the minimum, thereby, coincides with the equivalent radius of the cloud, meaning the radius a cloud of given mass would have if it would be a sphere.
Thus, in this context the VSF is an accurate tool to measure the size of a molecular cloud.
On smaller scales, which correspond to individual clumps and cores, one sees significant differences.

Furthermore, the examples illustrate how the clouds and their VSFs react to different scenarios that affect the turbulent structure of the entire clouds. 
In Fig.~\ref{pic:turb_vsf_example_M4_0012} one sees the standard case of VSFs where the turbulence is fully developed and decaying towards smaller scales within the clouds. 
This is the dominant case within the first $\sim$1.5~Myr of the simulations \citep[see][for more details]{Ibanez-Mejia2017,Seifried2017b}.
During this interval of time the clouds experience the effect of self-gravity for the first time in their evolution and first need to adjust to this new conditions.
Until this is the case, their VSFs are dominated by the freely cascading turbulence that previously dominated the kinetic structure of the clouds.
Furthermore, this also implies that I can only reliably examine turbulence within the simulations after 1.5~Myr and needs carefully be taken into account in the further discussion.

The other examples represent the clouds at later stages of their evolution when the VSFs are dominated by sources that drive the turbulence within the clouds in a more extreme way.
Fig.~\ref{pic:turb_vsf_example_M3_0035} shows the VSF of \texttt{M3} at a time when the cloud has just been hit by a supernova shock. 
One clearly sees how the amplitude of the VSFs is increased by one to two orders of magnitudes compared to the previous example.
Especially the power at small scale turbulence ($\ell \sim$ few parsecs) is highly amplified, which is the imprint of the shock front itself.
Furthermore, the shock reduces the equivalent radius of the cloud.
This results in a steeper scaling of the VSF despite the increase of turbulent power at small scales.
However, the effect of SN shocks last for only a short period of time (see below).

The last example, Fig.~\ref{pic:turb_vsf_example_M3_0040}, demonstrates the imprint of (self-)gravitational collapse.
Here, the VSF is almost flat, or even slightly increasing towards smaller separation scales. 
This kind of profile is typical for gas that is self-gravitationally contracting \citep{Boneberg2015,Burkhart2015} since gas moves into the inner regions of the cloud, reducing the average lag distances, but not necessarily the relative velocities.
The latter may even be accelerated by the infall.
As a consequence, large amounts of kinetic energy are transferred to smaller scales which flattens the corresponding VSF.

\begin{table}
	\begin{center}
	\begin{tabular}{l|c|c|c}
    	sample  / $\zeta(p)$ & p = 1 & p = 2 & p = 3 \\ \hline
        \citet{Boldyrev2002} & 0.42 & 0.74 & 1.0 \\ 
        \citet{She1994} & 0.36 & 0.70 & 1.0 \\ \hline
        4 cells per $\lambda_J$  & 0.29 $\pm$ 0.29 & 0.54 $\pm$ 0.55 & 0.78 $\pm$ 0.79 \\
        8 cells per $\lambda_J$  & 0.35 $\pm$ 0.19 & 0.65 $\pm$ 0.38 & 0.95 $\pm$ 0.61 \\
        32 cells per $\lambda_J$ & 0.52 $\pm$ 0.08 & 1.10 $\pm$ 0.17 & 1.89 $\pm$ 0.44 % \\ \hline
%        4 cells per $\lambda_J$, 100 cm$^{-3} \leq$ n $\leq$ 8,000 cm$^{-3}$ & 0.37 $\pm$ 0.20 & 0.66 $\pm$ 0.41 & 0.92 $\pm$ 0.63 
	\end{tabular}
	\end{center}
	\caption[Statistical summary of $\zeta(p)$]{Statistical summary of scaling exponents, $\zeta(p)$, of the velocity structure functions and comparison with values by theoretically predicted by \citet{Boldyrev2002} and \citet{She1994}.
    }
	\label{tab:turb_strucfunc_zeta}
\end{table}

\pagebreak

\begin{figure}
	\centering
	\includegraphics[width=\textwidth]{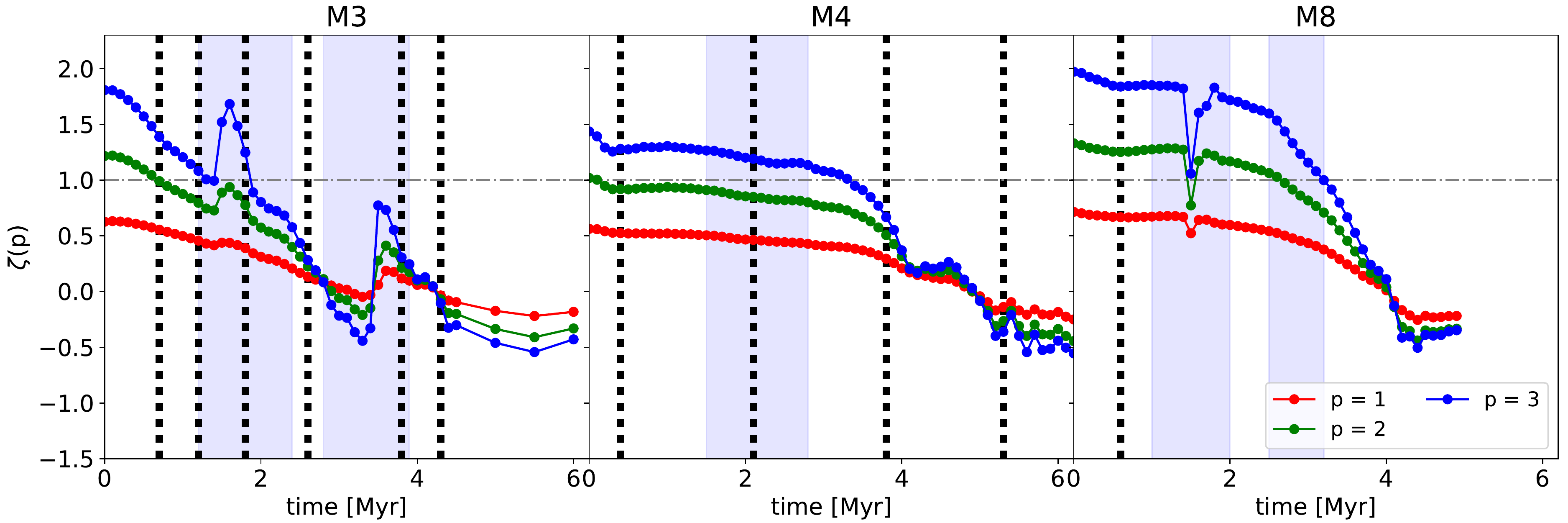}
    \caption[Time evolution of $\zeta$]{Time evolution of scaling exponent $\zeta$ of the $p^\mathrm{th}$ order structure function in \texttt{M3} (\textit{left}), \texttt{M4} (\textit{middle}), and \texttt{M8} (\textit{right}).
    The black dotted lines indicate the times when SNe occur, and the blue shaded areas do so when mass accretion peaks within the respective clouds.
    }
    \label{pic:turb_vsf_zeta_evol}
\end{figure}

\noindent The solid lines in the examples illustrate the fitted power-law relations as given in Eq.~(\ref{equ:turb_vsf_fitting}).
Table~\ref{tab:turb_strucfunc_zeta} (4~cells per $\lambda_J$, see Sect.~\ref{turb:vsf_resolution}) statistically summarises the measured values of $\zeta$ for $p$~=~1--3 for a better comparison.
Fig.~\ref{pic:turb_vsf_zeta_evol} plots the time evolution of the $\zeta$ obtained for all three clouds.
The figure shows several interesting features.
First, initially all calculated values of $\zeta$ are above the predicted values.
This means that the turbulence within the clouds is highly supersonic before the gas begins to react to the onset of self-gravity.
Second, all $\zeta$ decrease with time as the clouds gravitationally collapse.
Therefore, the gas transfers the turbulent power from large to small scales.
This process accelerates the relative motions between cells on all scales and causes a flat or even inverted profile in the VSF.
Third, occasionally one observes bumps and dips in all orders of VSFs (e.g., \texttt{M3} or \texttt{M8} around $t$~=~1.7~Myr). 
These features only last for short periods of time (up to 0.6~Myr), but set in quasi-instantly and represent a complete relocation of the turbulent power on all scales. 

The origin of these features can easily be traced back to the SNe occurring in the environment of the clouds.
To visualise this better, I add marks to Fig.~\ref{pic:turb_vsf_zeta_evol} that represent the times when the SNe explode (black dotted lines) and the periods of time when the clouds are heavily accreting gas (blue area).
The data are taken from \citet{Ibanez-Mejia2017}, where mean distances between the sites of SNe and the centres of the clouds can also be found.
Considering that the SNe shock fronts move at speeds of 50--100 km s$^{-1}$ through the ISM and with distances between 30--100~pc to the clouds, the shocks need around 1~Myr, on average, to reach the clouds.
Thus, one cannot only relate the major gas accretion events of the clouds to the arrival of those SNe that indeed affect the evolution of the clouds, but also all significant variations in the evolution of $\zeta$.
In most of the cases, the SNe cause an increase of VSF towards larger scales is reflected by a peaked $\zeta$.

\texttt{M8}, on the contrary, seems to develop differently.
At the time the SN, occurring at $t$~=~0.8~Myr, hits the cloud the $\zeta$ do not rise, as they have done within the other two clouds, but instead drop. 
After the shock all exponents grow to levels that are slightly above the pre-shock values, before they slowly decrease again.
This evolution appears to contradict the explanations I have given for the behaviour of turbulence within \texttt{M3} and \texttt{M4}.
However, looking at \texttt{M8} in detail this is not the case.
\citet{Ibanez-Mejia2017} show in their Fig.~5 that \texttt{M8} does not as strongly react to the onset of self-gravity as the other clouds do.
For example, the mass of \texttt{M8} remains almost constant for the first megayear while the other clouds strongly accrete gas from the ISM.
Similarly, the velocity dispersion within \texttt{M3} and \texttt{M4} increases within the first 1.5~Myr while it is close to constant in \texttt{M8} until the SN shock hits the cloud.
Thus, I can explain the evolution of $\zeta$ as follows:
In the beginning, \texttt{M8} evolves only slightly.
It loses a small fraction of gas \citep[Fig.~5, \textit{middle} panel]{Ibanez-Mejia2017}.
Therefore, the relative velocities increase towards larger separations.
At $t$~=~1.5~Myr the shock front of the $t$~=~0.8~Myr SN hits the cloud.
That causes a short periods of gravitational collapse, traced by the dip in $\zeta$.
As in the other clouds, the gas relaxes rapidly after the shock.
However, the cloud continues to gravitationally contract in a more regular way.
That is seen as slow decrease of $\zeta$ from $t$~=~1.8~Myr on. 

In summary, one can say that the scaling exponent, $\zeta$, that is obtained by fitting a power-law relation onto the measured velocity structure function, is a useful tool to understand and evaluate the time evolution of turbulence within molecular clouds. 
It is not only sensitive to both external (SNe) and internal (gravitational collapse) driving sources, but also reacts differently to the individual driving sources. 

However, this diagnostic requires a series of time steps to be significant.
According to \citet{She1994} and \citet{Boldyrev2002}, $\zeta$(3) is supposed to be equal or larger than unity whenever the gas experiences supersonic turbulence.
Although this is definitely the case in the model clouds \citep{IbanezMejia2016,Ibanez-Mejia2017}, one sees that $\zeta$(3) declines far below 1 due to gravitational collapse.
Thus, measuring $\zeta$ for individual moments in time cannot fully describe the turbulence of molecular clouds.

\begin{table}
	\centering
	\begin{tabular}{l|c|c|c}
    	sample  / $Z(p)$ & p = 1 & p = 2 & p = 3 \\ \hline
        \citet{Boldyrev2002} & 0.42 & 0.74 & 1.0 \\ 
        \citet{She1994} & 0.36 & 0.70 & 1.0 \\ \hline
        4 cells per $\lambda_J$  & 0.40 $\pm$ 0.12 & 0.72 $\pm$ 0.13 & 1.0 \\
        8 cells per $\lambda_J$  & 0.35 $\pm$ 0.09 & 0.67 $\pm$ 0.11 & 1.0 \\
        32 cells per $\lambda_J$ & 0.29 $\pm$ 0.06 & 0.59 $\pm$ 0.07 & 1.0 %\\ \hline
%        4 cells per $\lambda_J$, 100 cm$^{-3} \leq$ n $\leq$ 8,000 cm$^{-3}$ & 0.57 $\pm$ 0.85 & 0.87 $\pm$ 0.89 & 1.0
	\end{tabular}
	\caption[Statistical summary of $Z(p)$]{As Table~\ref{tab:turb_strucfunc_zeta} for $Z(p)$.
    }
	\label{tab:turb_strucfunc_zcomp}
\end{table}

\begin{figure}
	\centering
	\includegraphics[width=\textwidth]{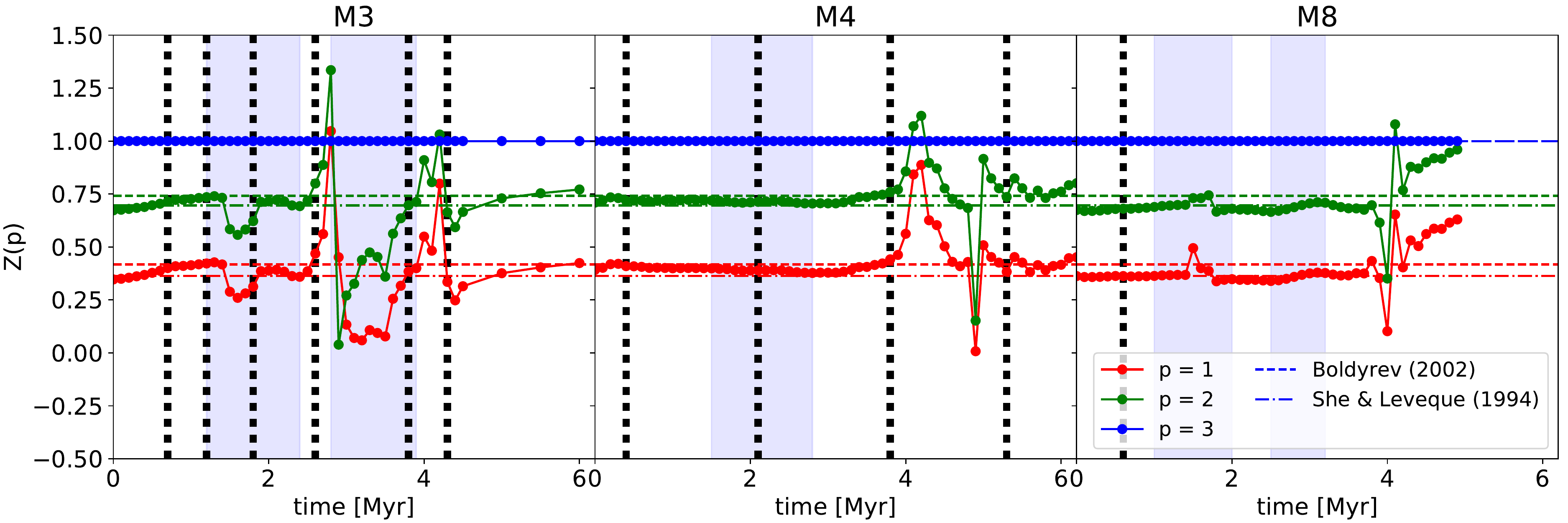}
    \caption[Time evolution of $Z$]{As Fig.~\ref{pic:turb_vsf_zeta_evol}, but with the measurements of $Z(p)$. The dashed and dashed-dotted horizontal lines mark the values that are theoretically predicted for supersonic, turbulent flows by \citet{Boldyrev2002} and \citet{She1994}, respectively.}
    \label{pic:turb_vsf_zeta_relevol}
\end{figure}

The principle of "extended self-similarity" \citep[Sect.~\ref{turb:vsf_theory}]{Benzi1993} offers a solution to this problem.
Therefore, I measure the ratio between the scaling of the individual VSF of $p^\mathrm{th}$ order relative to the 3$^\mathrm{rd}$ order VSF scaling, $Z(p) = \zeta(p)/\zeta(3)$. 
Table~\ref{tab:turb_strucfunc_zcomp} (4~cells per $\lambda_J$) provides a statistical summary of the measurements of $Z(p)$, as well as a comparison to literature values.
Fig.~\ref{pic:turb_vsf_zeta_relevol} plots the time evolution of $Z(p)$ based on the model clouds. 
One sees that most of the time the measured values of $Z(p)$ are in agreement or at least closely approaching the predicted values.
This reflects the supersonic nature of intercloud turbulence, even when it is dominated by gravitational collapse.
However, I also detect strong deviations that either reduce or increase the measured values of $Z(p)$.
Those derivations can be related to the physical forces that currently dominate the clouds.

The peaks in the $Z(p)$ (for example, in \texttt{M4} at $t$~=~4.1~Myr) occur at the times when the scaling exponents of the VSFs, $\zeta(p)$ reach values close or below 0.
That means that the VSF becomes flat ($\zeta(p)$~=~0) or increases forward smaller scales. 
In this case, self-gravitational contraction clearly dominates the cloud's evolution.
It transfers the majority of turbulent power from the large to the small scales, where filaments and fragments are forming and accreting gas.

The decrease in $Z(p)$ (for example, in \texttt{M3} around $t$~=~1.8~Myr), on the other hand, is caused by SN shocks that hit and heavily impact the clouds. 
This causes a sudden, but heavy increase of turbulent power on all scales, though on the larger scales more than on the smaller, resulting in a steepening of the VSF towards larger scales and an increase in $\zeta(p)$.
The VSFs become more sensitive to the shock the higher their order is.
Therefore, the $\zeta(3)$ increases more rapidly than $\zeta(2)$ and $\zeta(1)$, causing $Z(2)$ and $Z(1)$ to decrease.
\pagebreak

\noindent In summary, $Z$ is not as sensitive to gravitational contraction as $\zeta$, although it is very sensitive to inversion of turbulent power from large-scale to small-scale dominated due to gravity.
It also traces the impact of SN shocks on the clouds.
Whenever neither of these two extreme scenarios is acting on the clouds, the values of $Z$ are close to the predicted values (see discussion below).
This means that $Z$ is, as $\zeta$, a good tracer for the dominant forces driving the turbulence within the cloud, both internally and externally, as long as they are heavily impacting the cloud.
The disadvantage is that one cannot use $Z$ to distinguish between freely-floating fully-developed and moderate gravitational contraction, as one can do it with $\zeta$.
However, the advantage is that a single-epoch observation of $Z(p)$ is sufficient to trace extreme motions and driving sources within observed molecular clouds.

At the times when the clouds are not impacted by the above described extreme cases, one sees that the ratio of the VSF scaling exponents is mostly in agreement with the principle of "self-similarity", although the clouds are dominated by gravitationally contracting motions during their evolution.
The measured $Z(p)$, thereby, do not uniquely follow the predictions of only \citet{Boldyrev2002} or \citet{She1994}, but are normally between the predicted values of both theories. 
Recalling that both studies describe supersonic, fully developed turbulent flows, the only difference between them are the different geometries along which they allow the gas to flow.
In the case of \citet{Boldyrev2002}, the gas flows are sheet-like, while they are filamentary in the work by \citet{She1994}.

Since \texttt{M3} is heavily impacted by many SN shocks and heavy collapse motions, it does not allow us to make strong predictions about its 'steady-state' evolution.
The other two clouds, on the contrary, are less externally impacted.
This allows us to relate the developments of measured $Z(p)$ to the internal evolution of the clouds.

In \texttt{M8}, $Z(p)$ follows the predictions by \citet{She1994} for most of the cloud's evolution. 
This shows that \texttt{M8} mostly transfers gas along filamentary substructures which means that the cloud is highly hierarchically structured already at the beginning of the simulations and before self-gravity is active.
This also implies that the formation of filamentary structures does not require gravity \citep[e.g.,][]{Federrath2016}.
The fact that I do not detect fragments before the clouds have evolved due to the influence of self-gravity for at least one megayear (see Chapter~\ref{frag}), however, demonstrates that gravity is essential for the fragmentation of filaments and formation of further substructures.
\pagebreak

\noindent The evolution of $Z(p)$ in \texttt{M4} shows a slightly different picture that still contains significant conclusion.
While the values of $Z(p)$ in \texttt{M8} are mostly constant over time, they are decreasing in \texttt{M4}, starting at values that are in agreement with the sheet-like flows of \citet{Boldyrev2002} to those that are predicted for filamentary flows by \citet{She1994}. 
This demonstrates that \texttt{M4} develops its hierarchical structure as it evolves under the influence of self-gravity. 
As a consequence, its turbulent structure becomes more dominated by the filaments with time which causes the decline of $Z(p)$.
Considering that the gas of \texttt{M4} is indeed first flattened into more sheet-like geometry through the impact of the SNe \citep{Ibanez-Mejia2017} this observation is very interesting.
It agrees with the studies, for example by \citet{Lin1965} or \citet{McKee2007}, that claim that molecular clouds are supposed to collapse in a dimension-losing, outside-in fashion that does not require any initial substructures within the clouds.

In summary, $Z$ reflects the global geometry of turbulence within molecular clouds. 
Thereby, the measure values of $Z$ are in good agreement with theoretical predictions, as long as one carefully ensures that the dominating turbulent mode in the cloud matches the modes that are considered in the respective theory.
This makes $Z$ a reliable parameter for examining turbulent modes in observational studies, as well.
Furthermore, the time evolution of $Z$ shows how the geometry of turbulence is changed due to the gravitational contraction, from sheet-like to filamentary vortices, accompanying the change of basic parameters describing the turbulent structure of the entire cloud. 
$Z$ can also deviate from the predicted values. 
This, however, only occurs in time spans of extreme turbulence driving, such as SN shocks or when the distribution of turbulent power is inverted due to gravitational contraction.
Fortunately, these extreme driving sources affect $Z$ differently; SN shocks decrease $Z$, while gravity causes a quasi-momentary peak in $Z$.
In both cases, $Z$ relaxes to the pre-perturbation values within a short time.

\subsection{The Effect of Jeans Length Refinement}\label{turb:vsf_resolution}

The results I have discussed so far are based on data of \citet{IbanezMejia2016}'s simulations.
Due to the huge computational expense the variety of physical and numerical processes (fluid dynamics, adaptive mesh refinement, supernovae, magnetic fields, radiative heating and cooling, and many more) those simulations, though, have also demanded some compromises.

One of these compromises has been the Jeans refinement criterion that is part of the AMR mechanisms.
Therefore, the authors have resolved local Jeans lengths by only four cells ($\lambda_J$~=~$4\Delta{}x$).
This is the minimal requirement for modelling self-gravitating gas in order to avoid artificial fragmentation \citep{Truelove1998}. 
Other studies, for example by \citet{Turk2012}, yet have shown that a significant higher refinement is needed to reliably resolve turbulent structures and flows on scales of single cells.

In the appendix of \citet{Ibanez-Mejia2017}, the authors examine the effect the number of cells used for the Jeans refinement has on the measured kinetic energy.
For this, they have rerun the simulations of \texttt{M3} twice; 
once with a refinement of eight cells per Jeans length ($\lambda_J$~=~$8\Delta{}x$) for the first 3~Myr after self-gravity was activated, and once with 32 cells per Jeans length ($\lambda_J$~=~$32\Delta{}x$) for the first megayear of the cloud's evolution.
The authors show that the $\lambda_J$~=~$32\Delta{}x$ simulations smoothly reveal the energy power spectrum on all scales already after this first megayear.
The other two setups also do this.
However, they need more time to overcome the resonances in the respective power spectra that originate from the previous resolution steps. 
This is why one can only fully reliably trust the findings in this chapter after the clouds have evolved for approximately 1.5~Myr \citep[see also][]{Ibanez-Mejia2017,Seifried2017b}.
\pagebreak

\noindent More importantly, though, \citet{Ibanez-Mejia2017} have also calculated the difference in the cloud's total kinetic energy as function of time and refinement level.
They found that the $\lambda_J$~=~$4\Delta{}x$ simulations miss a significant amount of kinetic energy, namely up to 13\% compared to $\lambda_J$~=~$8\Delta{}x$ and 33\% compared to $\lambda_J$~=~$32\Delta{}x$.
However, they also observed that these differences peak around $t$~=~0.5~Myr and decrease afterwards again, as the $\lambda_J$~=~$4\Delta{}x$ and $\lambda_J$~=~$8\Delta{}x$ simulations adjust to the new refinement levels.
This, of course, means that the results I have derived from the $\lambda_J$~=~$4\Delta{}x$ simulations always needs to be evaluated with respect to this lack of energy, although the clouds' dynamics is dominated by gravitational collapse; yet it also means that the $\lambda_J$~=~$4\Delta{}x$ data become more reliable the longer the simulations have time to evolve.

In this section, I estimate how the lack in total kinetic energy influences the behaviour of the VSFs.
In order to do so, I analyse the data of the $\lambda_J$~=~$8\Delta{}x$ and $\lambda_J$~=~$32\Delta{}x$ simulations in the same way as I have done with the $\lambda_J$~=~$4\Delta{}x$ data: measure the VSFs and fit power-law relations onto them.
Tables~\ref{tab:turb_strucfunc_zeta} and \ref{tab:turb_strucfunc_zcomp} summarise the measurements statistically and in comparison with the previous measurements and the predicted values by \citet{Boldyrev2002} and \citet{She1994}.

\begin{figure}[h!t]
	\centering
    \begin{subfigure}{0.48\textwidth}
      \includegraphics[width=\textwidth]{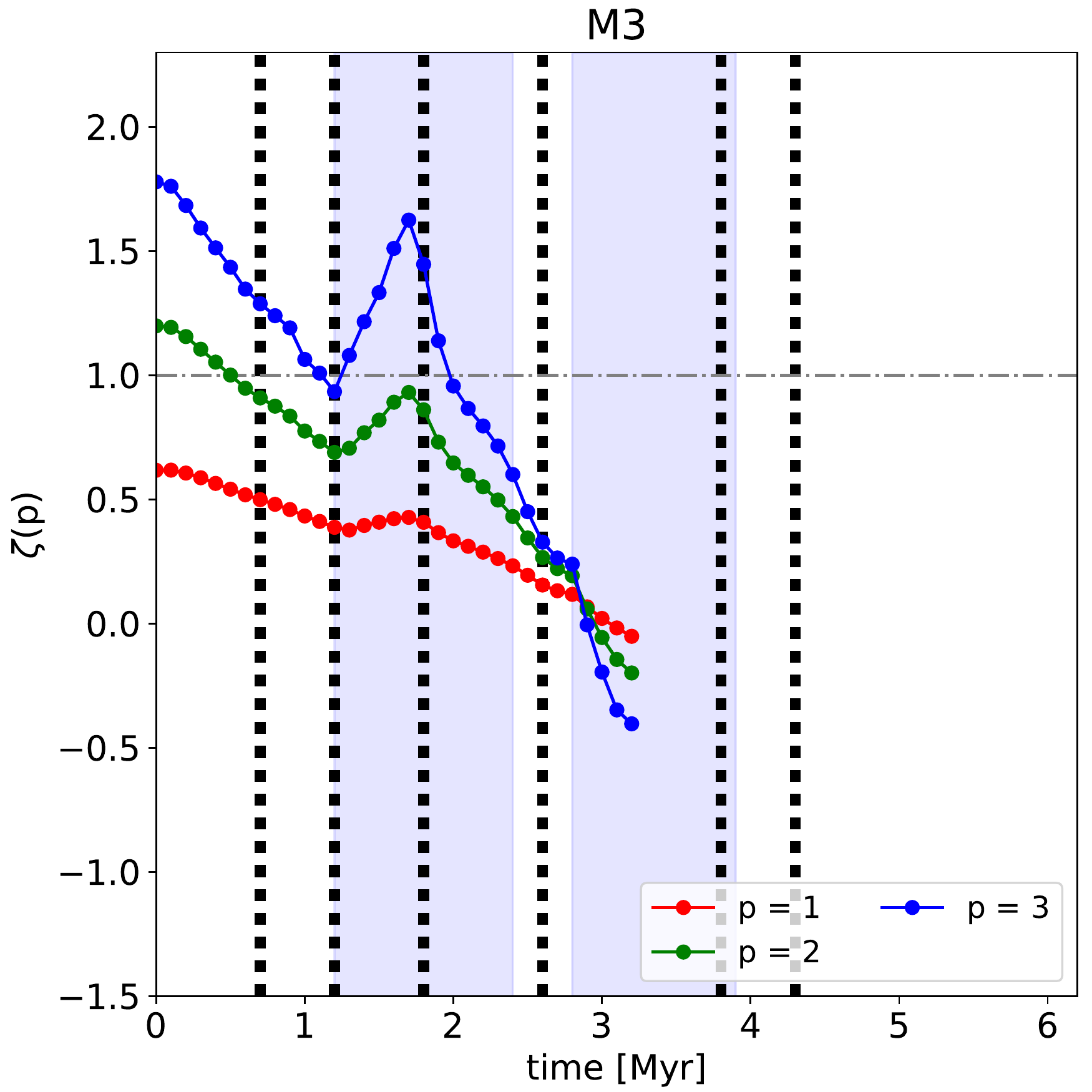}
      \caption{time evolution of $\zeta$}
      \label{pic:turb_vsf_zeta_evol_vsfhr08}
    \end{subfigure} 
    \hfill
    \begin{subfigure}{0.48\textwidth}
      \includegraphics[width=\textwidth]{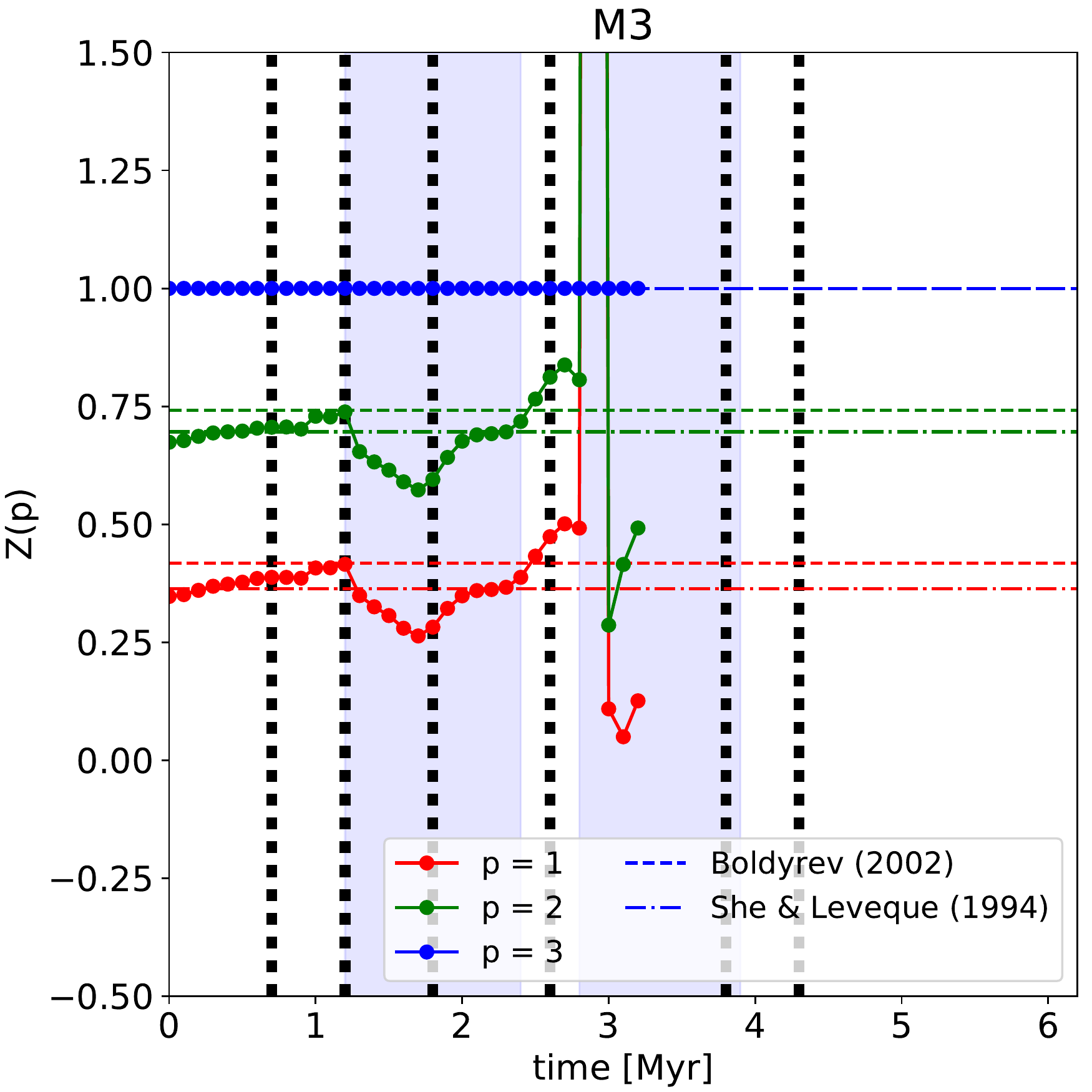}
      \caption{time evolution of $Z$}
      \label{pic:turb_vsf_zeta_relevol_vsfhr08}
    \end{subfigure}
    \caption[Time evolution of $\zeta$ and $Z$ for 8 cells per Jeans length]{As Fig.~\ref{pic:turb_vsf_zeta_evol} and Fig.~\ref{pic:turb_vsf_zeta_relevol} but based on data that resolve the local Jeans length with 8 cells.}
    \label{pic:turb_vsf_vsfhr08}

	\begin{subfigure}{0.48\textwidth}
	\includegraphics[width=\textwidth]{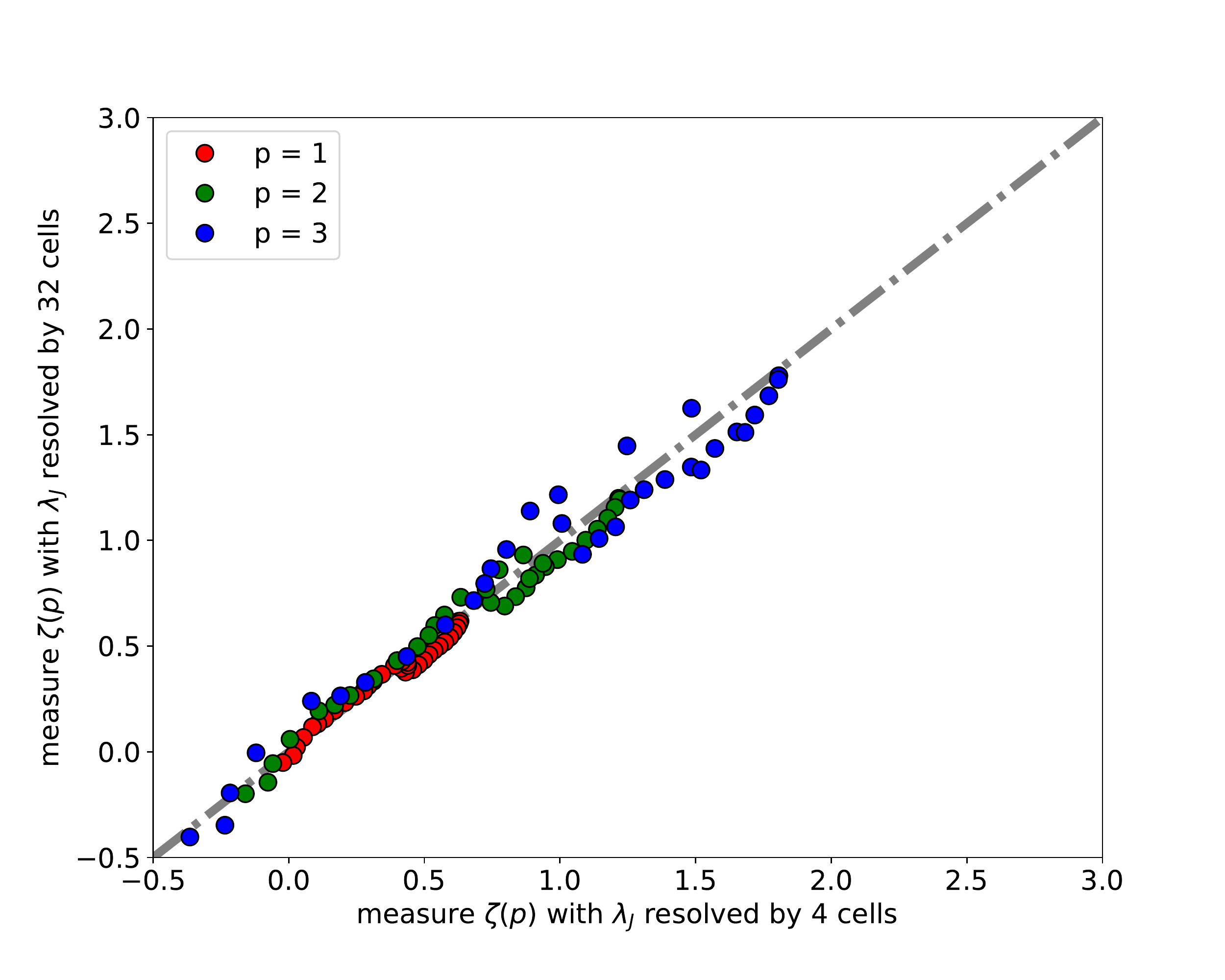}
    \caption{scaling exponents $\zeta$}
    \label{pic:turb_vsf_compres_vsfhr08_zeta}
    \end{subfigure}
    \hfill
    \begin{subfigure}{0.48\textwidth}
      \includegraphics[width=\textwidth]{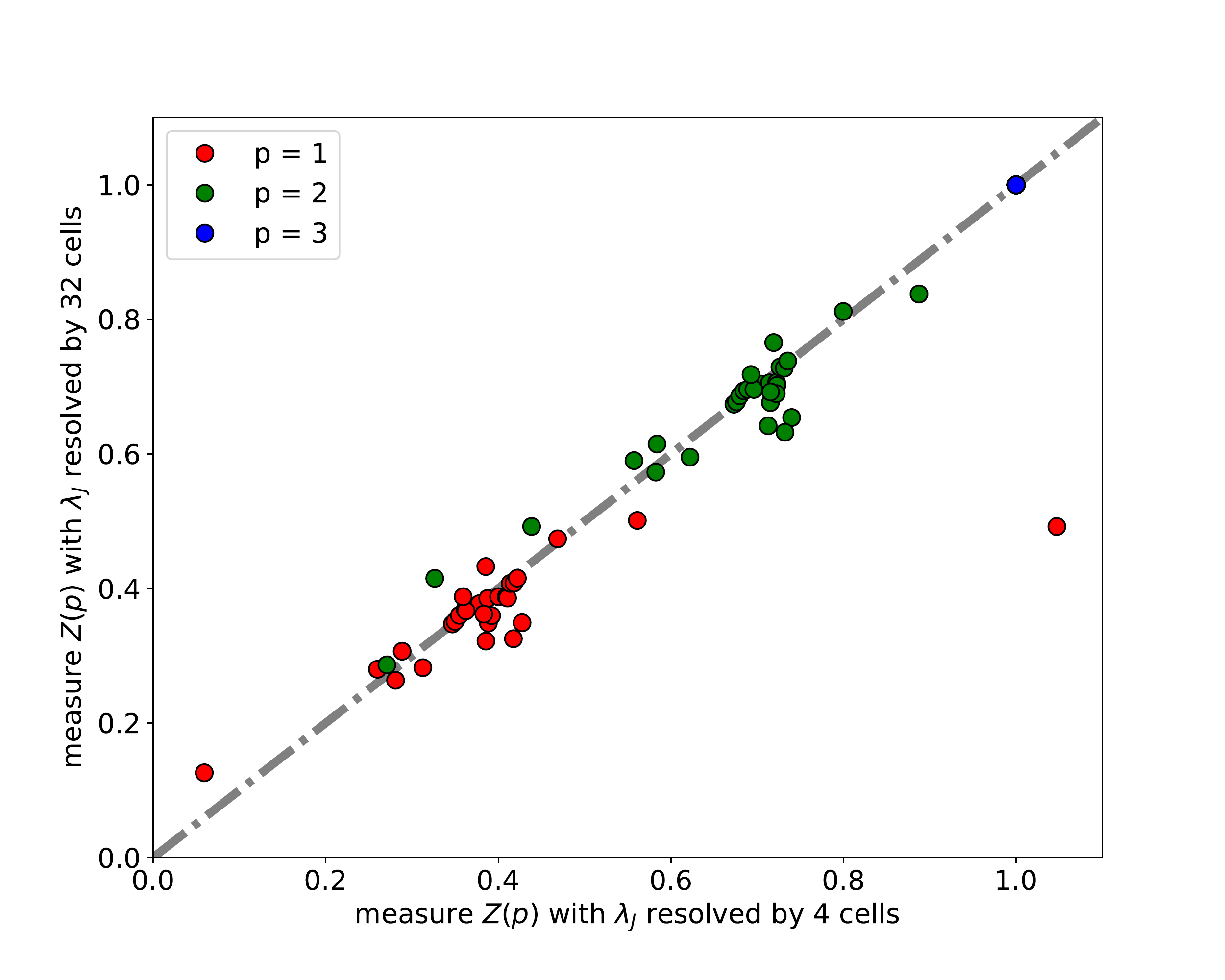}
      \caption{ratio of scaling exponents $Z$}
      \label{pic:turb_vsf_compres_vsfhr08_z}
    \end{subfigure}
    \caption[Comparision of $\zeta$ and $Z$ for 8 cells per Jeans length to 4 cells per Jeans length]{Comparison of measured $\zeta$ (\textit{left}) and $Z$ (\textit{right}) values based on data taking 4 cells per Jeans length (\textit{abscissa}) and 8 cells per Jeans length (\textit{ordinate}).
    The grey dashed-dotted line illustrates the line along which the measured values of both simulation samples are equal.
    }
    \label{pic:turb_vsf_compres_vsfhr08}
\end{figure}

\begin{figure}[h!t]
	\centering\begin{subfigure}{0.48\textwidth}
	\includegraphics[width=\textwidth]{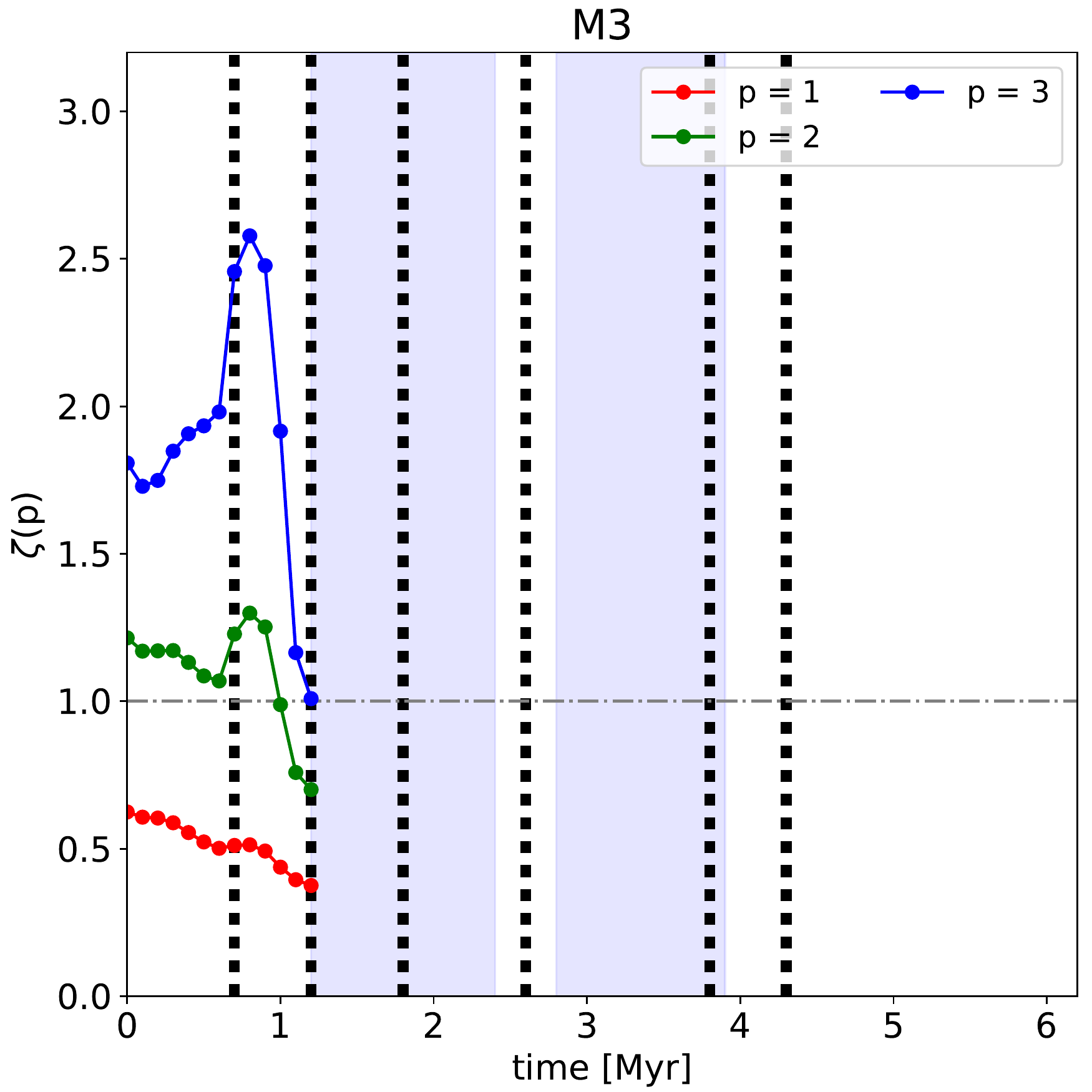}
    \caption{time evolution of $\zeta$}
    \label{pic:turb_vsf_zeta_evol_vsfhr32}
    \end{subfigure}
    \hfill
    \begin{subfigure}{0.48\textwidth}
      \includegraphics[width=\textwidth]{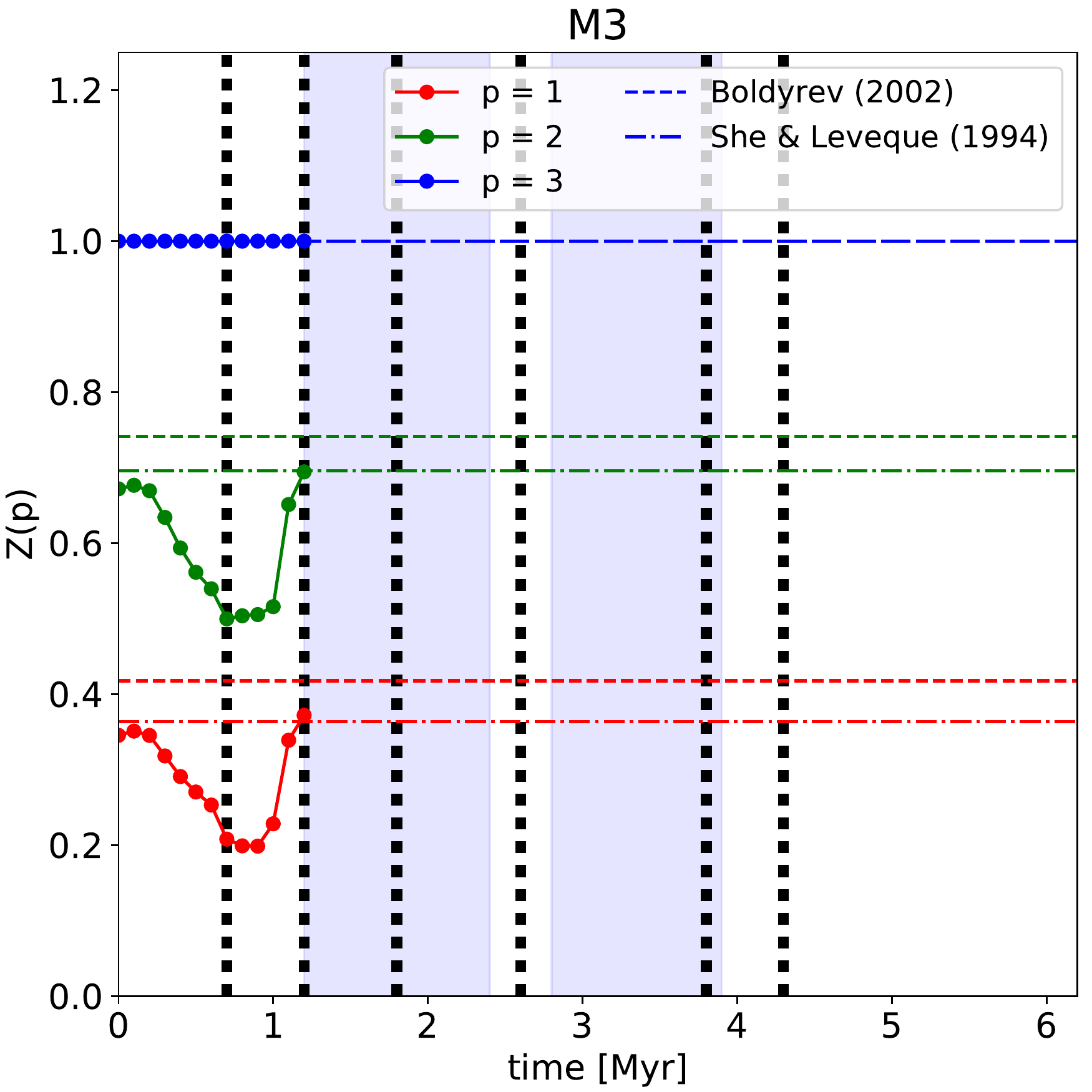}
      \caption{time evolution of $Z$}
      \label{pic:turb_vsf_zeta_relevol_vsfhr32}
    \end{subfigure}
    \caption[Time evolution of $\zeta$ and $Z$ for 32 cells per Jeans length]{As Fig.~\ref{pic:turb_vsf_vsfhr08} but based on data that resolve the local Jeans length with 32 cells.}
    \label{pic:turb_vsf_vsfhr32}

	\begin{subfigure}{0.495\textwidth}
		\includegraphics[width=\textwidth]{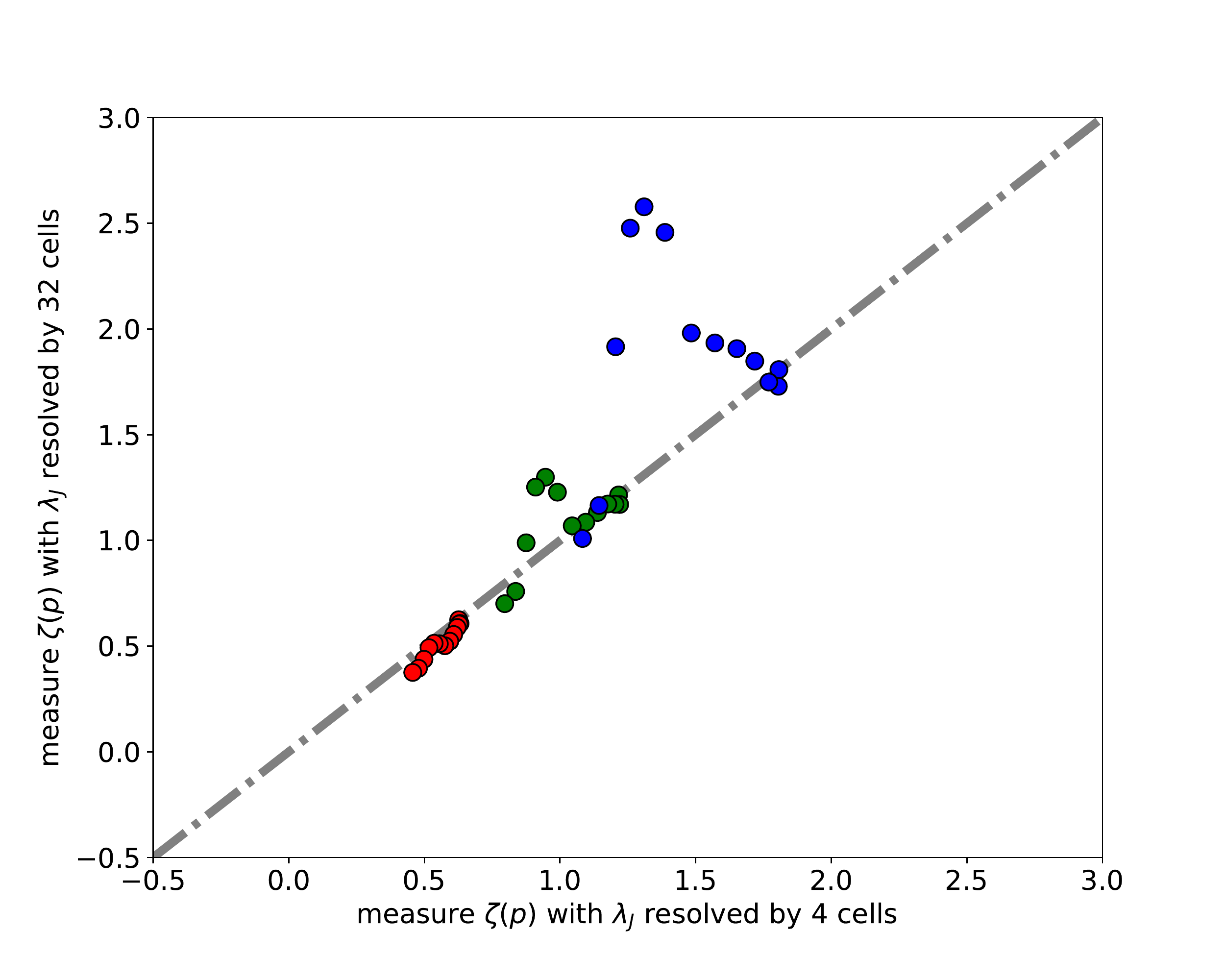}
		\caption{scaling exponents $\zeta$}
		\label{pic:turb_vsf_compres_vsfhr32_zeta}
	\end{subfigure}
%	\hfill
	\begin{subfigure}{0.495\textwidth}
		\includegraphics[width=\textwidth]{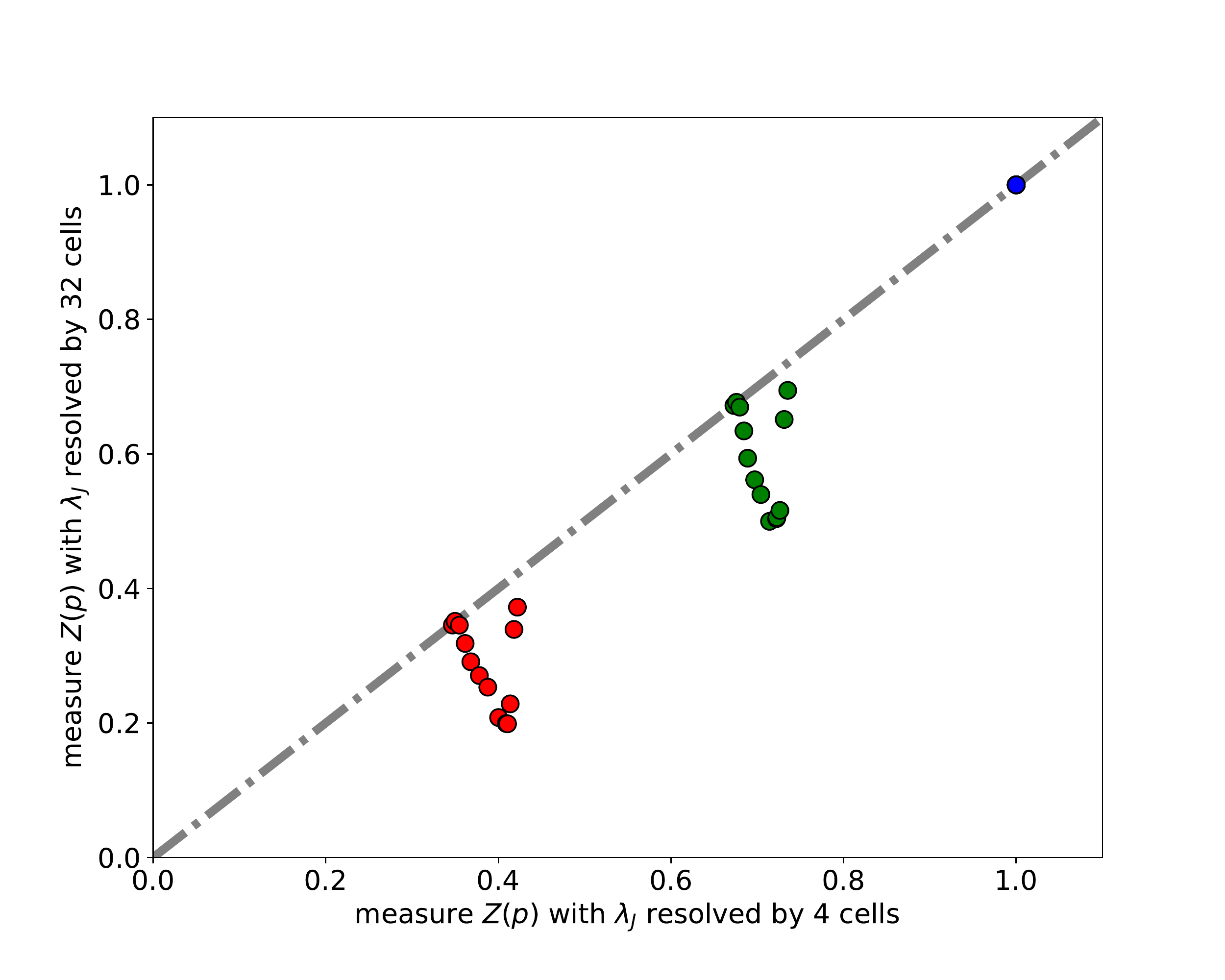}
		\caption{ratio of scaling exponents $Z$}
		\label{pic:turb_vsf_compres_vsfhr32_z}
	\end{subfigure}
	\caption[Comparision of $\zeta$ and $Z$ for 23 cells per Jeans length to 4 cells per Jeans length]{As Fig.~\ref{pic:turb_vsf_compres_vsfhr08} for measured $Z$ values.}
	\label{pic:turb_vsf_compres_vsfhr32}
\end{figure}

Figs.~\ref{pic:turb_vsf_vsfhr08} and \ref{pic:turb_vsf_vsfhr32} plot the measure values of $\zeta(p)$ and $Z(p)$ for the $\lambda_J$~=~$8\Delta{}x$ and $\lambda_J$~=~$32\Delta{}x$ runs, respectively.
$\lambda_J$~=~$8\Delta{}x$ (Fig.~\ref{pic:turb_vsf_zeta_evol_vsfhr08}) shows the same behaviour as $\lambda_J$~=~$4\Delta{}x$ previously. 
The VSFs steeply increase forward larger scales, causing values of $\zeta$ far higher than the values predicted by \citet{Boldyrev2002} and \citet{She1994}.
Yet, as the cloud contracts under the influence of self-gravity the kinetic energy is shifted to smaller scales, flattening the VSFs and reducing the corresponding values of $\zeta(p)$.
One also sees the peak in $\zeta(p)$ due to the SN exploding at $t$~=~0.8~Myr. 
However, the derivations in $\zeta$ are not as sharp as they have been measured for $\lambda_J$~=~$4\Delta{}x$.
Instead there is a continuous growth of $\zeta$ that lasts for 0.5~Myr, before the VSFs flatten again for $\sim$0.5~Myr.
This procedure lasts twice as long as the reaction in $\lambda_J$~=~$4\Delta{}x$ and overlaps with the response to the second SN (exploding at $t$~=~1.2~Myr).
In both simulations, the latter has not produced as significant a peak in $\zeta$ as the first SN, but its effect is still visible in $\lambda_J$~=~$4\Delta{}x$ as it reduces the rate at which $\zeta$ drops due to self-gravity.
In the $\lambda_J$~=~$8\Delta{}x$ simulations one cannot distinguish the contribution of the second SN from the peak produced by the first SN.
However, the smoother development of $\zeta$ in the $\lambda_J$~=~$8\Delta{}x$ simulation demonstrate that I can resolve the shock front better due to the higher refinement level as it moves from the ISM and the cloud.

This is also observable in Fig.~\ref{pic:turb_vsf_zeta_relevol_vsfhr08} where the sink of $Z$ due to the SN shock lasts longer than it has done in within the $\lambda_J$~=~$4\Delta{}x$ simulations. 
Besides this, the time evolution of $Z$ based on the $\lambda_J$~=~$8\Delta{}x$ simulations is as sensitive to the turbulence-related events as it has been for $\lambda_J$~=~$4\Delta{}x$.
Also the peak which is produced when gravity has transferred the majority of power to smaller scales occurs at the same time. 
The height of the peak is thereby a numerical artefact caused by $\zeta(3)$ being 0 at this very time step. 

Fig.~\ref{pic:turb_vsf_compres_vsfhr08} compares the measurements more directly by plotting the values of $\zeta$ (Fig.~\ref{pic:turb_vsf_compres_vsfhr08_zeta}) and $Z$ (Fig.~\ref{pic:turb_vsf_compres_vsfhr08_z}) measured for $\lambda_J$~=~$8\Delta{}x$ as function of the values measured in the $\lambda_J$~=~$4\Delta{}x$ simulations.
Each of the dots corresponds to the same time steps in both simulation samples. 
One sees that both simulation samples are not only qualitatively, but also quantitatively in good agreement.

This means that, although refining Jeans lengths with 4~cells only misses about 13\% of kinetic energy, the effect on the structure and behaviour of the turbulence is rather small and not traced by a VSF analysis.

The picture, however, changes when analysing the VSFs based on the $\lambda_J$~=~$32\Delta{}x$ runs.
Here one sees that the measured values of both $\zeta$ (Fig.~\ref{pic:turb_vsf_zeta_evol_vsfhr32}) and $Z$ (Fig.~\ref{pic:turb_vsf_zeta_relevol_vsfhr32}) are similar to those for $\lambda_J$~=~$4\Delta{}x$ for the first 0.2~Myr.
After this short period, though, the evolutions of $\zeta$ diverge. 
While $\zeta(1)$ and $\zeta(2)$ continue to decrease in similar, but lower rates compared to $\lambda_J$~=~$4\Delta{}x$, $\zeta(3)$ increases until it peaks at $t$~=~0.8~Myr, before it falls steeply down again.
At $t$~=~1.2~Myr, the last time step of this sample, the values of all $\zeta$ equal the measurements of $\lambda_J$~=~$4\Delta{}x$ again (see also Fig.~\ref{pic:turb_vsf_compres_vsfhr32_zeta}). 
However, since there is no information of how the $\lambda_J$~=~$32\Delta{}x$ simulations develop further I cannot predict whether this correspondence will continue.

Fig.~\ref{pic:turb_vsf_compres_vsfhr32} illustrates the different evolutions of measured $\zeta$ and $Z$ in the two samples of simulations more clearly.
One sees that the differences between the simulation samples follow the same pattern for all orders of $p$.
The order of difference, though, increases with the order; while the values for $\zeta(1)$ are still in good agreement, the measured values of $\zeta(2)$ and $\zeta(3)$ for $\lambda_J$~=~$32\Delta{}x$ are 40\% and 100\% higher than those measured for $\lambda_J$~=~$32\Delta{}x$, respectively.
Consequently, this causes differences in $Z(p)$ within 30--52\% between the simulations (also see Fig.~\ref{pic:turb_vsf_compres_vsfhr32_z}).
\pagebreak

\noindent Following the explanations before, this behaviour of $\zeta$ and $Z$ corresponds to the reaction of the cloud's gas to a shock wave running through the cloud; caused by a supernova that exploded before $t$~=~0~Myr. 
Indeed one sees a SN at $t$~=~-1.11~Myr at a distance of 172~pc. 
Due to the distance the shock front is too weak to effectively compress the gas within \texttt{M3}.
This is why it has not been detected previously in the less refined samples.
However, the SN explodes far below the mid-plane of the modelled disk galaxy, in a region without dense gas.
This means that the shock wave that has been injected by the explosion is less damned as it propagates through the ISM. 
By the time the front arrives at \texttt{M3} it is still energetic enough to drive strong winds, with velocity above 300~km~s$^{-1}$, at the closer edge of the cloud. 
This causes an increase of VSFs at longer lag scales and the increase of $\zeta$, as well as the drop in $Z$.

Thus, the derivations of $Z$ to the predicted do only trace external turbulent driving.
Whether the source is a SN shock front propagating through the cloud or a strong wind, cannot be distinguished by $Z$ only. 
Yet, $Z$ remains a fine probe for the geometry of turbulence and the scales at which turbulence is driven.

%%%%%%%%%%%%%%%%%%%%%%%%%%%%%%%%%%%%%%%%%%%%%%%%%%%%%%%%%%%%%%%%%%%%%%%%%%%%%%%%%%%%%%%%%%%%
\section{The Interplay of Turbulence and Fragmentation}\label{turb:veldir}

This section returns to the question how filaments fragment.
In Chapter~\ref{frag}, I have investigated this question by analysing the thermal stability of filaments, that had formed self-similarly in the model clouds.
I have seen that the fragments use to form early in the evolution of the filaments while those are supposed to still be thermally supported.
Hence, the fragmentation needs to be triggered by other mechanisms, like colliding flows.

Sect.~\ref{turb:vsf} demonstrates that the turbulence in the model clouds is dominated by gravitational contraction for most of their evolution. 
Occasionally, the clouds are affected by SN shocks, though they relax into the initial conditions within short times ($\sim$0.6~Myr).
However, this gravitational collapse is not the dominant mechanism that initiates the fragmentation of the filaments within the clouds (Chapter~\ref{frag}).
Due to the physics considered in the simulations the most promising candidates for triggering fragmentation are magnetic fields and turbulent flows.
As magnetic fields are generally supposed to stabilise structures, I investigate the interplay of fragmentation and turbulence in this section.
Sect.~\ref{turb:veldir_theory} introduces the parameter I use for the analysis in Sect.~\ref{turb:veldir_results}.

\subsection{Characterising Turbulent Flows}\label{turb:veldir_theory}

For analysis I focus on three parameters:
\begin{itemize}
	\item the 3D divergence of the velocity field, $\nabla \cdot \vec{u}$;
		\begin{equation}
			\nabla \cdot \vec{u} = \frac{\partial u_x}{\partial x} + \frac{\partial u_y}{\partial y} + \frac{\partial u_z}{\partial z} ,
		\end{equation}
	\item the angle between the local direction of the filament's long axis, $\vec{\ell}$, and the mean flow direction, $\vec{u}$, given by,
		\begin{equation}
			\varphi =  \arccos \left(\frac{\vec{\ell} \cdot \vec{u}}{|\vec{\ell}| \cdot |\vec{u}|}\right) ;
 			\label{equ:turb_def_relang_filvel}
		\end{equation}
	\item and the relative angle between the mean velocity vectors, $\vec{u}$, of two neighbouring filament segments, $\vec{x}_i$ and $\vec{x}_{i+1}$, defined as
		\begin{equation}
			\psi(\vec{x}_\mathrm{i}) =  \arccos \left(\frac{\vec{u}(\vec{x}_\mathrm{i}) \cdot \vec{u}(\vec{x}_\mathrm{i+1})}{|\vec{u}(\vec{x}_\mathrm{i})| \cdot |\vec{u}(\vec{x}_\mathrm{i+1})|}\right) .
			\label{equ:turb_def_relang_vels}
		\end{equation}
\end{itemize}

\noindent Note that, for this analysis, I cannot directly use the simulated velocity vectors.
There are two factors that the vectors need to be corrected for. 
First, I need to remove the contribution of the large-scale motions of the respective cloud.
Recall that our model clouds are part of multi-kpc box containing a segment of a spiral-like galaxy and move along its disk.
Thus, the measured velocities, $\vec{v}$, do not only contain information about the clouds' internal turbulence, but also a large-scale component that describes how the entire cloud moves within the ISM with respect to the galactic potential.
In order to characterise the turbulence within the clouds, represented by $\vec{v}_\mathrm{int}$ , I need to remove the external component which is approximated to first order by the mean motions of all cells belonging to the respective cloud, $\langle \vec{v}(\vec{\tilde{x}}) \rangle_{\vec{\tilde{x}} \, \in \, \mathrm{cloud}}$,
\begin{equation}
	\vec{v}_\mathrm{int}(\vec{x}) = \vec{v}(\vec{x}) - \langle \vec{v}(\vec{\tilde{x}}) \rangle_{\vec{\tilde{x}} \, \in \, \mathrm{cloud}}.
	\label{equ:turb_vsf_orig-mean}
\end{equation}
\pagebreak

\noindent This has not been a major concern in Sect.~\ref{turb:vsf} as the contribution of the large scale motions has been remove by the two-point-correlation ansatz.
Here, however, I have to remove the net contribution to not falsify the results.

Second, it is essential to take the influence of numerical diffusion on small scales into account.
As mentioned in Sect.~\ref{turb:vsf_resolution}, the Jeans length are refined by four cells only to reduce the computational demands. 
This is the limit to ensure that the clouds do not artificially fragment due to numerical effects.
However, for reliably resolving the turbulent structures at least ten zones are required. 
As a consequence, I can only trace the average motion of gas from one segment of the skeleton to another, but I cannot resolve the turbulence within individual segments.
In terms of implementation this means that the radius of the filament need to be redefined as $R_f$~=~0.5~pc, hereafter (0.3~pc in Sect.~\ref{frag}). 
Thus, for each cell $\vec{x}_\mathrm{i}$ along the filaments the average the velocity vectors, $\vec{u}(\vec{x}_\mathrm{i})$ is given by,
\begin{equation}
	\vec{u}(\vec{x}_\mathrm{i}) = \frac{\sum_{j \in |\vec{x}_i-\vec{x}_j| \leq R_f} \rho(\vec{x}_\mathrm{j}) \, \vec{v}_\mathrm{int}(\vec{x}_\mathrm{j})}{ \sum_{j \in |\vec{x}_i-\vec{x}_j| \leq R_f} \rho(\vec{x}_\mathrm{j}) } ,
    \label{equ:turb_vsf_velaver}
\end{equation}
with $\rho(\vec{x}_\mathrm{i})$ being the volume mass density.

\begin{figure}[h!t]
	\includegraphics[width=\textwidth]{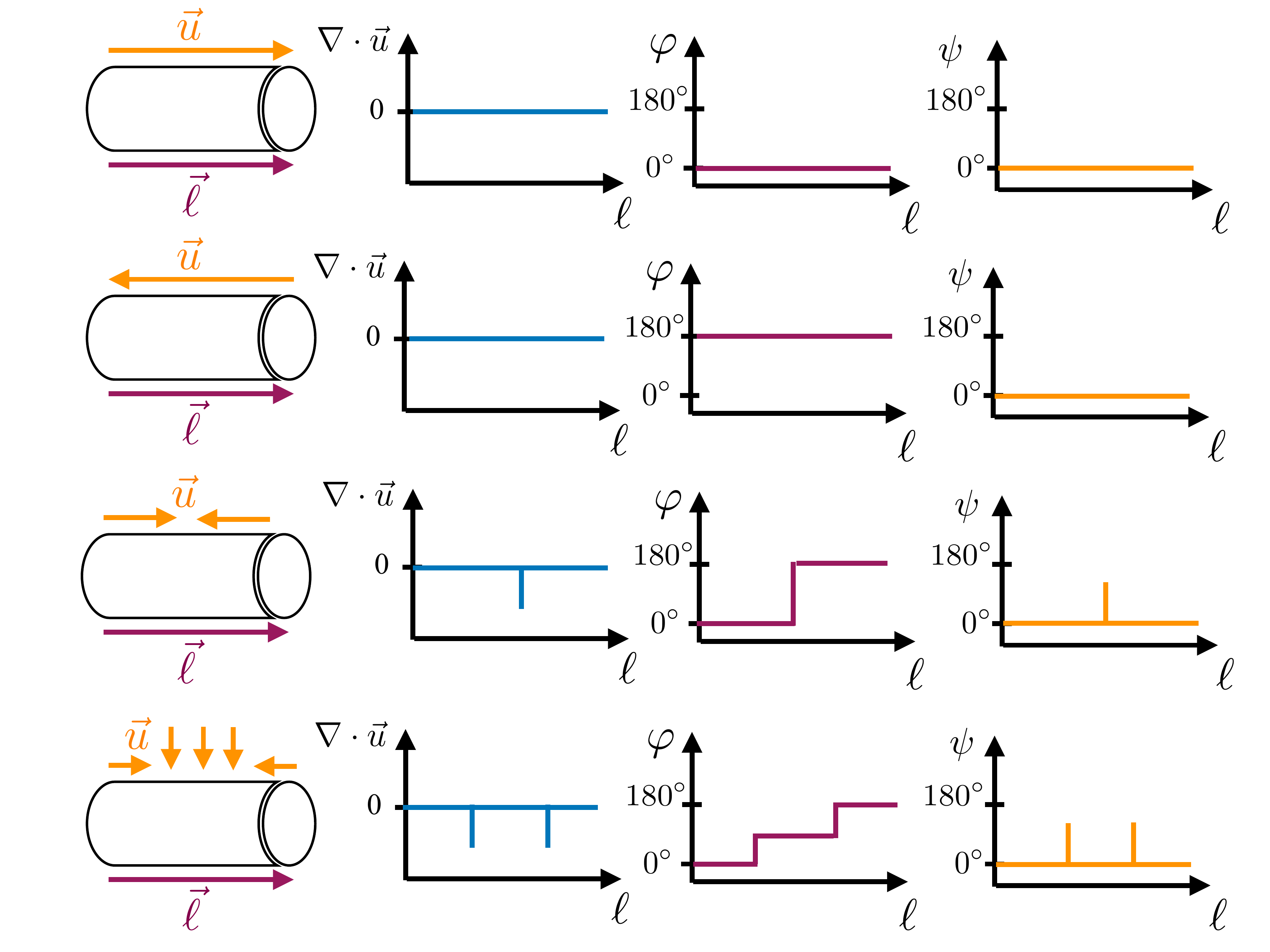}
	\caption[Sketch for methodology for examining turbulent flows in filaments]{Sketch demonstrating how relative angles resemble given flow patterns. Shown are four examples (\textit{from} top to \textit{bottom}): a gas flow that runs parallel to the filament, one that runs anti-parallel to the filament, a colliding flow within the filaments, and a segment of the filament that is locally mainly accreting gas radially from its surrounding. The \textit{left} plots show the behaviour of velocity divergence, $\nabla \cdot \vec{u}$. The \textit{middle} panels show the relative angle between the filament long axis and the mean velocity vector as function of position along the filament, $\phi$. The \textit{right} panels illustrate the relative angle between two neighbouring velocity vectors as function of position along the filament, $\psi$.
	}
    \label{pic:turb_sketch_veldir}
\end{figure}

Fig.~\ref{pic:turb_sketch_veldir} sketches the behaviour of the three parameters of interest, $\nabla \cdot \vec{u}$, $\varphi$ and $\psi$, for four simple, but commonly occurring scenarios.
The first two scenarios show gas flowing parallel or antiparallel to the direction of the respective filament segment.
The third example demonstrates two colliding flows.
The last model describes a more complex scenario that implies that the centre of the segments accretes gas perpendicularly from the surrounding of the filament, as well as along the skeleton.

Fig.~\ref{pic:turb_sketch_veldir} shows that all three quantities react differently in each of the situations, and that only observing their behaviour relative to each other can reveal what happens in the filament. 
For example, in $\nabla \cdot \vec{u}$ and $\psi$ and one clearly sees the location when the dominant velocity direction changes, from accretion along to perpendicularly the filament and vice versa. 
However, the condensation of matter and the formation of a fragment is possible along the whole slice between the divergence signals since there are at least two flows of gas converging onto one point. 
This is only reflected in $\varphi$.
Yet, $\varphi$ cannot reflect the dynamics relative to the filament's axis. 
Assuming that gas is always accreted perpendicularly to the filament, as in the last example in Fig.~\ref{pic:turb_sketch_veldir}, $\varphi$ cannot differentiate whether the accretion direction is always the same. 
For this, $\psi$ is essential as it provides the second necessary angle to resemble the full 3D relation between the filament axis and flow direction.

\subsection{Turbulent Flows within Filaments}\label{turb:veldir_results}

Table~\ref{tab:turb_veldir_statangle} summarises the measured values of $\nabla \cdot \vec{u}$, $\varphi$, and $\psi$ statistically.
Therefore, I group the filaments into two samples: those filaments that do not contain any fragment, and those filaments within which I detect filaments. 
In the latter group, I additionally mark the time steps when the individual fragments have been detected for the first time.
This subsample is important since the measurements here reflect the conditions under which the respective fragments have formed.

\begin{table}
	\begin{center}
	\begin{tabular}{l|c|c|c|c}
		Quantity	& minimum & mean & median & maximum \\ \hline
        \multicolumn{5}{l}{filaments without fragments} \\ \hline
		$\nabla \cdot \vec{u}$	& -144.7	& -1.5	& 0.1	& 127.3 \\
		$\varphi$ [deg]			&   48.9	& 90.7	& 91.3	& 159.1 \\
		$\psi$ [deg]			&    0.0	& 79.7	& 83.6	& 98.5 \\  \hline
        \multicolumn{5}{l}{filaments with fragments} \\ \hline
		$\nabla \cdot \vec{u}$	& -140.1	&  3.0	& 0.4	& 144.4 \\
		$\varphi$ [deg]			&   56.6	& 89.1	& 90.0	& 113.7 \\
		$\psi$ [deg]			&   42.5 	& 80.2	& 83.4	&  93.9 \\  \hline
        \multicolumn{5}{l}{filaments with fragments at the time of first detection} \\ \hline
		$\nabla \cdot \vec{u}$	& -39.2		& 11.0	& 0.6	& 144.4 \\
		$\varphi$ [deg]			&  56.6		& 89.7	& 91.1	& 110.0 \\
		$\psi$ [deg]			&  42.5 	& 78.4	& 83.5	& 89.78 \\  \hline
	\end{tabular}
	\end{center}
	\caption[Statistical summary of measurements of $\nabla \cdot \vec{u}$, $\varphi$, and $\psi$]{Statistical summary of measurements of the gas velocity divergence, $\nabla \cdot \vec{u}$, angle between the filaments' axis and gas flow direction, $\varphi$, and relative angle of gas velocity vectors within neighbouring filament segments, $\psi$. The filaments have been classified by whether or not they inhabit fragments. The last sample of filaments is a subset of those filaments with fragments, but only takes those time steps into account at which the fragments have been detected for the first time. }
	\label{tab:turb_veldir_statangle}
\end{table}

\begin{figure}
	\centering
    \includegraphics[width=\textwidth]{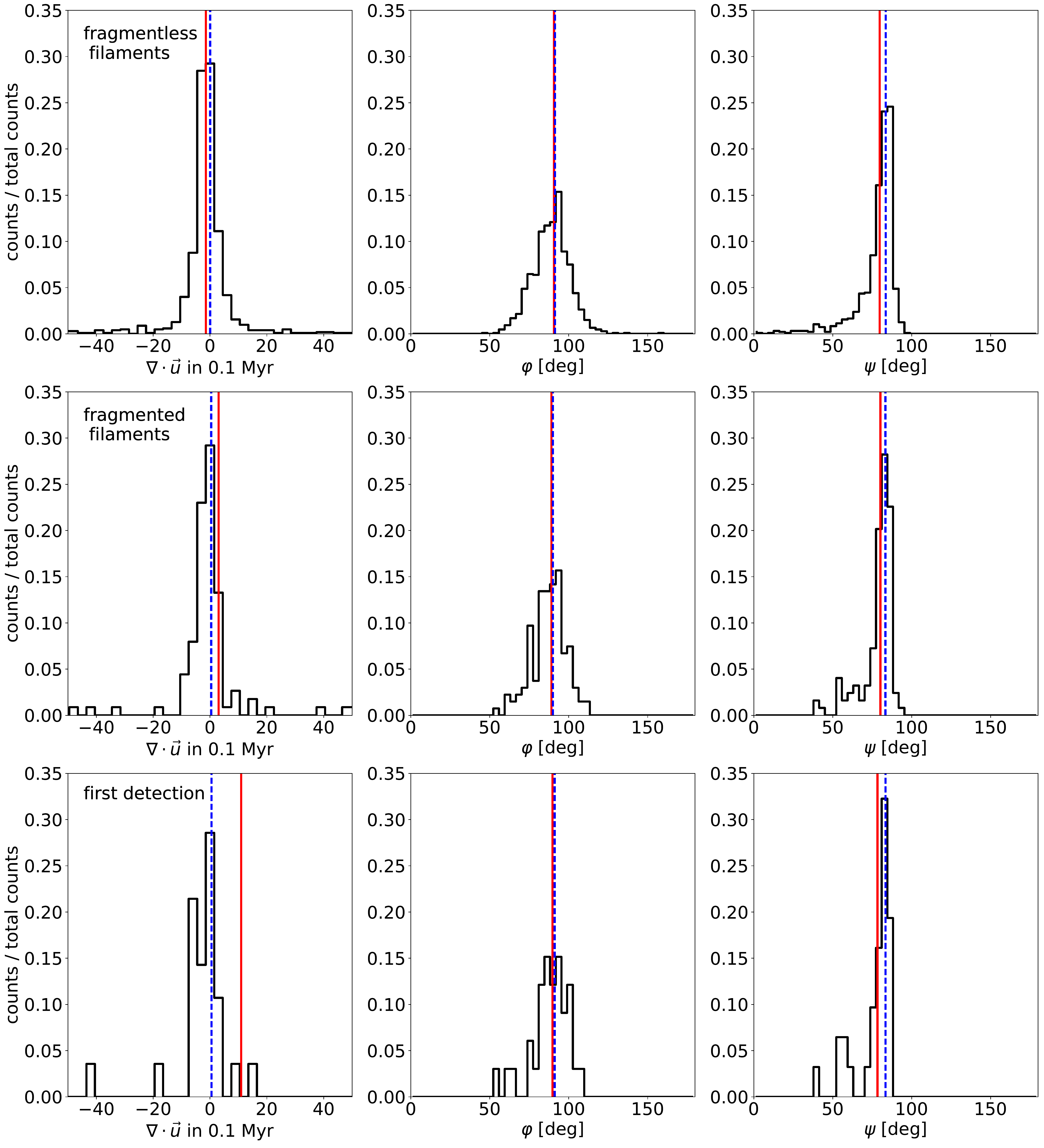}
    \caption[Histogram of $\nabla \cdot \vec{u}$, $\varphi$, and $\psi$ for fragment-less and fragment filaments]{Histograms of 3D velocity divergence, $\nabla \cdot \vec{u}$ (\textit{left}), the angles between the filament skeletons and velocity vectors, $\varphi$ (\textit{middle}), and angles between neighbouring velocity vectors, $\psi$ (\textit{right}). 
    The individual counts represent the measured mean value of an entire filament that does not or does inhabit any fragment (\textit{top} and \textit{middle}, respectively). 
    The histogram in the \textit{bottom} panel illustrates the measurements of fragmented filaments at the very time step when a new fragment has been detected.
    The solid red lines and blue dashed lines mark the mean and median values, respectively.
    }
    \label{pic:turb_veldir_stat}
\end{figure}

Fig.~\ref{pic:turb_veldir_stat} shows histograms of measured $\nabla \cdot \vec{u}$, $\varphi$, and $\psi$ for all samples.
Each count represents the average value of the respective quantity for an entire filament. 
One sees that all distributions are generally Gaussian with mean divergence around 0, while $\varphi$ and $\psi$ are distributed around 90$^\circ$. 
This means the filaments use to accrete gas from their surrounding, preferentially perpendicularly to their axes, represented by $\varphi$~=~90$^\circ$. 
This is in agreement with other studies \citep[e.g.,][]{ZamoraAviles2017} that demonstrate that the gas is transported from the surrounding onto the filament via filament spines and fibres that are perpendicular to the main filaments' axes.
$\psi$~=~90$^\circ$ shows that the gas does not have any favoured infall direction. 
Consequently, the filaments accrete gas from their entire surrounding, instead of being fed from a particular side (as it would be the case of shock waves).
This may be motivated by the gravitational contraction of the clouds that causes the formation and growth of substructures (see also Sect.~\ref{frag:thermsupport_mean}).
The low degree in velocity divergence confirms this scenario as there are on average no sharp changes in flow directions that dominate the dynamics of the filaments externally. 

\begin{figure}
	\centering
	
	\begin{subfigure}{\textwidth}
		\includegraphics[width=\textwidth]{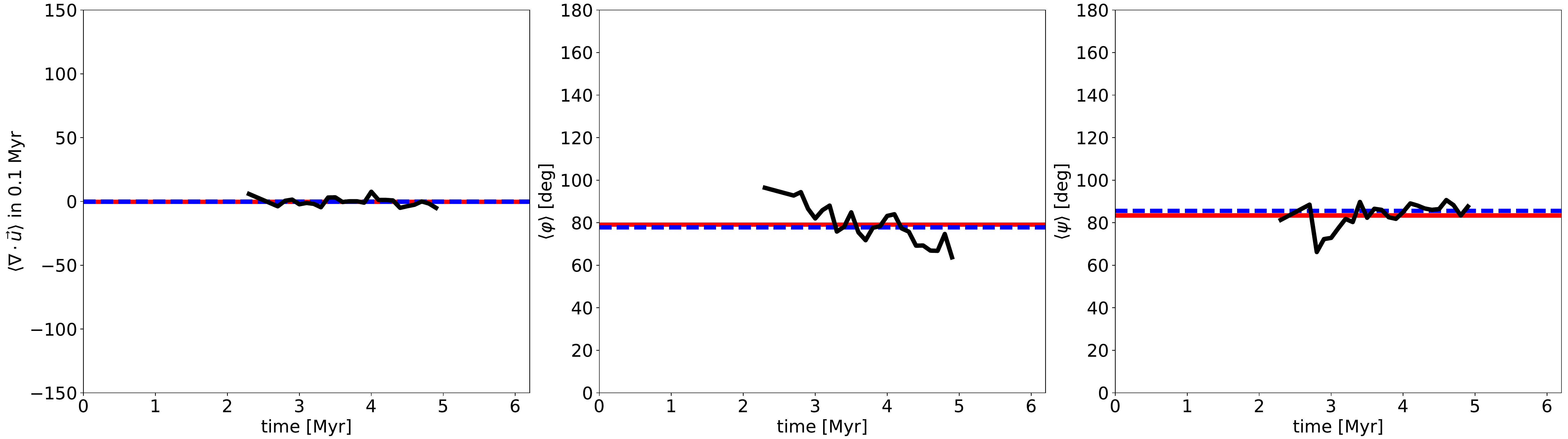}
		\caption{Filament \texttt{M8\#012}}
		\label{pic:turb_veldir_m8_f012_mean}
	\end{subfigure}
	
	\begin{subfigure}{\textwidth}
		\includegraphics[width=\textwidth]{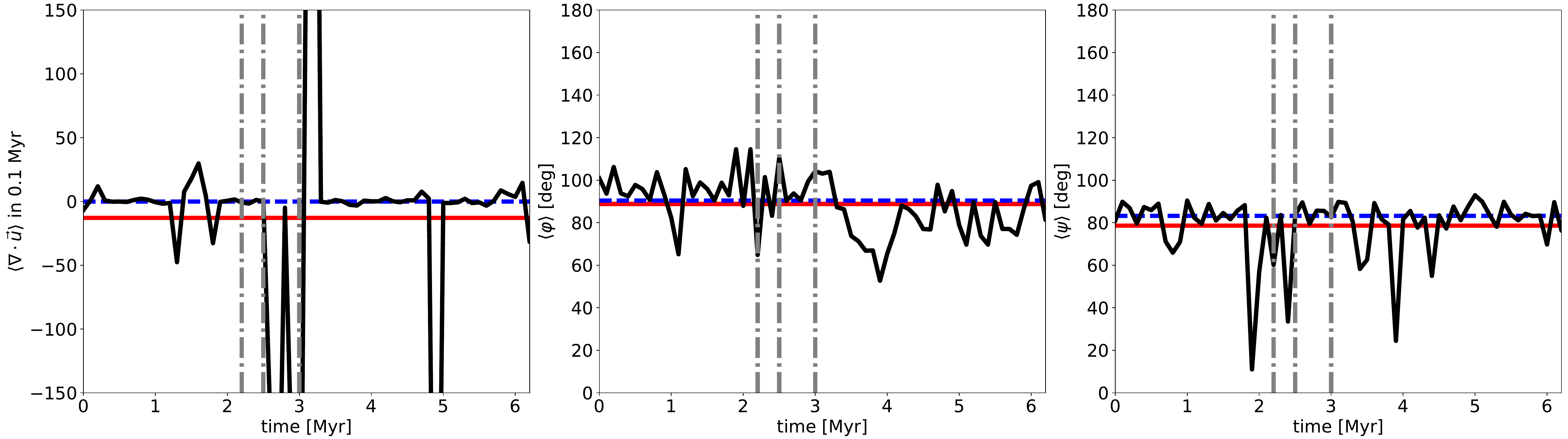}
		\caption{Filament \texttt{M4\#001}}
		\label{pic:turb_veldir_m4_f001_mean}
	\end{subfigure}
	
	\caption[Example of time evolution of mean $\langle \nabla \cdot \vec{u} \rangle$, $\langle \varphi  \rangle$, and $\langle \psi \rangle$ in individual filaments.]{Example of time evolution of mean $\langle \nabla \cdot \vec{u} \rangle$, $\langle \varphi  \rangle$, and $\langle \psi \rangle$ (black solid lines) in individual filaments. The red solid lines and the dashed blue line illustrate the all-time mean and median values of this filaments, respectively. The grey dash-dotted lines mark the time when the fragments are detected within the corresponding filaments for the first time.}
	\label{pic:turb_veldir_filmean}
\end{figure}

\begin{figure}
	\centering
	
	\begin{subfigure}{\textwidth}
		\centering
		\includegraphics[height=0.42\textheight]{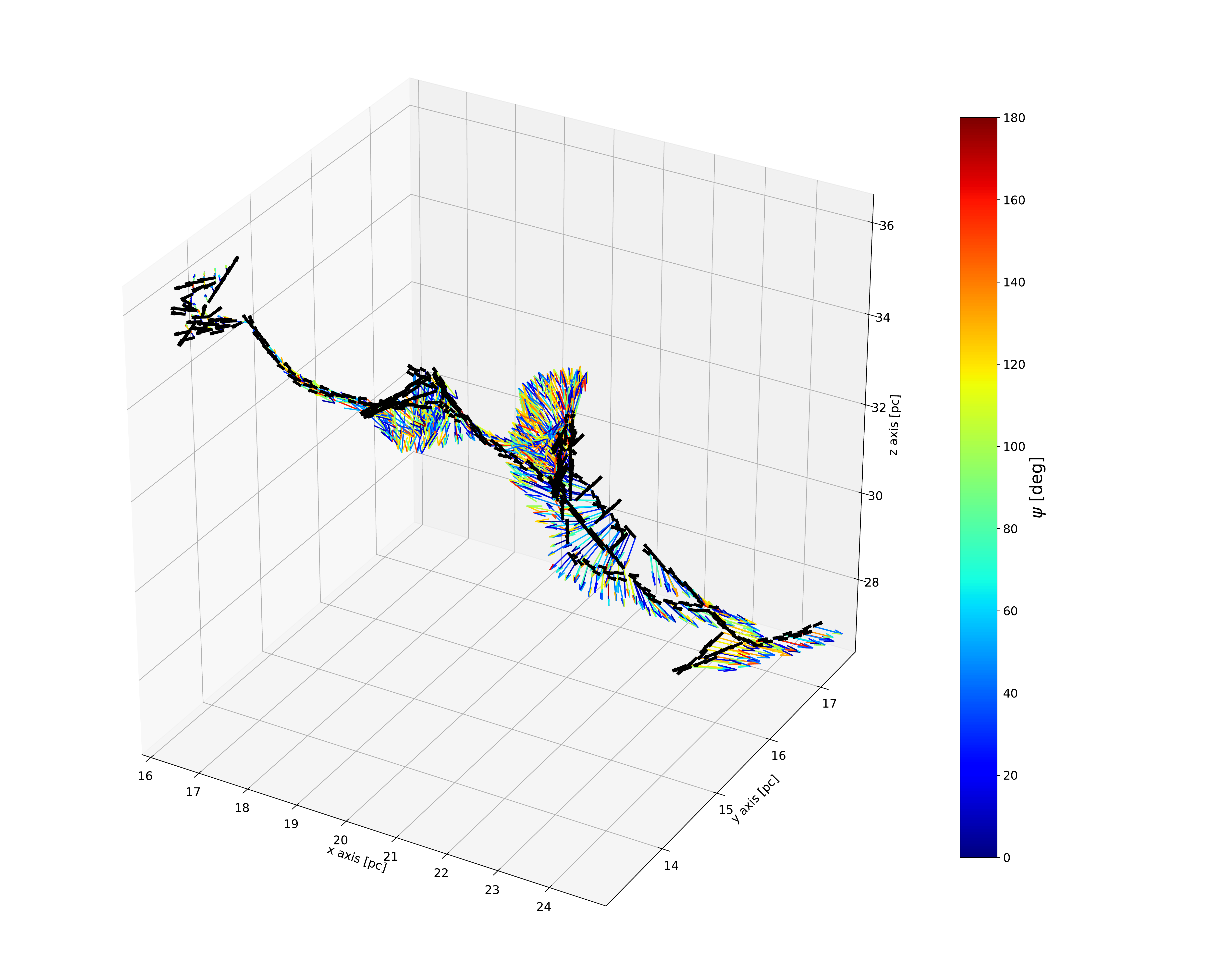}
		\caption{\texttt{M8\#012} at $t$~=~3.0~Myr}
		\label{pic:turb_veldir_3d_example_m8_f012}
	\end{subfigure}	
	
	\begin{subfigure}{\textwidth}
		\centering
		\includegraphics[height=0.42\textheight]{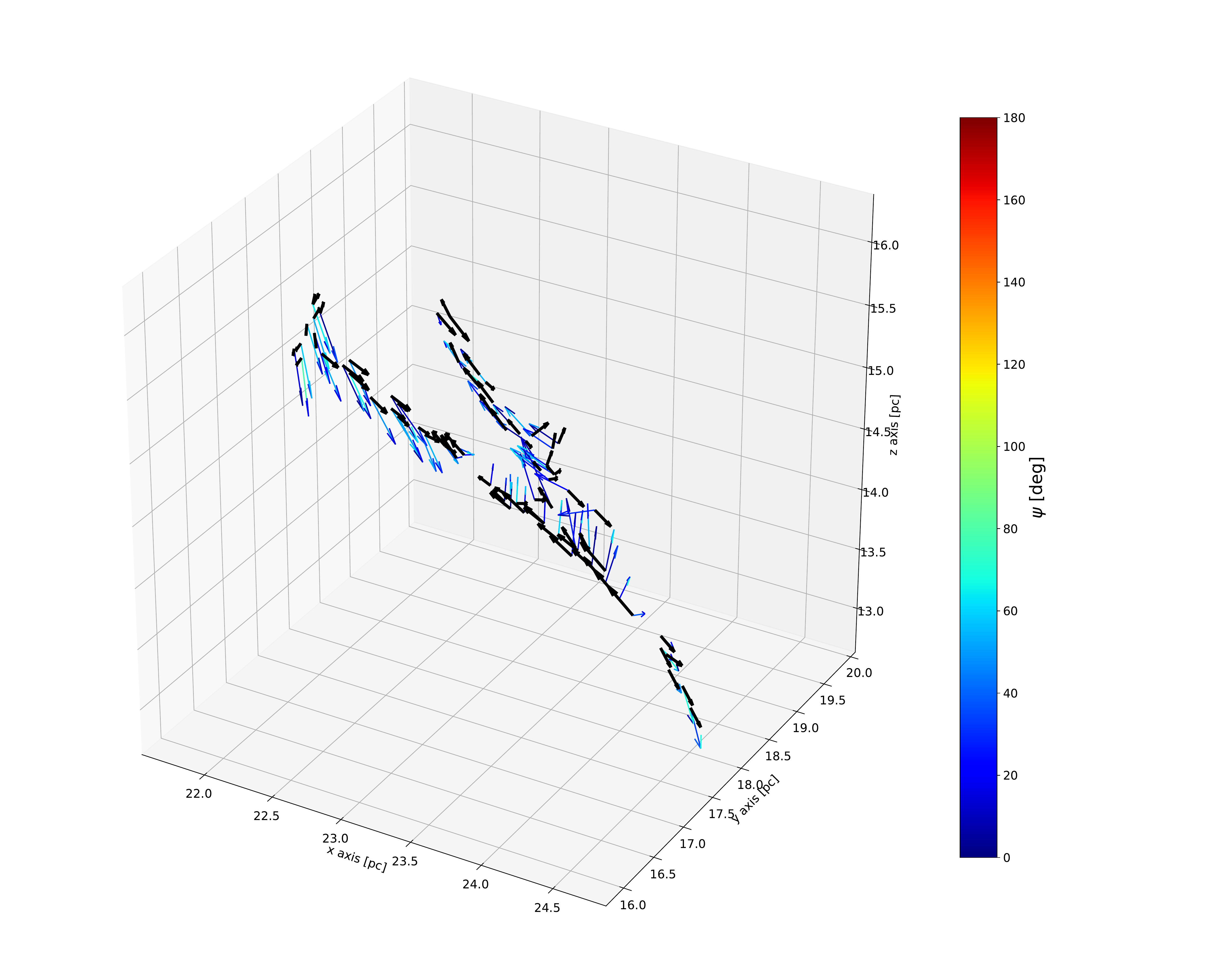}
		\caption{\texttt{M4\#001} at $t$~=~2.4~Myr}
		\label{pic:turb_veldir_3d_example_m4_f001}
	\end{subfigure}	
	
	\caption[3D plot of \texttt{M4\#001} at $t$~=~2.4~Myr and \texttt{M8\#012} at $t$~=~3.0~Myr.]{3D plot of \texttt{M4\#001} at $t$~=~2.4~Myr (\textit{left}) and \texttt{M8\#012} at $t$~=~3.0~Myr (\textit{right}). The black lines represent the filament's axis and the coloured vectors illustrate the mean velocity vector $\vec{u}$ for each of the segments. The colour bar shows the relative angle between neighbouring velocity vectors, $\psi$. }
	\label{pic:turb_veldir_3d_example}
\end{figure}

However, in absolute numbers (indicated in Table~\ref{tab:turb_veldir_statangle}) one sees that there are filaments with a high degree of divergence.
Fig.~\ref{pic:turb_veldir_filmean} plots the time evolution of mean $\langle \nabla \cdot \vec{u} \rangle$, $\langle \varphi \rangle$, and $\langle \psi \rangle$ in two filaments in \texttt{M4} and \texttt{M8}.
I have chosen these examples because they demonstrate the connection between gas flow and fragmentation.
Filament \texttt{M8\#012} is a quiescence filament that does not inhabit any detected fragment.
One sees that although there is a large amount of velocity divergence within the filament  (see Fig.~\ref{pic:turb_veldir_3d_example_m8_f012}), $\langle \varphi \rangle$ and $\langle \psi \rangle$ show only small derivation over the entire evolution of the filament. 
In this example the gas within the filament is highly turbulent, yet the motions are rather random that leads to no significant variations on scales of the entire filament.
\pagebreak

\noindent Filament \texttt{M4\#001} (Fig.~\ref{pic:turb_veldir_m4_f001_mean}), on the contrary, is located within the inner regions of \texttt{M4} and one of the first filaments that fragments there.
Both $\langle \nabla \cdot \vec{u} \rangle$ and $\langle \psi \rangle$ reflect that the region is dynamically active and that the first fragment forms $\sim$0.5~Myr after a sequence of significant divergence and 0.2~Myr after a major change in $\langle \psi \rangle$.
The $\langle \psi \rangle$ is reduced to less than 20$^\circ$ while $\langle \varphi \rangle$ is still around 90$^\circ$. 
Low values $\langle \psi \rangle$ mean that the majority of gas velocity vectors are parallel to each other.
As the vectors are mostly perpendicular to each other before this change implies that there needs to be an externally acting pressure that forces the gas flows to align.
Indeed, \texttt{M4} is affected by a SN shock wave at this time (see Sect.~\ref{turb:vsf_results}) that redirects the gas flows into the propagation direction of the shock front (see Fig.~\ref{pic:turb_veldir_3d_example_m4_f001} for illustration).
After the shock front has passed and the cloud has relaxed again, the filament accretes along its entire surface, which increases $\langle \psi \rangle$.

These examples indicate that it is more likely to form a fragment in an active environment than in a quiescent filament.
However, they do not explain what actually causes the formation of the fragments.
In the case of \texttt{M4\#001} the turbulence is driven by the shock front that hits the filament perpendicularly (otherwise, $\langle \varphi \rangle$ must deviate from 90$^\circ$).
This only causes that the gas in the filaments is strongly compressed along the long axis of the filament and for a short range of time. 
There are two possible explanations for this behaviour: either turbulence is not primarily involved in the formation of the fragments, or the mean evolution of the entire filament does not reflect the local, small scales motions that are involved in the fragment formation.
To unravel this ambiguity, I focus on the flow motions along the filament spine.
I continue to focus on \texttt{M4\#001} and the first fragment that forms within it. 
Note that one sees similar pattens in the other filaments across all clouds, as Figs.~\ref{pic:turb_veldir_focusdend_m3_f000} and~\ref{pic:turb_veldir_focusdend_m8_f000} demonstrate. 

\begin{figure}
	\centering
	\includegraphics[width=\textwidth]{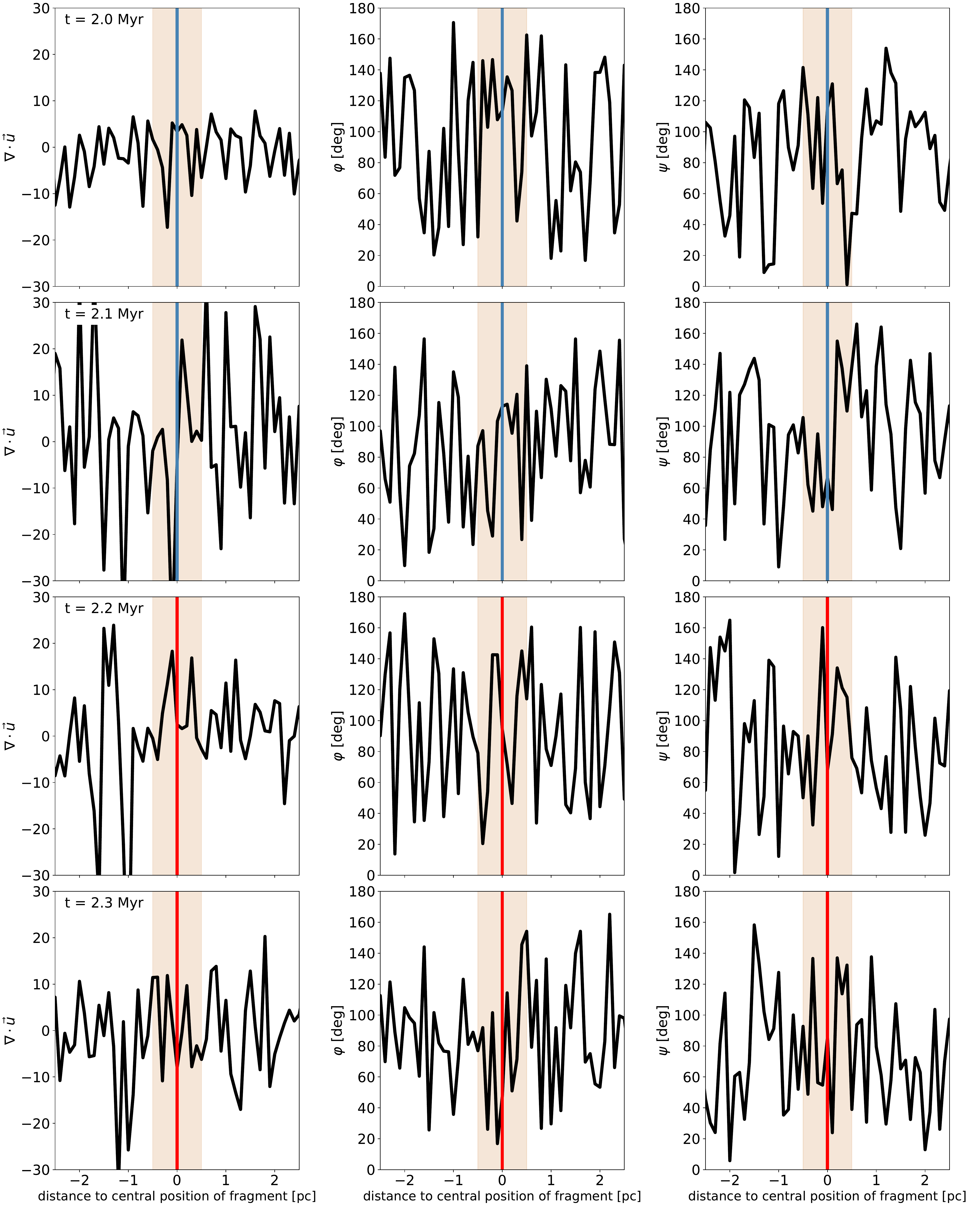}	
	
	\caption[Zoom on fragment \texttt{M4\_f001}.]{Zoom on fragment \texttt{M4\_f001} within $t$~=~2.0--2.3~Myr (from \textit{top} to \textit{bottom}). Shown are the measured values of $\nabla \cdot \vec{u}$ (\textit{left}), $\varphi$ (\textit{middle}), and $\psi$ (\textit{right}) along the filament \texttt{M4\#001} at the given time steps. The abscissa refers to individual segments of the filament at the given distance to the central position of the fragment of interest. The orange area illustrate the radius of the fragment. The blue vertical lines mark the future position of the not-yet formed fragment, whereas red vertical lines indicate the actual position. }
	\label{pic:turb_veldir_focusdend_m4_f001}
\end{figure}

\begin{figure}
	\centering
	\includegraphics[width=\textwidth]{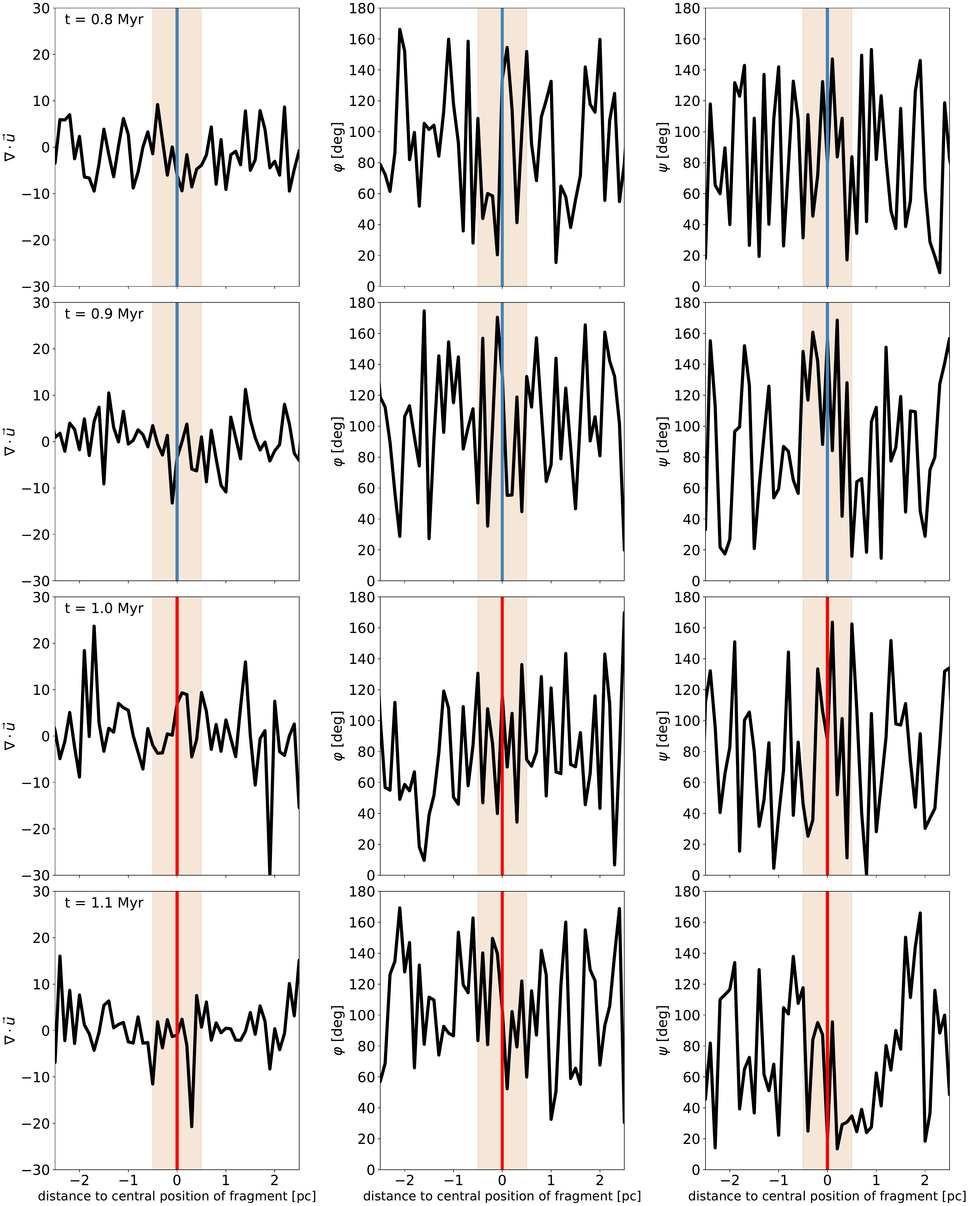}	
	
	\caption[Zoom on fragment \texttt{M3\_f000}.]{As Fig.~\ref{pic:turb_veldir_focusdend_m4_f001} but for the fragment \texttt{M3\_f000} within $t$~=~0.8--1.1~Myr (from \textit{top} to \textit{bottom}). }
	\label{pic:turb_veldir_focusdend_m3_f000}
\end{figure}

\begin{figure}
	\centering
	\includegraphics[width=\textwidth]{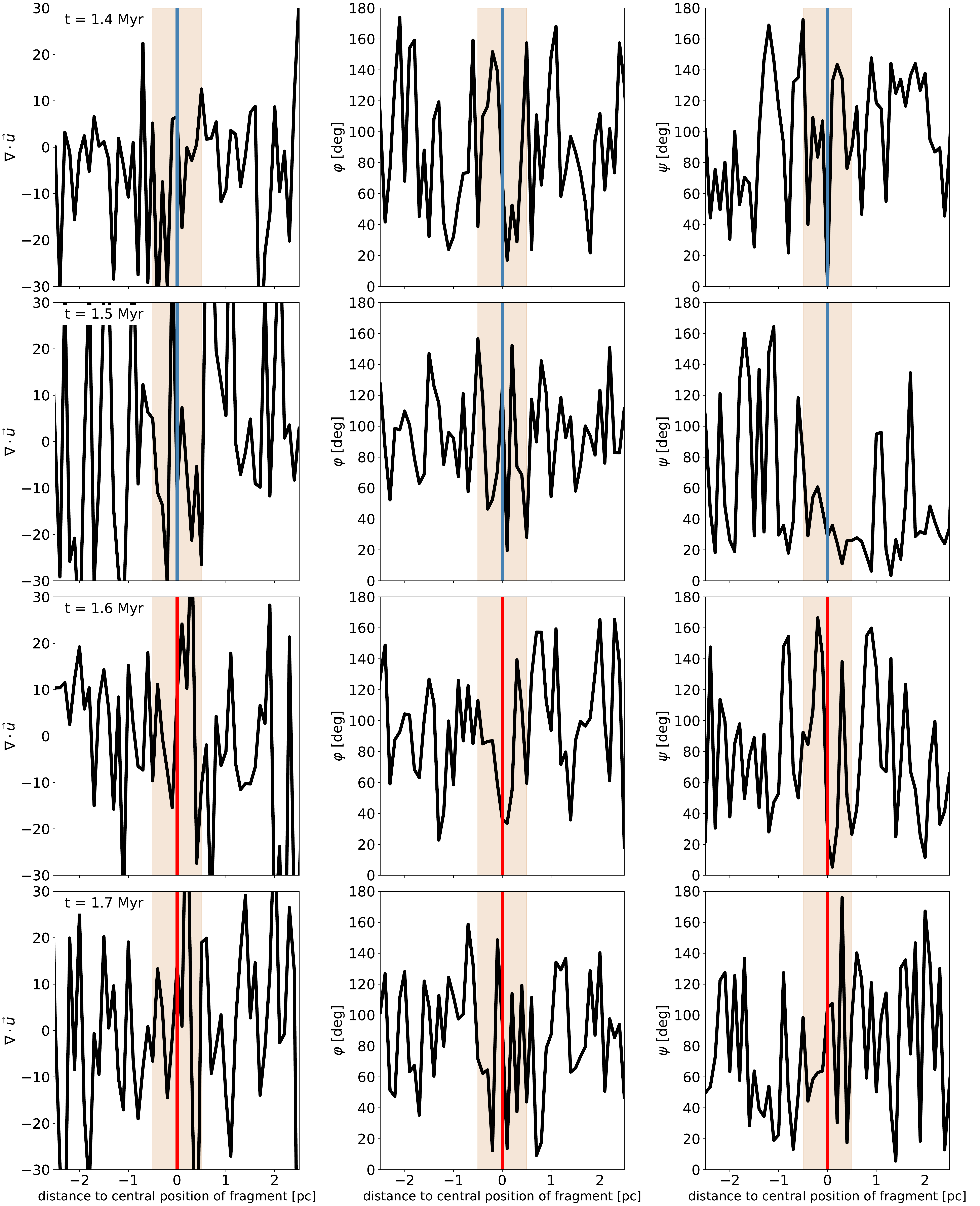}	
	
	\caption[Zoom on fragment \texttt{M8\_f000}.]{As Fig.~\ref{pic:turb_veldir_focusdend_m4_f001} but for the fragment \texttt{M8\_f000} within $t$~=~1.4--1.8~Myr (from \textit{top} to \textit{bottom}). }
	\label{pic:turb_veldir_focusdend_m8_f000}
\end{figure}

Fig.~\ref{pic:turb_veldir_focusdend_m4_f001} plots the measured values of $\nabla \cdot \vec{u}$, $\varphi$, and $\psi$ as function of the position along the spine (abscissa, relative to the central position of the -- future -- fragment) and time (from \textit{top} to \textit{bottom}). 
The first two plots show the measurements in the 0.2~Myr before the detection of the fragment, the last two within the 0.2~Myr after its first detection.
The orange area illustrate the radius of the fragment whose central position is marked with a blue and red vertical line before or after its detection, respectively.
Note that Figs.~\ref{pic:turb_veldir_focusdend_m3_f000} and~\ref{pic:turb_veldir_focusdend_m8_f000} show the same data for the first fragments formed in \texttt{M3} and \texttt{M8}, namely \texttt{M3\_f000} and \texttt{M8\_f000}, respectively.

One sees that the distributions in Fig.~\ref{pic:turb_veldir_focusdend_m4_f001} are more irregular than the mean distributions described above. 
On average, the values still fluctuate around 0 in the case of $\nabla \cdot \vec{u}$, and around 90$^\circ$ in the cases of $\varphi$ and $\psi$, respectively (see Fig.~\ref{pic:turb_veldir_focusdend_m4_f001}, \textit{top}).
However, in the time step before the first detection, at $t$~=~2.1~Myr (Fig.~\ref{pic:turb_veldir_focusdend_m4_f001}, \textit{second from top}), $\nabla \cdot \vec{u}$ falls significantly at the exact central position of the future fragment.
One also sees strong divergences, both positive and negative, close to its outer edges at separations between 0.5--1~pc from the central positions. 
This means that there are strong changes in flow directions at these parts of the filaments.
This behaviour is also indicated in $\varphi$ and $\psi$.
From the one side of the fragment to the other, $\varphi$ increases along the filament.
This implies that the flow direction of the gas relative to the filament axis changes from preponderantly parallel to antiparallel.
$\psi$ behaves similarly as $\nabla \cdot \vec{u}$ in the sense that it shows a sharper transition from values around 60$^\circ$ to 140$^\circ$ within the area of the future fragment. 
This strongly suggests that there are two flows collide with each other.
This collision does not occur face-on, as in the examples sketched in Fig.~\ref{pic:turb_sketch_veldir}.
The involved gas flows origin in the environment of the filaments are still dominantly accreted by the filament and, thus, tend to fall perpendicularly into the filament. 
As the gas approaches the filament, it is additionally dragged by the inner-filament flows, which reduce (or increase) the angle between the infalling gas velocity vectors and the axis of the filament. 

At the time when the fragment is detected for the first time ($t$~=~2.2~Myr, Fig.~\ref{pic:turb_veldir_focusdend_m4_f001}, \textit{second from bottom}), the velocity divergence within the fragment is less compared to the time step before.
Yet, there are two peaks close to the edges of the fragment that indicate that the turbulence is still actively developing within this part of the filament.
The distribution of $\psi$ peaks at $\sim$160$^\circ$ at the central position of the fragment while the values in the close environment of the fragments are mostly below 90$^\circ$.
This suggests that the fragment concentrates the flows in this surrounding and accretes them more focussed. 
However, this is not clearly displayed by $\varphi$ that oscillates between values of $\sim$50$^\circ$ and $\sim$150$^\circ$ both around and within the fragment. 

At $t$~=~2.3~Myr (Fig.~\ref{pic:turb_veldir_focusdend_m4_f001}, \textit{bottom}), $\varphi$ shows a modest increase from slightly below to slightly above 90$^\circ$ from the one edge of the fragment to the another one while the variations in $\nabla \cdot \vec{u}$ and $\psi$ cease.
This hints at the scenario that the turbulence in this region of the filament relaxes while the fragment commences to influence the gas dynamics. 
This means that the gas from the surrounding falls again perpendicularly onto the filament, which moderates $\nabla \cdot \vec{u}$ and let $\phi$ and $\psi$ return to values around 90$^\circ$. 
At the same time, the fragment condenses further and, due to its gravitational potential, accretes gas from the filament.
As a consequence, and considering that the fragment also accretes perpendicularly to its surface, the component of the mean velocity vectors that is parallel to the filaments axis becomes stronger the closer the segment is to the edges of the fragment. 
This behaviour becomes more intense as the fragment grows in mass and gains more gravitational potential that induces that more gas is accreted along filament.

This process continues as long as the filament provides enough mass. 
At later stages, when the fragments are massive enough, they become dynamically more dominating within the filament and begin to interact each other.
In some of the cases this may cause that fragments merge, which means that they approach each other until their separation cannot be resolved any more (see Sect.~\ref{frag:thermsupport_frags}).

%%%%%%%%%%%%%%%%%%%%%%%%%%%%%%%%%%%%%%%%%%%%%%%%%%%%%%%%%%%%%%%%%%%%%%%%%%%%%%%%%%%%%%%%%%%%
\section{Summary}\label{turb:conclusions}

In this chapter, I analyse the turbulent structures of molecular clouds and filaments that have formed within 3D AMR FLASH simulations of the self-gravitating, magnetised, supernova-driven ISM by \citet{IbanezMejia2016}.
The main results are as follows.
\medskip

\begin{itemize}
	\item The scaling of velocity structure functions is sensitive to both internal (gravitational contraction) and external (SN shocks, winds) driving sources of turbulence. 
		Applied on simulated data, the time evolution of the scaling exponent, $\zeta$, can reveal which driving mechanism dominates the turbulence of an entire molecular cloud. 
		The ratio of the $p^\mathrm{th}$-to-3$^\mathrm{rd}$ order VSF, $Z$, though, is not directly sensitive to gravitational contraction. 
		Yet, it can be used as observational tracer as it significantly reacts to SNe.
	\item As long as the molecular cloud is not affected by a shock, $Z$ is in good agreement with predicted values for supersonic flows. 
		This makes it a fine probe for the properties of dominant turbulent modes, such as the geometry, and their evolution in the context with the evolution of the cloud. 
	\item I test the influence of Jeans refinement on the VSFs. 
		I find that the absolute amount of kinetic energy does not influence the evolution of $\zeta$ and $Z$, as long as the power spectrum is properly resembled, or similarly resembled if different samples are compared.
	\item I investigate the connection between the gas flows and the formation of fragments within the simulated filaments. 
		I find that the sites where the fragments are detected show sights of colliding flows. 
		This strongly suggest that such flows that compress gas locally are required for the formation of core-like objects. 
\end{itemize}

%\chapterimage{head02.pdf} % Chapter heading image

\chapter[Final Summary and Conclusions]{Final Summary and Conclusions\footnotemark[8]}\label{conclusion}

\footnotetext[8]{The content presented in this chapter are partly published in \citet{Chira2016,Chira2017}.}

In this thesis, I investigate the nature of filamentary structure and their connection to star formation. 
Thereby, I focus on three major questions: 
How are the properties of filaments observable in dust emission?
Are filaments describable as quasi-static thermally supported cylinders?
Is the fragmentation of filaments triggered by internal turbulence?
My main results are the following:
\medskip

\begin{itemize}
	% Chapter 2
	\item One sees that the mean flux density of simple cylinders rises with increasing dust temperature, but it is not significantly influenced by changing the viewing angle.
	\item For the 3D models of the $\rho$~Ophiuchi cloud and the G11.11 \textit{Snake}, one observes that the mean effective dust temperature is approximately constant, whereas mean effective dust number density changes significantly, especially when I rotate the models into the direction where their long axis is parallel to the LoS. Since the dust emission is optically thin, the column densities strongly depend on the viewing angle. The dust temperature is determined by the local heating and cooling processes, which is unchanged when I rotate the models. 
    \item I investigate how sensitivity limits and noise correction procedures influence the results. I find that common data reduction processes reduce the level of variation in column density. The variations in effective dust temperature have increased, but have been still insignificant enough in the observational context.
    % chapter 3
	\item From three molecular cloud models that have been extracted from 3D FLASH ARM MHD simulations of the ISM in a disk galaxy I obtain that the dense gas mass fraction (DGMF) steadily grows as a function of time. Although these clouds grow in mass as they accrete material from their environment, the DGMF continues growing in time, consistent with runaway gravitational collapse.
	\item I find that the average line masses of the filaments always increase in time, with significant differences depending on the volume (or column) density thresholds adopted for their identification. This and the continuously increasing filament-to-cloud mass ratio confirm that the gas of the parental clouds collapses into smaller scale structures, as the evolution of the DGMF has already indicated.  \pagebreak
	\item Filaments already start to fragment well before their line masses reach the critical mass for the collapse of uniform density, self-gravitating, hydrostatic cylinders \citep{Ostriker1964b}. This is true both for the line masses of individual filaments, as well as for the the average of all identified filaments. This implies that the filaments in the simulation never resemble the isolated, hydrostatic configuration of \citet{Ostriker1964b} that is commonly used as the initial condition in analytic filament evolution and fragmentation models. Instead, they are embedded in the hierarchical collapse of the larger cloud, and thus subject to substantial surface pressures.
	\item I compare the performance of different filament finder codes. I find that different codes clearly identify different structures, and further, that the filament properties derived depend strongly on the choice of input parameters.
	\item I compare the properties of the filaments identified in 3D density distributions from the models with those identified in projected 2D column density distributions. I find that, for a given identification threshold density, all 3D filaments have counterparts in 2D column density data, but not vice versa. This is because the 2D filaments may also be composed of the overlap of more diffuse structures along the given LoS that do not fulfil the identification criteria in 3D. As a consequence, the average properties of a sample of filaments and how they evolve in time are not well recovered from column density data. However, since all 3D filaments have counterparts in 2D, the correspondence is better in the case of individual isolated filaments.
	%chapter 4
	\item The scaling of velocity structure functions is sensitive to both internal (gravitational contraction) and external (SN shocks, winds) driving sources of turbulence. 
		Applied on simulated data, the time evolution of the scaling exponent, $\zeta$, can reveal which driving mechanism dominates the turbulence of an entire molecular cloud. 
		The ratio of the $p^\mathrm{th}$-to-3$^\mathrm{rd}$ order VSF, $Z$, though, is not directly sensitive to gravitational contraction. 
		Yet, it can be used as observational tracer as it significantly reacts to SNe.
	\item As long as the molecular cloud is not affected by a shock, $Z$ is in good agreement with predicted values for supersonic flows. 
		This makes it a fine probe for the properties of dominant turbulent modes, such as the geometry, and their evolution in the context with the evolution of the cloud. 
	\item I test the influence of Jeans refinement on the VSFs. 
		I find that the absolute amount of kinetic energy does not influence the evolution of $\zeta$ and $Z$, as long as the power spectrum is properly resembled, or similarly resembled if different samples are compared.
	\item I investigate the connection between the gas flows and the formation of fragments within the simulated filaments. 
		I find that the sites where the fragments are detected show sights of colliding flows. 
		This strongly suggest that such flows that compress gas locally are required for the formation of core-like objects. 
\end{itemize}

\noindent I conclude that there is no quantity in the analysis related to dust emission that tracks the inclination of a filament uniquely.
A notably high column density at a given dust temperature can indicate that an observed object is elongated along the LoS direction.
For true inclinations and confirming masses, line observations are required.
However, with all the data obtained by dust surveys, it is possible to identify candidates of filaments, which may be elongated along the LoS.
It is important to learn more about the distribution and orientation of filaments in the Galactic plane, as it improves the understanding of the role of filaments within the star formation process.

My results on the simulated molecular clouds indicate that filament fragmentation is affected by the environment of the cloud they form in. 
In order to understand the onset and development of fragmentation, future theoretical studies likely need to abandon the hydrostatic initial condition and to consider the formation of filaments and their subsequent fragmentation together.

\noindent Furthermore, the results demonstrate that establishing common practices for how to define filaments in 3D and 2D data from simulations and observations is crucial for studying the properties and evolution of filaments and especially for comparing different filament studies with each other.
Studies using filament finders must thoroughly test applicability of the adopted algorithms to address the problem in question.

My analysis shows that velocity structure functions are fine tools for examining the driving source of turbulence within molecular clouds.
Therefore, I recommend its usage in future studies of molecular clouds.
For the model clouds, the VSFs illustrate that gravitational contraction dominates the evolution of the clouds for most their evolution, with short periods within which SN shock waves accelerate the turbulent powers on all scales. 

The flow patterns within the modelled filament indicate that the fragments form at sites where flows collide with each other.
This setting is preferentially given by the accretion of gas from the more diffuse parts of the parental clouds onto the filaments that occurs perpendicularly to the filament.
This mechanism locally compresses the gas until it forms a gravitationally bound core-like object that continues to accrete more mass from the surrounding filament. 

However, it requires further studies to verify this to be the common fragment formation scenario. 
Especially, a higher Jeans length refinement is needed to resolve the velocity structures on scales of individual grid cells (0.1~pc in this case).
This is crucial for following the local behaviour of the gas as neither the average behaviour of the filaments nor the dominant turbulence driving source of the entire molecular clouds mirror the underling flow patterns that are necessary for this scenario. 

Furthermore, I recommend extensive radiative transfer studies based on self-consistently formed and fragmenting filaments, like those I have presented.
It is necessary to explore to which extend my parameter space (the 3D divergence of the velocity field and the relative angles between the velocity field and the filament axis and consecutive resolution elements) can be applied to observed data as line observations primarily trace the velocity component along the LoS only. 
Since observations from the most inner part of the filaments are required, I propose to focus high-density molecular tracers, such as HCO$^+$ and HCN.

%-------------------------
% Appendix
%-----------------------

\appendix 
	%\chapterimage{head02.pdf} % Chapter heading image

%\chapter{Appendix}
%
%\renewcommand{\thechapter}{\Alph{chapter}}
%\setcounter{chapter}{1}
%\setcounter{figure}{0}
%\setcounter{table}{0}
%\counterwithin{figure}{chapter}
%\counterwithin{table}{chapter}
% 	\renewcommand{\thefigure}{A.\arabic{figure}}
% 	\renewcommand{\thetable}{A.\arabic{table}}

%\addcontentsline{toc}{chapter}{List of Figures}
\listoffigures

%\addcontentsline{toc}{chapter}{List of Tables}
\listoftables

\newpage

\addcontentsline{toc}{chapter}{List of Publications}
\chapter*{List of Publications}

\begin{large}
(* = These publications have been submitted and published during my studies and are partially or fully presented in this thesis)
\bigskip

\begin{enumerate}
	\item *Chira, R.-A., Siebenmorgen, R., Henning, Th., Kainulainen, J., 2016, \aap, 504, 883 \medskip
	\item *Chira, R.-A., Kainulainen, J., Ib\'a\~{n}ez-Mej\'{\i}a, J.~C., Henning, Th., Mac~Low, M.-M., subm., \aap
\end{enumerate}
\end{large}

\pagestyle{bibliography}
\addcontentsline{toc}{chapter}{Bibliography}
\singlespacing
\bibliographystyle{aa}
\begin{footnotesize}
	\bibliography{ref}
\end{footnotesize}

	\newpage
	\thispagestyle{empty}
	\cleardoublepage
    
	%\addcontentsline{toc}{chapter}{Acknowledgement}
\thispagestyle{empty}
\vspace*{2cm}
\begin{center}
    \textbf{Acknowledgements}
\end{center}

\vspace*{1cm}

\begin{doublespace}
\noindent At this point, I want to thank to all the people without whom I would not have come this far.

First of all, I would like to thank Prof.~Dr.~Thomas Henning who offered me the possibility to conduct my PhD the Max-Planck-Institut für Astronomie and motivated me whenever I needed it most. I enjoyed working and discussing with him.
Thanks also to Prof.~Dr. Henrik Beuther who has provided me a lot of help, as well as to Prof.~Dr.~Cornelis P.~Dullemond for reviewing this thesis and always having time for good advices or pleasant chats.
Special thanks also go to Dr.~Juan Ib\'{a}\~{n}ez-Mej\'{i}a, Prof.~Dr.~Mordecai Mac Low and Dr.~Jouni Kainulainen who took me under their wings and encouraged me in elaborating those fascinating projects.
Furthermore, I want to thank the European Southern Observatory for having been part of its Studentship Programme, as well as Dr.~Ralf Siebenmorgen who started this path with me.

A special thanks go to my friends, in particular Clio Bertelli Motta, Renate Hubele, Sarolta Zahorecz, Laura Inno, Franziska Brems, and Anne Klitsch, who never seriously complained about the extra hours they needed to work because of my "communicative" personality (Henning, 2017). 
Of course I will also miss all the other people at both the MPIA and ESO with whom I had the pleasure to spend these intensive years with. 

Last, but of course not least, I thank my parents for always supporting and driving me and my curiosity in Astronomy.
\end{doublespace}

	\thispagestyle{empty}
	\cleardoublepage
	
    %\addcontentsline{toc}{chapter}{Affidavit}
\thispagestyle{empty}
\begin{large}

\vspace*{2cm}

\begin{center}
    \textbf{Erkl\"arung / Affidavit}
\end{center}

\vspace*{1cm}

\noindent
Ich versichere hiermit, dass ich diese Dissertation selbst\"andig verfasst und nur die angegebenen Quellen und Hilfsmittel verwendet habe.

\vspace{2cm}

\noindent
With this, I assure that I have authored this Master thesis by my own and used only the named sources and aids.

\vspace{2cm}

\noindent
Heidelberg, November 13, 2017

\vspace{3cm}

\hspace*{7cm}%
\dotfill\\
\hspace*{8.5cm}%
\textit{(Roxana-Adela Chira)}

\end{large}

\end{document}